\documentclass[11pt]{cernrep}
\usepackage{psfig}
\usepackage{epsfig}
\usepackage{axodraw}
\usepackage{graphicx}
\usepackage{here}
\usepackage{a4}
\usepackage{graphics}
\usepackage{dcolumn}
\usepackage{amsmath}
\usepackage{psfrag}
\usepackage{cite}
\usepackage{rotating}
\usepackage{amssymb}
\usepackage{amsbsy}

\newcommand{\ssx}{\hspace*{0.6cm}}

\begin{document}

\pagestyle{myheadings}
\thispagestyle{empty}

\begin{center}

{\Large\sc \bf THE HIGGS WORKING GROUP: }

\vspace*{0.3cm}

{\Large\sc \bf Summary Report}

\vspace*{.7cm}

Conveners: \\[0.2cm]
{\sc D.\,Cavalli$^1$, A.\,Djouadi$^2$, K.\,Jakobs$^3$,
A.\,Nikitenko$^4$, M.\,Spira$^5$, C.E.M.\,Wagner$^6$} and {\sc W.-M.\,Yao$^7$}

\vspace*{0.5cm}

 Working Group: \\[0.2cm]
{\sc
K.A.\,Assamagan$^8$,
G.\,Azuelos$^9$,
S.\,Balatenychev$^{10}$,
G.\,B\'elanger$^{11}$,
M.\,Bisset$^{12}$,
A.\,Bocci$^{13}$,
F.\,Boudjema$^{11}$,
C.\,Buttar$^{14}$,
M.\,Carena$^{15}$,
S.\,Catani$^{16}$,
V.\,Cavasinni$^{17}$,
Y.\,Coadou$^{18}$,
D.\,Costanzo$^{17}$,
A.\,Cottrant$^{11}$,
A.K.\,Datta$^2$,
A.\,Deandrea$^{19}$,
D.\,de Florian$^{20}$,
V.\,Del Duca$^{21}$,
B.\,Di Girolamo$^{22}$,
V.\,Drollinger$^{23}$,
T.\,Figy$^{24}$,
M.\,Frank$^{25}$,
R.M.\,Godbole$^{26}$,
M.\,Grazzini$^{27}$,
M.\,Guchait$^{2,28}$,
R.\,Harper$^{14}$,
S.\,Heinemeyer$^8$,
J.\,Hobbs$^{29}$,
W.\,Hollik$^{25,30}$,
C.\,Hugonie$^{31}$,
V.I.\,Ilyin$^{10}$, 
W.B.\,Kilgore$^8$,
R.\,Kinnunen$^{32}$,
M.\,Klute$^{33}$,
R.\,Lafaye$^{34}$,
Y.\,Mambrini$^2$,
R.\,Mazini$^9$,
K.\,Mazumdar$^{35}$,
F.\,Moortgat$^{36}$,
S.\,Moretti$^{16,31}$,
G.\,Negri$^1$,
L.\,Neukermans$^{34}$,
C.\,Oleari$^{24}$,
A.\,Pukhov$^{10}$,
D.\,Rainwater$^{15}$,
E.\,Richter--Was$^{37}$,
D.P.\,Roy$^{35}$,
C.R.\,Schmidt$^{38}$,
A.\,Semenov$^{11}$,
J.\,Thomas$^3$,
I.\,Vivarelli$^{17}$,
G.\,Weiglein$^{31}$
and} {\sc 
D.\,Zeppenfeld$^{24}$. }
\vspace*{.7cm}

{\small
$^1$ INFN and Physics Department Milano University, Italy. \\
$^2$ LPMT, Universit\'e Montpellier II, F--34095 Montpellier Cedex 5, France. \\
$^3$ Institut f\"ur Physik, Universit\"at Mainz, Germany. \\
$^4$ Imperial College, London, UK. \\
$^5$ Paul Scherrer Institut, CH-5232 Villigen PSI, Switzerland. \\
$^6$ HEP Division, ANL, 9700 Cass Ave., Argonne, IL 60439 and
Enrico Fermi Institute, University of Chicago, 5640 Ellis Avenue, Chicago, IL60637, USA. \\
$^7$ LBNL, One Cyclotron Road, Berkeley, CA 94720, USA. \\
$^8$ Department of Physics, BNL, Upton, NY 11973, USA. \\
$^9$ University of Montreal, Canada. \\
$^{10}$ SINP, Moscow State University, Moscow, Russia. \\
$^{11}$ LAPTH, Chemin de Bellevue, B.P.\,110, F-74941 Annecy-le-Vieux, 
Cedex, France. \\
$^{12}$ Department of Physics, Tsinghua University, Beijing, P.R.\,China
100084. \\
$^{13}$ Rockefeller University, 1230 York Avenue, New York, NY 10021, USA. \\
$^{14}$ Department of Physics and Astronomy, University of Sheffield, UK. \\
$^{15}$ FNAL, Batavia, IL 60510, USA. \\
$^{16}$ CERN, Theory Division, CH--1211, Geneva, Switzerland. \\
$^{17}$ INFN and University of Pisa, Italy. \\
$^{18}$ University of Uppsala, Sweden. \\
$^{19}$ IPNL, Univ.\,de Lyon I, 4 rue E.\,Fermi, F--69622 Villeurbanne
Cedex, France.\\
$^{20}$ Departamento de F\'isica, Universidad de Buenos Aires, Argentina. \\
$^{21}$ I.N.F.N., Sezione di Torino via P.\,Giuria, 1 -- 10125 Torino, Italy. \\
$^{22}$ EP Division, CERN, CH--1211 Gen\`eve 23, Switzerland. \\
$^{23}$ Department of Physics and Astronomy, University of New Mexico, USA. \\
$^{24}$ Department of Physics, University of Wisconsin, Madison, WI 53706, USA. \\
$^{25}$ Institut f\"ur Theoretische Physik, Universit\"at Karlsruhe, D--76128 Karlsruhe, Germany. \\
$^{26}$ Centre for Theoretical Studies, Indian Institute of Science,
Bangalore 560 012, India. \\
$^{27}$ INFN, Sezione di Firenze, I--50019 Sesto Fiorentino, Florence, Italy. \\
$^{28}$ The Abdus Salam International Centre for Theoretical Physics, Strada Costieara 11,
I--34014 Trieste, Italy. \\
$^{29}$ SUNY at Stony Brook, Dept of Physics, Stony Brook, NY 11794, USA. \\
$^{30}$ Max--Planck--Institut f\"ur Physik, F\"ohringer Ring 6, D--80805 M\"unchen, Germany. \\
$^{31}$ IPPP, University of Durham, Durham DH1 3LR, UK. \\
$^{32}$ HIP, Helsinki, Finland. \\
$^{33}$ Physikalisches Institut, Universit\"at Bonn, Germany. \\
$^{34}$ LAPP, Chemin de Bellevue, B.P.\,110, F-74941 Annecy-le-Vieux, Cedex,
France. \\
$^{35}$ Tata Institute of Fundamental Research, Mumbai, 400 005, India. \\
$^{36}$ Physics Department Universitaire Instelling Antwerpen, Wilrijk, 
Belgium. \\
$^{37}$ Inst. of Computer Science, Jagellonian University; Inst. of Nuclear Physics, Cracow, Poland. \\
$^{38}$ Department of Physics and Astronomy Michigan State University
East Lansing, MI 48824, USA. \\
}

\vspace*{.5cm}

{\it Report of the HIGGS working group for the Workshop \\[0.1cm]
``Physics at TeV Colliders", Les Houches, France, 21 May -- 1 June 2001.}
\end{center}

\vspace*{.5cm}
\begin{center}
{\bf \large CONTENTS}
\end{center}

\vspace*{0.1cm}

\noindent {\bf \ssx Preface} \hfill 3 \\

\noindent {\bf A. Theoretical Developments}
\hfill 4 \\[0.2cm] \hspace*{0.5cm}
S.\,Balatenychev, G.\,B\'elanger, F.\,Boudjema, A.\,Cottrant, M.\,Carena,
S.\,Catani, V.\,Del Duca, \\ \ssx
D.\,de Florian, M.\,Frank, R.M.\,Godbole,
M.\,Grazzini, S.\,Heinemeyer, W.\,Hollik, C.\,Hugonie, \\
\ssx V.\,Ilyin, W.B.\,Kilgore, R.\,Lafaye, S.\,Moretti, C.\,Oleari, A.\,Pukhov, 
D.\,Rainwater, D.P. Roy, \\
\ssx C.R.\,Schmidt, A.\,Semenov, M.\,Spira, C.E.M.\,Wagner, G.\,Weiglein
and D.\,Zeppenfeld \\

\noindent {\bf B. Higgs Searches at the Tevatron}
\hfill 34 \\[0.2cm] \hspace*{0.5cm}
A.\,Bocci, J.\,Hobbs, and W.-M.\,Yao \\

\noindent {\bf C. Experimental Observation of an invisible Higgs Boson at LHC}
\hfill 42 \\[0.2cm]  \hspace*{0.5cm}
B.\,Di Girolamo, L.\,Neukermans, K.\,Mazumdar, A.\,Nikitenko and
D.\,Zeppenfeld \\

\noindent {\bf D. Search for the Standard Model Higgs Boson using Vector Boson
Fusion at the LHC}
\hfill 56 \\[0.2cm]  \hspace*{0.5cm}
G.\,Azuelos, C.\,Buttar, V.\,Cavasinni, D.\,Costanzo,
T.\,Figy, R.\,Harper, K.\,Jakobs, \\
\ssx M.\,Klute, R.\,Mazini, A.\,Nikitenko,
E\,.Richter--Was, I.\,Vivarelli and D.\,Zeppenfeld \\

\noindent {\bf E. Study of  the MSSM channel \mbox{$\rm{A/H \rightarrow \tau
\tau}$} at the LHC}
\hfill 67 \\[0.2cm] \hspace*{0.5cm}
D.\,Cavalli, R.\,Kinnunen, G.\,Negri, A.\,Nikitenko and J.\,Thomas \\

\noindent {\bf F. Searching for Higgs Bosons in $t\bar t H$ Production}
\hfill 80 \\[0.2cm] \hspace*{0.5cm}
V.\,Drollinger \\

\noindent {\bf G. Studies of Charged Higgs Boson Signals for the Tevatron
and the LHC}
\hfill 85 \\[0.2cm] \hspace*{0.5cm}
K.A.\,Assamagan, M.\,Bisset, Y.\,Coadou,
A.K.\,Datta, A.\,Deandrea, A.\,Djouadi, \\
\ssx M.\,Guchait, Y.\,Mambrini, F.\,Moortgat and S.\,Moretti \\

\newpage


\noindent
{\Large\bf PREFACE} \\

\noindent
In this working group we have investigated the propects for Higgs boson
searches at the Tevatron and LHC and, in particular, the potential of
these colliders to determine the Higgs properties once these particles
have been found. The analyses were done in the framework of the Standard
Model (SM) and its supersymmetric extensions as the minimal (MSSM) and
next-to-minimal (NMSSM) supersymmetric extensions. The work for the
discovery potential of the LHC mainly concentrated on the difficult
regions of previous analyses as those which are plagued by invisible
Higgs decays and Higgs decays into supersymmetric particles. Moreover,
the additional signatures provided by the weak vector-boson fusion
process (WBF) have been addressed and found to confirm the results of
previous analyses. A major experimental effort has been put
onto charged Higgs boson analyses. The final outcome was a significant
improvement of the discovery potential at the Tevatron and LHC than
previous analyses suggested.

For an accurate determination of Higgs boson couplings, the theoretical
predictions for the signal and background processes have to be improved.
A lot of progress has been made during and after this workshop for the
gluon-fusion $gg\to H+(0,1,2jets)$ and the associated $t\bar tH$
production process. A thorough study of the present theoretical
uncertainties of signal and background processes has been initialized,
culminating in a list of open theoretical problems. A problem
of major experimental interest is the proper treatment of processes
involving bottom quark densities, which is crucial for some important signal
and background processes. Further theoretical
improvements have been achieved for the MSSM Higgs boson masses and Higgs
bosons in the NMSSM.

This report summarizes our work. The first part deals with theoretical
developments for the signal and background processes. The second part
gives an overview of the present status of Higgs boson searches at the
Tevatron. The third part analyzes invisible Higgs boson decays at the LHC and
the forth part the Higgs boson search in the WBF channel. Part 5 summarizes
the progress that has been achieved for $A/H\to \tau^+\tau^-$ decays in
the MSSM. In part 6 the status of the Higgs boson search in $t\bar tH$
production is presented. Finally, part 7 describes the charged Higgs
boson analyses in detail. \\

\noindent
{\bf Acknowledgements.} \\
We thank the organizers of this workshop for the friendly and
stimulating atmosphere during the meeting. We also thank our colleagues
of the QCD/SM and SUSY working group for the very constructive
interactions we had. We are grateful to the ``personnel'' of the Les
Houches school for enabling us to work on physics during day and night
and their warm hospitality during our stay.

\newpage

\noindent
{\Large \bf A. Theoretical Developments} \\[0.5cm]
{\it S.\,Balatenychev, G.\,B\'elanger, F.\,Boudjema, A.\,Cottrant, M.\,Carena,
S.\,Catani, V.\,Del Duca, D.\,de Florian, M.\,Frank, R.\,Godbole,
M.\,Grazzini, S.\,Heinemeyer, W.\,Hollik, V.\,Ilyn, W.\,Kilgore, R.\,Lafaye,
S.\,Moretti, C.\,Oleari, A.\,Pukhov, D.\,Rainwater, DP\,Roy, C.\,Schmidt,
A.\,Semenov, M.\,Spira, C.\,Wagner, G.\,Weiglein and D.\,Zeppenfeld}

\begin{abstract}
New theoretical progress in Higgs boson production and background
processes at hadron colliders and the relations between the MSSM Higgs
boson masses is discussed. In this context new proposals for benchmark
points in the MSSM are presented. Additional emphasis is put on
theoretical issues of invisible SUSY Higgs decays and multiple Higgs boson
production within the NMSSM.
\end{abstract}

{
\catcode`@=11
\def\citer{\@ifnextchar
[{\@tempswatrue\@citexr}{\@tempswafalse\@citexr[]}}

%

\def\@citexr[#1]#2{\if@filesw\immediate\write\@auxout{\string\citation{#2}}\fi
  \def\@citea{}\@cite{\@for\@citeb:=#2\do
    {\@citea\def\@citea{--\penalty\@m}\@ifundefined
       {b@\@citeb}{{\bf ?}\@warning
       {Citation `\@citeb' on page \thepage \space undefined}}%
\hbox{\csname b@\@citeb\endcsname}}}{#1}}
\catcode`@=12

\newcommand{\beq}{\begin{eqnarray}}
\newcommand{\eeq}{\end{eqnarray}}
\newcommand{\nn}{\noindent}
\newcommand{\non}{\nonumber}
\newcommand{\ra}{\rightarrow}
\newcommand{\s}{\\ \vspace*{-3mm} }
\newcommand{\tgb}{\tan\beta}

\def\sla#1{\ifmmode%
\setbox0=\hbox{$#1$}%
\setbox1=\hbox to\wd0{\hss$/$\hss}\else%
\setbox0=\hbox{#1}%
\setbox1=\hbox to\wd0{\hss/\hss}\fi%
#1\hskip-\wd0\box1 } 

\newcommand{\lsim}{\raisebox{-0.13cm}{~\shortstack{$<$ \\[-0.07cm] $\sim$}}~}
\newcommand{\gsim}{\raisebox{-0.13cm}{~\shortstack{$>$ \\[-0.07cm] $\sim$}}~}
\newcommand{\fbi}{~fb$^{-1}\;$}
 
\section[]{Higgs boson production at hadron colliders: signal and
background processes%
\footnote{\it D.\,Rainwater, M.\,Spira and D.\,Zeppenfeld}}

\subsection{Introduction}
The Higgs mechanism is a cornerstone of the Standard Model (SM) and its
supersymmetric extensions. Thus, the search for Higgs bosons is one of the
most important endeavors at future high-energy experiments.  In the SM
one Higgs doublet has to be introduced in order to break the
electroweak symmetry, leading to the existence of one elementary Higgs
boson, $H$ \cite{higgs}. The scalar sector of the SM is uniquely fixed by
the vacuum expectation value $v$ of the Higgs doublet and the mass $m_H$
of the physical Higgs boson \cite{hhg}.
The negative direct search for the Higgsstrahlung process $e^+e^-\to ZH$
at the LEP2 collider poses a lower bound of $114.1$ GeV on the SM
Higgs mass \cite{LEP-limit,mhLEP2001}, while triviality arguments force
the Higgs mass to be smaller than $\sim 1$ TeV \cite{triviality}.

Since the minimal supersymmetric extension of the Standard Model (MSSM)
requires the introduction of two Higgs doublets in order to preserve
supersymmetry, there are five elementary Higgs particles, two CP-even
($h,H$), one CP-odd ($A$) and two charged ones ($H^\pm$). At lowest order
all couplings and masses of the MSSM Higgs sector are fixed by two
independent input parameters, which are generally chosen as
$\tgb=v_2/v_1$, the ratio of the two vacuum expectation values $v_{1,2}$,
and the pseudoscalar Higgs-boson mass $m_A$. At LO the light scalar
Higgs mass $m_h$ has to be smaller than the $Z$-boson mass $m_Z$.
Including the
one-loop and dominant two-loop corrections the upper bound is increased
to $m_h\lsim 135$ GeV \citer{mhiggsRG2a,mhalphatsq}. The negative direct
searches for the
Higgsstrahlung processes $e^+e^-\to Zh,ZH$ and the associated production
$e^+e^-\to Ah,AH$ yield lower bounds of $m_{h,H} > 91.0$ GeV and
$m_A > 91.9$ GeV. The range $0.5 < \tgb < 2.4$ in the MSSM is
excluded by the Higgs searches at the LEP2 experiments
\cite{LEP-limit,mhLEP2001}.

The intermediate mass range, $m_H<196$~GeV at 95\% CL, is also favored
by a SM analysis of electroweak precision data~\cite{LEP-limit,mhLEP2001}.
In this 
contribution we will therefore concentrate on searches and measurements
for $m_H\lsim 200$~GeV. The Tevatron has a good chance to find 
evidence for such a Higgs boson, provided that sufficient integrated 
luminosity can be accumulated~\cite{Carena:2000yx}. The Higgs boson,
if it exists, 
can certainly be seen at the LHC, and the LHC can provide measurements of the
Higgs boson mass at the $10^{-3}$ level~\cite{CMSTDR,ATLASTDR}, and
measurements of 
Higgs boson couplings at the 5 to 10\% level~\cite{Zeppenfeld:2000td}.
Both tasks, discovery and measurement of Higgs properties, require accurate
theoretical predictions of cross sections at the LHC, but these requirements
become particularly demanding for accurate coupling measurements.

In this contribution we review the present status of QCD calculations of signal
and background cross sections encountered in Higgs physics at hadron colliders.
Desired accuracy levels can be estimated by comparing to the statistical errors
in the determination of signal cross sections at the LHC. For processes like
$H\to\gamma\gamma$, where a very narrow mass peak will be observed, backgrounds
can be accurately determined directly from data. For other decay channels, like
$H\to b\bar b$ or $H\to\tau\tau$, mass resolutions of order 10\% require modest
interpolation from sidebands, for which reliable QCD calculations are needed.
Most demanding are channels like $H\to W^+W^-\to l^+l^-\sla p_T$, for which
broad transverse mass peaks reduce Higgs observation to, essentially, a
counting experiment. Consequently, requirements on theory predictions vary
significantly between channels.  In the following we discuss production and
decay channels in turn and focus on theory requirements for the prediction of
signal and background cross sections. Because our main interest is in coupling
measurements, we will not consider diffractive channels in the following, which
are model-dependent and have large rate uncertainties~\cite{difractH};
potentially, they might contribute to Higgs discovery if, indeed, cross
sections are sufficiently large.  

\subsection{Gluon fusion}
The gluon fusion mechanism $gg\to \phi$ provides the dominant production
mechanism of Higgs bosons at the LHC in the entire relevant mass range
up to about 1 TeV in the SM and for small and moderate values of $\tgb$
in the MSSM \cite{habil}. At the Tevatron this process plays the relevant role
for Higgs masses between about 130 GeV and about 190 GeV
\cite{Carena:2000yx}. The gluon
fusion process is mediated by heavy quark triangle loops and, in the case
of supersymmetric theories, by squark loops in addition, if the squark
masses are smaller than about 400 GeV \cite{squark}, see 
Fig.~\ref{fg:gghlodia}.
\begin{figure}[hbt]
\vspace*{-.5cm}
\begin{center}
\setlength{\unitlength}{1pt}
\begin{picture}(180,100)(0,0)

\Gluon(0,20)(50,20){-3}{5}
\Gluon(0,80)(50,80){3}{5}
\ArrowLine(50,20)(50,80)
\ArrowLine(50,80)(100,50)
\ArrowLine(100,50)(50,20)
\DashLine(100,50)(150,50){5}
\put(155,46){$\phi$}
\put(20,46){$t,b,\tilde q$}
\put(-15,18){$g$}
\put(-15,78){$g$}
\end{picture}
\setlength{\unitlength}{1pt}
\caption[ ]{\label{fg:gghlodia} \it Typical diagram contributing to
$gg\to \phi$ at lowest order.}
\end{center}
\vspace*{-.5cm}
\end{figure}
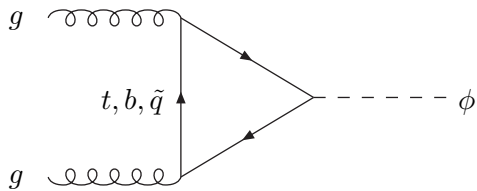

In the past the full two-loop QCD corrections have been determined. They
increase the production cross sections by 10--80\% \cite{glufusnlo},
thus leading to a significant change of the theoretical predictions. 
Very recently, Harlander and Kilgore have finished the full NNLO calculation, 
in the heavy top quark limit~\cite{3loop,Harlander:2002wh}.
This limit has been demonstrated to
approximate the full massive $K$ factor at NLO within 10\% for the SM Higgs
boson in the entire mass range up to 1 TeV \cite{softgluon}. Thus, a similar
situation can
be expected at NNLO. The reason for the quality of this approximation is
that the QCD corrections to the gluon fusion mechanism are dominated by
soft gluon effects, which do not resolve the one-loop Higgs coupling to gluons.
Fig.~\ref{fig:all14murf} shows the resulting $K$-factors at the LHC and
the scale variation of the $K$-factor. The calculation stabilizes at NNLO,
with remaining scale variations at the 10 to 15\% level. 
These uncertainties are comparable to the experimental errors which can 
be achieved with 200\fbi of data at the LHC, see solid lines in 
Fig.~\ref{fig:delsigh}. 
The full NNLO results confirm earlier estimates which were obtained 
in the frame work of soft gluon resummation
\cite{softgluon} and soft approximations
\cite{Catani:2001ic,Harlander:2001is} of the full
three-loop result. The full soft gluon resummation has been performed in
Ref.\,\cite{soft2}. The resummation effects enhance the NNLO result by
about 10\% thus signaling a perturbative stabilization of the
theoretical prediction for the gluon-fusion cross section.

\begin{figure}[hbt]
\hspace*{2.5cm} \includegraphics[width=10cm]{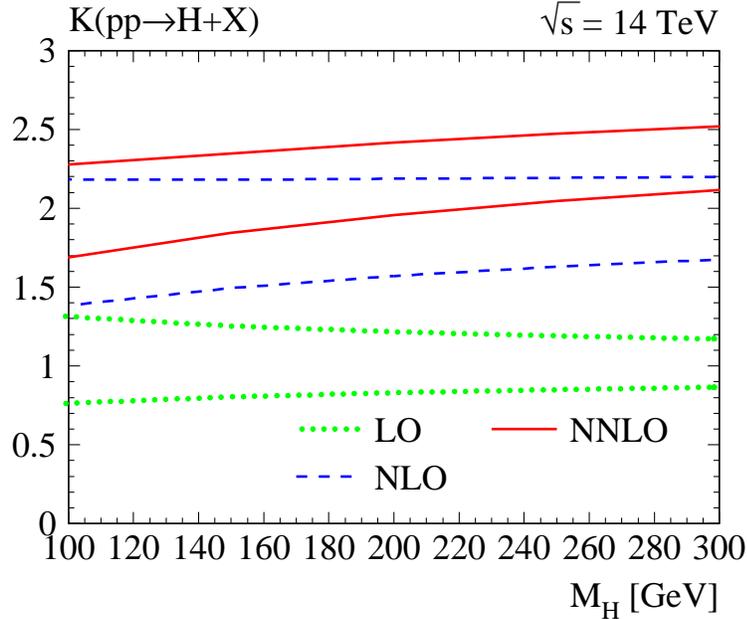}
      \caption[]{\it \label{fig:all14murf} Scale dependence of the $K$-factor 
      at the LHC. Lower curves for each pair are for 
      $\mu_R = 2m_H$, $\mu_F=m_H/2$, upper curves are for 
      $\mu_R =m_H/2$, $\mu_F=2m_H$.  The $K$-factor is
      computed with respect to the LO cross section at
      $\mu_R = \mu_F =m_H$. From Ref.~\cite{Harlander:2002wh}.}
\vspace*{-.2cm}
\end{figure}

In supersymmetric theories the gluon fusion cross sections for
the heavy Higgs, $H$, and, for small $m_A$, also for the light Higgs, $h$, 
may be dominated by
bottom quark loops for large values of $\tgb\gsim 10$ so that the heavy
top quark limit is not applicable. This can be clearly seen in the NLO
results, which show a decrease of the $K$ factor down to about 1.1 for
large $\tgb$ \cite{glufusnlo}. This decrease originates from an interplay
between the large
positive soft gluon effects and large negative double logarithms of the
ratio between the Higgs and bottom masses. 
In addition, the shape of the $p_T$ distribution of the Higgs boson may be 
altered; if the bottom loop is dominant, the $p_T$
spectrum becomes softer than in the case of top-loop dominance. These effects
lead to some model dependence of predicted cross sections.

%

 \begin{figure}
\hspace*{2cm} \includegraphics[width=8.5cm, angle=90]{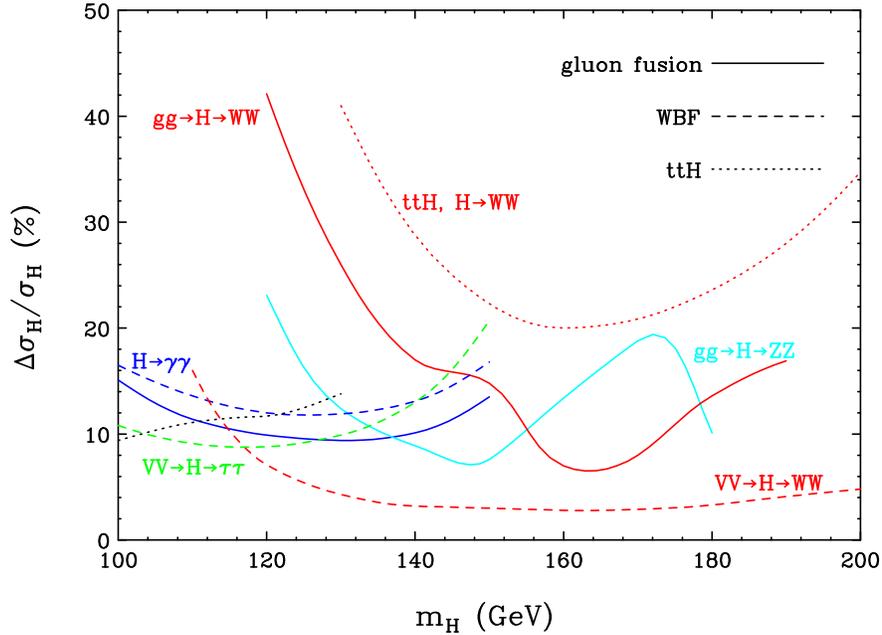}
\vspace*{-0.5cm}

 \caption{\it Expected relative error on the determination of 
$B\sigma$ for various Higgs search channels at the LHC 
with 200~\fbi of data~\cite{Zeppenfeld:2000td}. Solid 
lines are for inclusive Higgs production channels which 
are dominated by gluon fusion. Expectations for weak boson 
fusion are given by the dashed lines. Dotted lines are for 
$t\bar tH$ production with $H\to b\bar b$~\cite{drollinger:2001ym}
(black) and $H\to W^+W^-$~\cite{tth2ww}  (red) and assume 
300\fbi of data.  }
 \label{fig:delsigh}
\vspace*{-.5cm}
 \end{figure}

Let us now turn to a discussion of backgrounds for individual decay modes.

\noindent
\underline{\it (i) $\phi\to \gamma\gamma$}:
At the LHC the SM Higgs boson can be found in the mass range up to 
about 150~GeV by means of the rare photonic decay 
mode $H\to\gamma\gamma$~\cite{CMSTDR,ATLASTDR}. The dominant Higgs decays 
$H\to b\bar b, \tau^+\tau^-$ are overwhelmed by large 
QCD backgrounds in inclusive searches. 
The QCD $\gamma\gamma$ background is known at NLO,
including all relevant fragmentation effects. The present status is
contained in the program DIPHOX \cite{diphox}. 
The loop mediated process $gg\to\gamma\gamma$ contributes about 50\% to
the $\gamma\gamma$ background and has been calculated at NLO very
recently \cite{gggamgam}. However, a numerical analysis of the two-loop
result is still missing.

Once the experiment is performed, the diphoton background can be determined
precisely from the data, by a measurement of $d\sigma/dm_{\gamma\gamma}$
on both sides of the resonance peak. The NLO calculations are useful, 
nevertheless, for an accurate prediction of expected accuracies and for a
quantitative understanding of detector performance.
\vspace{0.1in}

\noindent
\underline{\it (ii) $H\to W^+W^-$}:
This mode is very important for Higgs masses above $W$-pair but below
$Z$-pair threshold, where $B(H\to WW)$ is close to 100\%.
In order to suppress the $t\bar t\to b\bar bW^+W^-$ background for 
$W^+W^-$ final states, a jet veto is crucial. However, gluon fusion 
receives sizeable contributions from real gluon bremsstrahlung at NLO,
which will also be affected by the jet veto. These effects have recently
been analyzed in Ref.~\cite{Catani:2001cr}, in the soft approximation 
to the full NNLO calculation. A veto of additional jets 
with $p_{Tj}>15$~GeV, as e.g. envisioned by ATLAS~\cite{ATLASTDR}, 
reduces the NNLO $K$-factor
to about $K=0.8$\footnote{It should be noted that for this strong cut in
$p_{Tj}$ the NNLO result may be plagued by large logarithms of this cut,
which have to be resummed, see \cite{soft2}.}, i.e. one loses more than
60\% of signal events. 
In addition the scale dependence of the cross section starts to grow with
such stringent veto criteria. These effects need to be modeled with a
NLO Monte Carlo program for $H+jet$ production in order to reach a 
reliable quantitative result for the signal rate. Since stop and sbottom 
loops are sizeable in supersymmetric theories for squark
masses below about 400 GeV, their inclusion is important in these
investigations.

From the perspective of background calculations, $H\to WW$ is the most 
challenging channel. Backgrounds are of the order of the signal rate or larger,
which requires a 5\% determination or better for the dominant background cross
sections in order to match the statistical power of LHC experiments. In
fact, the large errors at $m_H\lsim 150$~GeV depicted in Fig.~\ref{fig:delsigh}
($gg\to H\to WW$ curve) are dominated by an assumed 5\% background 
uncertainty. Clearly, such small errors cannot be achieved by NLO calculations
alone, but require input from LHC data. 
Because of two missing neutrinos in the 
$W^+W^-\to l^+l^-\sla p_T$ final state, the Higgs mass cannot be reconstructed
directly. Rather, only wide $(l^+l^-;\sla p_T)$ transverse mass distributions
can be measured, which do not permit straightforward sideband 
measurements of the 
backgrounds. Instead one needs to measure the normalization of the backgrounds
in signal poor regions and then extrapolate these, with the help of 
differential cross sections predicted in perturbative QCD, to the signal
region. The theory problem is the uncertainty in the shape of the
distributions used for the extrapolation, which will depend on an 
appropriate choice of the ``signal poor region''. No analysis of the 
concomitant uncertainties, at LO or NLO QCD, is available to date.

After the jet veto discussed above, the dominant background processes
are $pp\to W^+W^-$ and (off-shell) $t\bar t$
production~\cite{CMSTDR,ATLASTDR}. $W^+W^-$ production is known 
at NLO~\cite{vvnlo} and available in terms of parton level Monte Carlo
programs. In addition, a full NLO calculation including spin correlations
of the leptonic $W,Z$ decays, in the narrow width approximation, 
is available \cite{kunszt}. For Higgs boson
masses below the $W^+W^-(ZZ)$ threshold, decays into $WW^*(ZZ^*)$ are
important \cite{offshell,habil}. Since hadron colliders will be sensitive
to these off-shell
tails, too, the backgrounds from $VV^*$ production become relevant. There
is no NLO calculation of $VV^*$ background processes available so far, so
that it is not clear if NLO effects will be significant in the tails of
distributions needed for the Higgs search in these cases. Moreover, for
$WW^*$ production the inclusion of spin correlations among the final state
leptons is mandatory \cite{Dittmar:1997ss}. 

Top quark backgrounds arise from top-pair and $tWb$ production.
Recently, a new theoretical analysis of $pp\to t^{(*)}\bar t^{(*)}$ has
become available including full lepton correlations and off-shell effects 
of the final state top quarks arising from the non-zero top decay 
width~\cite{ttoff}. This calculation automatically
includes $pp\to tbW$ and those contributions to $pp\to b\bar b W^+W^-$,
which are gauge-related to $tbW$ couplings and describes
the relevant tails for the Higgs search at LO. It is now necessary to
investigate the theoretical uncertainties of this background.
A NLO calculation of off-shell top-pair production may well be needed
to reach the required 5\% accuracy for extrapolation to the Higgs search
region.

Other important reducible backgrounds are the $Wt\bar t,Zt\bar t,Wb\bar b$ 
and $Zb\bar b$ production processes. While $Vt\bar t$ ($V=W,Z$) production 
is only known
at LO, the associated vector boson production with $b\bar b$ pairs is
known at NLO including a soft gluon resummation \cite{veseli}. Thus $Vb\bar b$
production can be considered as reliable from the theoretical point of
view, while a full NLO calculation for $Vt\bar t$ production is highly
desirable, since top mass effects will play a significant role. In
addition, the background from $gb \to tH^-, g\bar b\to \bar t H^+$ has to
be taken into account within the MSSM framework. The full LO matrix 
elements are included in the
ISAJET Monte Carlo program, which can easily be used for experimental
analyses.

\vspace{0.1in}
\noindent
\underline{\it (iii) $H\to ZZ \to 4\ell^\pm$}:
A sharp Higgs peak can be observed in the four lepton invariant mass 
distribution. Hence, the $ZZ\to 4\ell^\pm$ backgrounds are directly 
measurable in the sidebands and can safely be interpolated to the 
signal region. 

\subsection{$qq\to qqH$}
In the SM the $WW,ZZ$ fusion processes $qq\to qqV^*V^*\to qqH$ play a
significant role at the LHC for the entire Higgs mass range up to 1 TeV. 
We refer to them as weak boson fusion (WBF). The WBF
cross section becomes comparable to the gluon fusion cross
section for Higgs masses beyond $\sim 600$ GeV \cite{habil} and is sizable,
of order 20\% of $\sigma(gg\to H)$, also in the intermediate mass region. 
The energetic quark jets in the
forward and backward directions allow for additional cuts to suppress the
background processes to WBF. The NLO QCD corrections
can be expressed in terms of the conventional corrections to the DIS
structure functions, since there is no color exchange between the two
quark lines at LO and NLO. NLO corrections increase the production cross 
section by about 10\% and are thus small and under theoretical 
control~\cite{fusion-nlo,SUSYQCDcor}. These small theory uncertainties make 
WBF a very promising tool for precise coupling measurements. However, 
additional studies are needed to assess the
theoretical uncertainties associated with a central jet veto. This veto
enhances the color singlet exchange of the signal over color octet 
exchange QCD backgrounds~\citer{zeppenfeld,wbf.wwlomh}. 

In the MSSM, first parton level
analyses show that it should be possible to cover the
full MSSM parameter range by looking for the light Higgs decay 
$h\to \tau^+\tau^-$ (for $m_A\gsim 150$~GeV) and/or the heavy Higgs 
$H\to\tau^+\tau^-$resonance (for a relatively small $m_A$) 
in the vector-boson fusion processes~\cite{vvhtau}.
Although these two production processes are suppressed with respect
to the SM cross section, their sum is of SM strength.

For the extraction of Higgs couplings it is important to distinguish 
between WBF and gluon fusion processes which lead to $H+jj$ final states.
With typical WBF cuts, including a central jet veto, gluon fusion 
contributions are expected at order 10\% of the WBF cross section, i.e. 
the contamination is modest~\cite{DelDuca:2001eu,wbf.exp}.
The gluon fusion processes are mediated by heavy top and
bottom quark loops, in analogy to the LO gluon fusion diagram of 
Fig.~\ref{fg:gghlodia}. The full massive cross section for $H+jj$ 
production via gluon fusion has been obtained only
recently~\cite{DelDuca:2001eu},
while former analyses were performed in the heavy top quark 
limit~\cite{hjjlimit}. Since stop and sbottom loops yield a sizeable 
contribution to the inclusive gluon fusion cross section, a similar 
feature is expected for $H+jj$
production. Thus, it is important to compute the effects of stop and
sbottom loops in $H+jj$ gluon fusion processes, which has not been done 
so far. 

\vspace{0.1in}
\noindent
\underline{\it (i) $H\to\gamma\gamma$}:
Parton level analyses show that $H\to \gamma\gamma$ decays in WBF Higgs
production can be isolated with signal to background ratios of order 
one~\cite{zeppenfeld} and with statistical errors of about 15\%, for 200\fbi of 
data (see Fig.~\ref{fig:delsigh}). Like for the inclusive $H\to\gamma\gamma$ 
search, background levels can be precisely determined from a sideband analysis
of the data. Prior to data taking, however, full detector simulations are 
needed to confirm the parton level results and improve on the search 
strategies. 

Improved background calculations are desirable as well. In particular,
the $pp\to \gamma \gamma jj$ background via quark loops (see
Fig.~\ref{fg:gamgamjj}) has not been calculated so far.
\begin{figure}[hbt]
\vspace*{-.2cm}
\begin{center}
\setlength{\unitlength}{1pt}
\begin{picture}(200,170)(5,-30)
\Gluon(0,20)(50,20){-3}{5}
\Gluon(0,80)(50,80){3}{5}
\Gluon(90,-5)(140,-35){-3}{5}
\Gluon(90,105)(140,135){3}{5}
\Photon(130,80)(180,80){3}{5}
\Photon(130,20)(180,20){3}{5}
\ArrowLine(50,20)(50,80)
\ArrowLine(50,80)(90,105)
\ArrowLine(90,105)(130,80)
\ArrowLine(130,80)(130,20)
\ArrowLine(130,20)(90,-5)
\ArrowLine(90,-5)(50,20)
\put(55,46){$q$}
\put(-15,18){$g$}
\put(-15,78){$g$}
\put(145,133){$g$}
\put(145,-37){$g$}
\put(185,78){$\gamma$}
\put(185,18){$\gamma$}
\put(215,48){$+$}
\end{picture}
\begin{picture}(180,180)(-45,-30)
\Gluon(0,20)(50,20){-3}{5}
\Gluon(0,80)(50,80){3}{5}
\Gluon(25,20)(75,-10){-3}{5}
\Gluon(25,80)(75,120){3}{5}
\Photon(100,80)(150,80){3}{5}
\Photon(100,20)(150,20){3}{5}
\ArrowLine(50,20)(50,80)
\ArrowLine(50,80)(100,80)
\ArrowLine(100,80)(100,20)
\ArrowLine(100,20)(50,20)
\put(55,46){$q$}
\put(-15,18){$g$}
\put(-15,78){$g$}
\put(80,118){$g$}
\put(80,-12){$g$}
\put(155,78){$\gamma$}
\put(155,18){$\gamma$}
\end{picture}
\setlength{\unitlength}{1pt}
\caption[ ]{\label{fg:gamgamjj} \it Typical diagrams contributing to
$gg\to \gamma\gamma jj$ at lowest order.}
\end{center}
\vspace*{-.2cm}
\end{figure}
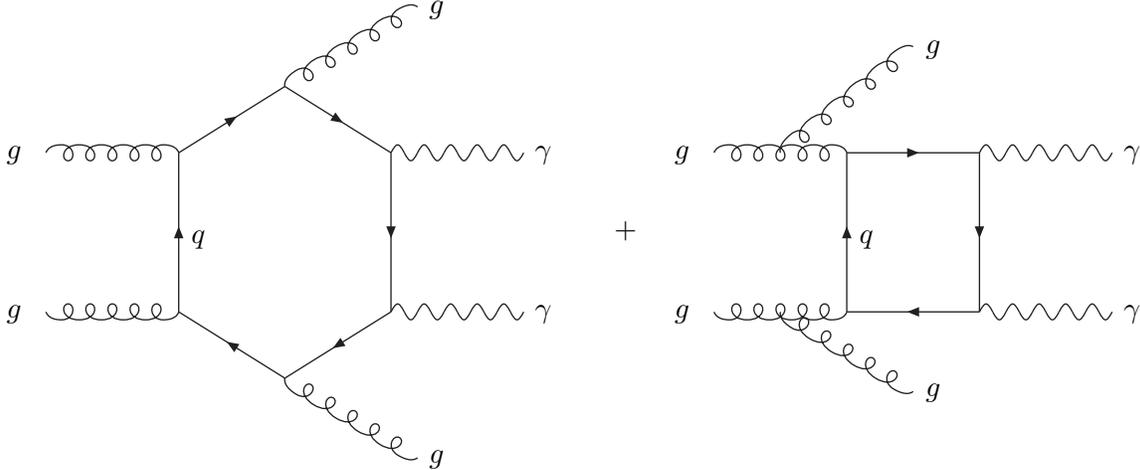

\noindent
\underline{\it (ii) $H\to\tau^+ \tau^-$}:
The observation of $H\to\tau\tau$ decays in WBF will provide crucial 
information on Higgs couplings to fermions~\cite{Zeppenfeld:2000td}
and this channel alone guarantees Higgs observation within the 
MSSM~\cite{vvhtau} and may be an important discovery channel
at low pseudoscalar mass, $m_A$. Recent detector 
simulations~\cite{wbf.exp} confirm parton level
results~\cite{zeppenfeld-tau} for 
the observability of this channel. (See Fig.~\ref{fig:delsigh} for parton 
level estimates of statistical errors.)

\begin{figure}[thb]
\hspace*{2cm} \includegraphics[width=9.0cm, angle=90]{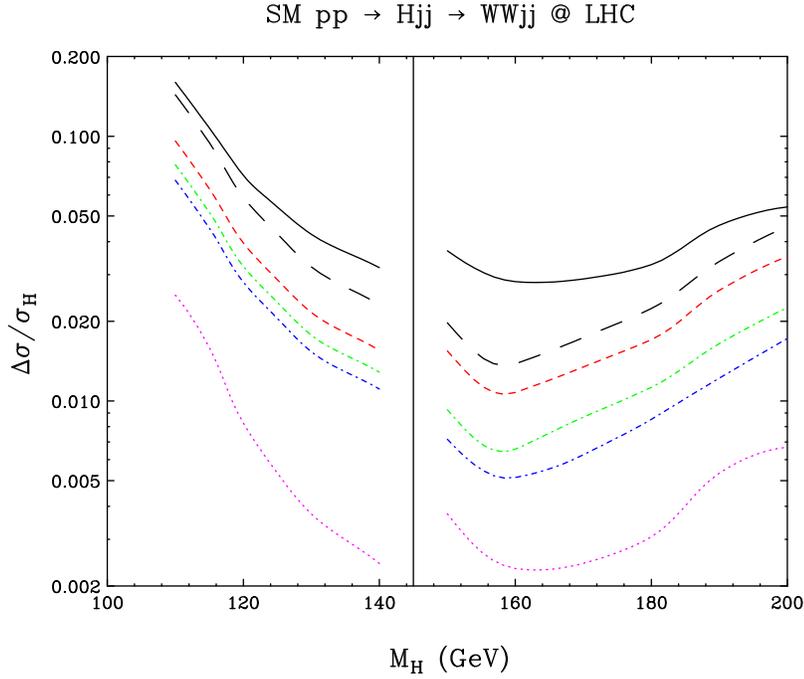}%
 \caption{\it Contributions of background systematic errors $\Delta\sigma$ 
to a measurement of $\sigma_H=\sigma B(H\to WW)$ in WBF. 
Shown, from bottom to top, are the effects of 
a 10\% uncertainty of the $\tau\tau jj$ rate (dotted line),
a 50\% error on the QCD WWjj rate (blue dash-dotted),
a 30\% error on the electroweak WWjj rate (green dash-dotted), and 
a 10\% error on $\sigma(t\bar t+$jets) (red dashes). 
The long-dashed line adds these errors in quadrature. 
For comparison, the solid line shows the expected statistical 
error for 200\fbi.
The vertical line at 145 GeV separates analyses optimized 
for small~\cite{wbf.wwlomh}  and large~\cite{zeppenfeld-ww} Higgs masses.
} \label{fig:errorbugetWW}
 \end{figure}

The $\tau^+ \tau^-$-invariant mass can be reconstructed at the LHC with
a resolution of order 10\%. This is possible in the $qq\to qqH$ mode 
because of the large transverse momentum of the Higgs. 
In turn this means a sideband analysis can be used,
in principle, to directly measure backgrounds. The most important of
these backgrounds is QCD $Zjj$ production (from QCD corrections to Drell-Yan)
or electroweak $Zjj$ production via WBF~\cite{zeppenfeld-tau}. The (virtual)
Z (or photon) then decays into a $\tau^+\tau^-$ pair. These $Zjj$ backgrounds,
with their highly nontrivial shape around $m_{\tau\tau}\approx m_Z$, can be 
precisely determined be observing $Z\to e^+e^-,\mu^+\mu^-$ events in identical
phase space regions. Theoretically the QCD $Zjj$ background is under control
also, after the recent calculation of the full NLO corrections~\cite{zjjNLO}.
For the $\tau^+\tau^-$ backgrounds the inclusion of $\tau$ polarization
effects is important in order to obtain reliable tau-decay distributions
which discriminate between signal processes
($h,H\to \tau^+\tau^-$) and backgrounds. This can be achieved by linking
the TAUOLA program \cite{tauola} to existing Monte Carlo programs.

\vspace{0.1in}
\noindent
\underline{\it (iii) $H\to WW\to \ell^+ \ell^- \sla p_T$}:
The most challenging WBF channel is $H\to WW^{(*)}$ decay which does not
allow for direct Higgs mass reconstruction and, hence, precludes a simple
sideband determination of backgrounds. The important 
backgrounds~\cite{zeppenfeld-ww,wbf.wwlomh} involve
(virtual) $W$ pairs, namely top decays 
in $t\bar t+$jets production, and QCD and electroweak $WWjj$ production.
QCD and EW $\tau\tau jj$ production are subdominant after cuts, they are
known at NLO~\cite{zjjNLO}, and they can be determined directly, in phase 
space regions for jets which are identical to the signal region and with
high statistics, by studying
$e^+e^-$ or $\mu^+\mu^-$ pairs instead of $\tau^+\tau^-$. 

Demands on QCD
calculations can be estimated by comparing the effects of systematic 
background errors on the measurement of the signal rate with statistical 
errors achievable at the LHC with 200\fbi of data. Results are shown in
Fig.~\ref{fig:errorbugetWW} for an assumed 10\% error on 
$\sigma(t\bar t+$jets), a 50\% error on the QCD WWjj rate, and
a 30\% error on the electroweak WWjj rate. The latter two should be achievable
from a LO extrapolation from signal poor to signal rich regions of phase space.
A 10\% error of $\sigma(t\bar t+$jets), on the other hand, may require a NLO
calculation, in particular of the on-shell $t\bar t+1$~jet cross section
which dominates the $t\bar t$ background. 
Off-shell effects have recently been studied at 
LO~\cite{ttoff} and a ${\cal O}(20\%)$ increase of the $t\bar t$ background 
was found, which, presumably,
is small enough to permit the inclusion of off-shell effects at LO only.
However, a dedicated study is needed to devise optimal techniques for a 
reliable background determination for $H\to WW$ searches in WBF, for all major 
backgrounds.

\vspace{0.1in}
\noindent
\underline{\it (iv) Jet veto and Jet Tagging}:
Background suppression in the WBF channels relies on double forward jet 
tagging to identify the scattered quark jets of the $qq\to qqH$ signal
and it employs a veto of relatively soft central jets 
(typically of $p_T>20$~GeV) to exploit the different gluon radiation patterns
and QCD scales of $t$-channel color singlet versus color octet exchange. 
Transverse momenta of these tagging or veto jets are relatively small for 
fixed order perturbative calculations of hard processes at the LHC. Thus,
dedicated studies will be needed to assess the applicability of NLO QCD for 
the modeling of tagging jets in WBF and for the efficiency of a central jet 
veto in the Higgs signal. First such studies have been performed in the past 
at LO, for $Wjj$ or $Zjj$ events~\cite{Chehime:1992ub}. While NLO Monte Carlos
for QCD $Vjj$ production are now available~\cite{zjjNLO,MCFM}, 
the corresponding
NLO determination of electroweak $Vjj$ cross sections would be highly 
desirable. This would allow a comparison of calculated and measured 
veto efficiencies in a WBF process. These efficiencies must be known at the 
few percent level for the signal in order to extract Higgs couplings without
loss of precision.

At present, the veto efficiencies for signal and background processes are 
the most uncertain aspect of WBF Higgs production at the LHC. Any improvement
in their understanding, from QCD calculations, from improved Monte Carlo tools,
or from hadron collider data would be very valuable.

\subsection{$t\bar t\phi$ production}
SM Higgs boson production in association with $t\bar t$ pairs plays a
significant role at the LHC for Higgs masses below about 130 GeV, since
this production mechanism makes the observation of 
$H\to b\bar b$ possible
\cite{CMSTDR,ATLASTDR,drollinger:2001ym,tth2bb.atlas,tth2bb.cms}. 
The decay $H\to\gamma\gamma$ is 
potentially visible in this channel at high integrated luminosity. For Higgs 
masses above about 130 GeV, the decay $H\to W^+W^-$ can be 
observed~\cite{tth2ww}. 
$t\bar tH$ production could conceivably be used to determine the top Yukawa 
coupling directly from the cross section, but this requires either assumptions 
about the branching ratio for $H\to b\bar{b}$, which are not justified in 
extensions of the SM, or observability of decay to either $\gamma\gamma$
or $W^+W^-$. 
Recently, the
NLO QCD corrections have become available. They decrease the cross section
at the Tevatron by about 20--30\% \cite{tthnlo,tthnloq}, while they
increase the signal rate at the LHC by about 20--40\% \cite{tthnlo}. The
scale dependence of the production cross
section is significantly reduced, to a level of about 15\%, which can be
considered as an estimate of the theoretical uncertainty. Thus, the signal
rate is under proper theoretical control now. In the MSSM, $t\bar th$
production with $h\to \gamma\gamma,b\bar b$ is important at the LHC in
the decoupling regime, where the light scalar $h$ behaves as the SM
Higgs boson \cite{CMSTDR,ATLASTDR,drollinger:2001ym,tth2bb.atlas,tth2bb.cms}. 
Thus, the SM results can also be used for $t\bar{t}h$ 
production in this regime. \\

\noindent
\underline{\it (i) $t\bar t\phi\to t\bar t b\bar b$}:
The major backgrounds to the $\phi\to b\bar b$ signal in associated $t\bar
t\phi$ production come from $t\bar t jj$ and $t\bar t b\bar b$ production,
where in the first case the jets may be misidentified as $b$ jets. A full
LO calculation is available for these backgrounds and will be included in
the conventional Monte Carlo programs. However, an analysis of the
theoretical uncertainties is still missing. A first step can be made by
studying the scale dependence at LO in order to investigate the effects on
the total normalization and the event shapes. But for a more sophisticated
picture a full NLO calculation is highly desirable. A second question is
whether these backgrounds can be measured in the experiments off the Higgs
resonance and extrapolated to the signal region. \\

\noindent
\underline{\it (ii) $t\bar t\phi\to t\bar t \gamma\gamma$}:
The $t\bar t \gamma\gamma$ final states develop a narrow resonance in the
invariant $\gamma\gamma$ mass distribution, which enables a measurement
of the $t\bar t \gamma\gamma$ background directly from the sidebands.\\

\noindent
\underline{\it (iii) $t\bar t\phi\to t\bar t W^+W^-$}:
This channel does not allow reconstruction of the Higgs. Instead, it relies 
on a counting experiment of multiplepton final states where the background 
is of approximately the same size as the signal~\cite{tth2ww}. The
principal backgrounds are 
$t\bar{t}Wjj$ and $t\bar{t}\ell^+\ell^- (jj)$, with minor backgrounds of 
$t\bar{t}W^+W^-$ and $t\bar{t}t\bar{t}$. For the $3\ell$ channel, the 
largest background is $t\bar{t}\ell^+\ell^-$ where one lepton is lost. 
It is possible that this rate could be measured directly for the lepton pair 
at the $Z$ pole and the result extrapolated to the signal region of phase 
space. However, for $t\bar{t}Vjj$ backgrounds the QCD uncertainties 
become large and unknown, due to the presence of two additional soft jets 
in the event. Further investigation of these backgrounds is essential, and 
will probably require comparison with data, which is not expected to be 
trivial.

\subsection{$b\bar b\phi$ production}
In supersymmetric theories $b\bar b\phi$ production becomes the dominant
Higgs boson production mechanism for large values of $\tgb$ \cite{habil},
where the
bottom Yukawa coupling is strongly enhanced. In contrast to $t\bar t\phi$
production, however, this process develops potentially large logarithms,
$\log m_\phi^2/m_b^2$, in the high-energy limit due to the smallness of the
bottom quark mass, which are
related to the development of $b$ densities in the initial state. They can
be resummed by evolving the $b$ densities according to the
Altarelli--Parisi equations and introducing them in the production
process \cite{willenbrock}. The introduction of conventional $b$ densities
requires an
approximation of the kinematics of the hard process, i.e.~the initial $b$
quarks are assumed to be massless, have negligible transverse momentum
and travel
predominantly in forward and backward direction. These approximations can be
tested in the full $gg\to b\bar b\phi$ process. At the Tevatron it turns
out that they are not valid so that the effective cross
section for $b\bar b\to \phi$ has to be considered as an overestimate of
the resummed result. An improvement of this resummation requires an
approach which describes the kinematics of the hard process in a
better way. Moreover, since the experimental analyses require 3 or 4 $b$
tags \cite{Carena:2000yx,CMSTDR,ATLASTDR}, the spectator $b$ quarks need
to have a sizeable
transverse momentum of at least 15--20 GeV. Thus a resummation of a
different type of potentially arising logarithms, namely
$\log m_\phi^2/(m_b^2+p_{tmin}^2)$ is necessary. This
can be achieved by the introduction of e.g.~unintegrated parton
densities \cite{uninpdf} or an extension of the available resummation
techniques.

As a first step, however, we have to investigate if the energy of the
Tevatron and LHC is sufficiently large to develop the factorization of
bottom densities. This factorization requires that the transverse
momentum distribution of the (anti)bottom quark scales like
$d\sigma/dp_{Tb}\propto p_{Tb}/(m_b^2+p_{Tb}^2)$ for transverse momenta
up to the factorization scale of the (anti)bottom density.
\begin{figure}[hbt]
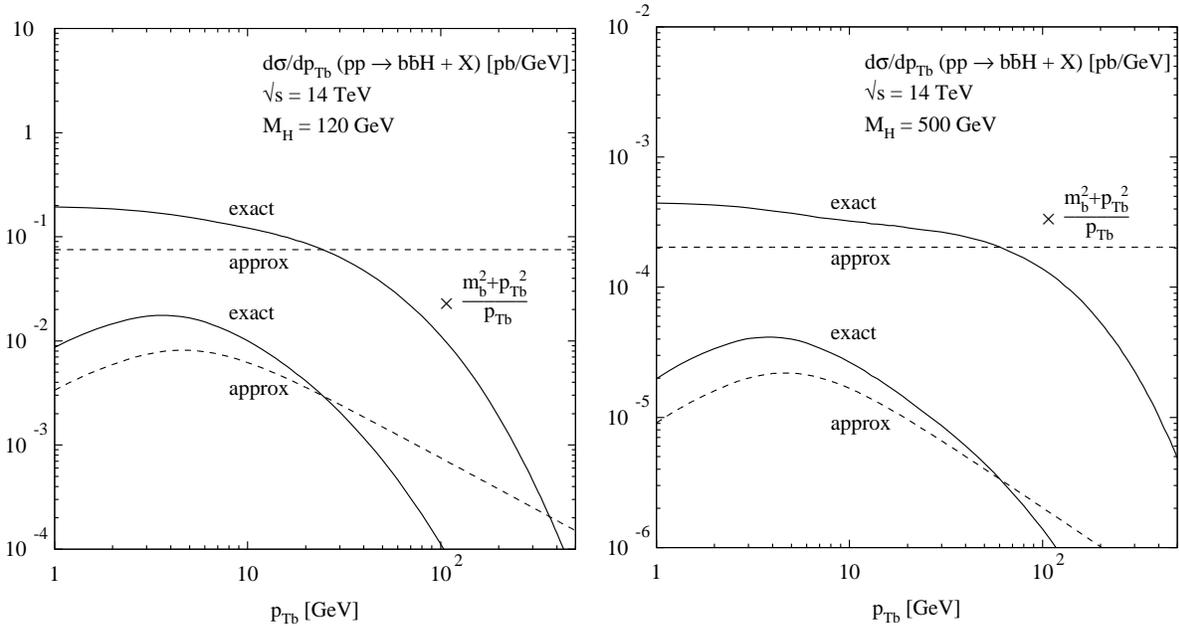

\vspace*{-0.5cm}
\hspace*{0.5cm}
\epsfxsize=7cm \epsfbox{ptlhc.120}
\vspace*{-9.75cm}

\hspace*{8.5cm}
\epsfxsize=7cm \epsfbox{ptlhc.500}
\vspace*{-1.5cm}
\caption[]{\label{fg:bbh} \it Transverse momentum distributions of the
bottom quark in $b\bar bH$ production for two Higgs masses at
the LHC. We have adopted CTEQ5M1 parton densities and a bottom mass of
$m_b=4.62$ GeV. The solid lines show the full LO result from
$q\bar q,gg\to b\bar bH$ and the dashed lines the factorized collinear
part, which is absorbed in the bottom parton density. The upper curves
are divided by the factor $p_{Tb}/(m_b^2+p_{Tb}^2)$ of the asymptotic
behavior, which is required by factorizing bottom densities.}
\end{figure} 
The transverse momentum distributions at the LHC
are shown in Fig.~\ref{fg:bbh}, for two different Higgs
masses. The solid curves show the full distributions of the $q\bar
q,gg\to b\bar b\phi$ processes, while the dashed lines exhibit the factorized
collinear part, which is absorbed in the bottom density. For a proper
factorization, these pairs of curves have to coincide approximately up to
transverse momenta of the order of the factorization scale, which is
usually chosen to be $\mu_F={\cal O}(m_H)$. It is clearly visible that
there are sizeable differences between the full result and the
factorized part, which originate from sizeable bottom mass and phase
space effects, that are not accounted for by an active bottom
parton density. Moreover, the full result falls quickly
below the approximate factorized part for transverse momenta of the
order of $m_H/10$, which is much smaller than the usual factorization
scale used for the bottom densities. We conclude from these plots that
$b\bar b\phi$ production at the LHC develops sizeable bottom mass effects,
so that the use of bottom densities in the process $b\bar b\to \phi$ may
lead to an overestimate of the correct theoretical result due to too
crude approximations in the kinematics of the hard process. The full NLO
calculation of the $gg\to b\bar b\phi$ will
yield much more insight into this problem, since the large logarithms
related to the evolution of bottom densities have to appear in the
NLO corrections, if the picture of active bottom quarks in the proton is
correct.

\subsection{$ZH,WH$ production}
Higgsstrahlung in $q\bar q\to WH,ZH$ plays a crucial role for the Higgs
search at the Tevatron, while it is only marginal at the LHC. At the
Tevatron it provides the relevant production mechanism for Higgs masses
below about 130 GeV, where $H\to b\bar b$ decays are
dominant~\cite{Carena:2000yx}. 
The NLO QCD corrections have been analyzed in the past. They are identical 
to the QCD corrections to the Drell--Yan processes $q\bar q\to W,Z$, if the
LO matrix elements are replaced accordingly. QCD corrections increase
the production cross sections by about 30--40\%~\cite{vhnlo,SUSYQCDcor}.

The most important backgrounds at the Tevatron are $Wjj$ and in particular
$Wb\bar b$ production. Both are known at NLO and are contained in a NLO
Monte Carlo program~\cite{MCFM}.  The
same applies also to the $Zjj$ and in particular $Zb\bar b$ 
backgrounds~\cite{veseli,zjjNLO}.
In addition, the $t\bar t$ background is relevant.

\subsection{Conclusions}
Considerable progress has been made recently in improving QCD calculations
for Higgs signal and background cross sections at hadron colliders.
Noteworthy examples are the NNLO corrections to the gluon fusion cross 
section~\cite{Harlander:2002wh}, the QCD $Zjj$ cross section at 
NLO~\cite{zjjNLO} and the determination of full finite top and $W$ width 
corrections to $t\bar t$ and $t\bar tj$ production at LO~\cite{ttoff}.
These improvements are crucial for precise coupling determinations of the
Higgs boson.

Much additional work is needed to match the statistical power of the LHC.
Largely, QCD systematic errors for coupling measurements have not been 
analyzed yet. Additional NLO tools need to be provided as
well, and these include NLO corrections to $t\bar t$ production with 
finite width effects and $t\bar tj$ production at zero top width.
A better understanding of central jet veto efficiencies is crucial for 
the study of WBF channels. These are a few examples where theoretical 
work is needed. Many more have been highlighted in this review. 
Higgs physics at the LHC remains a very rich field 
for \mbox{phenomenology.}}

{
\def\ltap{\raisebox{-.4ex}{\rlap{$\,\sim\,$}} \raisebox{.4ex}{$\,<\,$}}
\def\gtap{\raisebox{-.4ex}{\rlap{$\,\sim\,$}} \raisebox{.4ex}{$\,>\,$}}
\def\lra{\leftrightarrow}
\def\naive{na\"{\i}ve}
\newcommand\as{\alpha_{\mathrm{S}}}
\newcommand\f[2]{\frac{#1}{#2}}
\def\xcut{\f{p_T^{2cut}}{M_H^2}}
\def\dO{{\cal D}_{0}}
\def\dl{{\cal D}_{1}}
\def\dll{{\cal D}_{2}}
\def\dlll{{\cal D}_{3}}
\def\ee{$e^+e^-$}
\def\la{\lambda}
\def\beq{\begin{equation}}
\def\eeq{\end{equation}}
\def\beeq{\begin{eqnarray}}
\def\eeeq{\end{eqnarray}}
\def\bom#1{{\mbox{\boldmath $#1$}}}
\def\to{\rightarrow}
\def\nn{\nonumber}
\def\arrowlimit#1{\mathrel{\mathop{\longrightarrow}\limits_{#1}}}
\def\qt{q_{\perp}}
\def\res{{\rm res.}}
\def\ms{${\overline {\rm MS}}$}
\def\msbar{{\overline {\rm MS}}}
\def\asp{{\alpha_s}\over{\pi}}

\def\sqr{\sqrt{1 - \pitcut^2 }}
\def\pitcut{\pi_T}

\section[]{Direct Higgs production and jet veto%
\footnote{\it S.\,Catani, D.\,de Florian and M.\,Grazzini}}

\noindent
Direct Higgs production through gluon--gluon fusion, followed by
the decay $H\to W^*W^*,Z^*Z^*$ is a relevant channel to discover a Higgs boson
with mass $140\ltap M_H \ltap 190$ GeV both at the Tevatron and at the LHC.
In particular, 
the decay mode $W^*W^*\to l^+l^-\nu {\bar \nu}$ is quite important
\cite{Carena:2000yx,ATLASTDR,CMSTDR,Dittmar:1997ss},
since it is cleaner than $W^*W^*\to l\nu jj$, and the decay rate $H\to W^*W^*$
is higher than $H\to Z^*Z^*$ by about one order of magnitude.

An important background for the direct Higgs signal 
$H\to W^*W^*\to l^+l^-\nu {\bar \nu}$
is $t {\bar t}$ production ($tW$ production is also important at the LHC), where
$t \to l{\bar \nu} b$, thus leading to $b$ jets with high $p_T$ in the final 
state. If the $b$ quarks are not identified, 
a veto cut on the transverse momenta of the jets accompanying
the final-state leptons
can be applied to enhance the signal/background ratio.
Imposing a jet veto turns out to be essential, both at
the Tevatron \cite{Carena:2000yx,Han:1999ma} and at the LHC \cite{ATLASTDR,CMSTDR,Dittmar:1997ss},
to cut the hard $b$ jets arising from this background process.

Here we study the effect of a jet veto on direct Higgs
production. More details can be found in Ref.~\cite{Catani:2001cr}.
The events that pass the veto selection are those with
$p_T^{\rm jet} < p_T^{\rm veto}$, where $p_T^{\rm jet}$ is the transverse
momentum of any final-state jets, defined by a cone algorithm.
The cone size $R$ of the jets will be fixed
at the
value $R=0.4$.

The vetoed cross section $\sigma^{\rm veto}(s,M_H^2;p_T^{\rm veto},R)$ can be
written as 
\begin{equation}
\label{sigmaveto}
\sigma^{\rm veto}(s,M_H^2;p_T^{\rm veto},R) = \sigma(s,M_H^2) 
- \Delta \sigma(s,M_H^2;p_T^{\rm veto},R) \;\;,
\end{equation}
where $\sigma(s,M_H^2)$ is the inclusive cross section, and $\Delta \sigma$ is
the `loss' in cross section due to the jet-veto procedure.  The jet-vetoed
cross section is evaluated by using the large-$M_{\rm top}$ limit.  At NLO
(NNLO) the calculation is performed by subtracting the LO (NLO) cross section
for the production of Higgs plus jet(s) from the inclusive NLO (NNLO) result.

The NLO calculation is exact: apart from using the 
large-$M_{\rm top}$ limit, we do not perform any further approximations.
At the NNLO, the contribution $\Delta \sigma$ to Eq.~(\ref{sigmaveto})
is again evaluated exactly, by using the numerical program of Ref.~\cite{deFlorian:1999zd}. 
To evaluate the contribution of the inclusive
cross section we use the recent result of Ref.~\cite{Catani:2001ic,Harlander:2001is}, and in particular, we
rely on our approximate estimate
NNLO-SVC \cite{Catani:2001ic}.
In the following we present both NLO and NNLO numerical results for the vetoed
cross section. 
The results are obtained by using the
parton distributions of the MRST2000 set \cite{mrst2000},
with densities and coupling constant evaluated at each corresponding order.
The MRST2000 set includes (approximate) NNLO parton densities.

\begin{figure}[ht]
\begin{center}
\begin{tabular}{c}
\epsfxsize=11truecm
\epsffile{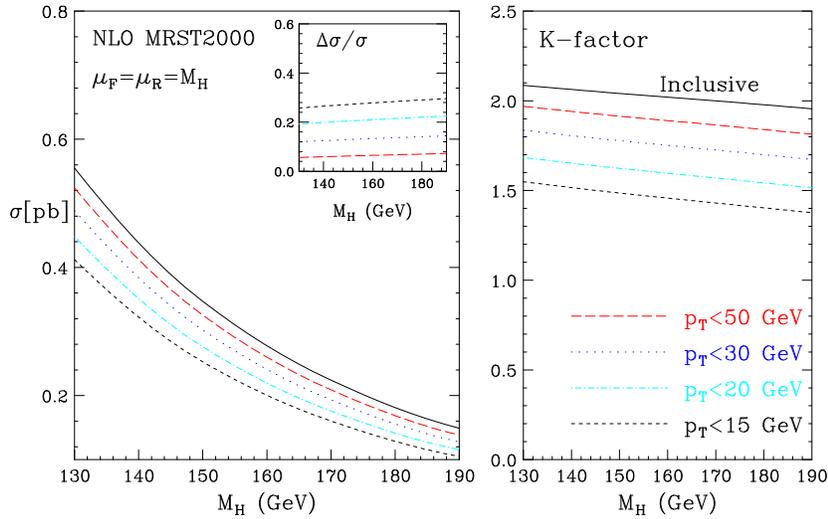}\\[-0.7cm]
\end{tabular}
\end{center}
\caption{\label{fig:nlotev}{\em Vetoed cross section
and K-factors: NLO results at the Tevatron Run II.}}
\vspace*{-0.2cm}
\end{figure}
We first present the vetoed cross section
at the Tevatron Run II.
In Fig.~\ref{fig:nlotev} we show the dependence of the NLO results 
on the Higgs mass for different values of 
 $p_T^{\rm veto}$ (15, 20, 30 and 50~GeV).
The vetoed cross sections $\sigma^{\rm veto}(s,M_H^2;p_T^{\rm veto},R)$
and the inclusive cross section $\sigma(s,M_H^2)$ are given in the plot 
on the left-hand side. The inset plot gives an idea of
the `loss' in cross section once the veto is applied,
by showing the ratio between the cross section difference $\Delta \sigma$ 
in Eq.~(\ref{sigmaveto}) and the inclusive cross section at the same
perturbative order.
As can be observed, for large values of the cut, say $p_T^{\rm veto}=50$~GeV,
less than 10\% of the inclusive cross section is vetoed. The veto effect
increases by decreasing $p_T^{\rm veto}$, but it is still smaller than 30\%
when $p_T^{\rm veto}=15$~GeV.
On the right-hand side of Fig.~\ref{fig:nlotev},
we show the corresponding K-factors, i.e. the vetoed
cross sections normalized to the LO result,
which is independent of the value of the cut.
Figure~\ref{fig:nnlotev} shows the analogous results at NNLO.
\begin{figure}[htb]
\begin{center}
\begin{tabular}{c}
\epsfxsize=11truecm
\epsffile{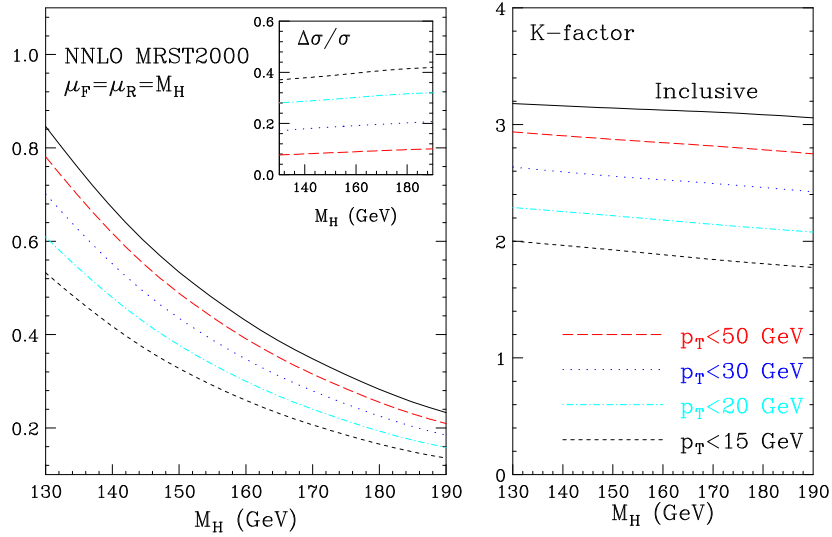}\\[-0.7cm]
\end{tabular}
\end{center}
\caption{\label{fig:nnlotev}{\em Vetoed cross section and K-factors:
NNLO results at the Tevatron Run II.}}
\vspace*{-0.2cm}
\end{figure}
\begin{figure}[htb]
\begin{center}
\begin{tabular}{c}
\epsfxsize=7.4truecm
\epsffile{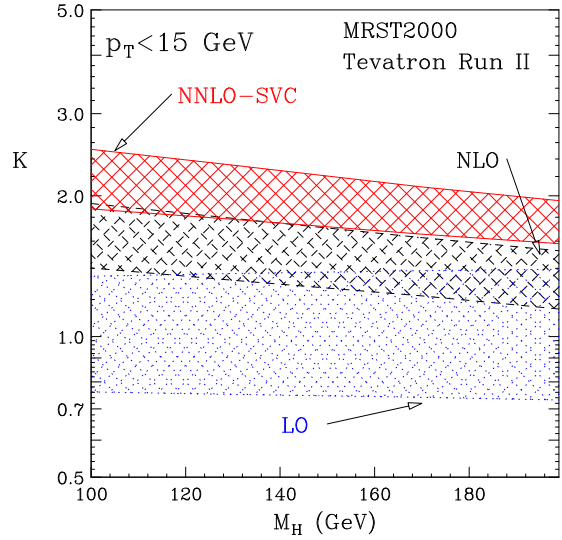}\\[-0.9cm]
\end{tabular}
\end{center}
\caption{\label{fig:kv}{\em  K-factors for Higgs production
at the Tevatron for a veto of $p_T^{\rm veto}=15$~GeV at LO, NLO
and NNLO-SVC.}}
\vspace*{-0.4cm}
\end{figure}
In Fig.~\ref{fig:kv} we show the LO, NLO and NNLO-SVC K-factor bands,
computed by varing renormalization ($\mu_R$) and factorization ($\mu_F$) scales in the range $1/2 M_H<\mu_F,\mu_R<2M_H$ and normalizing to the LO contribution at $\mu_F=\mu_R=M_H$.
The calculation is done with $p_T^{\rm veto}=15$~GeV.
Comparing Fig.~\ref{fig:kv} with the inclusive case (see Ref.~\cite{Catani:2001cr}),
we see that the effect of the veto is to partially reduce the
relative difference between the NLO and NNLO results; the increase of 
the corresponding K-factors
can be estimated to about $25\%$. 
\begin{figure}[htb]
\vspace*{-0.0cm}
\begin{center}
\begin{tabular}{c}
\epsfxsize=11truecm
\epsffile{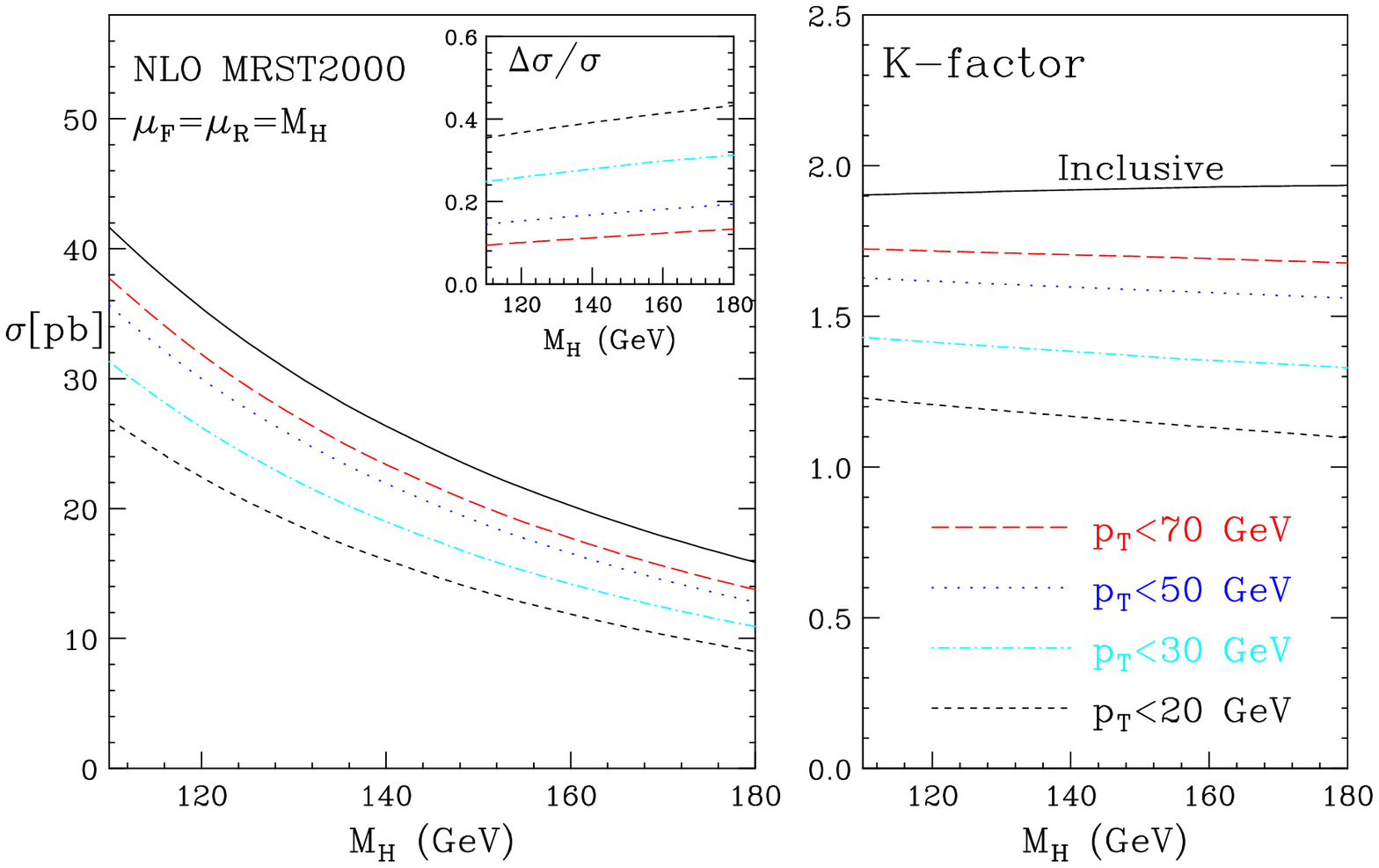}\\[-0.8cm]
\end{tabular}
\end{center}
\caption{\label{fig:nlolhc}{\em Vetoed cross sections and
K-factors at NLO at the LHC. }}
\vspace*{-0.4cm}
\end{figure}
\begin{figure}[htb]
\begin{center}
\begin{tabular}{c}
\epsfxsize=11truecm
\epsffile{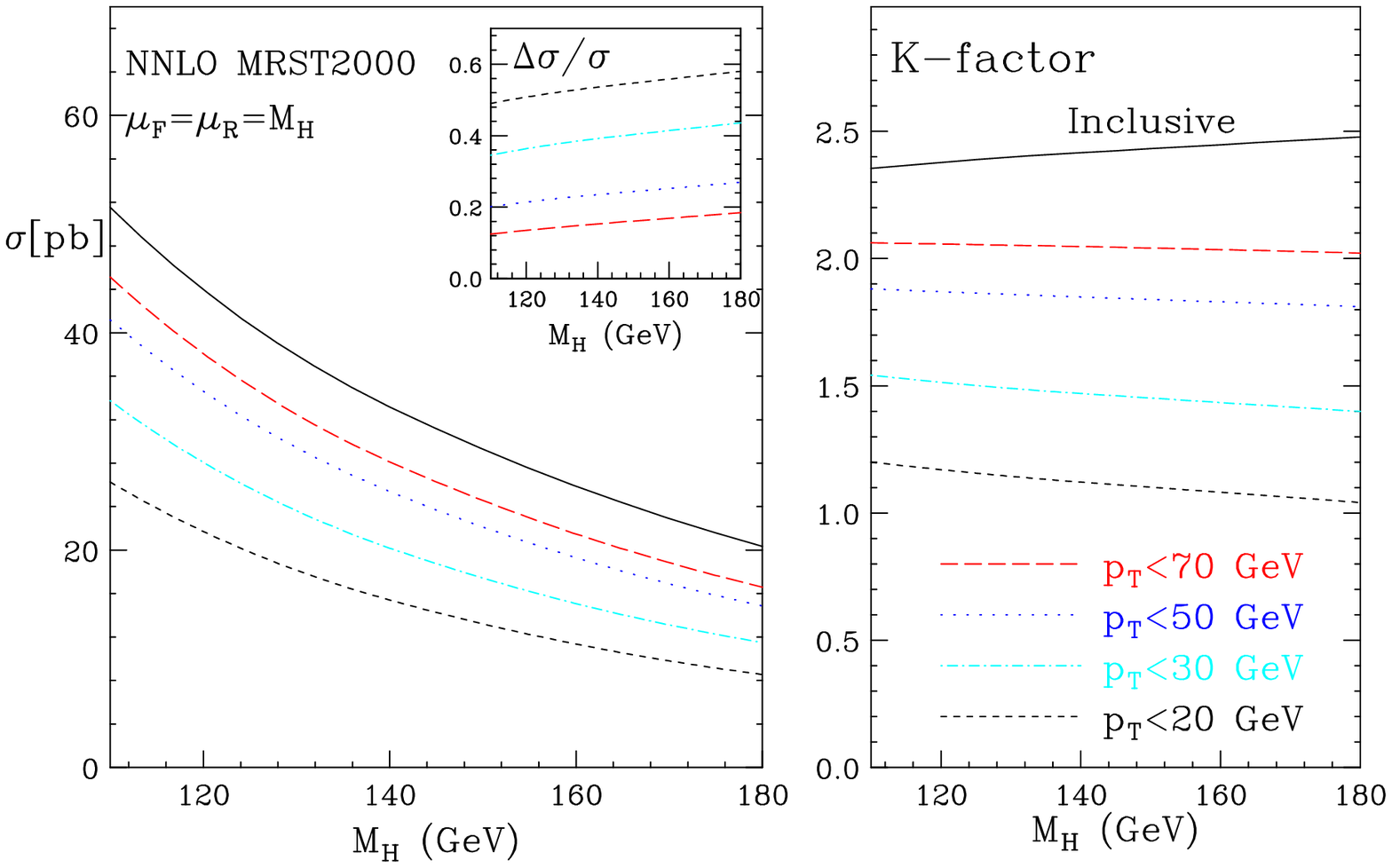}\\[-0.7cm]
\end{tabular}
\end{center}
\caption{\label{fig:nnlolhc}{\em Vetoed cross sections and
K-factors at NNLO at the LHC.  }}
\vspace*{-0.2cm}
\end{figure}
 
The results for the vetoed cross sections at the LHC are presented in
Figs.~\ref{fig:nlolhc} and \ref{fig:nnlolhc} for 
$p_T^{\rm veto}=20$, 30, 50 and 70~GeV.
At fixed value of the cut, the impact of the jet veto, 
both in the `loss' of cross section and in the reduction of the K-factors,
is larger at the LHC than at the Tevatron Run II. This effect can also be 
appreciated by comparing Fig.~\ref{fig:kvlhc} and Fig.~\ref{fig:kv}.
At the LHC, the value of $p_T^{\rm veto}=30$~GeV is already sufficient to reduce
the difference between the NNLO and NLO results
to less than $10\%$. 
\begin{figure}[htb]
\vspace*{-0.0cm}
\begin{center}
\begin{tabular}{c}
\epsfxsize=7.4truecm
\epsffile{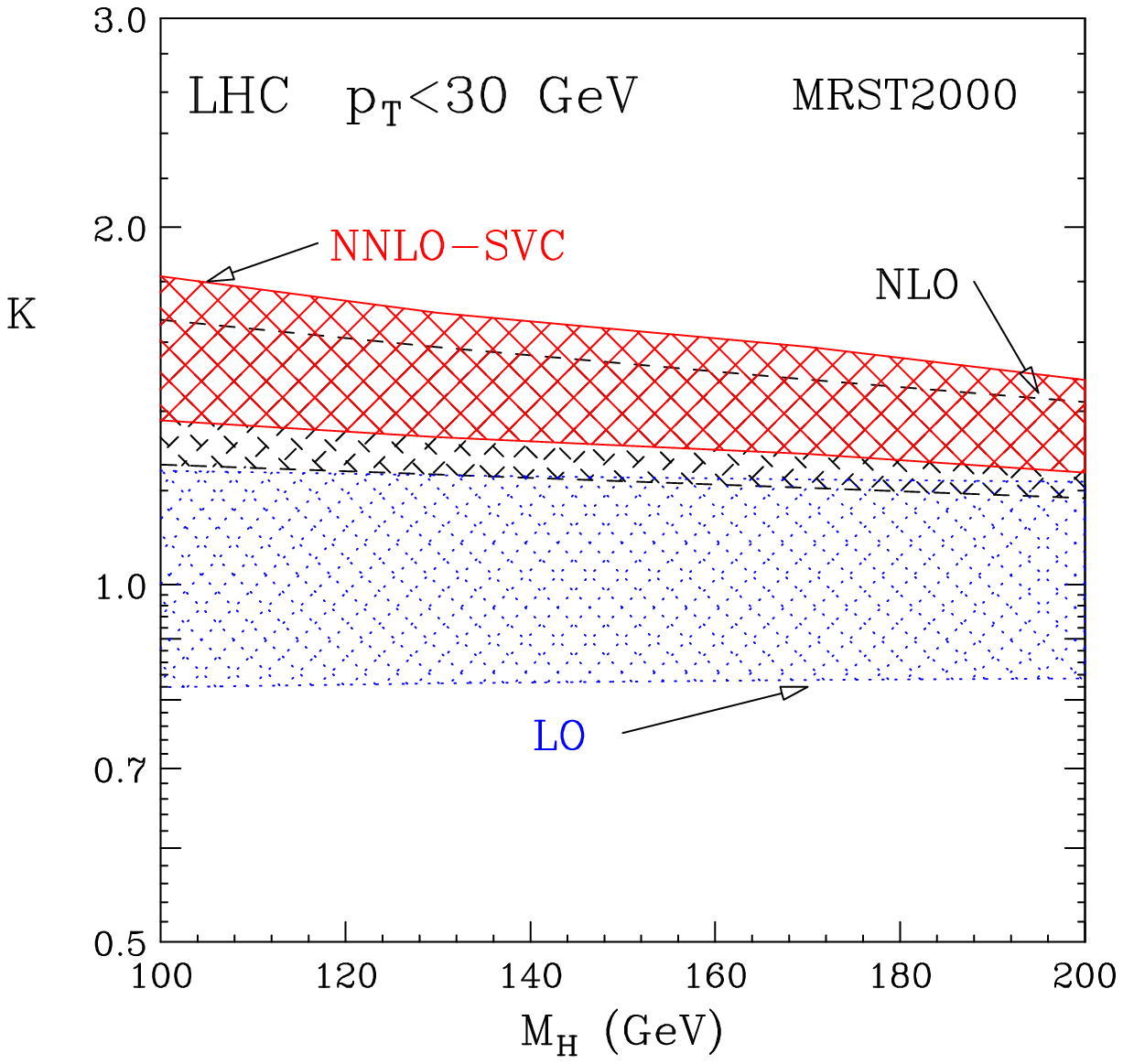}\\[-0.8cm]
\end{tabular}
\end{center}
\caption{\label{fig:kvlhc}{\em  The same as in Fig.~\ref{fig:kv}, but at the LHC
and with $p_T^{cut}=30$~GeV.}}
\vspace*{-0.4cm}
\end{figure}

The results presented above can be interpreted according to a simple physical picture.
The dominant part of QCD corrections 
is due to soft and collinear radiation \cite{Catani:2001ic}.
The characteristic scale of the highest
transverse momentum $p_T^{\rm max}$ of the accompanying jets is 
$p_T^{\rm max}\sim \langle 1- z \rangle M_H$, 
where the average value 
$\langle 1- z \rangle = \langle 1 - M_H^2/{\hat s}\rangle$ of the distance
from the partonic threshold is small. As a consequence the jet veto procedure
is weakly effective unless the value of $p_T^{\rm veto}$ is
substantially smaller than $p_T^{\rm max}$.
Decreasing $p_T^{\rm veto}$,
the enhancement of the inclusive cross section due to soft radiation at higher
orders is reduced, and the jet veto procedure tends to
improve the convergence of the perturbative series.
At the LHC Higgs production is less
close to threshold than at the Tevatron and, therefore, the accompanying jets
are harder. This is the reason why, at fixed $p_T^{\rm veto}$, the effect of the jet veto is
stronger at the LHC than at the Tevatron.

When $p_T^{\rm veto}$ is much smaller than the characteristic scale
$p_T^{\rm max}\sim \langle 1- z \rangle M_H$, the
perturbative expansion
of the vetoed cross section contains large logarithmic
contributions that can spoil the convergence
of the fixed-order expansion in $\as$. Since $\langle 1- z \rangle M_H$ is
larger at the LHC than at the Tevatron, the value of $p_T^{\rm veto}$ at which
these effects become visible is larger at the LHC. Whereas at the Tevatron the perturbative calculation for $p_T^{\rm veto}=15$ GeV seems still to be reliable, at the LHC, with the same value of $p_T^{\rm veto}$, the perturbative result suggests that the effect of these logarithmic contributions is large \cite{Catani:2001cr}.
 

\noindent {\bf Note added.} After the completion of this work, the full
NNLO
QCD contribution to {\em inclusive} Higgs boson production has been
computed
\cite{Harlander:2002wh}. These results influence
those in the present paper through Eq.~(\ref{sigmaveto}), since in our
NNLO
calculation the inclusive cross section $\sigma(s,M_H^2)$ is evaluated
by
using
the approximate (soft-collinear) estimate (named NNLO-SVC) of
Ref.~\cite{Catani:2001ic}. We have considered the effect of the
additional
hard corrections of Ref.~\cite{Harlander:2002wh} and, as expected
\cite{Catani:2001ic,Catani:2001cr}, we find that they are relatively
small.
The inclusive cross section at full NNLO is smaller than its NNLO-SVC
approximation by about $5\%$ ($7\%$) at the LHC (Tevatron Run II).
This correction can directly be applied to our results. For instance,
the NNLO K-factors in Figs.~\ref{fig:nnlotev}, \ref{fig:kv} and
\ref{fig:nnlolhc} can be modified as $K \rightarrow K - \Delta K$,
where  $\Delta K=0.20$-0.21 at the Tevatron
and $\Delta K=0.11$-0.13 at the LHC (the variations of $\Delta K$
correspond
to variations of the Higgs mass in the range considered in the Figures).
}

{
\newcommand{\beq}{\begin{equation}}
\newcommand{\eeq}{\end{equation}}
\newcommand{\bea}{\begin{eqnarray}}
\newcommand{\eea}{\end{eqnarray}}
\newcommand{\nn}{\nonumber}

\def\eqn#1{Eq.~(\ref{#1})}
\def\eqns#1#2{Eqs.~(\ref{#1}) and~(\ref{#2})}
\def\eqnss#1#2{Eqs.~(\ref{#1})-(\ref{#2})}
\def\fig#1{Fig.~{\ref{#1}}}
\def\sec#1{Section~{\ref{#1}}}
\def\app#1{Appendix~\ref{#1}}
\def\tab#1{Table~\ref{#1}}


\def\cM{{\cal M}}
\def\sep{\mbox{$\,|\;$}}

\newcommand\sss{\scriptscriptstyle}
\newcommand\as{\alpha_{\sss S}}
\newcommand\mh{M_{\sss {\rm H}}}
\newcommand\mt{M_{t}}
\newcommand\pt{p_{{}\sss \perp}}
\newcommand\ph{p_{\sss {\rm H}}}
\newcommand\mhp{m_{\sss {{\rm H}_\perp}}}
\newcommand\sah{s_{j_1\sss {\rm H}}}
\newcommand\sbh{s_{j_2\sss {\rm H}}}
\newcommand\yh{y_{\sss {\rm H}}}
\newcommand\qip{q_{i\sss \perp}}


\section[]{The high-energy limit of $\boldsymbol{H+2}$ jet production via 
gluon fusion%
\footnote{\it V.\,Del Duca, W.B.\,Kilgore, C.\,Oleari, C.R.\,Schmidt
and D.\,Zeppenfeld}}


At the Large Hadron Collider (LHC), the main production channels
of a Higgs boson are gluon fusion and weak-boson fusion 
(WBF)~\cite{CMS,ATLASTDR}. 
The WBF process, $q q\to q q H$, occurs through the exchange of a
$W$ or a $Z$ boson in the $t$ channel, and is characterized by
the production of two forward quark 
jets~\cite{Dokshitzer:1987nc}. 
Even though it is smaller than the gluon fusion 
channel by about a factor of 5 for an intermediate mass Higgs boson,
it is interesting because it is expected to provide information
on Higgs boson couplings~\cite{Zeppenfeld:2000td}. 
In this respect, $H + 2$ jet production
via gluon-gluon fusion, which has a larger production rate before cuts, can be 
considered a background; it
has the same final-state topology, and thus may hide the features of the
WBF process.

In Higgs production via gluon fusion, the Higgs boson is produced 
mostly via a top quark loop. The computation of $H + 2$ jet production
involves up to pentagon quark loops~\cite{DelDuca:2001eu}. However,
if the Higgs mass is smaller than the threshold for the creation of a 
top-quark pair, $\mh \lesssim 2 \mt$, the coupling of the Higgs to the
gluons via a top-quark loop can be replaced by an effective 
coupling~\cite{Shifman:1979eb}: this is called the {\it large-$\mt$ limit}.
It simplifies the calculation, because it reduces the number of loops 
in a given diagram by one. 
In $H + 2$ jet production, the large-$\mt$ limit yields a good
approximation to the exact calculation if, in addition to the
condition $\mh \lesssim 2 \mt$, we require that
the jet transverse energies are smaller than the top-quark mass, 
$\pt \lesssim \mt$~\cite{DelDuca:2001eu}.
However, 
the large $\mt$ approximation is quite insensitive to the value of the
Higgs--jet and/or dijet invariant masses. The last issue
is not academic, because Higgs
production via WBF, to which we should like to compare, features
typically two forward quark jets, and thus a large dijet invariant mass.

In this contribution, we consider $H + 2$ jet production 
when Higgs--jet and/or dijet 
invariant masses become much larger than the typical momentum 
transfers in the scattering. We term these conditions
the {\it high-energy limit}. 
In this limit the scattering amplitude factorizes into 
{\it impact factors} connected by a gluon exchanged in the $t$ channel.
Assembling together different impact factors, the amplitudes for different
sub-processes can be obtained. Thus the high-energy factorization 
constitutes a stringent consistency check on any amplitude for
the production of a Higgs plus one or more jets.

In the high-energy limit of $H + 2$ jet production,
the relevant (squared) energy scales are the parton center-of-mass energy $s$, 
the Higgs mass $\mh^2$, the dijet invariant mass $s_{j_1j_2}$, and 
the jet-Higgs invariant masses $\sah$ and $\sbh$. At leading order they
are related through momentum conservation,
\begin{equation}
s = s_{j_1j_2} + \sah + \sbh - \mh^2\, .\label{Hjjmtmcons}
\end{equation}
There are two possible high-energy limits to consider: 
$s_{j_1j_2}\gg\sah,\sbh\gg\mh^2$ and
$s_{j_1j_2},\sbh\gg\sah,\mh^2$.
In the first case the Higgs boson is centrally located in rapidity between
the two jets, and very far from either jet.  In the second case the Higgs boson
is close to one jet, say to jet $j_1$, in rapidity, and both of these are 
very far from jet $j_2$. In both cases
the amplitudes will factorize, and the relevant Higgs vertex in case 1
and the Higgs--gluon and Higgs--quark impact factors in case 2 can be obtained 
from the amplitudes for $q\,Q\to q\,Q\,H$ and $q\,g\to q\,g\,H$ scattering.

\subsection*{The high-energy limit $\boldsymbol{ s_{j_1j_2} \gg \sah,\, 
\sbh\gg \mh^2}$ }

We consider the production of two partons of momenta $p_1$ and $p_3$
and a Higgs boson of momentum $\ph$,
in the scattering between two partons of momenta $p_2$ and $p_4$, where 
all momenta are taken as outgoing.
We consider the limit in which the Higgs boson is produced 
centrally in rapidity, and very far from either jet,
$s_{j_1j_2} \gg \sah,\, \sbh\gg \mh^2$,
which is equivalent to require that
\begin{equation}
p_1^+ \gg \ph^+ \gg p_3^+ \, ,\qquad
p_1^- \ll \ph^- \ll p_3^- \, ,\label{mrk}
\end{equation}
where we have introduced the light-cone coordinates 
$p^{\pm}= p_0\pm p_z $, and complex transverse coordinates 
$p_{\perp} = p^x + i p^y$.
In the limit~(\ref{mrk}), the amplitudes
are dominated by gluon exchange in the $t$ channel, with emission
of the Higgs boson from the $t$-channel gluon. 
We can write the amplitude for $q\, Q\to q\, Q\, H$ scattering 
in the high-energy limit as~\cite{us}
\begin{eqnarray}
\lefteqn{i\ \cM^{qq\to Hqq}(p_2^{-\nu_1},p_1^{\nu_1} \sep H \sep 
p_3^{\nu_3}, p_4^{-\nu_3}) } \nonumber\\ &=& 2s \left[ g\,
T^c_{a_1 \bar a_2}\, C^{\bar q;q}(p_2^{-\nu_1};p_1^{\nu_1})\right]
{1\over t_1} \left[ \delta^{cc'} C^{\sss H}(q_1,\ph,q_2)\right] {1\over t_2}
\left[ g\, T^{c'}_{a_3 \bar a_4}\, 
C^{\bar q;q}(p_4^{-\nu_3};p_3^{\nu_3})\right]\, ,\label{HqqqqHE}
\end{eqnarray}
where $q_1 = - (p_1+p_2)$, $q_2 = p_3+p_4$, $t_i\simeq - |\qip|^2$,
$i = 1, 2$, and the $\nu$'s are the quark helicities. 
In \eqn{HqqqqHE} we have made explicit the helicity conservation
along the massless quark lines.
The effective vertex $C^{\bar q;q}$ for the production of a quark jet,
$q\, g^* \rightarrow q$, 
contributes a phase factor~\cite{DelDuca:2000ha}: its square is 1.
The effective vertex for Higgs production
along the gluon ladder, $g^* g^* \to H$,
with and off-shell $g^*$ is
\begin{equation}
C^{\sss H}(q_1,\ph,q_2) = 2 g^2 \mt^2/v \cdot \left( \mhp^2 
A_1(q_1,q_2) - 2A_2(q_1,q_2) \right)\, .\label{hif}
\end{equation} 
The scalar coefficients of the triangle vertex with two off-shell gluons,
$A_{1,2}$, are defined in terms of the form factors $F_T$ and $F_L$ of 
Ref.~\cite{DelDuca:2001eu} as
\begin{equation}
A_1 = i F_T / (4\pi)^2 \;, \qquad
A_2 = i \left( F_T\ q_1\cdot q_2 + F_L\ q_1^2 q_2^2
\right)/ (4\pi)^2\; .
\end{equation} 

We have checked analytically that the amplitude for 
$q\, g\to q\, g\, H$ scattering can also be written as
\eqn{HqqqqHE}, provided we perform on one of the two effective vertices 
$C^{\bar q;q}$ the substitution (for the sake of illustration, we display 
it here for the lower vertex)
\begin{equation}
i g\, f^{bb'c}\, C^{g;g}(p_b^{\nu_b};p_{b'}^{\nu_{b'}}) \leftrightarrow g\, 
T^c_{b' \bar b}\, C^{\bar q;q}(p_b^{-\nu_{b'}};p_{b'}^{\nu_{b'}})\, 
,\label{qlrag}
\end{equation}
and use the effective vertices $g^*\, g \rightarrow g$ for the production
of a gluon jet~\cite{DelDuca:2000ha} (which contribute a phase factor as 
well). 
The same check on the (squared) amplitude for
$g\, g\to g\, g\, H$ scattering has been performed numerically.
Thus, in the high-energy limit~(\ref{mrk}), 
the amplitudes for $q\, Q\to q\, Q\, H$,
$q\, g\to q\, g\, H$ and $g\, g\to g\, g\, H$ scattering only
differ by the color strength in the jet-production vertex. Therefore, 
in a production rate it is enough to consider one of them and include 
the others through the effective parton distribution 
function~\cite{Combridge:1984jn},
$f_{\rm eff}(x,\mu_F^2) = G(x,\mu_F^2) + (C_F/C_A)\sum_f
\left[Q_f(x,\mu_F^2) + \bar Q_f(x,\mu_F^2)\right]$,
where $x$ is the momentum fraction of the incoming parton,
$\mu_F^2$ is the collinear factorization scale, 
and where the sum is over the quark flavors.

\begin{figure}[t]
\vspace*{-0.5cm}
\begin{center}
\begin{turn}{-90}
\epsfig{figure=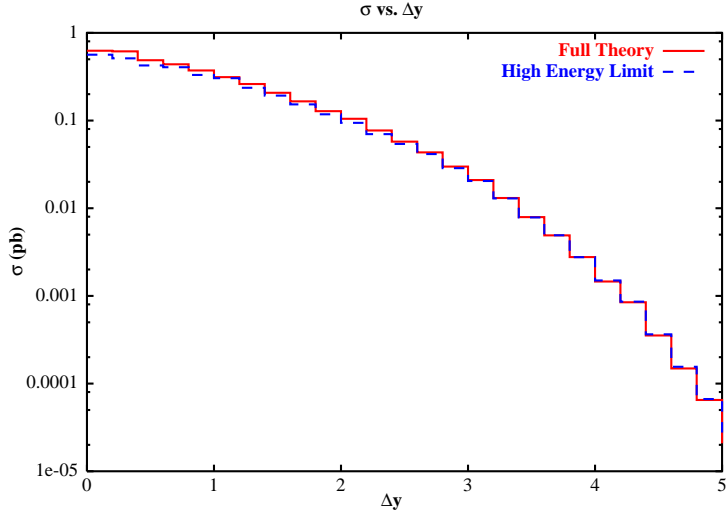,width=7cm}
\end{turn}
\end{center}
\vspace*{-0.5cm}
\caption{Cross section in $H+2$~jet production in $pp$ collisions 
at the LHC energy $\sqrt{s}=14$~TeV as a function of $\Delta y$,
with $\mh= 120$~GeV and $\mt = 175$~GeV. The dijet invariant mass
fulfills the constraint $\sqrt{s_{j_1j_2}} \ge 600$~GeV.
The rapidity interval $\Delta y$
is defined as $\Delta y = {\rm min}(|y_{j_1} - \yh|, |y_{j_2} - \yh|)$,
with the kinematical constraint $y_{j_1} > \yh > y_{j_2}$.
The solid line is the exact production rate; the dashed line is
the rate in the high-energy limit.} 
\label{fig:dydistra}
\vspace*{-0.5cm}
\end{figure}

In \fig{fig:dydistra} we plot the cross section in $H+2$~jet production 
at the LHC energy $\sqrt{s}=14$~TeV, as a function 
of $\Delta y$, which is defined as the smallest rapidity difference 
between the Higgs and
the jets, $\Delta y = {\rm min}(|y_{j_1} - \yh|, |y_{j_2} - \yh|)$,
with the kinematical constraint $y_{j_1} > \yh > y_{j_2}$.
The solid line is the exact production rate,
with the amplitudes evaluated in Ref.~\cite{DelDuca:2001eu}; 
the dashed line is the rate in the high-energy limit~(\ref{mrk}),
with the amplitudes evaluated using Eqns.~(\ref{HqqqqHE})--(\ref{qlrag}).
It is apparent that the high-energy limit works very well over the whole
$\Delta y$ spectrum. However, in the evaluation of the effective 
vertex~(\ref{hif}), 
we used the exact value of the scalar coefficients $A_{1,2}$.
A more conservative statement is to say that when any kinematic quantity
involved in the amplitude~(\ref{HqqqqHE}) is evaluated in the 
limit~(\ref{mrk}),
we expect the high-energy limit to represent a good approximation of 
the exact calculation when $\Delta y\gtrsim 2$.

\subsection*{The high-energy limit 
$\boldsymbol{ s_{j_1j_2},\, \sbh\gg \sah,\, \mh^2}$ }
\label{sec:limit2}

Next, we consider the limit in which the Higgs is produced 
forward in rapidity, and close to one of the jets, say to jet $j_1$,
and both are very far from jet $j_2$, {\it i.e.}
$s_{j_1j_2},\, \sbh\gg \sah,\, \mh^2$. This limit implies that
\begin{equation}
p_1^+ \simeq \ph^+ \gg p_3^+ \, ,\qquad
p_1^- \simeq \ph^- \ll p_3^- \, .\label{mrk2}
\end{equation}
In this limit, the amplitudes are again dominated by gluon exchange in 
the $t$ channel, and factorize into in effective vertex for the
production of a jet and another for the production of a Higgs plus a jet.
For example, in the limit~(\ref{mrk2}) the amplitude for 
$q\ g\to q\ g\ H$ scattering~\cite{DelDuca:2001eu}
with the incoming gluon (quark) of momentum $p_2$ ($p_4$),
can be written as~\cite{us}
\bea
\lefteqn{ i\ \cM^{gq\to gHq}(p_2^{\nu_2}; p_1^{\nu_1}, H \sep 
p_3^{\nu_3}; p_4^{-\nu_3}) } \nn\\ &=& 2 s \left[i g\, f^{a_2a_1c}\, 
C^{g; {\sss H} g}(p_2^{\nu_2}; p_1^{\nu_1},\ph)\right] 
{1\over t} \left[i g\, T^c_{a_3 \bar a_4}\, 
C^{\bar q;q}(p_4^{-\nu_3};p_3^{\nu_3}) \right]\, ,\label{HqgqgHE}
\eea
where $C^{g; {\sss H} g}(p_2^{\nu_2}; p_1^{\nu_1},\ph)$ is the
effective vertex for the production of a Higgs boson and a gluon jet,
$g^* g\to g H$. It has two independent helicity configurations, which
we can take to be $C^{g; {\sss H} g}(p_2^-; p_1^+,\ph)$ and
$C^{g; {\sss H} g}(p_2^+; p_1^+,\ph)$~\cite{us}. 
High-energy factorization also implies that the amplitude for
$g\ g\to g\ g\ H$ scattering can be put in the form~(\ref{HqgqgHE}),
up to replacing the incoming quark with a gluon 
via the substitution~(\ref{qlrag}). Likewise,
the amplitude for $q\ Q\to q\ Q\ H$ scattering can be written as
\bea
\lefteqn{i\ \cM^{q Q\to q H Q}(p_2^{-\nu_1};p_1^{\nu_1}, \ph \sep 
p_3^{\nu_3}; p_4^{-\nu_3}) } \nonumber\\ &=& 
2s \left[ g\, T^c_{a_1 \bar a_2}\, 
C^{\bar q; {\sss H} q}(p_2^{-\nu_1};p_1^{\nu_1},\ph)\right] {1\over t}
\left[ g\, T^{c'}_{a_3 \bar a_4}\, 
C^{\bar q;q}(p_4^{-\nu_3};p_3^{\nu_3})\right]\, ,\label{HqqqqHE2}
\eea
where $C^{\bar q; {\sss H} q}(p_2^{-\nu_1};p_1^{\nu_1},\ph)$ is the 
effective vertex for the production of a Higgs and a quark jet,
$g^* q\to q H$. There is only one independent helicity configuration, which
we can take to be $C^{\bar q; {\sss H} q}(p_2^-;p_1^+,\ph)$, and its
expression
is given in Ref.~\cite{us}, where an analysis of the limit (\ref{mrk2})
with the kinematic parameters of \fig{fig:dydistra} can also be found.

In conclusion, we have considered $H + 2$ jet
production via gluon fusion, when either
one of the Higgs-jet or the dijet invariant masses become much larger
than the typical momentum transfers in the scattering. These limits
also occur naturally in Higgs production via WBF.
We have shown that
we can write the scattering amplitudes in accordance to high-energy
factorization, Eqns.~(\ref{HqqqqHE}), (\ref{HqgqgHE}) and (\ref{HqqqqHE2}).
The corresponding effective vertices, whose squares are the impact factors, 
can be found in Ref.~\cite{us}.
}

{
\catcode`@=11
\def\citer{\@ifnextchar
[{\@tempswatrue\@citexr}{\@tempswafalse\@citexr[]}}

\def\@citexr[#1]#2{\if@filesw\immediate\write\@auxout{\string\citation{#2}}\fi
  \def\@citea{}\@cite{\@for\@citeb:=#2\do
    {\@citea\def\@citea{--\penalty\@m}\@ifundefined
       {b@\@citeb}{{\bf ?}\@warning
       {Citation `\@citeb' on page \thepage \space undefined}}%
\hbox{\csname b@\@citeb\endcsname}}}{#1}}
\catcode`@=12

\newcommand{\nn}{\noindent}
\newcommand{\ra}{\rightarrow}
\newcommand{\s}{\\ \vspace*{-3mm}}
\newcommand{\gae}{\stackrel{\scriptscriptstyle>}{\scriptscriptstyle\sim}}
\newcommand{\beq}{\begin{eqnarray}}
\newcommand{\eeq}{\end{eqnarray}}
\newcommand{\bq}{\begin{equation}}
\newcommand{\eq}{\end{equation}}
\newcommand{\be}{\begin{equation}}
\newcommand{\ee}{\end{equation}}
\newcommand{\sla}[1]{/\!\!\!#1}



\input paperdef1

\newcommand{\psfragtextscale}{0.75}
\psfrag{Mh0}[][][\psfragtextscale]{$m_h [\mathrm{GeV}]$}
\psfrag{MA0}[][][\psfragtextscale]{$M_A [\mathrm{GeV}]$}
\psfrag{TB}[][][\psfragtextscale]{$\tan \beta$}
\psfrag{5x-1}[][][\psfragtextscale]{0.5}
\psfrag{1x0}[][][\psfragtextscale]{1}
\psfrag{5x0}[][][\psfragtextscale]{5}
\psfrag{1x1}[][][\psfragtextscale]{10}
\psfrag{5x1}[][][\psfragtextscale]{50}
\psfrag{0}[][][\psfragtextscale]{0}
\psfrag{1}[][][\psfragtextscale]{1}
\psfrag{2}[][][\psfragtextscale]{2}
\psfrag{3}[][][\psfragtextscale]{3}
\psfrag{4}[][][\psfragtextscale]{4}
\psfrag{5}[][][\psfragtextscale]{5}
\psfrag{10}[][][\psfragtextscale]{10}
\psfrag{20}[][][\psfragtextscale]{20}
\psfrag{30}[][][\psfragtextscale]{30}
\psfrag{40}[][][\psfragtextscale]{40}
\psfrag{50}[][][\psfragtextscale]{50}
\psfrag{60}[][][\psfragtextscale]{60}
\psfrag{70}[][][\psfragtextscale]{70}
\psfrag{80}[][][\psfragtextscale]{80}
\psfrag{90}[][][\psfragtextscale]{90}
\psfrag{100}[][][\psfragtextscale]{100}
\psfrag{110}[][][\psfragtextscale]{110}
\psfrag{120}[][][\psfragtextscale]{120}
\psfrag{130}[][][\psfragtextscale]{130}
\psfrag{140}[][][\psfragtextscale]{140}
\psfrag{150}[][][\psfragtextscale]{150}
\psfrag{160}[][][\psfragtextscale]{160}
\psfrag{170}[][][\psfragtextscale]{170}
\psfrag{180}[][][\psfragtextscale]{180}
\psfrag{190}[][][\psfragtextscale]{190}
\psfrag{200}[][][\psfragtextscale]{200}
\psfrag{300}[][][\psfragtextscale]{300}
\psfrag{400}[][][\psfragtextscale]{400}
\psfrag{500}[][][\psfragtextscale]{500}
\psfrag{600}[][][\psfragtextscale]{600}
\psfrag{700}[][][\psfragtextscale]{700}
\psfrag{800}[][][\psfragtextscale]{800}
\psfrag{900}[][][\psfragtextscale]{900}
\psfrag{1000}[][][\psfragtextscale]{1000}
\psfrag{1500}[][][\psfragtextscale]{1500}
\psfrag{-1}[][][\psfragtextscale]{-1}
\psfrag{-2}[][][\psfragtextscale]{-2}
\psfrag{-3}[][][\psfragtextscale]{-3}
\psfrag{-4}[][][\psfragtextscale]{-4}
\psfrag{-5}[][][\psfragtextscale]{-5}
\psfrag{-10}[][][\psfragtextscale]{-10}
\psfrag{-20}[][][\psfragtextscale]{-20}
\psfrag{-30}[][][\psfragtextscale]{-30}
\psfrag{-40}[][][\psfragtextscale]{-40}
\psfrag{-50}[][][\psfragtextscale]{-50}
\psfrag{-60}[][][\psfragtextscale]{-60}
\psfrag{-70}[][][\psfragtextscale]{-70}
\psfrag{-80}[][][\psfragtextscale]{-80}
\psfrag{-90}[][][\psfragtextscale]{-90}
\psfrag{-100}[][][\psfragtextscale]{-100}
\psfrag{-110}[][][\psfragtextscale]{-110}
\psfrag{-120}[][][\psfragtextscale]{-120}
\psfrag{-130}[][][\psfragtextscale]{-130}
\psfrag{-140}[][][\psfragtextscale]{-140}
\psfrag{-150}[][][\psfragtextscale]{-150}
\psfrag{-160}[][][\psfragtextscale]{-160}
\psfrag{-170}[][][\psfragtextscale]{-170}
\psfrag{-180}[][][\psfragtextscale]{-180}
\psfrag{-190}[][][\psfragtextscale]{-190}
\psfrag{-200}[][][\psfragtextscale]{-200}
\psfrag{-300}[][][\psfragtextscale]{-300}
\psfrag{-400}[][][\psfragtextscale]{-400}
\psfrag{-500}[][][\psfragtextscale]{-500}
\psfrag{-600}[][][\psfragtextscale]{-600}
\psfrag{-700}[][][\psfragtextscale]{-700}
\psfrag{-800}[][][\psfragtextscale]{-800}
\psfrag{-900}[][][\psfragtextscale]{-900}
\psfrag{-1000}[][][\psfragtextscale]{-1000}
\psfrag{-1500}[][][\psfragtextscale]{-1500}


\section[]{FeynHiggs1.2: Hybrid \msbarbf /on-shell Renormalization
 for the MSSM Higgs%
\footnote{\it M.\,Frank, S.\,Heinemeyer, W.\,Hollik and G.\,Weiglein}}


\subsection{Introduction}


In this section we present an updated version of the Fortran code
\fh~\cite{feynhiggs} that evaluates the neutral $\cp$-even Higgs
sector masses and mixing angles~\cite{mhiggsletter,mhiggslong}. It
differs from the previous version as presented in 
\citere{feynhiggs} 
by a modification of the  renormalization scheme 
concerning the treatment of subleading terms at the \onel\
level; the two-loop corrections, for which the leading contributions
of \order{\alt\als} and \order{\alt^2} are implemented,
are not affected.
In particular, an \msbar\ 
renormalization for $\tb$ and the field renormalization constants has
been used (where the \msbar\ quantities are evaluated 
at the scale $\mt$). The renormalization in the new version
of \fh\ does no longer involve the derivative 
of the $A$~boson self-energy and the $AZ$~mixing self-energy. This
leads to a more stable behavior around 
thresholds, e.g.\ at $\MA \approx 2\,\mt$, and avoids unphysically large
contributions in certain regions of the MSSM parameter space.
Thus, the new renormalization scheme stabilizes the prediction of the
masses and mixing angles in the $\cp$-even Higgs sector of the MSSM.


\subsection{Renormalization schemes}

At the tree-level, The MSSM Higgs boson masses $\mh$ and
$\mH$ can be evaluated in terms of the SM gauge
couplings and two MSSM parameters, conventionally chosen as
$\MA$ and $\tb$. 
Beyond lowest order, the Feynman-diagrammatic (FD) approach allows to
obtain in principle the most precise evaluation of the neutral
$\cp$-even Higgs boson sector, since in this way
the effect of different mass scales of the supersymmetric
particles and of 
the external momentum can consistently be included. 
The masses of the two $\cp$-even Higgs bosons are obtained in this
approach by
determining the poles of the $h-H$-propagator
matrix, which is equivalent to solving the equation
\begin{equation}
[ \; q^2 - m_{h,{\rm tree}}^2 + \hat\Sigma_{h}(q^2) \; ]
[ \; q^2 - m_{H,{\rm tree}}^2 + \hat\Sigma_{H}(q^2) \; ] -
[ \; \hat\Sigma_{hH}(q^2) \; ]^2 = 0 ,
\label{eq:propmatrix}
\end{equation}
where $\hat\Sigma_s, s = h, H, hH$, denote the renormalized Higgs
boson self-energies. For the renormalization within the FD approach
usually the on-shell scheme is
applied~\cite{mhiggslong}. This means in particular that all the masses
in the FD result are the physical ones, i.e.\ they correspond to
physical observables. Since \refeq{eq:propmatrix} is solved iteratively,
the result for $m_h$ and $m_H$ contains a dependence on the field
renormalization constants of $h$ and $H$, which is 
formally of higher order. Accordingly, there is some freedom in choosing 
appropriate renormalization conditions for fixing the field
renormalization constants (this can also be interpreted as affecting the
renormalization of $\tb$). Different renormalization conditions have
been considered, e.g.\ ($\hSip$ denotes the derivative with respect to
the squared momentum):
\begin{enumerate}
\item
on-shell renormalization for $\hSi_Z, \hSi_A,
\hSip_A, \hSi_{AZ}$, and  
$\de v_1/v_1 = \de v_2/v_2$~\cite{mhiggs1lfull}
\item
on-shell renormalization for 
$\hSi_Z, \hSi_A, \hSi_{AZ}$, and 
$\de v_i = \de v_{i, {\rm div}}, i = 1,2$~\cite{mhiggs1lfullb}
\item
on-shell renormalization for $\hSi_Z, \hSi_A$~\cite{mhiggs1lfull},
\msbar\ renormalization for $\de Z_h, \de Z_H$, $\tb$~\cite{mhiggsrenorm}.
\end{enumerate}
The previous version of \fh\ is based on renormalization~1, involving
the derivative of the $A$~boson self-energy. The new version of \fh,
see {\tt www.feynhiggs.de}, is based on renormalization~3 (a detailed 
discussion can be found in \citere{mhiggsrenorm}).


\subsection{Numerical comparison}

In this section we numerically compare the output of 
the previous version (based on renormalization~1) and the
new version (based on renormalization~3) of \fh.
We also show results for the recently obtained
non-logarithmic \order{\alt^2} corrections~\cite{mhalphatsq,maulpaul}
that are also included in the new version of \fh.
The comparison is performed for the parameters of the three LEP 
benchmark scenarios~\cite{benchmark}. In this way, the effect of the
new renormalization and the non-logarithmic \order{\alt^2} corrections
on the analysis of the LEP Higgs-boson searches can easily be read off.

In \reffis{fig:mhmax}--\ref{fig:largemu} we show the results in the
``$\mhmax$'', ``no-mixing'' and ``large $\mu$'' scenario as a function
of $\MA$ (left column) and of $\tb$ (right column) for two values of
$\tb$ ($\tb = 3, 50$) and $\MA$ ($\MA = 100, 1000 \gev$ for the $\mhmax$
and the no-mixing scenario, $\MA = 100, 400 \gev$ for the large~$\mu$
scenario), respectively. The solid lines correspond to the new 
result while the dashed lines show the old results. The dotted lines
correspond to the new result including the non-logarithmic
\order{\alt^2} contributions. Concerning the new renormalization
scheme, in the $\mhmax$
(\reffi{fig:mhmax}) and the no-mixing scenario (\reffi{fig:nomix}) the
new result is larger by $\approx$1--2~GeV for 
not too small $\MA$ and $\tb$. For small $\tb$ and large $\MA$ the
enhancement can be 
even larger. In the large $\mu$ scenario (\reffi{fig:largemu}) the largest
deviations appear for small $\tb$ for both large and small $\MA$. 
While the previous prescription for the field renormalization constants 
leads to
unphysically large threshold effects in some regions of the parameter
space, which arise from the $AZ$ mixing self-energy and the 
derivative of the $A$~boson
self-energy, no threshold kinks are visible for the result based on the
new renormalization.
The shift in $\mh$ of $\approx$1--2~GeV related to the modification of the
renormalization prescription lies in the range of the anticipated
theoretical uncertainty from unknown non-leading electroweak two-loop
corrections~\cite{mhiggsstatus}. 
The new \order{\alt^2} corrections can further increase $\mh$
by up to $\approx$3~GeV for large $\Stop$~mixing (a detailed analysis
will be presented elsewhere~\cite{mhalphatsqAnal}).

\smallskip
The new version of \fh\ can be obtained from {\tt www.feynhiggs.de} .


%

\begin{figure}[htb!]
\begin{center}
\mbox{
\epsfig{figure=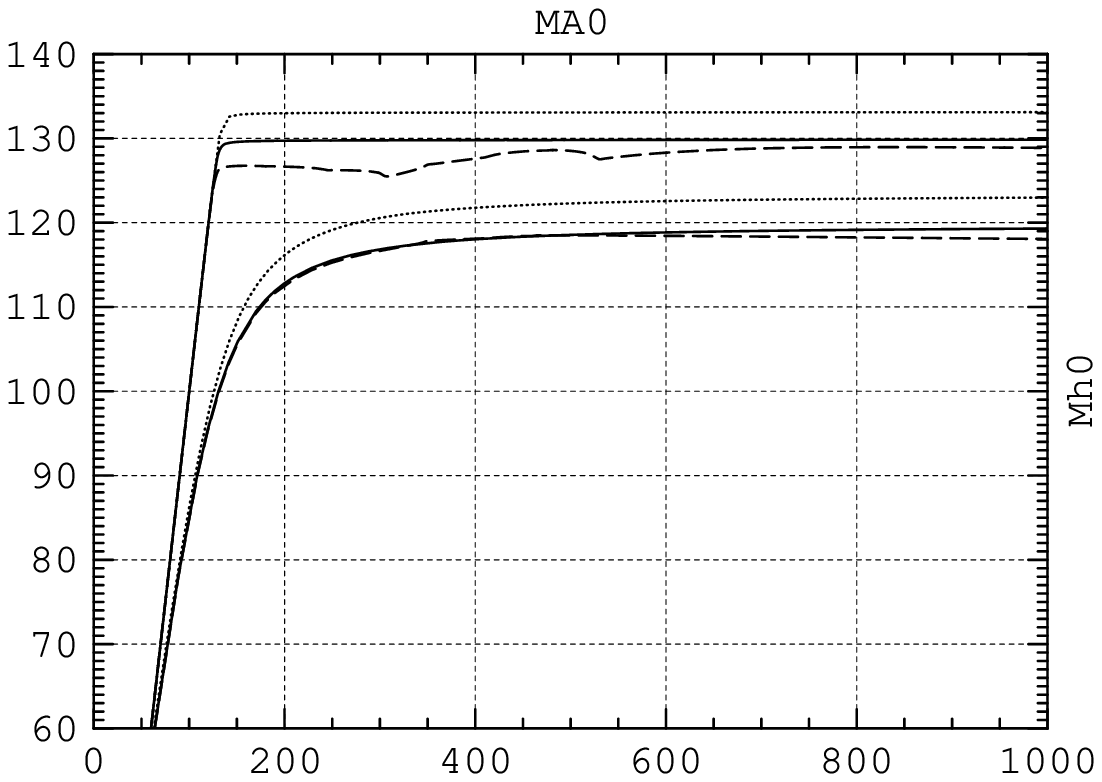,width=7cm,height=5.5cm} 
\hspace{1em}
\epsfig{figure=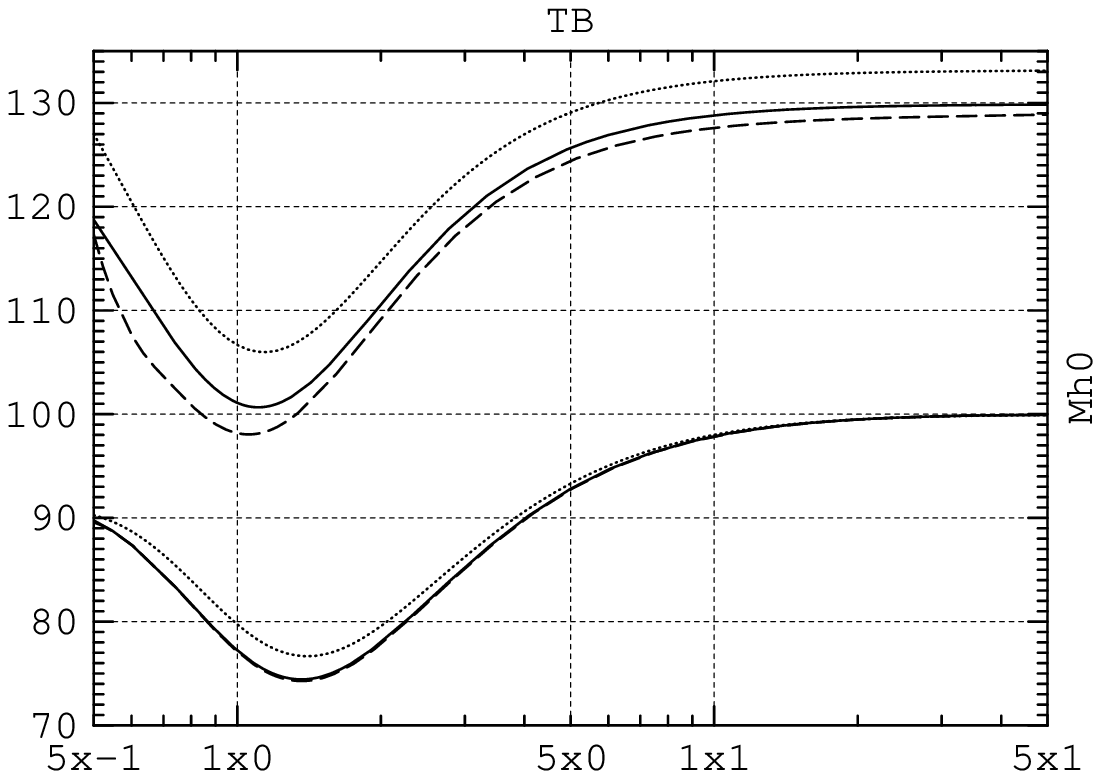,width=7cm,height=5.5cm}}
\end{center}
\vspace{-2.1em}
\caption[]{The new renormalization (3, solid) and the old scheme (1,
dashed) are compared in the $\mhmax$ scenario. The dotted line shows
the inclusion of the non-logarithmic \order{\alt^2} corrections. The 
lower curves are for 
$\tb = 3$ (left plot) or $\MA = 100 \gev$ (right). The upper curves are
for $\tb = 50$ (left) or $\MA = 1000 \gev$ (right).}
\label{fig:mhmax}
\vspace{-0.3cm}
\end{figure}
\begin{figure}[htb!]
\begin{center}
\mbox{
\epsfig{figure=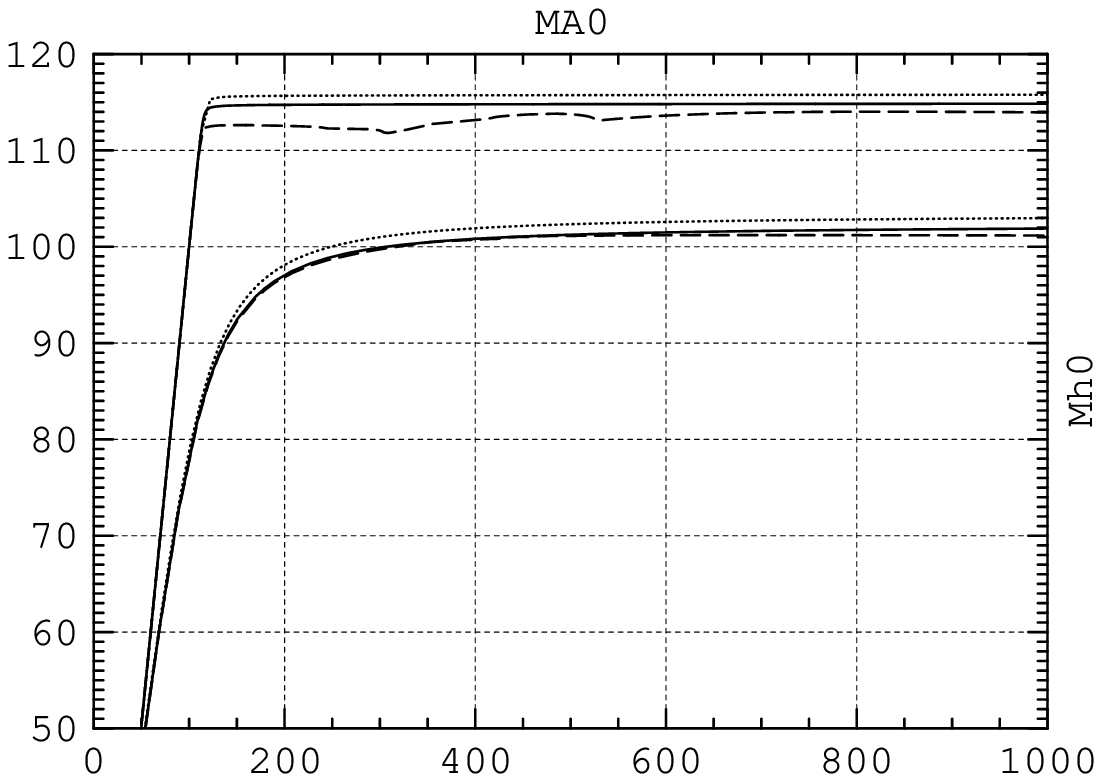,width=7cm,height=5.5cm} 
\hspace{1em}
\epsfig{figure=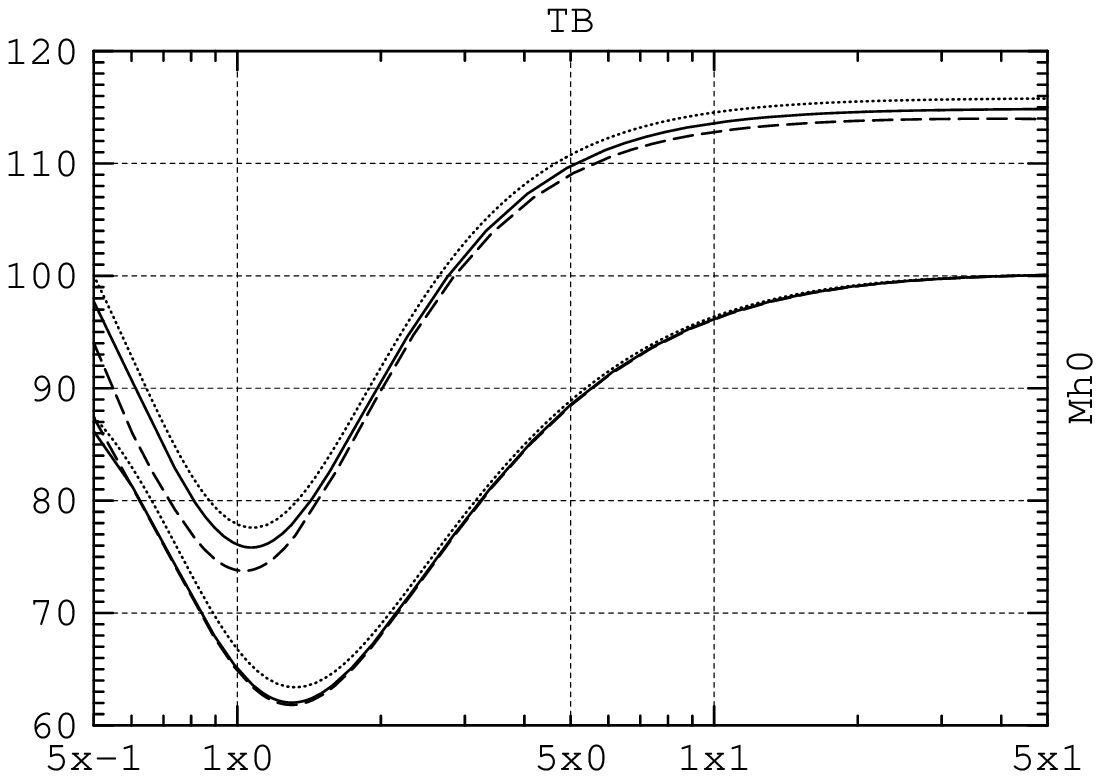,width=7cm,height=5.5cm}}
\end{center}
\vspace{-2.1em}
\caption[]{The new renormalization (3, solid) and the old scheme (1,
dashed) are compared in the no-mixing scenario. The dotted line shows
the inclusion of the non-logarithmic \order{\alt^2} corrections. The lower
curves are for 
$\tb = 3$ (left plot) or $\MA = 100 \gev$ (right). The upper curves are
for $\tb = 50$ (left) or $\MA = 1000 \gev$ (right).}
\label{fig:nomix}
\vspace{-0.3cm}
\end{figure}
\begin{figure}[htb!]
\begin{center}
\mbox{
\epsfig{figure=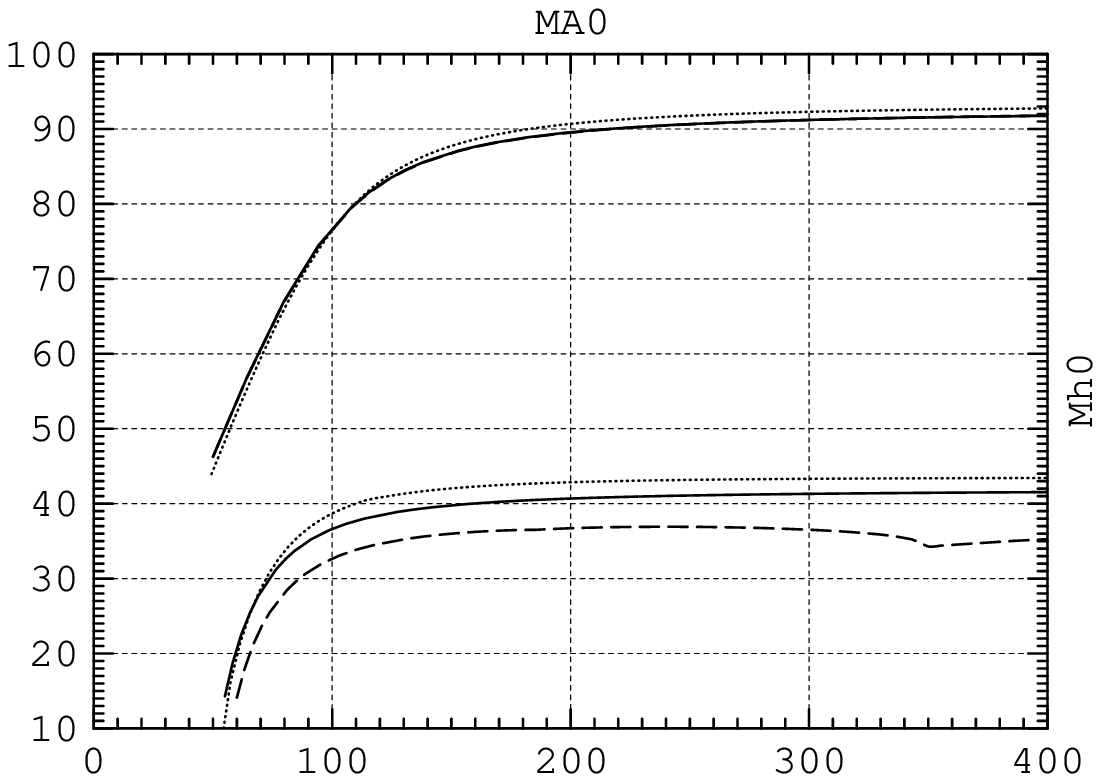,width=7cm,height=5.5cm} 
\hspace{1em}
\epsfig{figure=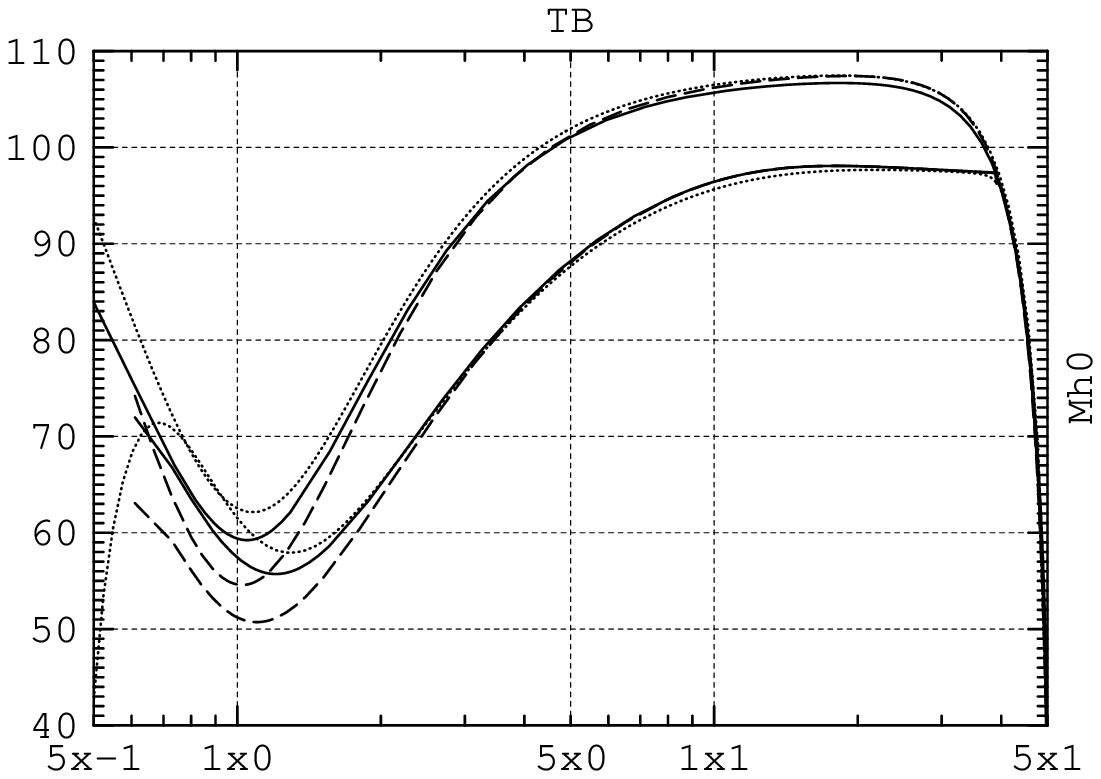,width=7cm,height=5.5cm}}
\end{center}
\vspace{-2.1em}
\caption[]{The new renormalization (3, solid) and the old scheme (1,
dashed) are compared in the large $\mu$ scenario. The dotted line shows
the inclusion of the non-logarithmic \order{\alt^2} corrections. The lower
curves are for 
$\tb = 50$ (left plot) or $\MA = 100 \gev$ (right). The upper curves are
for $\tb = 3$ (left) or $\MA = 400 \gev$ (right).}
\vspace{-0.6em}
\label{fig:largemu}
\vspace{-0.3cm}
\end{figure}
}

{
\catcode`@=11
\def\citer{\@ifnextchar
[{\@tempswatrue\@citexr}{\@tempswafalse\@citexr[]}}

%

\def\@citexr[#1]#2{\if@filesw\immediate\write\@auxout{\string\citation{#2}}\fi
  \def\@citea{}\@cite{\@for\@citeb:=#2\do
    {\@citea\def\@citea{--\penalty\@m}\@ifundefined
       {b@\@citeb}{{\bf ?}\@warning
       {Citation `\@citeb' on page \thepage \space undefined}}%
\hbox{\csname b@\@citeb\endcsname}}}{#1}}
\catcode`@=12

\newcommand{\nn}{\noindent}
\newcommand{\ra}{\rightarrow}
\newcommand{\s}{\\ \vspace*{-3mm}}
\newcommand{\gae}{\stackrel{\scriptscriptstyle>}{\scriptscriptstyle\sim}}
\newcommand{\beq}{\begin{eqnarray}}
\newcommand{\eeq}{\end{eqnarray}}
\newcommand{\bq}{\begin{equation}}
\newcommand{\eq}{\end{equation}}
\newcommand{\be}{\begin{equation}}
\newcommand{\ee}{\end{equation}}
\newcommand{\sla}[1]{/\!\!\!#1}



\input paperdef2


\section[]{Suggestions for MSSM Benchmark Scenarios
for Higgs Boson Searches at Hadron Colliders%
\footnote{\it M.\,Carena, S.\,Heinemeyer, C.E.M.\,Wagner and G.\,Weiglein}}


\subsection{Introduction and theoretical basis}

Within the MSSM the masses of the $\cp$-even neutral Higgs bosons are
calculable in terms of the other MSSM parameters. The lightest Higgs
boson has been of particular interest, since its mass, $\mh$, is
bounded from above according to $\mh \leq \MZ$ at the tree level.
The radiative corrections at \onel\ order have 
been supplemented in the last years with the leading \twol\ corrections, 
performed by renormalization group (RG) 
methods~\cite{mhiggsRG1}, by renormalization 
group improvement of the  one-loop effective potential 
calculation~\cite{mhiggsRG2a}, 
by two-loop effective  potential calculations~\cite{mhalphatsq,maulpaul},   
and in the Feynman-diagrammatic (FD)
approach~\cite{mhiggsletter,mhiggslong}. These calculations predict an
upper bound for $\mh$ of about $\mh \lsim 135 \gev$.%
\footnote{
This value holds for $\mt = 175 \gev$ and $\msusy = 1 \tev$. If $\mt$
is raised by $5 \gev$ then the $\mh$ limit is increased by about $5 \gev$; 
using $\msusy = 2 \tev$ increases the limit by about $2 \gev$.
}

After the termination of LEP, the Higgs boson search has now shifted to
the Tevatron and will later be continued to the LHC. Due to the large
number of 
free parameters, a complete scan of the MSSM parameter space is too
involved. Therefore at LEP the search has been performed in three
benchmark scenarios~\cite{benchmark}. Besides the $\mhmax$~scenario,
which has been used to obtain conservative bounds on
$\tb$~\cite{tbexcl}, and the no-mixing scenario, the
large-$\mu$~scenario had been designed to encourage the investigation
of flavor and decay-mode independent decay channels
(instead of focusing on the $\hbb$ channel). The investigation 
of these channels has lead to exclusion
bounds~\cite{mhLEP2001} that finally completely ruled out the
large-$\mu$~scenario.

The different environment at hadron colliders implies different Higgs
boson production channels and also different relevant decay channels
as compared to LEP. The main production modes at the Tevatron will be
$V^* \to V \phi$ ($V = W, Z, \phi = h, H, A$) 
and also $b\bar b \to b\bar b \phi$, while the
relevant decay modes will be $\phi \to b\bar b$ 
and $\phi \to \tau^+\tau^-$~\cite{Carena:2000yx}.
At the LHC, on the other hand, the most relevant process for a Higgs
boson with $\mh \le 135 \gev$ will be $gg \to h \to \ga\ga$,
supplemented by $t \bar t \to t \bar t h \to t \bar t b \bar b$.
In order to investigate these different modes, we propose new
benchmark scenarios for the Higgs boson searches at hadron colliders.
Contrary to the new ``SPS''~benchmark scenarios proposed in
\citere{sps} for general SUSY
searches, the scenarios proposed here are designed specifically to
study the MSSM Higgs sector without assuming any particular soft
SUSY-breaking scenario and 
taking into account constraints only from the Higgs boson sector~itself.


The tree-level value for $\mh$ within the MSSM is determined by $\tb$, 
the $\cp$-odd Higgs-boson mass $\MA$, and the $Z$-boson mass $\MZ$. 
Beyond the tree-level, the main correction to $\mh$ stems from the 
$t$--$\Stop$-sector, and for large values of $\tb$ also from the 
$b$--$\Sbot$-sector (see \citere{benchmark} for our notations.)
Accordingly, the most important parameters for the corrections to $\mh$
are $\mt$, $\msusy$ (in this work we assume that the soft
SUSY-breaking parameters for sfermions are equal:
$\msusy := \MstL = \MstR = \MsbL = \MsbR$), 
$\Xt$~($\equiv \At - \mu/\tb$), 
and $\Xb$~($\equiv \Ab - \mu\tb$) ($A_{t,b}$ are
the trilinear Higgs sfermion couplings, $\mu$ is the Higgs mixing
parameter.)
$\mh$ depends furthermore on the SU(2) gaugino mass
parameter, $M_2$ (the U(1) gaugino mass parameter is given by
$M_1 = 5/3\, \sw^2/\cw^2\, M_2$.) 
At the two-loop level also the gluino mass, $\mgl$, enters the
prediction for $\mh$.

It should be noted in this context that the FD result has been obtained
in the on-shell (OS) renormalization scheme (the corresponding Fortran
code, that has been used for the studies by the LEP collaborations, is
\fh~\cite{feynhiggs,mhiggsrenorm}), whereas the RG result has been 
calculated using the \msbar\ scheme; see \citere{bse} for details (the
corresponding Fortran code, also used by the LEP collaborations, is
\subh~\cite{mhiggsRG1,bse}). 
While the corresponding shift in the parameter $\msusy$ turns out to be 
relatively small in general, sizable differences can occur between the 
numerical values of 
$\Xt$ in the two schemes; see \citeres{mhiggslong,bse,maulpaul}. For this reason
we specify below different values for $\Xt$ within the two approaches.


\subsection{The benchmark scenarios}

In this section we define four benchmark scenarios suitable for the
MSSM Higgs boson search at hadron colliders\footnote{Here, we will
comment only on the phenomenology of the lightest $h$ boson. The couplings of 
the $H,A$ and $H^\pm$ bosons are also subject to important radiative 
corrections in the large tanbeta regime, see 
for instance ref.~\cite{heavy-couplings,bsganlo}. It 
is customary to define the searches for these 
particles in terms of their tree-level couplings. Since the corrections to these
couplings are strongly dependent on the value of the supersymmetry
breaking parameters, a precise interpretation of these search 
analyses within the MSSM will demand a knowledge of the characteristic
supersymmetry breaking parameters. In the case of the $h$ boson, 
see~\cite{benchmarkhad,deltamb2} for more details.}. 
In these scenarios the values of the $\Stop$~and $\Sbot$~sector as
well as the gaugino masses will be fixed, while $\tb$ and $\MA$ are
the parameters that are varied.%
\footnote{
Plots that show the behavior of different Higgs production and decay
channels in the four scenarios can be found at {\tt www.feynhiggs.de}~.
The numerical evaluation is based on
\citeres{hdecay,deltamb1,deltamb2,hff}. 
}%
~It has been checked that
the scenarios evade the LEP2 bounds~\cite{mhLEP2001} over a wide range
of the $\MA$-$\tb$-plane, where the variation should be chosen
according to: $ 0.5 \le \tb \le 50, \quad \MA \le 1000 \gev $.


\noindent {\it (i) The $\mhmax$ scenario}:
This scenario is kept as presented in \citere{benchmark},
since it allows for conservative $\tb$ exclusion
bounds~\cite{tbexcl}
(only the sign of $\mu$ is switched to a positive value.)
The parameters are chosen such that the maximum possible 
Higgs-boson mass as a function of $\tb$ is obtained
(for fixed $\msusy$, 
and $\MA$ set to its maximal value, $\MA = 1 \tev$).
The parameters are%
\footnote{Better agreement with $\br(b \to s \ga)$ constraints is obtained for
the other sign of $\Xt$ (called the ``constrained $\mhmax$''
scenario)~\cite{bsganlo}. However, this lowers the maximum $\mh$ values by 
$\sim 5 \gev$.}:
\BEA
&& \mt = 174.3 \gev, \quad \msusy = 1 \tev, \quad
\mu = 200 \gev, \quad M_2 = 200 \gev, \ \mgl = 0.8\,\msusy, \non \\
\label{mhmax}
&& \Xt^{\OS} = 2\, \msusy  \; \mbox{(FD calculation)}, \quad
   \Xt^{\MS} = \sqrt{6}\, \msusy \; \mbox{(RG calculation)} , \quad \Ab = \At~.
\EEA


\noindent {\it (ii) The no-mixing scenario}:
This benchmark scenario is the same as the $\mhmax$ scenario, but with
vanishing mixing in the $\Stop$~sector and with a higher SUSY mass
scale to avoid the LEP Higgs bounds:
\BEA
&& \mt = 174.3 \gev, \quad \msusy = 2 \tev, \quad
\mu = 200 \gev, \quad M_2 = 200 \gev, \ \  \mgl =0.8\msusy, \non \\
&& \Xt = 0  \; \mbox{(FD/RG calculation)}, \quad
   \Ab = \At~.
\label{nomix}
\EEA


\noindent {\it (iii) The gluophobic Higgs scenario}:
In this scenario the main production cross section for the light Higgs
boson at 
the LHC, $gg \to h$, is strongly suppressed (see \citere{ggsuppr}).
The parameters are:
\BEA
&& \mt = 174.3 \gev, \  \msusy = 350 \gev, \ 
\mu = 300 \gev, \ M_2 = 300 \gev, \ \mgl = 500 \gev , \non \\
\label{ggsup}
&& \Xt^{\OS} = -750 \gev \; \mbox{(FD calculation)}, \quad
   \Xt^{\MS} = -770 \gev \; \mbox{(RG calculation)}, \quad 
\Ab = \At~.
\EEA


\noindent {\it (iv) The small $\aeff$ scenario}:
Besides the channel $gg \to h \to \ga\ga$ at the LHC, the other
channels for light Higgs 
searches at the Tevatron and at the LHC mostly rely on the decays $\hbb$
and $\htautau$. In comparison to the Standard Model, both $hf\bar f$ 
couplings have an additional factor of $\sin\aeff/\cos\be$, where
$\aeff$ is the mixing angle of the neutral $\cp$-even Higgs sector,
including radiative corrections (see e.g. \citeres{deltamb2,hff}). 
If $\aeff$ is 
small, these two decay channels can be heavily suppressed ($\hbb$ can
receive also large corrections from
$\Sbot$-$\gl$~loops~\cite{deltamb1,deltamb2}). 
This case is realized for large $\tb$ and not too large $\MA$ (in a
similar way as in the large-$\mu$ scenario~\cite{benchmark}) for the
following parameters: 
\BEA
&& \mt = 174.3 \gev, \  \msusy = 800 \gev, \ 
\mu = 2.5 \, \msusy, \ M_2 = 500 \gev, \  \mgl = 500 \gev \non \\
\label{smallaeff}
&& \Xt^{\OS} = -1100 \gev \; \mbox{(FD calculation)}, \ 
   X_t^{\MS} = -1200 \gev \; \mbox{(RG calculation)}, \quad \Ab = \At.~ 
\EEA


\subsection{Conclusions}

We have presented four benchmark scenarios for the MSSM Higgs boson
search at hadron colliders, evading the exclusion bounds obtained at
LEP2. These scenarios exemplify 
different features of the MSSM parameter space, such as large $\mh$
values and significant $gg \to h$ or $\hbb$, $\htautau$
suppression. 
In analyzing the new benchmark scenarios, it will be helpful to make use
of the complementarity of different channels accessible at the
Tevatron and the LHC (see e.g.\ \citere{deltamb2} for details).


%
}

{
\newcommand{\beqn}{\begin{eqnarray}}
\newcommand{\eeqn}{\end{eqnarray}}
\newcommand{\ra}{\rightarrow}

\newcommand{\np}{Nucl.\,Phys.\,}
\newcommand{\pl}{Phys.\,Lett.\,}
\newcommand{\pr}{Phys.\,Rev.\,}
\newcommand{\prl}{Phys.\,Rev.\,Lett.\,}
\newcommand{\prep}{Phys.\,Rep.\,}
\newcommand{\nuclinst}{{\em Nucl.\ Instrum.\ Meth.\ }}
\newcommand{\annp}{{\em Ann.\ Phys.\ }}
\newcommand{\intjmp}{{\em Int.\ J.\ of Mod.\  Phys.\ }}


\newcommand{\mw}{M_{W}}
\newcommand{\mww}{M_{W}^{2}}
\newcommand{\mwmw}{M_{W}^{2}}

\newcommand{\mz}{M_{Z}}
\newcommand{\mzz}{M_{Z}^{2}}

\newcommand{\cw}{\cos\theta_W}
\newcommand{\sw}{\sin\theta_W}
\newcommand{\tw}{\tan\theta_W}
\def\cww{\cos^2\theta_W}
\def\sww{\sin^2\theta_W}
\def\tww{\tan^2\theta_W}

\def\noi{\noindent}
\def\nn{\noindent}

\def\sinb{\sin\beta}
\def\cosb{\cos\beta}
\def\sinbb{\sin (2\beta)}
\def\cosbb{\cos (2 \beta)}
\def\tgb{\tan \beta}
\def\tgbt{$\tan \beta\;\;$}
\def\tgbsq{\tan^2 \beta}
\def\sel{\tilde{e}_L}
\def\ser{\tilde{e}_R}
\def\msel{m_{\sel}}
\def\mser{m_{\ser}}


\def\neuto{\tilde{\chi}_1^0}
\def\mneuto{m_{\tilde{\chi}_1^0}}

\def\mh{m_h}

\section[]{The invisible SUSY Higgs and Dark Matter%
\footnote{\it G.\,B\'elanger, F.\,Boudjema, A.\,Cottrant,
R.M.\,Godbole, A.\,Pukhov and A.\,Semenov}}




\subsection{Introduction}
 Current limits\cite{leplimit2001} on both the
Higgs and the neutralino in a general SUSY model are such that it
is kinematically possible for the light Higgs to decay into the lightest
neutralino. If the decay rate is substantial the Higgs will be
mainly invisible, while its usual branching ratios will be
dramatically reduced preventing a detection in the much studied
channels at the LHC and the Tevatron.
Some theoretical
studies\cite{Kane-invisible-lhc,DP-invisible-lhc,Zeppenfeld-h-invisible}
have addressed the issue of how to hunt an invisibly decaying
Higgs at a hadronic machine.
For the LHC it has
been suggested to use $WH/ZH$ production , $t\bar t
h$\cite{DP-invisible-lhc} or more recently the 
the $W$ fusion process \cite{Zeppenfeld-h-invisible}. 
The results for the latter are quite promising
since for a luminosity of $100fb^{-1}$ a branching ratio into
invisibles as low as $5\%$ is enough for Higgs discovery. 
 The aim of the
present report  is to summarize our findings on the size
of the  branching ratio of the Higgs into
neutralinos, taking into account the latest data from colliders as well
as from cosmology. 
\subsection{MSSM parameters and $h\ra \chi\chi$}

For a substantial branching fraction of the  Higgs into invisible to occur 
 one needs both enough phase space for the decay 
  as well as a large enough coupling of the Higgs to neutralinos. 
Considering that the present experimental 
and theoretical limits on  the lightest MSSM Higgs implies
that its mass lies in the interval
$113-135$ GeV, the maximum LSP mass must be below $55-65$GeV.
In models with gaugino unification where $M_1 \approx M_2/2$ , the 
lower limit on the chargino mass (which depends essentially on $M_2$ and $\mu$)
 turns into a lower limit on the neutralino mass, leaving only a small window for the
Higgs into neutralinos. In fact in this type
of models we found that the branching is never above
20\%\cite{nous_hinvisible_lhc} (see also ref.\cite{hlspBR}). 
For this reason we will relax the relation between $M_1$ and
$M_2$ and consider these as independent parameters.
In order that  the coupling of the LSP to the Higgs be large, 
 it can be shown that the LSP has to be a mixture of gaugino and Higgsino
 \cite{nous_hinvisible_lhc}.
  However a light LSP, which corresponds to $M_1$ small,
 is mostly a Bino. To have a non negligible Higgsino component
 implies that $\mu$ should be small as well.
However $\mu$ is bounded below by  the chargino mass constraint.
From these arguments, we can already expect that if the Higgs invisible
decay is large then the 
chargino and next to lightest neutralino should not be far above the 
present LEP limit.
One also finds \cite{nous_hinvisible_lhc} that positive $\mu$ values lead to
larger couplings.
Large $\tan\beta$ values also lead to  large Higgs mass  and more phase space
for the invisible decays, however  the LSP mass increases even
faster with $\tan\beta$, and  we found that their coupling to the 
Higgs get smaller with increasing $\tan\beta$ \cite{nous_hinvisible_lhc}.
Therefore the largest effect for the Higgs occurs for moderate $\tan\beta$
and we will consider $\tan\beta=5$.

We take a model with a 
common scalar mass $m_0$ (defined at the GUT scale) for the SUSY
breaking sfermion mass terms of both left and right sleptons of
all three generations. As for the gaugino masses, we take $M_1=r
M_2$ at the weak scale.  For $r<1/3$ or
so, this scheme leads to almost no running of the right slepton
mass, since the contribution from the running is of order $M_1^2$,
while left sleptons have an added $M_2^2$ contribution and would
be ``much heavier". Indeed, neglecting Yukawa couplings one has
\beqn
\label{m0running} \mser^2&=&m_0^2\;+\; .88 \;r^2 M_2^2\;-\;\sww
\mzz \cosbb \nonumber
\\ \msel^2&=&m_0^2\;+ (0.72+.22\; r^2) M_2^2\; -\;(.5-\sww)\mzz\cosbb 
\;\;\;\;\;\;
\eeqn
Even with a common scalar mass
squarks are much heavier than sleptons, since they receive a large 
contribution from  the SU(3) gaugino mass.
For simplicity, we then assume all squarks to be heavy (1TeV).
In any case heavy squarks especially stops would be required in order to get
a heavy enough light Higgs.   
 Of course, to 
allow for a low $\mu$ in this scenario one needs to appropriately
choose the soft SUSY Higgs scalar masses at high scale. It is
important to stress that the kind of models we investigate in this
report are quite plausible. The GUT-scale relation which equates
all the gaugino masses at high scale need not be valid in a more
general scheme of SUSY breaking. 
SUGRA models with general kinetic terms\cite{nmSUGRA,nonuni-24}, superstring models
with moduli-dominated or with a mixture of
moduli and dilaton fields, as well as anomaly-mediated SUSY breaking
mechanisms, all lead to non-universality of the gaugino
masses\cite{nonuniversal-strings}.
\subsection{Constraints}

Our scenario requires as large a Higgs mass as possible without
a too large value for  \tgbt. We will then only consider the MSSM in the
decoupling limit with $M_A \sim 1$TeV and choose large enough stop
masses ($m_{\tilde{t}}=$1TeV) and large mixing ($A_t=2.4$TeV).
With these 
parameters we have $m_h=125$GeV  for $\tgb=5$
and we are never in conflict with the lower limit on the Higgs mass
$m_h>113$GeV.

The limits on $M_1,M_2,\mu$, the key ingredients for this
analysis, are set from the chargino mass limit at LEP2,
$m_{\chi_1^\pm}>103$GeV\cite{leplimit2001}. This bound can be
slightly relaxed depending on $\tgb$ and the sneutrino mass,
however we prefer to take the strongest constraint so that our
results are more robust.
In addition to this, one must include the limits
from LEP2 on pair production of neutralinos,
as well as the limit on the invisible width of the Z.
For the parameters we have studied these two constraints are
weaker than the chargino mass constraint.
We will also take $m_{\tilde{l}}>96$GeV, for all sleptons $\tilde
l$, even though the limit on the lightest stau is slightly
lower\cite{leplimit2001}.

Apart from the chargino mass limit, the most important constraint
comes from the relic density of the LSP. In the models
we are considering the LSP is {\em mainly} (but not totally) a
bino. Since it is rather light  the main annihilation channels are into
the light fermions. The largest contributions are
from processes involving ``right-handed" sleptons since they 
 have the largest hypercharge. In this case the relic
density may be approximated as $\Omega h^2 \sim 10^{-3}
m_{\tilde{l}_R}^4/ \mneuto^2$ (all masses in GeV) which imposes
a  strong constraint on $m_{\tilde{l}_R}$.
However this approximation does not hold if
the neutralino mass is such that 
 annihilation through the $Z$ pole, $\neuto \neuto \ra Z$,  
 occurs. In this case the contribution of this channel alone is enough
 to bring the relic density in the relevant range irrespective of the slepton 
 mass.

We use a new code\cite{OmegaComphep} for the calculation of
the relic density that tackles all $s$-channels poles, threshold
effects and includes all co-annihilations channels. The program
extracts all {\em exact} matrix elements 
from {\tt CompHEP} \cite{CompHEP} and is linked to {\tt FeynHiggs}
\cite{feynhiggs,mhiggsletter}
 for the Higgs mass. Radiative corrections to Higgs partial widths are
extracted from HDECAY \cite{hdecay}.

Fig.~1a shows the allowed parameter space in
the $M_2, \mu$ plane with $\tgb=5$ and $M_1=M_2/5$ for 
a light slepton, $m_0=100$GeV.
  The chargino mass limit
from LEP2 is delimited by a line. It does not depend on $m_0$. The
direct LEP2 limits, expectedly, cut on the lowest $\mu,M_2$
region. This is in contrast to the relic density requirement which
depends sensitively on $m_0$. We delineate three regions set by the
relic density: a) the overclosure region $ \Omega h^2>.3$ which we
consider as being definitely ruled out\cite{omega}, b) $ .1 <\Omega h^2<.3$
which is the preferred region and c) $ \Omega h^2<.1$ where one 
needs other form of Dark Matter than the SUSY Dark Matter
considered here. As $m_0$ increases the
allowed region for the relic density shrinks. However there
always remain 
allowed regions that  correspond essentially to the pole annihilation
$\neuto \neuto \ra Z$\cite{nous_hinvisible_lhc}. 
For $M_1=M_2/10$ (Fig.~1b),
the effect of the Z-pole
would be seen only at much larger values of $M_2$.
For $m_0=94$GeV, the relic density constraint leaves a sizeable
allowed region, however as soon as $m_0$ increases the region allowed
is restricted to the regions of parameter space where $m_\chi\approx M_z/2$.

In view of the latest theoretical calculations of the 
muon anomalous magnetic moment, showing consistency between the
experimental limit and the SM within 
$1.6\sigma$\cite{knecht}, all constraints
that were previously thought to play an important
role, (in particular the preference for light smuons) 
disappear in the range of parameters considered here.
Finally,
we note that $b\ra s\gamma$ is irrelevant since the squarks and
gluinos are assumed heavy and that we are choosing $\mu>0$ anyway.

\subsection{Results}
The branching ratio into invisible due to neutralinos will be
denoted by $B_{\chi \chi}$. The opening up of this channel will
not have any effect on any of the Higgs production mechanisms.
This is in contrast to other SUSY effects on the production and
decay of the Higgs, like those due to a light stop \cite{nous_Rggstophiggs_lhc}.
 Thus the Higgs discovery
significances of the different channels at the LHC (and the
Tevatron) are only affected by the reduction in the branching
ratio into $b\bar b$ and $\gamma \gamma$. We define $R_{bb}$ ($R_{\gamma
\gamma}$)as
the reduction factor of the branching ratio of $h \ra b \bar b$ ($h \ra \gamma
\gamma$)
due to invisible compared to the same branching ratio of a
standard model Higgs with the same Higgs mass.
Since in the absence of light neutralinos
the width of the Higgs is dominated by that into $b\bar b$, one
has roughly $R_{b b} \sim R_{\gamma \gamma} \sim 1-B_{\chi \chi}$. 
This is well supported by our full analysis and therefore we will
only show the behaviour of the branching into invisible.

\begin{figure*}[htb]
\vspace*{-0.2cm}
\begin{center}
\mbox{\includegraphics[height=9cm,width=.5\textwidth]{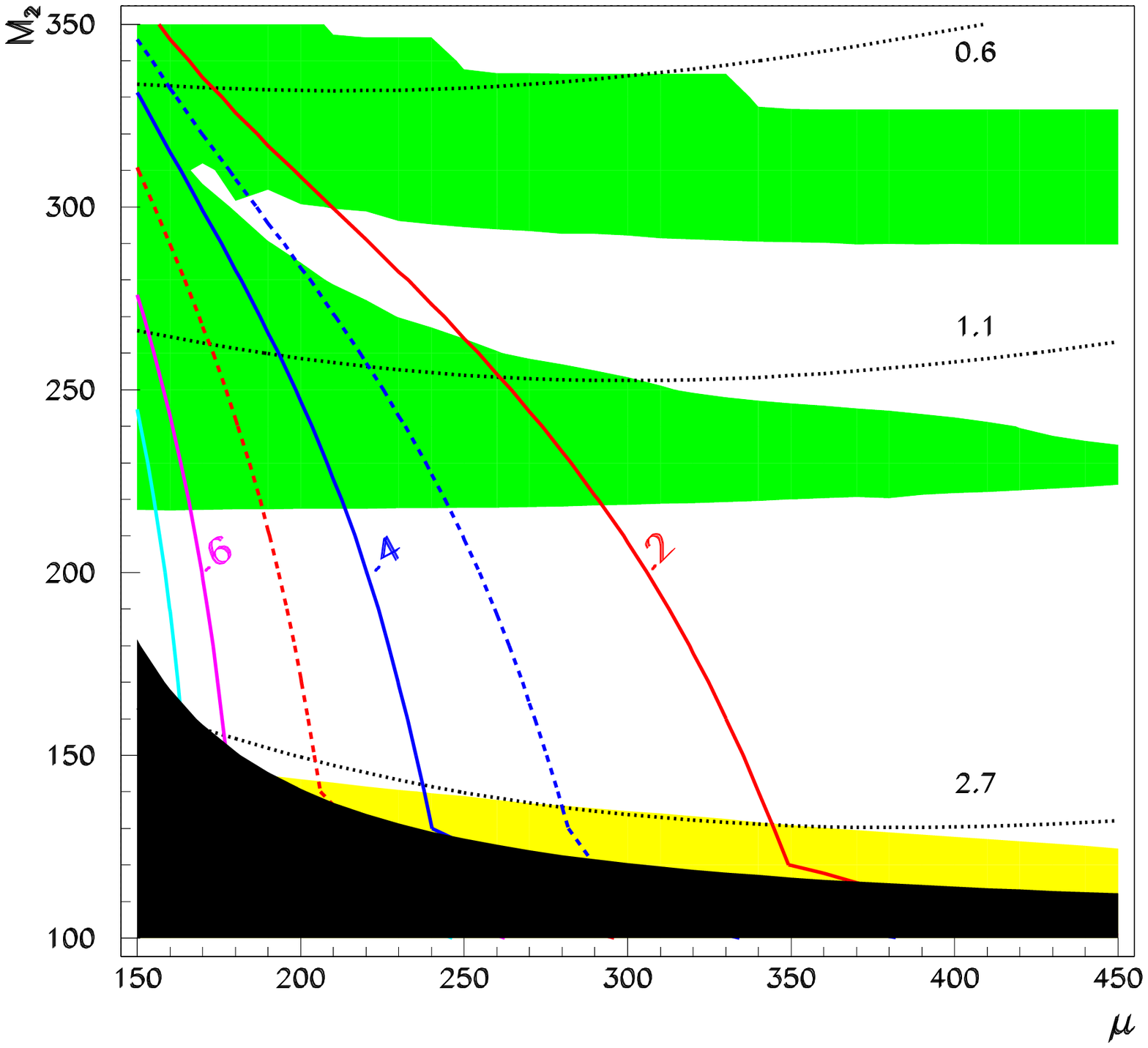}
\includegraphics[height=9cm,width=.5\textwidth]{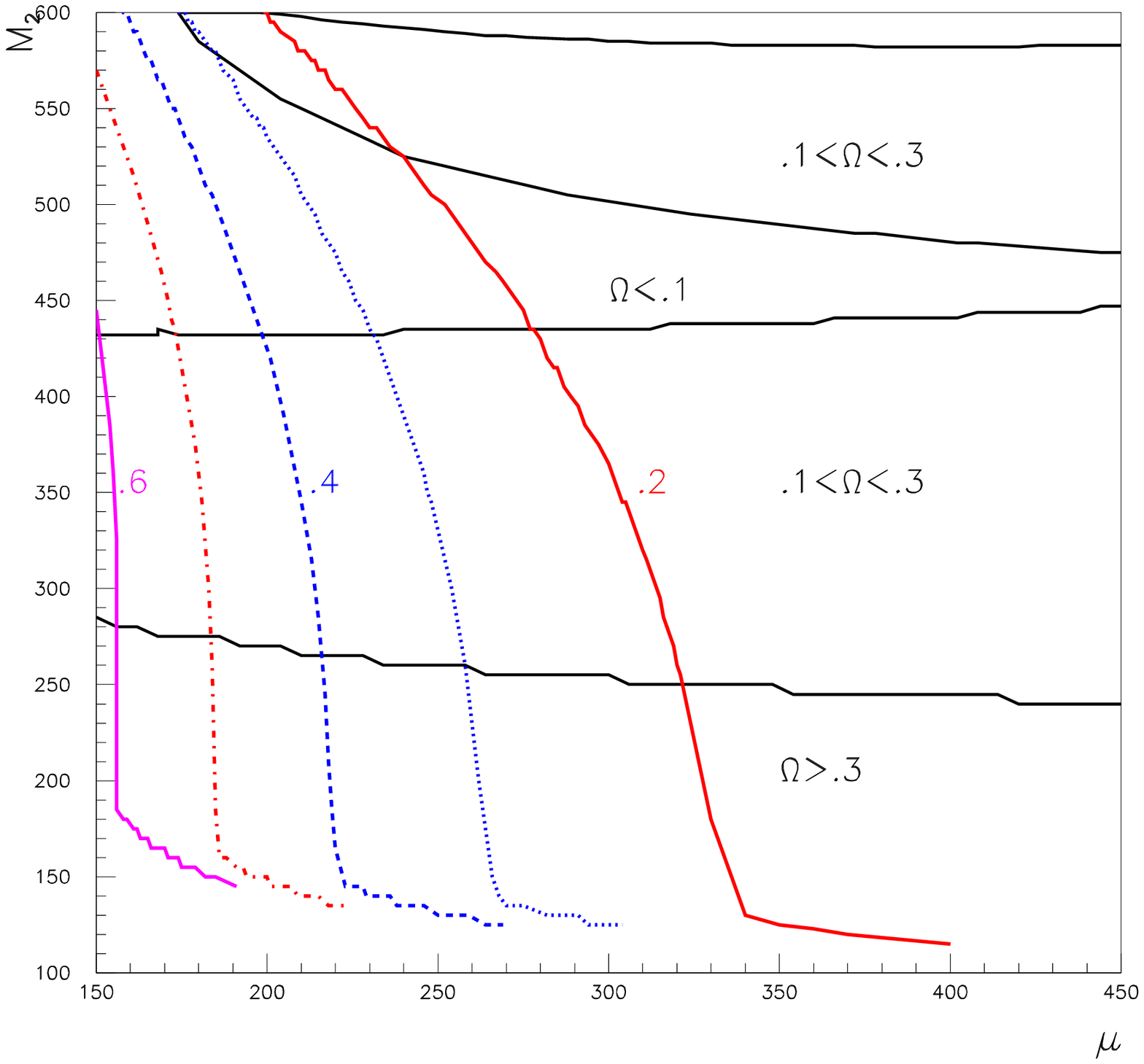}}
\vspace*{-1.0cm}
\caption{\label{fig_constraints}{\em 
 a) Contours of constant $Br_{\chi \chi}$ from .2
(far right) to .65 (far left) 
for  $M_1/M_2=1/5,m_0=100$GeV 
 We have also superimposed  the various constraints.  
The black area is excluded by the
chargino mass at LEP. The other shadings refer to the relic
density, with the allowed region (white), the overclosure region
(light grey) and the region with $\Omega<.1$ (medium grey).
The dotted lines are constant $a_\mu$
lines in units of $10^{-9}$\/.
b)Similarly  for $M_1/M_2=1/10,m_0=94$GeV. }}
\end{center}
\vspace*{-0.5cm}
\end{figure*}

Fig.~\ref{fig_constraints}a shows the
different contours in the $M_2-\mu$ plane of $B_{\chi \chi}$ for $M_1=M_2/5$
and for a  light slepton.
 We see that, even after  taking all constraints, we still find
large branching ratio of the lightest SUSY Higgs into neutralinos.
The largest branchings correspond to the
smallest $\mu$ values, which as argued before maximises the Higgsino content.
  It is also worth stressing
that even in these general models, the branching ratio into
invisible is never larger than $70\%$.
 For a lower ratio $M_1=M_2/10$, the invisible branching ratio can reach
over 60\% (Fig.~\ref{fig_constraints}b).
Even in the case of heavier sleptons large branching ratio into invisible are
possible although the allowed region of in the
$M_2-\mu$ plane corresponds to a narrow region around the Z  pole.

We have also searched, by making a large scan over $M_1,M_2,\mu$
and  $m_0$, but for fixed $\tgb=5$, which minimum value of $M_1$
one can entertain. The parameters were varied in the range
$10<M_1<100$GeV, $100<M_2,\mu<500$GeV, $70<m_0<300$GeV as given
by Eq.~1. We find that,
in order not to have too large a relic density, 
 $M_1$ must be above $20$GeV independently of $M_2$ and $\mu$, as
seen in  Fig.~\ref{tgb5largescan}. However, this is not a value that gives
the largest branching into invisibles, largest values 
are in the range $40<M_1<60$ GeV where one has both a
significant  Higgsino-gaugino mixing and a LSP light enough
for the Higgs to decay into it.
Note
that this lower bound on $M_1$ is more or less independent on
$\tan\beta$\cite{lowneutralino}.
We also show the relic density as a function of $M_1$. Note that one hits both
the Z pole and the Higgs pole. However for the latter configurations,
$B_{\chi\chi}$ is negligible.

To conclude we have found that there are  regions of
parameter space that give a substantial branching fraction of the
lightest SUSY Higgs into invisibles that can account 
 for the dark matter in the
universe. We also find that these scenarios do not always require
a very light slepton since we can obtain an acceptable amount of
LSP relic density through an efficient annihilation at the $Z$
pole. However scenarios with the largest branching ratio into LSP
do entail that the lightest chargino and at the least the next LSP
are light enough that they could be produced at the Tevatron. The
phenomenology at the Tevatron should somehow be similar to the
Sugra $SU(5)$ based ``24-model'' which was
studied in \cite{nonuni-24}. Among other things, due to the fact
that  one has a larger splitting between the LSP and the NLSP, as
compared to the usual unified scenario, one expects an excess of
events containing many isolated leptons originating, for example,
from  a real $Z$ coming from the decay of the NLSP. However to
make definite statements about observability of these states at
the Tevatron requires a thorough simulation. 

\begin{figure*}[htb]
\vspace*{-0.2cm}
\begin{center}
\includegraphics[width=16cm,height=8cm]{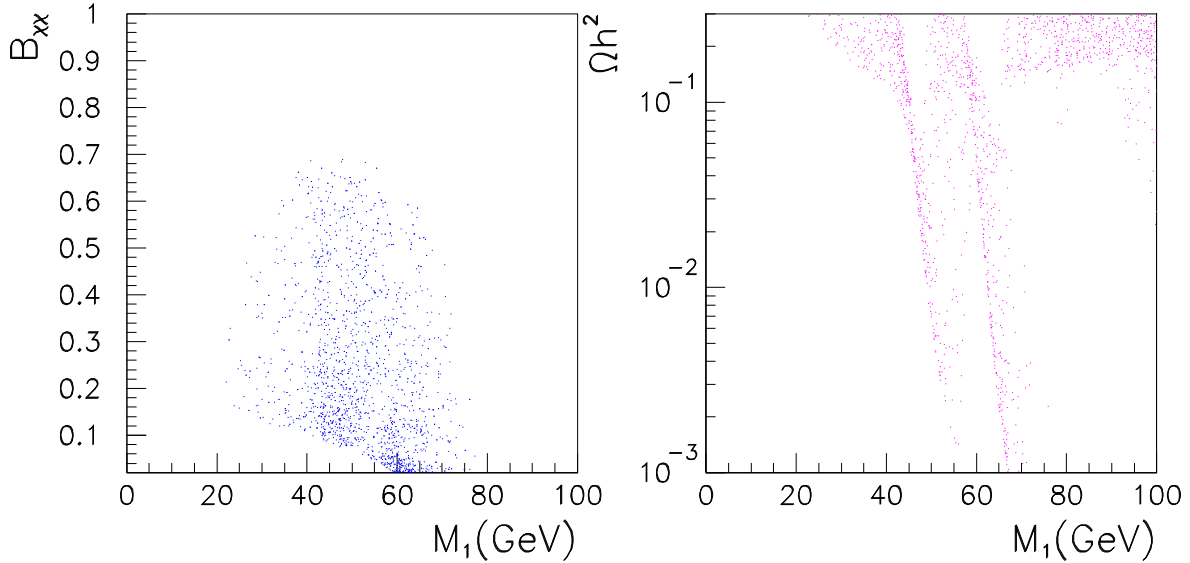}
\vspace*{-0.9cm}
\caption{\label{tgb5largescan}{\em Large scan over
$M_1,M_2,\mu,m_0$ for $\tgb=5$. The first panel shows the
branching ratio into invisibles {\it vs} $M_1$. The second panel
shows the relic density as a function of $M_1$. \/}}
\end{center}
\vspace*{-0.5cm}
\end{figure*}

%
%
}

{
\section[]{Search for the invisible Higgs in the $Wh/Zh$ channel at the LHC%
\footnote{\it S.\,Balatenychev, G.\,B\'elanger, F.\,Boudjema, R.M.\,Godbole, 
V.A.\,Ilyin and D.P.\,Roy}}  

One of the various methods to search for 
an invisible Higgs at the LHC is the associated production 
process $pp \rightarrow Wh(Zh)$ followed by the invisible decay of h is one of
them. The signatures being a single lepton with large transverse momentum and 
missing $P_T$, from the higgs as well as from the neutrino from the W decay, 
and lepton pair whose mass constructs to $Z$ and missing $P_T$, for the 
$Wh$ and $Zh$ production respectively.
 A parton level study, which took into account the dominant irreducible 
background caused by the $WZ (ZZ)$ production followed by the invisible
$Z \rightarrow \nu \bar \nu$ decay, had been made~\cite{DP-invisible-lhc}. This had shown 
that that it is possible to have $S/\sqrt{B} \sim 5.9$ for the process 
for $m_h = 120$ GeV and 100$\%$ B.R. into the invisible channel. The issue 
of the reach of these channels in terms of $m_h$ and 
the $B (h \rightarrow {\rm invisibles})$, is being revisited here.

{\bf In the $Wh$ process}: The backgrounds relevant for this signal are:
(1) The irreducible background due to the $WZ$ production followed 
by $Z \rightarrow \nu \bar \nu$.
(2) $WW$  production followed by  leptonic decays of both the 
$W$'s, one lepton being lost, due to low $P_T$ or too large a rapidity.  
(3) The large QCD  backgrounds caused by the production of $W$ with jets 
which are lost. The lost jets can add on to the missing $P_T$ of the 
decay $\nu$  from the $W$ and thus possibly give substantial missing $P_T$.
(4)  Another source of background would be the $t \bar t$ production
with their decay  producing a $W$ pair with two $b$ jets.  This can cause
a background if the $b$ jets are  lost along with one of the decay leptons. 
(5)  $Z + {\rm jets}$ production will also give a background if the jet(s) 
are missidentified as a lepton. 

Before discussing the separation of the signal from the mentionned QCD  
backgrounds, we report the result of a repetition of the 
of the calculation of the signal and the irreducible background, 
with more modern parton densities.  A calculation of the $Wh$ signal, using
the LO formulae, with $|\eta_l| < 2.5, P_T^\ell > 100 $ GeV, gives 28 fb. The
irreducible $WZ$ background for the same cuts is 40 fb.  The higher order 
corrections to the signal give rise to a moderate $K$ factor and might just 
compensate for lepton detection effeciencies of $70\%$. Thus the above 
numbers, though calculated from a LO formula can be considered representative.
All the numbers are for $m_h = 120$  GeV and $100 \%$ B.R. into the 
invisible channel.  

The $WW$ background mentioned in point (2) above has also been evaluated and 
is 9 fb.  The canonical values for LHC used here for the soft and forward 
leptons that lost, giving rise to a final state similar to the signal are, 
i) a $P_T^e   < 10$  GeV, ii) $P_T^\mu < 5 $  GeV or $|\eta_\mu| > 2.5$,
iii) $P_T^\tau < $ 20 GeV. Incidentally the background also has been evaluated
only at the LO.

The big discriminant between the QCD background and the signal is the
hadronically quiet nature of the signal.
Hence one  has to tune cuts and jet vetos such that we get rid of 
the background at low cost to the signal. Towards this end we have to first 
look at the effect of the initial state radiations on the kinematical 
distributions in the following variables:
1) Missing $P_T$ which no longer is just the lepton  $P_T$.
2)rapidity and $P_T$ of the jet with the largest $P_T$ in the event.
3) the same for the jet with the second largest $P_T$. 
The knowledge of  2 and 3 can help determine the vetos for the two $b$ jets 
that will be produced in $t \bar t$ process. 
Results of a priliminary calculation are very encouraging and suggest that
with a cut on the transverse mass $M_T$ of the lepton and the missing $P_T$,
as well as appropriate jet vetos, it should be possible to reduce the QCD 
background substantially without any harm to the signal. The last background
due to $Z + {\rm jets}$ has been evaluated. With a rejection factor of 
$10^{-5}$ against a misidentification of the jet as a lepton, this background
can at the most be 1.2 fb, even for a missing $P_T$ cut of $100$ GeV.

The NLO corrections to a pair of gauge bosons 
production~\cite{Zeppenfeld-h-invisible} shows
that a veto on the jet with $P_T > 50 $ GeV and $\eta < 3$  reduces them
to within $20 \%$ of the Born cross-section. This also gives an
indication that these kinds of jet vetos will work well to reduce 
the  background.  We will have to optimise these cuts once we 
after taking into account the effect of the initial state radiation on the 
signal. 

{\bf In the $Zh$ process}: This signal in this case is of course much smaller. 
The possible backgrounds in this case are:
(1) $ZZ$ production followed by the invisible $Z \rightarrow \nu \bar \nu$
decay of the $Z$.
(2)  $WZ$  production with leptonic decay of the  $W$  lepton 
getting  lost as outlined in the consideration of the background due to the
$WW$ production in the earlier caser. This can give rise to a final state 
with $l^+l^-$ coming from the $Z$ decay and missing $P_T$.  
(3) Production of  $Z$ + 1 jet where the  jet gets lost can also cause a 
small background.

In the $Zh$ case one would want to use a cut on the  missing $P_T$  unlike the 
case of the $Wh$ signal where demanding a large $P_T$ lepton automatically
guranteed a large missing $P_T$.  The signal and $ZZ$ background
as well as the $WZ$ background mentioned in (2)  has been calculated
at the parton level. The numbers  are for a missing $P_T$ cut of 100 GeV,
with  $P_T^\ell < 20 GeV$,  and  $|\eta^\ell| < 2.5$.  The size of the signal, 
$ZZ$ and $WZ$  background are  about 8 fb, 21 fb and 3 fb respectively. 
The cut on the missing $P_T$ can be increased without harming the signal but 
can bring down the QCD backgrounds effectively. Again to decide on the cuts 
to optimise the signal, one needs to know the effect of the initial state 
radiation on the $Zh$ signal.These calculations are in progress.
}

{
\noindent
\section[]{Simulation of neutral Higgs Pair Production in PYTHIA
using HPAIR Matrix Elements%
\footnote{\it R.\,Lafaye}}

\subsection{Introduction} 

Scalar Higgs boson pair production at LHC allows to study the trilinear Higgs
self couplings in the Minimal Supersymmetric Standard Model (MSSM) scheme.  In
$pp$ collisions the dominating process is the gluon fusion $gg\rightarrow HH$,
where $H$ can be any of $A$, $H$ or $h$. In PYTHIA those processes can be
generated through the resonance channel $H\rightarrow hh$ or via
$f_i\bar{f_i}\rightarrow AH, Ah$.  But for Standard Model (SM) and MSSM
scenarios with high values of $\tan\beta$ the contribution of $s$-channels
becomes negligible, thus PYTHIA alone can not be used to explore this region of
the parameters space where the cross section can rise above 1~pb for values of
$m_A$ up to 150~GeV.  
At LHC measuring trilinear self couplings in production modes dominated by
the gluon fusion would require a huge amount of data. Nonetheless 
the MSSM high $\tan\beta$ values cases are also interesting as a discovery 
channel, as their cross section is large enough.

This note presents the implementation of LO matrix elements calculated with
HPAIR\footnote{HPAIR is a program written by M.\,Spira.} into PYTHIA
6.1\cite{pythia}.  This implementation will allow a more complete simulation of
$hh$ production in resonance region as well as all neutral Higgs pairs in
continuum production.  After a brief description of the implementation steps,
we will present a comparison between cross sections values obtained with PYTHIA
Monte Carlo generation and those computed with VEGAS.  For a more complete
version of this work, refer to the ATLAS note about to be published.

\subsection{Neutral Higgs pair production in PYTHIA}


The already implemented processes into PYTHIA are $f_i\bar{f}_i\rightarrow
Ah^0$ and $f_i\bar{f}_i\rightarrow AH^0$.  Although those processes are
dominant in most MSSM scenarios, above $\tan\beta=30$ they contributes for only
10\% of the total cross section. In those cases, the dominant process is
$gg\rightarrow AA$ and all others processes contributes for 50\% of the total
cross section, as can be seen on figure \ref{fig1}.  
Thus for a more complete study of those scenarios we have added
into PYTHIA, the following processes\cite{SPIRA}: 
$gg\rightarrow h^0h^0$, 
$gg\rightarrow H^0h^0$,
$gg\rightarrow H^0H^0$,
$gg\rightarrow Ah^0$,
$gg\rightarrow AH^0$ and
$gg\rightarrow AA$.
All processes described above were implemented in a private 
version of PYTHIA, as standard $2\rightarrow 2$ PYTHIA processes. 
We tried to keep the modifications into PYTHIA to the minimum.
\begin{figure}
\vspace*{-1.1cm}
\begin{center}
  \includegraphics[width=10cm]{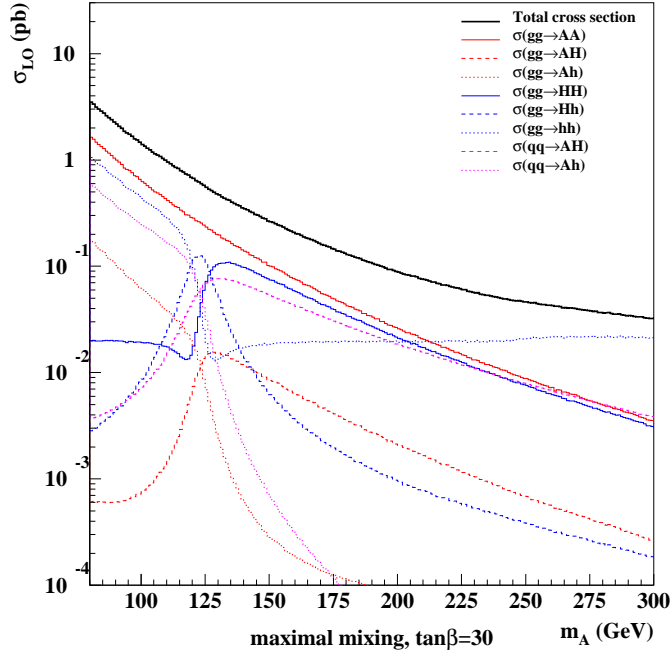} 
\vspace*{-0.4cm}
  \caption{\em Neutral Higgs pair LO cross sections as a function of
$m_A$ in maximal mixing for $\tan\beta=30$. 
Below the transition limit, the cross section
is dominated by $AA$, $Ah$ and $hh$ production, while above $AA$, $AH$ and
$HH$ dominate. The transition region is special as $A$, $H$ and $h$ have
all very similar masses and all possible Higgs pairs can be produced.
Higgs pair production studies can not disentangle different Higgs, as
dominating processes are made of degenerated Higgs with similar masses
and branching ratios.
}
  \label{fig1}
\end{center}
\vspace*{-0.8cm}
\end{figure}

For each event tried, the kinematic is chosen by PYTHIA, taking into account
resonance and continuum production in the shape of the phase space.
The PYSIGH routine then call the PYHPAIR function to get the differential cross section.
This differential cross section is computed with HPAIR Matrix Elements using
PYTHIA parameters. 
Finally, the event is
selected or rejected according to the PYTHIA standard Monte-Carlo procedure.
In SM and MSSM large $\tan\beta$ scenarios, the Higgs width might
be non negligible and Higgs particles might be produced off-shell. 
Thus in order to reproduce correctly the kinematic of the events,
this should be taken into account.

When $MSTP(42)=1$ Higgs are produced off-shell by PYTHIA. Those off-shell
values are then used to compute weight factors for the cross section like
in $WW$ production processes. The generation of events for those cases is much
slower and can sometimes leads to unphysical and very high differential cross sections.
Thus, except in the purpose of studying the differences in kinematics,
one should rely on on-shell bosons production.

For each off-shell Higgs the cross section is weighted by a factor
$B_H^*/B_H$ from the propagators of the virtual Higgs, defined as
follow:
$$
 B_H^* =
\frac{m_H^{*}\Gamma_H^*}{(m_H^{*2}-m_H^2)^2+m_H^{*2}\Gamma_H^{*2}} \ , \
 B_H = \frac{m_H\Gamma_H}{(m_H^{*2}-m_H^2)^2+m_H^2\Gamma_H^2}
$$
where $H$ can be any of $A$, $H$ and $h$. 
The total cross section suppression factor is then
($i=1,2$ denotes the two identical Higgs in the final state)
$$
B_{H_1}^*/B_{H_1}\cdot B_{H_2}^*/B_{H_1}\cdot
 m_{H^*_1}m_{H^*_2} / m^2_{H}
$$

\subsection{Monte-Carlo results}


The stand alone HPAIR program uses VEGAS to integrate the total cross section
of Higgs pair production processes.  It can also gives NLO order results with
QCD corrections.  Although, those corrections are not valid for $m_A$ larger
than 200~GeV.  As a check, one can compare cross section obtained using PYTHIA
and VEGAS with the same matrix elements. Comparisons were made with 1000 VEGAS
iterations and 100 PYTHIA events per bin. Most of the cross sections obtained
are very similar.  But unfortunately, the stand alone HPAIR code and PYTHIA
have some small differences in their Higgs mass spectrum algorithms. Thus the
cross section obtained might be slightly different, especially in cases where
it is sensitive to the $H,h$ mass difference like for the resonance
$gg\rightarrow H \rightarrow hh$ process at low $\tan\beta$.

Kinematic distributions have been investigated for the 4 $b$ final state,
that is when both Higgs disintegrate to $b\bar{b}$. The most important
for this final state is the $p_T$ distribution of the $b$, as current
selection algorithms ask for $b$ with a $p_T$ above 40~GeV. Furthermore,
as there is no $b$-trigger in ATLAS, high $p_T$ jets will be more likely
to pass the jet trigger thresholds.

First of all the distributions in $\eta$ and $p_T$ of the $b$ for the existing 
$gg\rightarrow H \rightarrow hh$ PYTHIA process 152 and our implemented
process $gg\rightarrow hh$ process in the resonance region ($\tan\beta=3$
and $m_A=300$~GeV) are very similar.
We then take a look at the $p_T$ distributions of the $b$ and mass resolution
of the $b$ pairs for $\tan\beta=50$ and $m_H=150$~GeV for on-shell and off-shell
Higgs bosons in the maximal mixing scenario. All Higgs pair production processes were turned on, including 
the already existing $f\bar{f}\rightarrow AH$
and $f\bar{f}\rightarrow Ah$. For this set of MSSM parameters, Higgs pair
production are largely dominated by $A$ and $H$ pairs.
The following table summarize the MSSM parameters and cross sections
obtained for those processes:

\begin{center}
\begin{tabular}{|c|c|c|c|c|c|c|c|} \hline
 $\tan\beta$  & $m_A$ & $m_H$   & $\Gamma_H$ & $\sigma$ & Number of events & Number of events \\
        &             & (GeV)   &            &   (pb)   &     generated    &      tried       \\ \hline
 50     & 148.9       &  150.0   &  9.1 GeV   &   1.46   &       20000      &     113186     \\ \hline
 50     & 148.9       &  150.0   &  9.1 GeV   &   1.45   &       20000      &     334236     \\ \hline
\end{tabular}
\end{center}

Events were then analyzed with the ATLAS fast simulation, ATLFAST, to take into
account the detector resolution. The
$b$ pairs invariant mass resolution of 15~GeV is convoluted by the Higgs width
of 9.1~GeV,
when bosons are produced off-shell, as expected. The distribution of
the $p_T$ of the jets is very similar and makes very little difference.

\subsection{Conclusion and prospects}

Higgs pair production processes have been successfully implemented
in PYTHIA 6.1. Although in some rare points of the parameters space 
when using off-shell bosons the behavior of the matrix elements 
should be investigated.
Neutral Higgs bosons production has two main interests\cite{DJOUADI}\cite{HOUCHES}.
The first is Higgs discovery and this could be achieved
through the study of the resonance production where the cross section
according to NLO predictions is of the order of 2~pb or for high
values of $\tan\beta$ where the cross section for $\tan\beta=50$
can reach 30~pb. The first case as already been studied in \cite{ELZBIETA}
while the second is under analysis.
An other interest is the reconstruction of the Higgs potential
which implies the measurement of the Higgs self couplings like
$\lambda_{Hhh}$. Whether this coupling can be measured for double Higgs
resonance production with sufficient integrated luminosity at the 
LHC is under investigation.
}

{
\tolerance=100000

\def\cO#1{{\cal{O}}\left(#1\right)}
\def\bea{\begin{eqnarray}}
\def\eea{\end{eqnarray}}
\def\nn {\nonumber}
\newcommand{\be}{\begin{equation}}
\newcommand{\ee}{\end{equation}}
\newcommand{\br}{\begin{eqnarray}}
\newcommand{\er}{\end{eqnarray}}
\newcommand{\ba}{\begin{array}}
\newcommand{\ea}{\end{array}}
\newcommand{\bi}{\begin{itemize}}
\newcommand{\ei}{\end{itemize}}
\newcommand{\bn}{\begin{enumerate}}
\newcommand{\en}{\end{enumerate}}
\newcommand{\bc}{\begin{center}}
\newcommand{\ec}{\end{center}}
\newcommand{\ul}{\underline}
\newcommand{\ol}{\overline}
\newcommand{\ar}{\rightarrow}
\newcommand{\sm}{${\cal {SM}}$}
\newcommand{\as}{\alpha_s}
\newcommand{\aem}{\alpha_{em}}
\newcommand{\ycut}{y_{\mathrm{cut}}}
\newcommand{\susy}{{{SUSY}}}
\newcommand{\Dir}{\kern -6.4pt\Big{/}}
\newcommand{\Dirin}{\kern -10.4pt\Big{/}\kern 4.4pt}
\newcommand{\DDir}{\kern -10.6pt\Big{/}}
\newcommand{\DGir}{\kern -6.0pt\Big{/}}
\def\Ecm{\ifmmode{E_{\mathrm{cm}}}\else{$E_{\mathrm{cm}}$}\fi}
\def\gluino{\ifmmode{\mathaccent"7E g}\else{$\mathaccent"7E g$}\fi}
\def\photino{\ifmmode{\mathaccent"7E \gamma}\else{$\mathaccent"7E \gamma$}\fi}
\def\mgluino{\ifmmode{m_{\mathaccent"7E g}}
             \else{$m_{\mathaccent"7E g}$}\fi}
\def\taugluino{\ifmmode{\tau_{\mathaccent"7E g}}
             \else{$\tau_{\mathaccent"7E g}$}\fi}
\def\mphotino{\ifmmode{m_{\mathaccent"7E \gamma}}
             \else{$m_{\mathaccent"7E \gamma}$}\fi}
\def\ML{\ifmmode{{\mathaccent"7E M}_L}
             \else{${\mathaccent"7E M}_L$}\fi}
\def\MR{\ifmmode{{\mathaccent"7E M}_R}
             \else{${\mathaccent"7E M}_R$}\fi}

\def\lsim{\buildrel{\scriptscriptstyle <}\over{\scriptscriptstyle\sim}}
\def\gsim{\buildrel{\scriptscriptstyle >}\over{\scriptscriptstyle\sim}}
\def\Jnl #1#2#3#4 {#1 {\bf #2}, (#3) #4}
\def\NPB {{\rm Nucl. Phys.} {\bf B}}
\def\PLB {{\rm Phys. Lett.}  {\bf B}}
\def\PRL {\rm Phys. Rev. Lett.}
\def\PRD {{\rm Phys. Rev.} {\bf D}}
\def\ZPC {{\rm Z. Phys.} {\bf C}}
\def\EPJC {{\rm Eur. Phys. J.} {\bf C}}
\def\Ord{\lower .7ex\hbox{$\;\stackrel{\textstyle <}{\sim}\;$}}
\def\OOrd{\lower .7ex\hbox{$\;\stackrel{\textstyle >}{\sim}\;$}}

\section[]{Multiple NMSSM Higgs boson signals at the LHC%
\footnote{\it C.\,Hugonie and S.\,Moretti}}

\noindent
In Ref.\,\cite{EGH} a
no-loose theorem was established for the NMSSM \cite{model}, guaranteeing
that the LHC will discover {\em at least one} neutral NMSSM Higgs boson
(unless there are large branching ratios for decays to SUSY particles 
and/or to other Higgs bosons). Here, we try to establish the plausibility
of a NMSSM scenario in which {\em multiple} neutral Higgs boson detection 
is possible at the LHC, with a number of available Higgs states in excess  
of those pertaining to the MSSM.
Similarly to what done there, we only consider the `direct' production
channels (at the accuracy described in  \cite{direct}), namely
($V=W^\pm,Z$, $Q=b,t$ and $q^{(')}$ refers to any possible quark
flavour):
\bea\label{procs}
gg \to \mathrm{Higgs} ~ ({\mathrm{gluon-gluon~fusion}}),
& &
q\bar q^{(')} \to V~\mathrm{Higgs} ~ ({\mathrm{Higgs-strahlung}}),  \\
q q^{(')}\to q q^{(')}~\mathrm{Higgs} ~ (VV-{\mathrm{fusion}}),
& &
gg,q\bar q\to Q\bar Q~\mathrm{Higgs} ~
({\mathrm{quark~associated~production}}). \nn
\eea
We neglect `indirect' Higgs production via 
decays/bremsstrahlung off SUSY particles \cite{indirect} and 
Higgs production in association with squarks \cite{Hsquark}. 

The parameter scan performed here is somewhat different though. 
By using the program described of Ref.~\cite{NMSSM},
we have first constrained the soft terms of the NMSSM
by requiring {\em universality} at the GUT scale. The independent parameters of
the model are then: a universal gaugino mass $M_{1/2}$, a universal mass for
the scalars $m_0$, a universal trilinear coupling $A_0$, the Yukawa coupling
$\lambda$ and the singlet self-coupling $\kappa$: see eqs.~(2.1)--(2.2)
of \cite{EGH}. The (well-known) value of the $Z$-boson mass fixes
one of these parameters with respect to the others, so that we end up with {\em
four} free parameters at the GUT scale. 
As independent inputs characterising
the NMSSM, we adopt here: $m_0/M_{1/2}$, $A_0/M_{1/2}$, $\lambda$ and $\kappa$.
We then integrate numerically the Renormalisation Group Equations
(RGEs) between the GUT and the weak scale and
minimised the two-loop effective potential.  Furthermore, we impose the current
experimental bounds on (s)particle masses and couplings, especially the 
LEP limits
on the Higgs mass vs. its coupling to gauge bosons, see
\cite{LEP1}. Finally, we assume the existence of one neutral CP-even Higgs
boson with mass 115 GeV and sufficient coupling to gauge bosons, as hinted
by LEP \cite{LEP2}.

The main result of this numerical analysis, as already pointed out in
Ref.~\cite{NMSSM}, is that the additional couplings appearing in 
the Superpotential are always small: $\lambda (\kappa) < 10^{-2}$. 
The mixing angles of the additional singlet
states to the non-singlet sector, being
proportional to these couplings, are also small  
and the singlet sector of the
{\em universal} NMSSM is then quasi decoupled. (In the non-universal 
scenario of the previous section, 
the outcome was quite different: see also Ref.~\cite{EGH}). 
Hence, the neutral
Higgs sector consists of a quasi pure (qp) CP-even Higgs singlet state, $S_r$,
a qp CP-odd singlet, $S_i$, and the doublet sector is basically MSSM-like,
apart from small perturbations of order $\sim \lambda^2$, so that results known
for the Higgs sector of the MSSM are also valid in our case.

Fixing the mass of the lightest visible (non-singlet) CP-even Higgs at 115 GeV
puts further constraints on the parameter space of the model: we find that
$\tan\beta$ is always larger than 4, the CP-odd doublet Higgs mass $M_A$ is
larger than 160 GeV and $M_{\mathrm{SUSY}}$ is larger than 350 GeV. In this
limit, the CP-even  doublet states are the qp interaction eigenstates. The 
Higgs 
state with mass 115 GeV is a qp $H_u$, and the qp $H_d$ is heavy (with mass
larger than 300 GeV). On the other hand, the masses of the singlet Higgs
states, $S_r$ and $S_i$, can vary from a few GeV to 1 TeV.
For each of the five neutral Higgs bosons of the NMSSM, 
we have computed the total number of
events obtained by summing the rates of all production processes in
(\ref{procs}), assuming 300 fb$^{-1}$
 as integrated luminosity, at the  LHC. We have plotted these rates versus
the mass of the given Higgs states in Fig.~\ref{fig:LHC}.
If, as tentative threshold of detectability of a signal, we assume 
100 events, the conclusions are quite encouraging\footnote{However, we 
emphasise 
that this is not intended to be a definite claim of visibility, as the 
evaluation of such thresholds 
would require hadron-level simulations and detector-dependent 
considerations which 
are beyond the scope of this preliminary study.}.

\begin{figure}[t]
\begin{minipage}[b]{.495\linewidth}
\hskip1.25cm{\small Total number of $S_r$ events, $N_{S_r}$}
\centering\epsfig{file=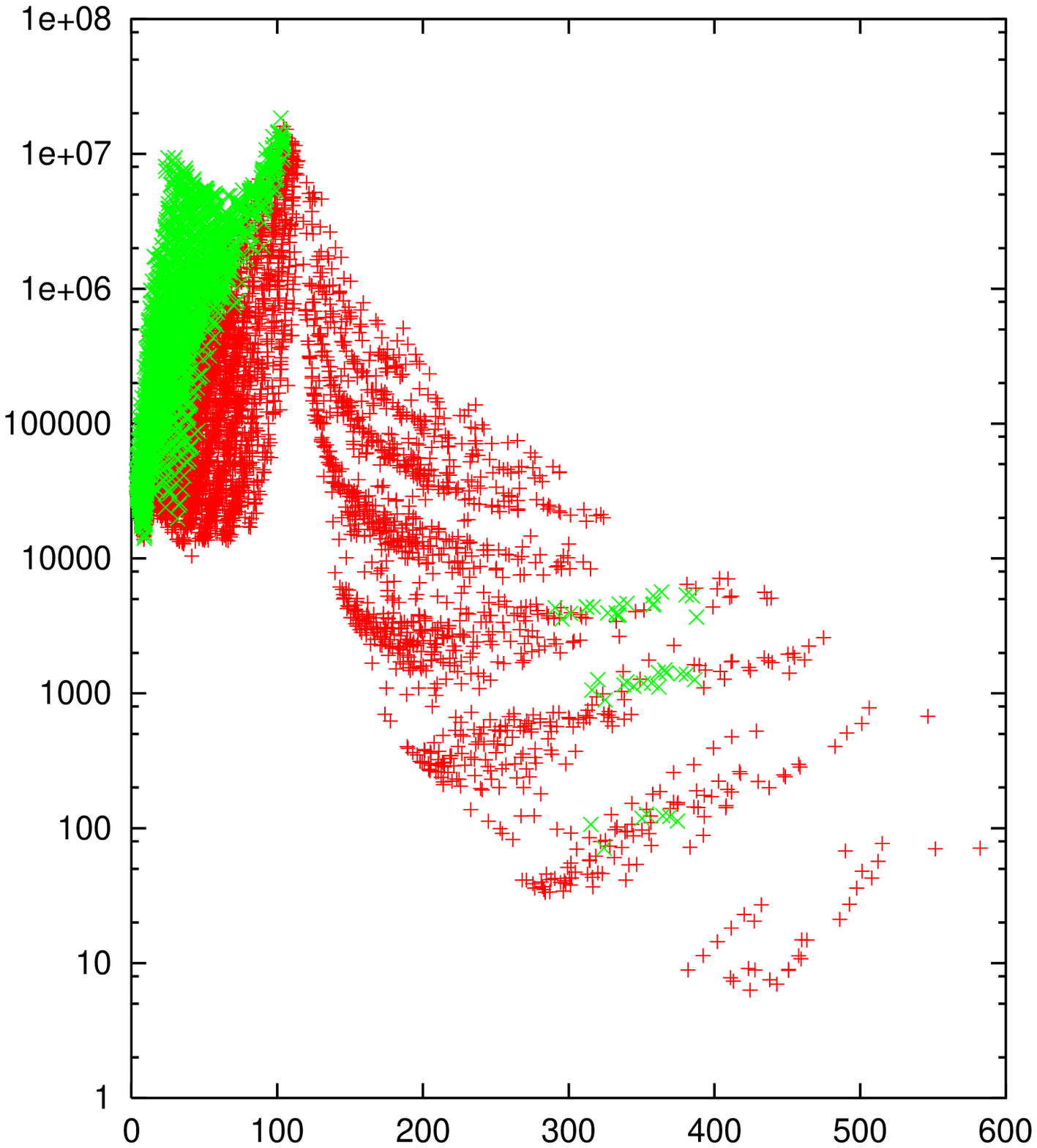,angle=0,height=5cm,width=\linewidth}
\end{minipage}\hfill\hfill
\begin{minipage}[b]{.495\linewidth}
\hskip1.25cm{\small Total number of $S_i$ events, $N_{S_i}$}
\centering\epsfig{file=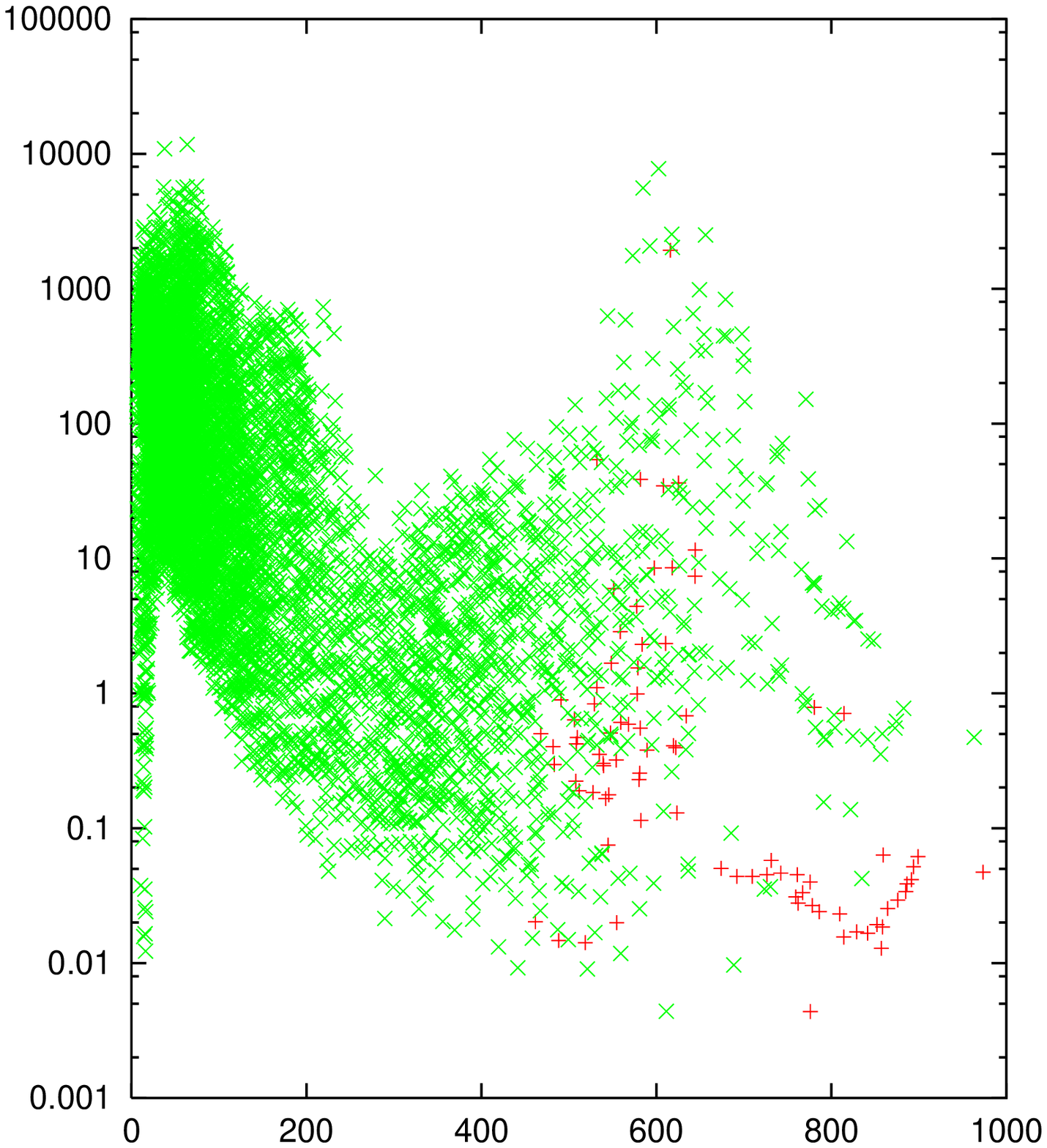,angle=0,height=5cm,width=\linewidth}
\end{minipage}\hfill\hfill

\hskip4.0cm{\small $M_{S_r}$ (GeV)}\hskip6.75cm{\small $M_{S_i}$ (GeV)}
\caption{Total number of events produced through processes (\ref{procs})
at the LHC  after 300 fb$^{-1}$ in the NMSSM,
for the CP-even singlet $S_r$ (left plot) and the CP-odd singlet $S_i$
(right plot), at LHC after 300 fb$^{-1}$.
(For an explanation of the colour coding, see the text.)}
\vspace*{-3mm}
\label{fig:LHC}
\end{figure}

At the LHC, all three non-singlet Higgs 
states, $H_u$, $H_d$ and $A$, might be 
visible at the same time, as they are MSSM-like. In the
singlet sector,  $S_r$ could be visible if its mass is $\Ord 600$ GeV and
$\lambda$ is not too small. In the NMSSM, this covers most of the parameter
space. Moreover, the CP-odd singlet, $S_i$, might be visible too, 
for an appreciable
part of the parameter space (when its mass is below 200 GeV or so). 
To render this
manifest, we have plotted in Fig.~\ref{fig:LHC} the total number of events
produced at the LHC with $S_r$ in the final state, $N_{S_r}$ (left plot)
in green (light) when the corresponding $S_i$ state is also visible 
($N_{S_i}>100$)
and in red (dark) when it is not ($N_{S_i}<100$). Similarly, we did 
for $S_i$ (right plot), 
with green (light) when the corresponding $S_r$ is visible
($N_{S_r}>100$) and red (dark) when it is not ($N_{S_r}<100$). 

Notice that  the discovery areas of multiple Higgs boson states
identified in Fig.~\ref{fig:LHC} are indeed associated
to the same regions of parameter space. 
In fact, a first glance at the total number of CP-odd singlet $S_i$ 
produced at the LHC 
might indicate that nearly all the parameter space of the model is already
covered by the CP-even singlet $S_r$ search, as all the plotted
points are in green (light). This is however not the case, as one can check
from the left-hand plot ($N_{S_r}$ vs. $M_{S_r}$), where a lot of
points are still under the 100 events threshold. The fact that one sees only
green (light) points on the right-hand plot is due to the very high density of
points considered, green (light) points being plotted after red (dark) ones.
Hence, there are red (dark) areas, uncovered by the $S_r$ searches, behind
green (light), covered, ones. 

The conclusions of this preliminary study are that, although the singlet sector
of the NMSSM tends to decouple from the rest of the neutral Higgs spectrum in
the universal case, quasi pure singlet states could still be found at the
LHC. In fact, one has to remember that a very light CP-even Higgs state is
not excluded by LEP searches if its coupling to gauge bosons is small enough.
Such a Higgs state could be visible at the LHC in the form of a CP-even
singlet Higgs state of the NMSSM (even with rather heavy masses), alongside a 
(light) singlet CP-odd state. Besides, often this scenario occurs where 
the MSSM-like non-singlet 
Higgses ($H_u$, $H_d$ and $A$) should also be visible, 
hence  making the whole 
neutral Higgs spectrum of the NMSSM 
in principle accessible at the CERN machine. 
(Rather similar conclusions also apply
the the next-to-MSSM (nMSSM) of Ref.~\cite{nMSSM}: see \cite{snowmass}.)

The caveat of our analysis is that we have not performed a
full Higgs decay analysis in the NMSSM. 
One may question whether the additional Higgs states would actually 
be visible. For example, they would certainly couple to 
singlinos -- $\tilde S$ is always the Lightest Supersymmetric Particles (LSP)
in our context -- hence decay into the latter and thus remain undetected.
This should however not be the case. In fact, the coupling of the singlet 
states to ordinary matter are generally stronger in comparison
 (of order $\lambda$, whereas those to two 
singlinos are $\sim\lambda^3$). So that, in the end, the main decay channels
of singlet Higgs states  should be those into detectable 
fermions and gauge bosons.
}

\vspace*{0.5cm}

\noindent
{\large \bf Acknowledgements.} \\
We would like to thank the organizers of the Les Houches workshop for
their invitation, warm hospitality and financial support. The work of
M.S. has been supported in part by the
Swiss Bundesamt f\"ur Bildung und Wissenschaft.
The work of D.Z. was partially supported by WARF and under DOE grant
No.~DE--FG02--95ER40896.
Fermilab is operated by URA under DOE contract No.~DE-AC02-76CH03000.
This work was supported in part by the EU Fourth Framework Programme
``Training and Mobility of Researchers'',
Network ``Quantum Chromodynamics and the Deep Structure of
Elementary Particles'', contract FMRX--CT98--0194 (DG 12 -- MIHT),
by the European Community's Human
Potential Programme under contract HPRN--CT--2000--00149 Physics at
Colliders and by the Indo--French
Collaboration IFCPAR--1701--1 {\em Collider Physics}.
The work of A.~Semenov and A.~Pukhov was supported in part by the
 CERN-INTAS grant
99-0377 and by RFFR grant 01-02-16710.
R.G. acknowledges the
hospitality of LAPTH where some of this work was done.

\setcounter{figure}{0}
\setcounter{table}{0}
\setcounter{section}{0}
\setcounter{equation}{0}
\newpage

{
\noindent
{\Large \bf B. Higgs Searches at the Tevatron} \\[0.5cm]
{\it A.\,Bocci, J.\,Hobbs and W.--M.\,Yao} 

\begin{abstract} 
Studies of the discovery reach for the SM and supersymmetric 
Higgs in Run II have been summarized. Combining the results from all possible
decay channels, and combining the data from both experiments, with 15 fb$^{-1}$
the Tevatron experiments can exclude a SM Higgs at the masses up to about 
190 GeV at 95 \% C.L. or discover it up to 120 GeV at the 5$\sigma$ 
level. A great deal of effort remains in order to raise 
the performance of the accelerator and bring the detectors on line and fully 
operational to the level demanded by the Higgs search. 
\end{abstract} 

\section{Introduction}

The search for the Higgs boson and the dynamics responsible for electroweak
symmetry breaking is the central goal of high energy physics today.
The Tevatron luminosity increase provided by the  
Main Injector and Recycler, along with the upgrades of both CDF and 
D0 detectors, will provide unique 
opportunities to search for the Higgs both in the Standard Model (SM)
and in supersymmetry model (SUSY). 
The Tevatron is expected to deliver an integrated luminosity of 2 
fb$^{-1}$ in the first two years (Run IIa) and 13 fb$^{-1}$ in the 
subsequent years (Run IIb) until the LHC starts.
Most of CDF and D0 Run IIa upgrades\cite{run2a} 
were successfully installed in spring of 
2001 and are now collecting data from 
$p\bar p$ collisions. Since the design of Run IIa 
upgrades is for the initial goal of 2 fb$^{-1}$ and will not survive the 
course of Run IIb, it is now anticipated that the Run 2b upgrades~\cite{run2b},
and in particular the replacement for the Run 2a silicon vertex detector, are 
necessary to carrying out this exciting program.

\section{Tevatron Run II SUSY/Higgs Workshop} 
 A year long workshop on the Tevatron Run II Higgs physics was held at 
Fermilab during 1998, a joint venture between CDF, D0 and theory group 
at FermiLab. The aim is to explore the discovery sensitivities for the 
Standard Model and MSSM Higgs bosons in Run II at the Tevatron. The results
is ultimately expressed in terms of the integrated luminosity required 
to either exclude the Higgs with 95\% 
confidence level, or discover it with 3-$\sigma$ or 5-$sigma$ 
statistical significance at a given mass. The details can be found in 
the report of the Higgs Working Group of the Tevatron Run II SUSY/Higgs 
Workshop~\cite{Carena:2000yx}.

At the time of the Workshop, neither CDF nor
D0 has had full Run II detector simulation package available. 
Two complementary approaches were adopted.
The first approach was based on a CDF Run I detector simulation with the 
geometrical acceptance extended to correspond to the Run IIa CDF detector. 
The second approach was based purely on SHW, a simple simulation package 
that uses an average of the expected CDF and D0 detector performances as a 
set of parametrized resolutions and acceptances to perform simple 
reconstruction of tracking, jets, vertices and trigger objects.  
In addition, a multivariate analysis using neural network\cite{nn} 
has been pursued and leads to a potential gain of Higgs sensitivity 
above the conventional analysis. 
 
Further challenges must be met in bringing the detectors online and fully 
operational, and in developing the techniques and understanding, particularly 
in $b\bar b$ jet-jet mass reconstruction and $b$ jet tagging, necessary to 
extract the small signal of the Higgs boson from the larger Standard Model
background. Here, we will quote the results with reasonably optimistic 
projections for what we might achieve after a great deal of hard work in 
the coming years. 

\subsection{$b\bar b$ mass resolution} 

The $b\bar{b}$ mass resolution assumed in making these estimates is
10\% in the central part of the distribution.  This represents a
significant improvement over the 14-15\% resolution obtained in Run 1.
One can improve upon the jet energy corrections
by making the best possible use of all detector information, 
including tracking, shower max, calorimeter, and muon chambers. 
Figure\ref{higgs-mass-res}
shows the improvement of jet energy resolution possible by determining jet 
energy from an optimum combination of all jet information. 
A great deal of effort, presently underway, is needed to understand
the jet energy corrections to the level required to attain 10\%
resolution.  The required integrated luminosity for Higgs discovery
scales linearly with this resolution.
\begin{figure}
\vspace*{-0.3cm}
\begin{center}
\includegraphics[width=10.0cm]{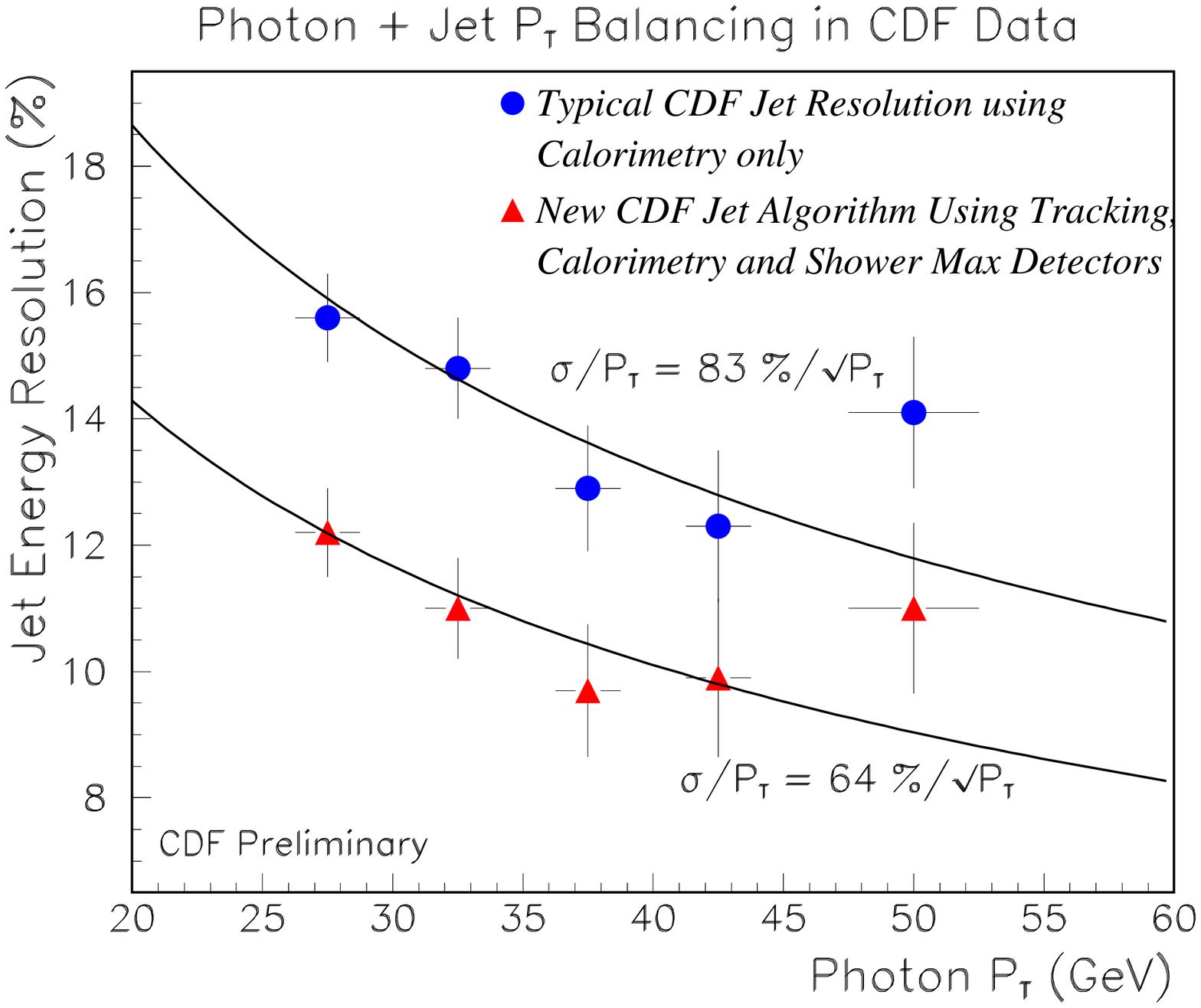}
\vspace*{-0.5cm}
\caption{Jet energy resolution as a function of jet $E_T$, comparing standard
corrections based on calorimeter only with energy determination combining
information from tracking, calorimetry, and shower max.} 
\label{higgs-mass-res}
\end{center}
\vspace*{-0.8cm}
\end{figure}

\subsection{$b$ tagging efficiency} 

The estimates of required integrated luminosity assume that the $b$
tagging efficiency and purity are essentially the same as in Run 1
in CDF, per taggable jet, shown in Figure~\ref{higgs-btag-eff}.  
The better geometric coverage of the 
Run 2a and 2b silicon systems, however, is taken into account and
leads to a much larger taggable jet efficiency.  Unlike the Run 1 
detector, the CDF Run II detector has a silicon vertex detector 
covering the entire luminous region, and has a 3D vertexing capability.
Since the required integrated luminosity scales inversely with the 
{\em square} of the tagging efficiency (assuming constant mistagging rates),
there is a potentially great improvement for developing high-efficiency
algorithms for $b$-tagging. 
\begin{figure}[htpb]
\begin{center}
\includegraphics[width=11.cm]{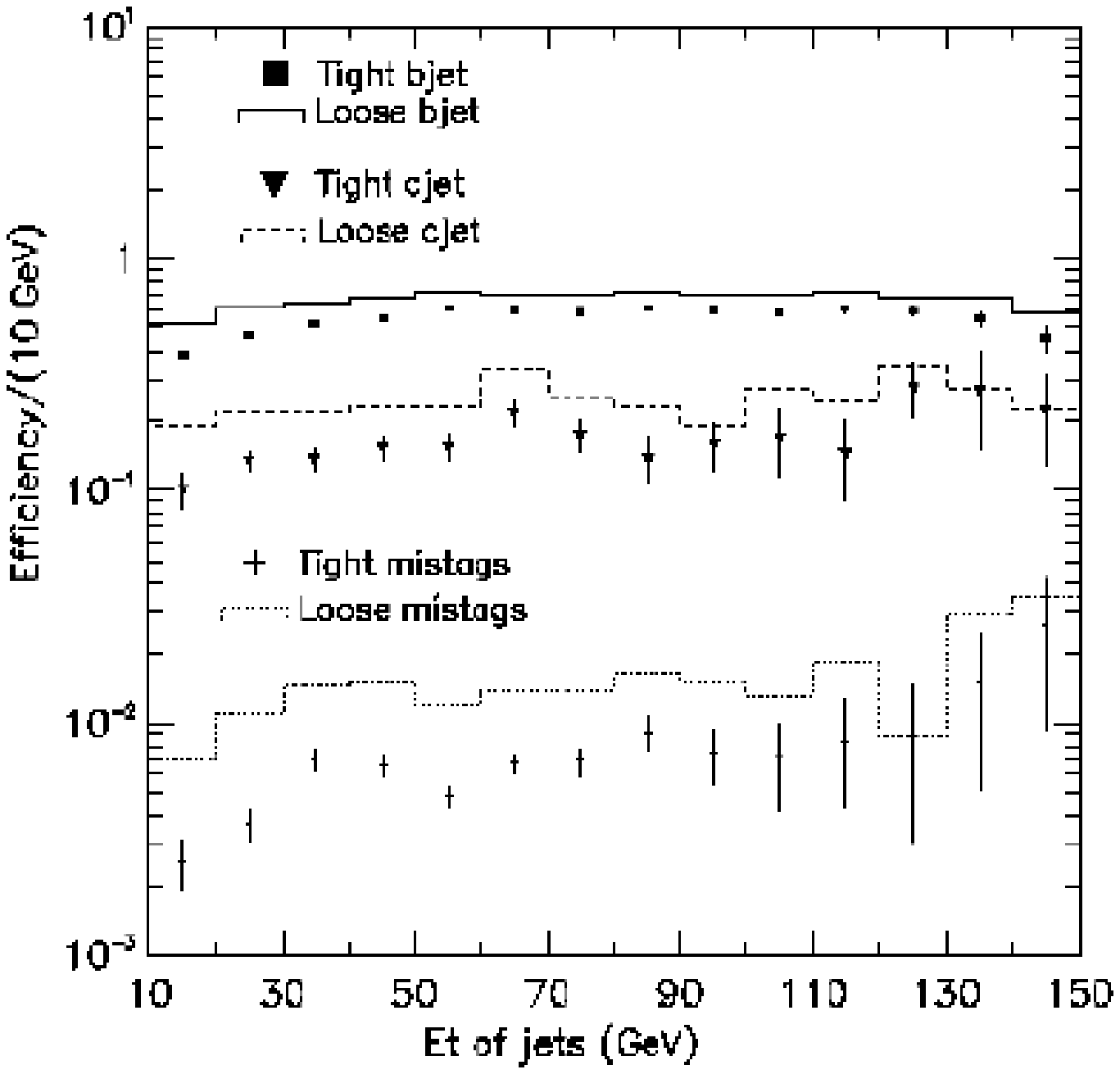}
\vspace*{-0.5cm}
  \caption{
 b tag efficiencies as a function of jet $E_T$, determined using CDF 
run1 taggable jets}
  \label{higgs-btag-eff}
\end{center}
\vspace*{-1.cm}
\end{figure}[htpb]
\begin{figure}
\begin{center}
\includegraphics[width=11.cm]{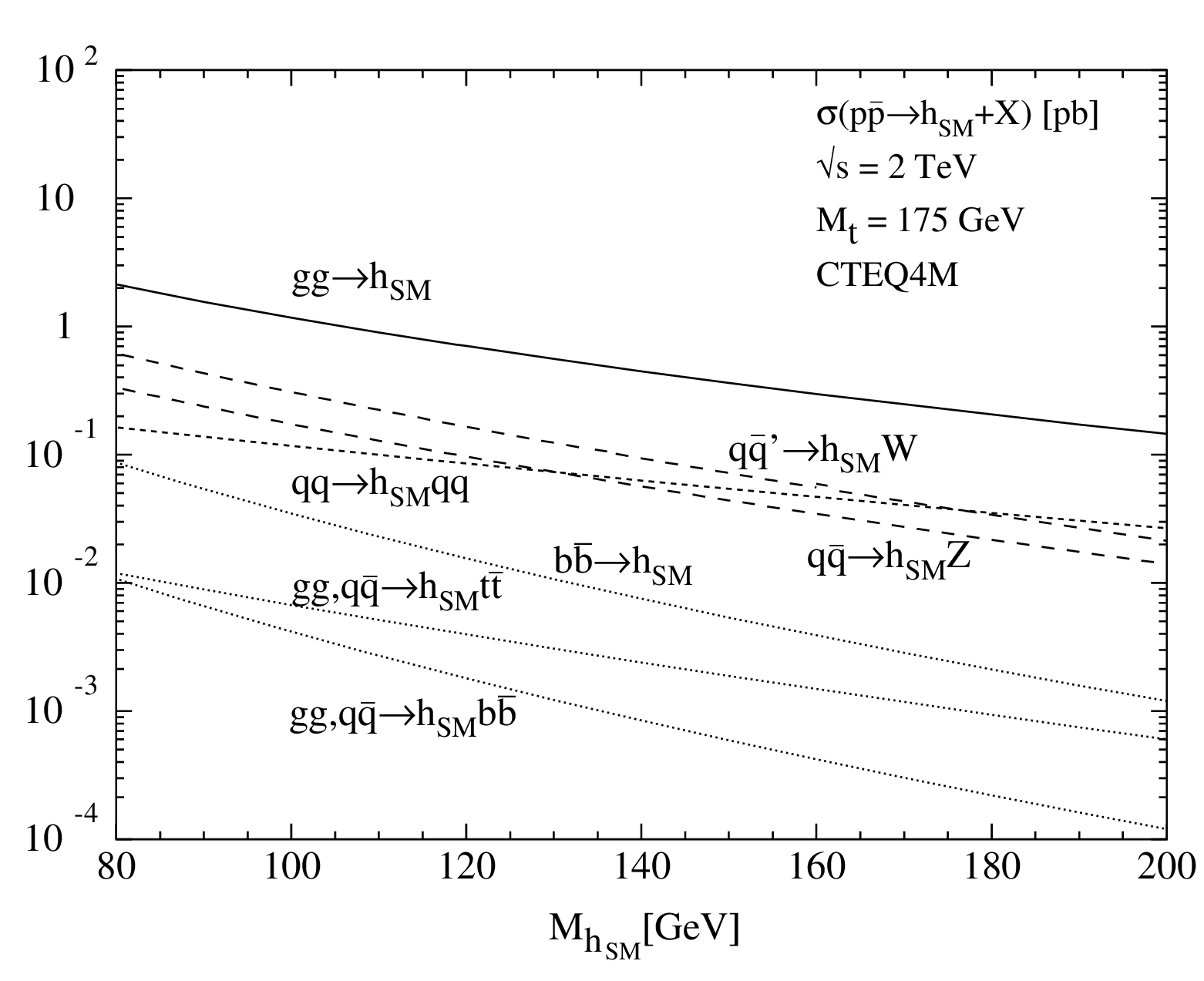}
\vspace*{-0.5cm}
  \caption{Production cross section for Standard Model Higgs at the Tevatron
           as a function of Higgs mass.}
  \label{higgs-xsec-fig}
\end{center}
\vspace*{-1.cm}
\end{figure}

\subsection{Background Systematic Studies}

Most particle searches are designed to have a small background, typically
only a few events.  The effective fractional statistical precision of the
background is of the same order as the background size, implying that
(fractional) systematic errors can be rather large.  Systematic uncertainties
of 30\% of the background are common in searches.  However, the Run II Higgs
search will have hundreds of background events, which come predominantly 
from the direct production of vector bosons plus heavy flavors, top and 
single top, di-boson, and QCD jets production. In some cases, 
the magnitude or the shape of the Standard Model backgrounds 
are not known at the required level of accuracy. For example, 
in the $\nu \bar {\nu} b\bar b$ channel analysis no estimates were made of 
the generic $b\bar b$ dijet background. This process has a very large cross 
section but tiny acceptance, and is thus not modeled reliably. In the CDF Run 1 
analysis, this background was about half of the total, and estimated from the 
data. Uncertainties which affect the dijet mass spectrum are particularly 
important to control because the components have different spectra. 
With collider data in hand, the understanding will be improved using dedicated 
studies and by tuning the event simulations to match collider data control 
samples.

  The remainder of this section is an estimate, in advance of significant data
taking, of the uncertainty in background estimates arising from systematic
uncertainties in simulations.  One of the possible sources of uncertainty has
been chosen for this study.  The effect from $q^2$ dependence on the background
inside a mass window of $\pm2\sigma$ around a target Higgs mass is
explored. Signal and background events for the channel $p\bar p\rightarrow WH
\rightarrow e\nu bb$ were generated using the Pythia generator and passed
through the detector simulation used during the Run II SUSY/Higgs workshop.
The basic event selection outlined in the SUSY/Higgs report was applied, and
the acceptance within the mass window was computed.  This process was repeated
with the $q^2$ scale in Pythia changed to $2q^2$ and $q^2/2$, ranges commonly
chosen when assessing systematic uncertainties from simulation.  The results,
expressed as fractional change in acceptance, are shown in table~\ref{t-q2}.
Systematic effects from varying $q^2$ scale become important when the overall
(fractional) uncertainty approaches 5\%.  As expected, the largest uncertainty
comes from the steeply falling $W+b\bar b$ background.  This is particularly
important, because $W+b\bar b$ is the dominant component of the background.
\smallskip

\begin{table}
\begin{center}
\caption{Fractional change in backgrounds to the $WH$ signal as a function of
  $q^2$.  The result shows the fractional change of each component of the
  background as $q^2$ changes from the nominal to $q^2/2$ and to $2q^2$.  In
  the average, the absolute values of percent changes for the two $q^2$ choices
  were used.  \label{t-q2}}
\begin{tabular}{|c|c|} \hline
  Source     & Percent change   \\ \hline
  $W+b\bar b$ &  $ 6\pm1\% $     \\
  $t\bar t$   &  $ 1\pm1\% $ \\
 Single top  &  $ 3.8\pm0.2\% $ \\
   $WZ$      &  $ 2.4\pm0.1\% $ \\
 $WH$ signal &  $ 0.4\pm0.1\% $ \\
\hline
\end{tabular}
\end{center}
\vspace*{-0.8cm}
\end{table}

\section{Standard Model Higgs}

The dominant SM Higgs production at the Tevatron is gluon-gluon fusion
via a heavy quark loop, giving a single Higgs produced.  The Higgs can
also be produced in association with a $W$ or $Z$ boson via its
couplings to the vector bosons.  Figure~\ref{higgs-xsec-fig} shows the
production cross section for various modes as a function of Higgs
mass. In the range below 135 GeV Higgs mass, the decay of $b\bar b$ dominates, 
and for larger masses the decay to W pairs dominates.

In the gluon fusion case, since the Higgs decays predominantly to
$b\bar{b}$ (for Higgs masses below 135 GeV), there is an
overwhelming background from QCD production of $b\bar{b}$ pairs.  The
$WH$ and $ZH$ modes, however, have been extensively studied~\cite{Carena:2000yx}
and lead to several distinct signatures in which a Higgs signal can be
observed with sufficient integrated luminosity.  

\subsection{Low-mass Higgs} 

For low mass ($ < 135 $GeV) Higgs, the most sensitive signatures arise
from the leptonic decays of the $W$ and $Z$, and are denoted $\ell\nu
b\bar{b}$, $\nu\bar{\nu} b\bar{b}$, and $\ell^+\ell^- b\bar{b}$.
Hadronic decays of the $W$ and $Z$ lead to the $q\bar{q}b\bar{b}$
final state which suffers from large backgrounds from QCD multijet
production.

In Run 1 in CDF, all four of these channels were studied, and led to 
limits on the Higgs cross section times branching ratio to $b\bar b$
as depicted in Figure~\ref{higgs-xsec-run1}. As the plots shown, the 
Run 1 limits are more than an order of magnitude above the expected
Standard Model cross section. Improvements to the detector,
coupled with much higher luminosity in Run II lead to the greatly 
enhanced sensitivity in the Standard Model search. 

Maximizing the sensitivity of the search for the Higgs in these
channels depends most critically on three things as mentioned above:
attaining the best
possible $b\bar{b}$ mass resolution, attaining the best possible $b$ jet
tagging efficiency and purity, and attaining as large a data sample as
possible.  
\begin{figure}
\vspace*{-0.3cm}
\begin{center}
\includegraphics[width=8.5cm]{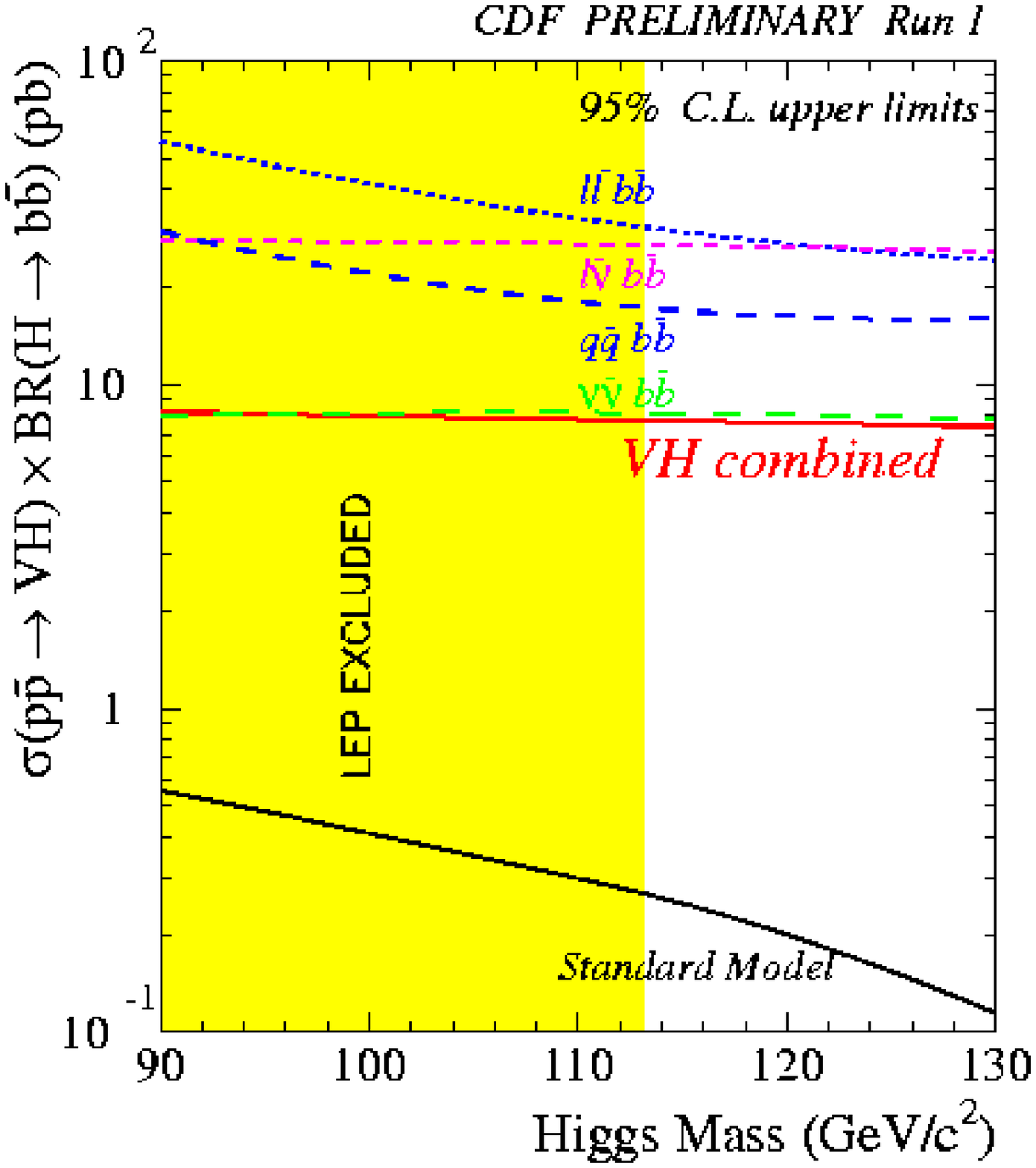}
\vspace*{-0.7cm}
  \caption{Limits on SM Higgs cross section times branching ratio to 
$b\bar b$ from CDF run1.}
  \label{higgs-xsec-run1}
\end{center}
\vspace*{-0.8cm}
\end{figure}

Figure~\ref{higgs-lvbb-mass} shows the two $b$-tagged dijet
 mass distribution and
Figure~\ref{higgs-lvbb-fig} shows the background-subtracted
signal in the $\ell\nu b\bar{b}$ case, for a 120 GeV SM Higgs, combining
data from both CDF and D\O\ representing 15 fb$^{-1}$ integrated 
luminosity, which clearly illustrates that even with the best resolution 
attainable, discovering the Higgs at Tevatron remains a major challenge. 

\subsection{High-mass Higgs} 
For larger Higgs masses ($>$ 135 GeV), the Higgs decays predominantly
to $WW^{(*)}$.  Two modes have been shown to be sensitive in this mass
range: $\ell\nu\bar{\ell}\bar{\nu}$ (from gluon fusion production of
single Higgs) and $\ell^\pm\ell^\pm jj$ (from tri-vector-boson final
states)~\cite{Han:1999ma}.  
The critical issues in these search modes are accurate
estimation of the $WW$ background in the former channel and estimation
of the $W/Z$+jets background in the latter.
\begin{figure}
\begin{center}
\includegraphics[width=10.cm]{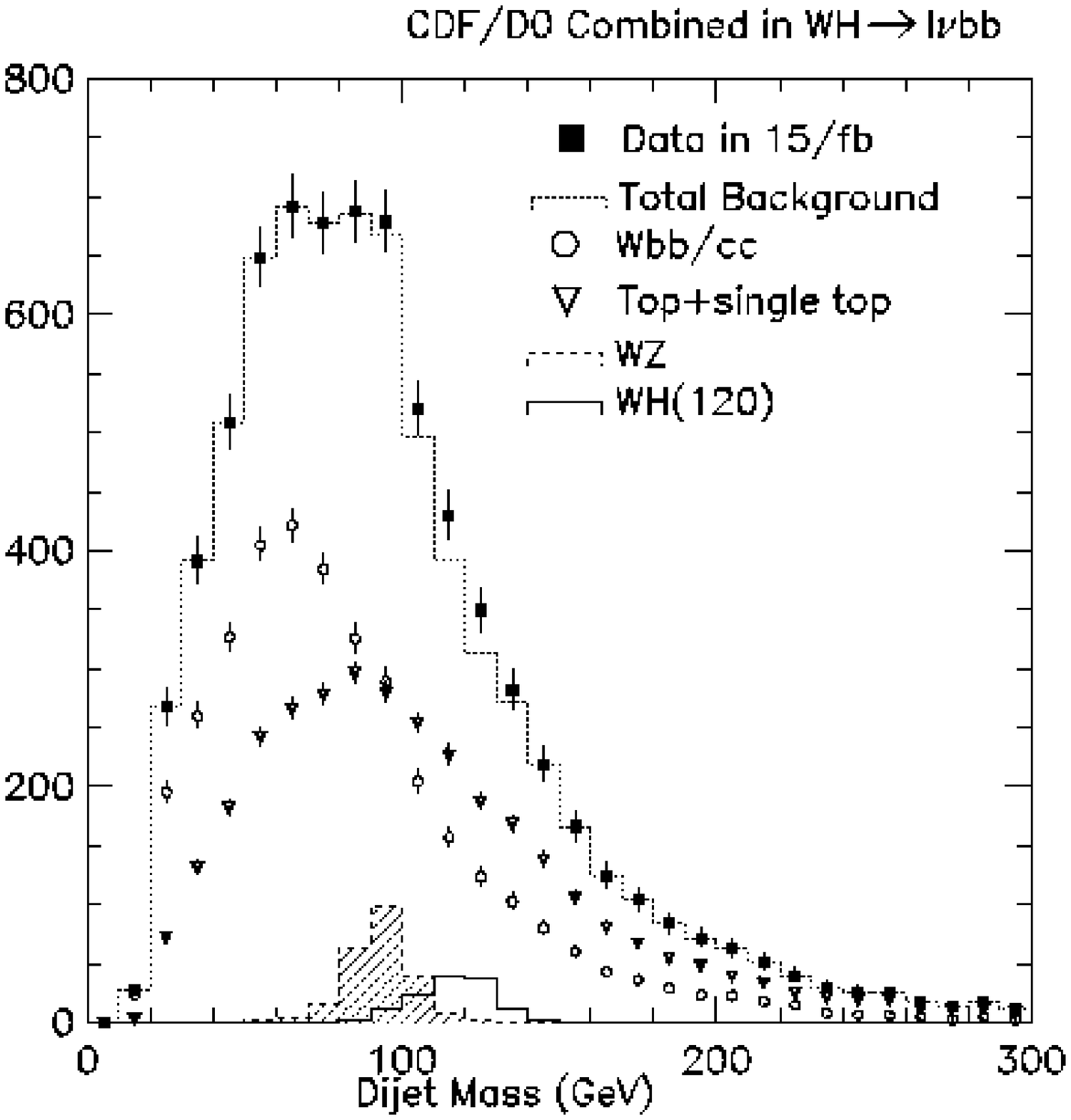}
\vspace*{-0.5cm}
  \caption{Distribution of $b\bar{b}$ mass distribution in the 
           $\ell\nu b\bar{b}$ channel, showing expected signal from 
           120 GeV SM Higgs, combining 15 fb$^{-1}$ of data 
           from CDF and D\O\ .}
  \label{higgs-lvbb-mass}
\end{center}
\vspace*{-0.8cm}
\end{figure}
\begin{figure}
\begin{center}
\includegraphics[width=10.cm]{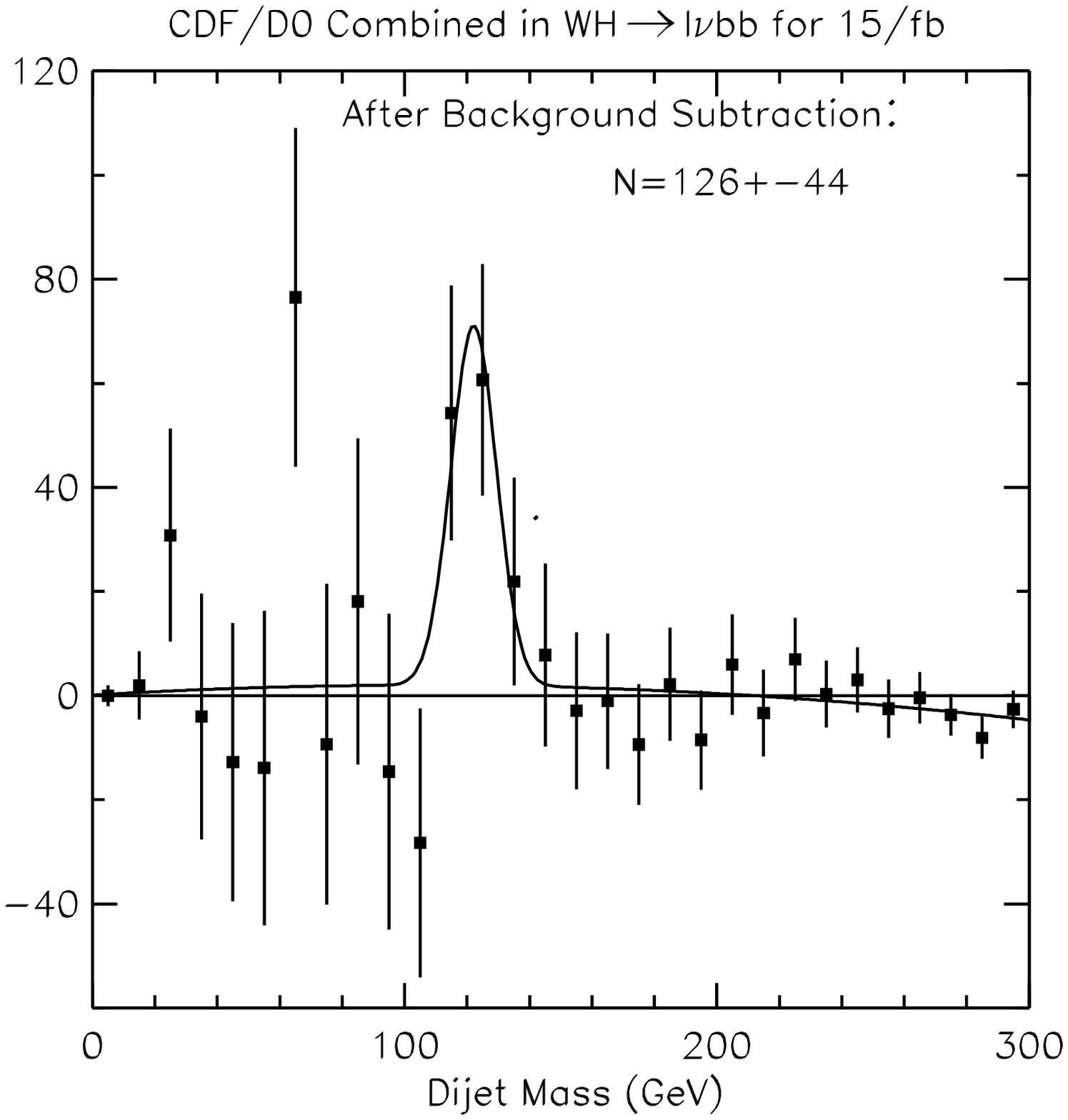}
\vspace*{-0.3cm}
  \caption{Background subtracted $b\bar{b}$ mass distribution in the 
           $\ell\nu b\bar{b}$ channel, showing expected signal from 
           120 GeV SM Higgs, combining 15 fb$^{-1}$ of data 
           from CDF and D\O\ .}
  \label{higgs-lvbb-fig}
\end{center}
\vspace*{-0.8cm}
\end{figure}

\subsection{Standard Model Higgs Reach in Run II} 

The integrated luminosity required to discover or exclude the Standard
Model Higgs, combining all search channels and combining the data from
CDF and D0 , is shown in figure \ref{higgs-final-fig}.  The lower
edge of the bands is the nominal estimate of the Run 2 study, and 
the bands extend upward with a width of about 30\%, indicating the
systematic uncertainty in attainable mass resolution, $b$ tagging
efficiency, and other parameters.  

The figure clearly shows that discovering a SM (or SM-like) Higgs at 
the 5-sigma level requires a very large data sample: even with 
15 fb$^{-1}$, the mass reach is about 120 GeV at best.  A 95\% CL
exclusion can, however, be attained over the entire mass range 
115-190 GeV with the integrated luminosity foreseen in Run 2b.
\begin{figure}
\begin{center}
\includegraphics[width=10.cm]{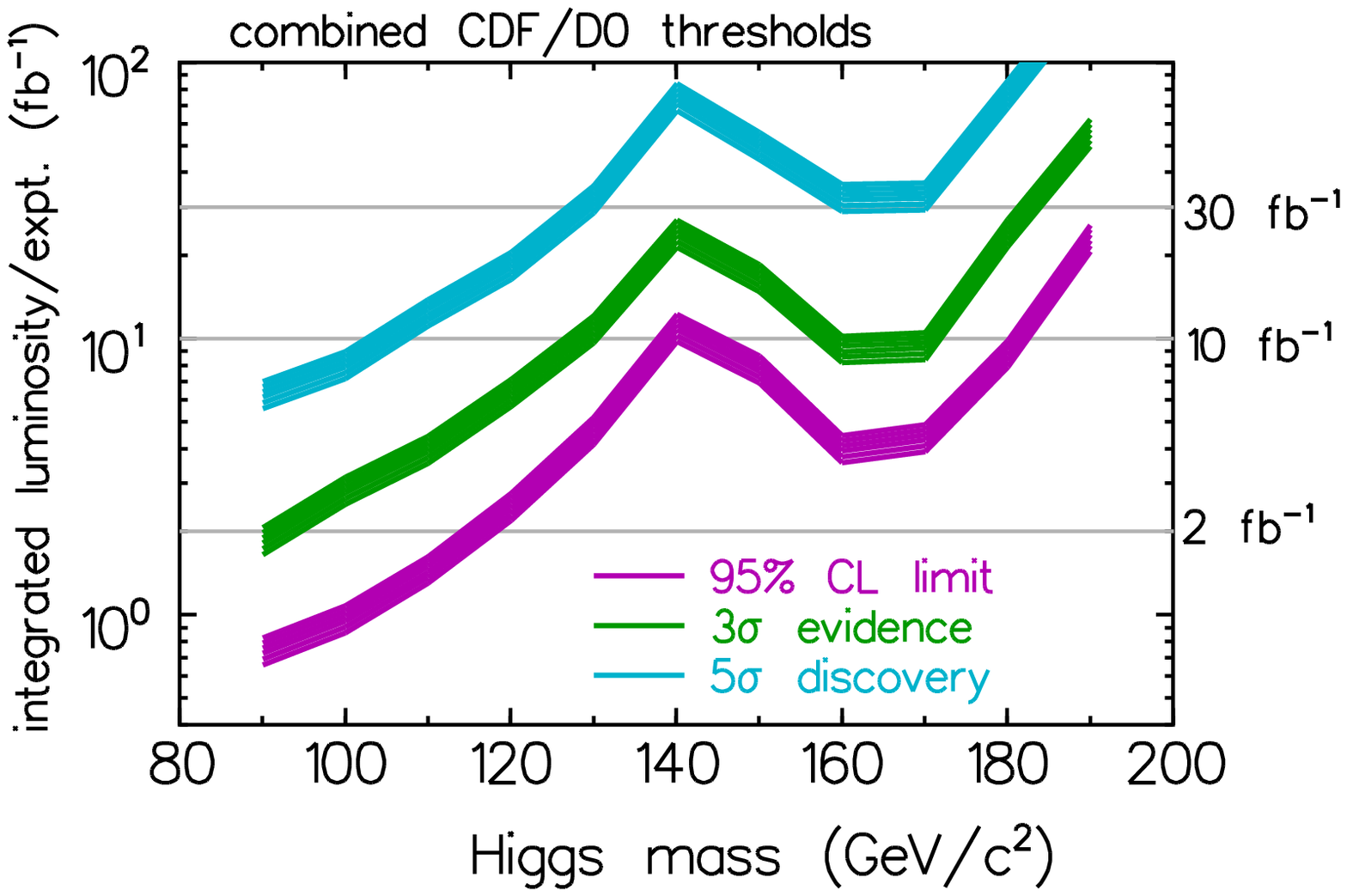}
\vspace*{-0.5cm}
  \caption{The integrated luminosity required per experiment, to
            either exclude a SM Higgs boson at 95\% CL or discover it at the
            $3\sigma$ or $5\sigma$ level, as a function of the Higgs mass.  
            These results are based on the combined statistical power of 
            both CDF and D\O\ and combining all search channels.}
  \label{higgs-final-fig}
\end{center}
\vspace*{-0.8cm}
\end{figure}

\section{SUSY Higgs} 

In the context of the minimal supersymmetric standard model (MSSM) the
Higgs sector has two doublets, one coupling to up-type quarks and the
other to down-type quarks and leptons.  There are five physical Higgs
boson states, denoted $h$, $A$, $H$, and $H^\pm$.  The masses and
couplings of the Higgses are determined by two parameters, usually
taken to be $m_A$ and $\tan\beta$ (the ratio of the vacuum expectation
value of the two Higgs doublets), with corrections from the scalar top
mixing parameters.  

The light scalar $h$ can appear very Standard-Model-like or nearly so
over a larger range of MSSM parameter space.  In this scenario the
results of the search for the SM Higgs produced in the $WH$ and $ZH$
modes are directly interpretable.  

More interesting is the case of large $\tan\beta$.  Since the coupling
of the neutral Higgses ($h/A/H$) to down-type quarks is proportional
to $\tan\beta$, there is an enhancement factor of $\tan^2\beta$ for
the production of $b\bar{b}\phi, \phi=h, A, H$ relative to the SM rate
appearing in figure \ref{higgs-xsec-fig}.  This leads to distinct final
states with four $b$ jets; if we demand that at least three of the jets
be tagged, the background from QCD multijet processes is relatively 
small.  In Run 1, CDF searched for this process, and from the null
result excluded a large swath of MSSM parameter space inaccessible to
LEP, as shown in figure \ref{higgs-bbbb-fig}.
\begin{figure}
\begin{center}
\includegraphics[width=10.cm]{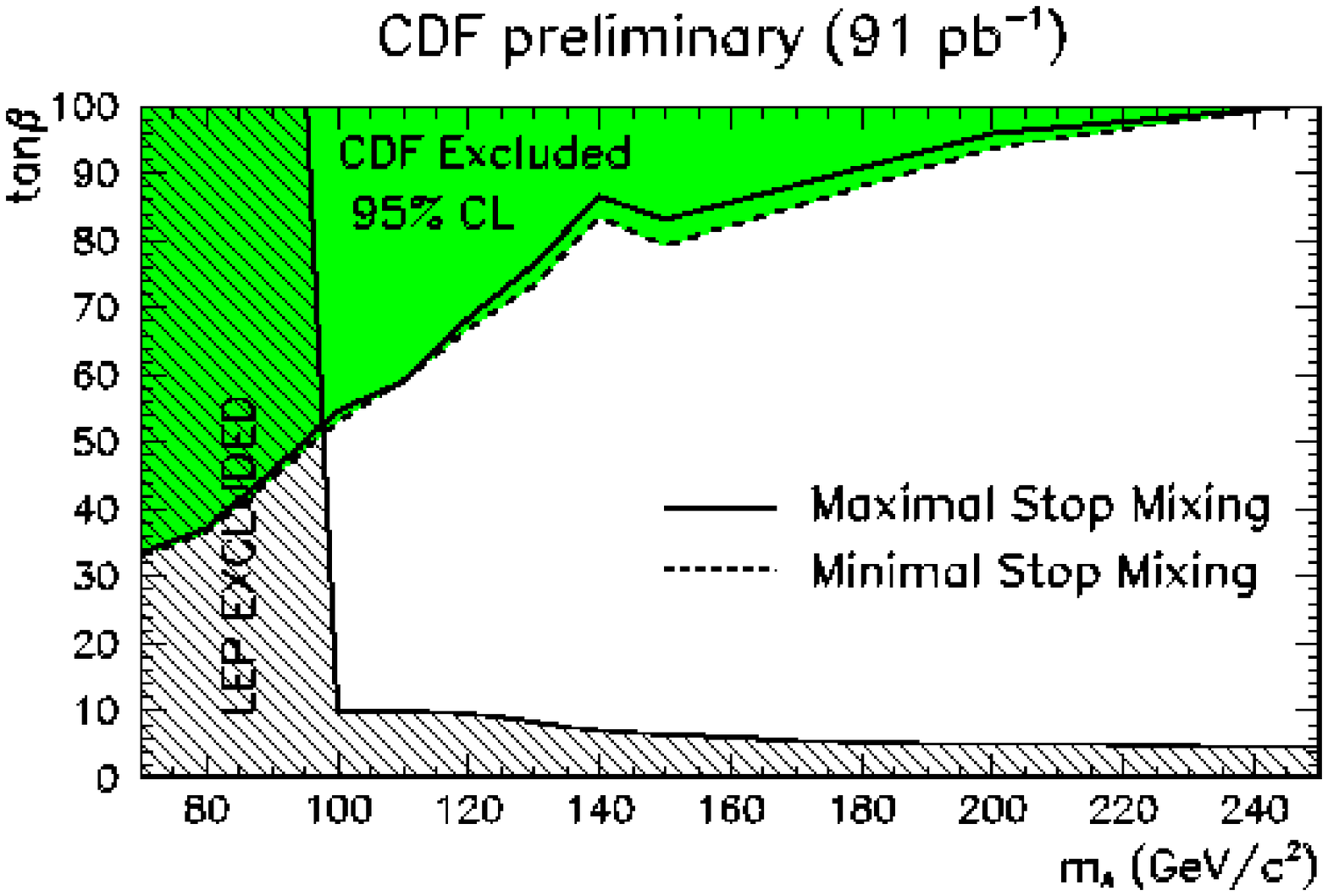}
\vspace*{-0.5cm}
\caption{CDF limits on MSSM Higgs using $b\bar{bb}\bar{b}$ final state.}
\label{higgs-bbbb-fig}
\end{center}
\vspace*{-0.8cm}
\end{figure}

Based on the Run 1 analysis, and taking into account the improved
$b$-tagging efficiency, figure \ref{higgs-mssm-fig} shows the 
regions of $m_A$ versus $\tan\beta$ that the Tevatron can cover for different
integrated luminosities. 
\begin{figure}
\begin{center}
\includegraphics[width=10.cm]{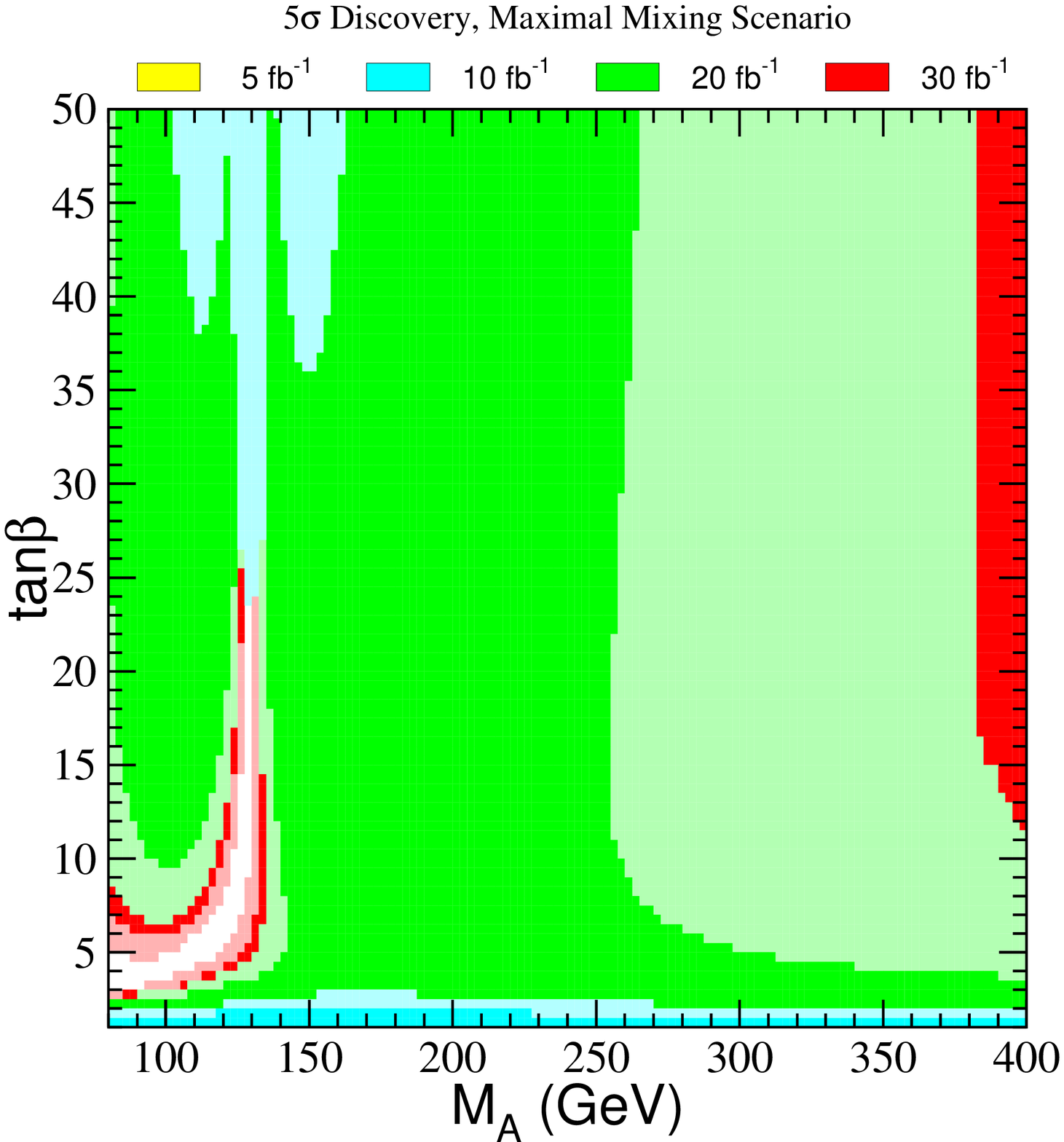}
\vspace*{-0.5cm}
  \caption{Regions of MSSM Higgs parameter space where 5$\sigma$ 
           discovery is possible, using SM Higgs search results.}
  \label{higgs-mssm-fig}
\end{center}
\vspace*{-0.8cm}
\end{figure}

\section{Conclusion} 

Studies of the discovery reach for the SM and supersymmetric 
Higgs in Run II have been summarized. Combining the results from all possible
decay channels, and combining the data from both experiments, with 15 fb$^{-1}$
the Tevatron experiments can exclude a SM Higgs at the masses up to about 
190 GeV at 95\% C.L. or discover it up to 120 GeV at the 5$\sigma$ 
level. A great deal of effort remains in order to raise 
the performance of the accelerator and bring the detectors on line and fully 
operational to the level demanded by the Higgs search. 
}

\setcounter{figure}{0}
\setcounter{table}{0}
\setcounter{section}{0}
\setcounter{equation}{0}
\newpage

{
\newcommand{\sla}[1]{/\!\!\!#1}

\newenvironment{2figures}[1]{\begin{figure}[#1]
  \begin{center}
    \begin{tabular}{p{.47\textwidth}p{.47\textwidth}} }
 {  \end{tabular}
  \end{center}
\vspace*{-1cm}
 \end{figure}
}

\noindent
{\Large\bf C. Experimental Observation of an invisible Higgs Boson at LHC}
\\[0.5cm]
{\it 
    B.\,Di Girolamo, L.\,Neukermans, K.\,Mazumdar, A.\,Nikitenko and
    D.\,Zeppenfeld}

  \begin{abstract}
    We present ATLAS and CMS simulation studies on the observability
    of an invisible Higgs boson produced via weak boson fusion at the LHC.
    With full and fast detector simulations we have checked the selection
    efficiency of the basic cuts proposed to search for such an object. The 
    Level 1 and High Level trigger strategies for this purely jet and 
    missing $E_{T}$ final state are discussed.
  \end{abstract} 

\section{Introduction}

Some extensions of the Standard Model (SM) exhibit Higgs bosons which can decay
into stable neutral weakly interacting particles, therefore giving rise to
invisible final states. In supersymmetric models, the Higgs bosons can decay
with a large branching ratio into the lightest neutralinos or gravitinos in 
some region of parameter space \cite{c1}, leading to an invisible final 
state 
if R parity is conserved. Invisible Higgs decay also happens in models with
an enlarged symmetry breaking sector, $e.g.$ in Majoron models \cite{c2},
\cite{c3}, where the Higgs disintegrates into light weakly interacting 
scalars. Another possibility arises in models with 
large extra dimensions~\cite{dimo}. In Randall-Sundrum type models~\cite{c4a},
the Higgs boson can mix
with the scalar radion field which then predominantly decays, invisibly, 
to graviton states~\cite{c5}. The invisible decay of the Higgs boson is a 
possibility which needs to be addressed in collider searches~\cite{c5a}.

Presently, the LEP II collaborations exclude invisible Higgs masses up to
114.4 GeV \cite{c6}. The presence of invisible Higgs decays makes much
more difficult the Higgs boson search at hadron colliders. Phenomenological
studies have been done on observability of the invisible Higgs in $ZH$ and
$WH$ associated production \cite{DP-invisible-lhc} and $t \bar{t} H$ production 
\cite{c8}. Assuming that the Higgs boson is produced with SM strength,
while decaying with an invisible branching fraction of $\approx 100$ \%, 
associated $ZH$ production 
was estimated to be sensitive to Higgs masses $\leq$ 150 GeV \cite{DP-invisible-lhc} at 
the LHC, while
$t \bar{t}H$ production might extend the Higgs mass range to 250 GeV
\cite{c8}. In recent work \cite{Zeppenfeld-h-invisible} 
it was shown that the LHC potential
for the search of an invisibly decaying Higgs boson can be considerably 
extended by studying Higgs production via weak boson fusion. According to
these parton level studies, 10 $fb^{-1}$ of data should allow 
to discover these particles with 
masses up to 480 GeV, at the 5$\sigma$ level, provided their invisible 
branching ratio is 1. A method for background estimation directly from 
the data has also been proposed.

The search strategy for an invisible Higgs described in \cite{Zeppenfeld-h-invisible} 
heavily 
relies on 
the performance of the ATLAS/CMS calorimetry for jets and $\sla{p}_{T}$
reconstruction as well as on a dedicated calorimeter trigger. In the study
presented here we basically repeat the analysis done in \cite{Zeppenfeld-h-invisible}, but
with a more dedicated simulation of the detectors. We also discuss possible
Level 1 and High Level triggers for the most efficient on-line selection
of invisible Higgs events. The efficiency of the basic selections proposed in 
\cite{Zeppenfeld-h-invisible} 
has been checked. Below we will refer to the following cuts on the
$E_{T}$ of tagging jets, a rapidity gap 
between two tagging jets, an effective mass of tagging jets ($M_{jj}$), 
missing transverse momentum ($\sla{p}_{T}$) and the azimuthal angle between 
two jets in the transverse plane ($\phi _{jj}$) :
\begin{eqnarray}
&E_{T}^{j1,~j2} >~40~GeV,~~|\eta _{j}| < 5.0,~~
|\eta _{j1} - \eta _{j2}|~>~4.4,~~\eta _{j1} \eta _{j2}~<~0, \\
&\sla{p}_{T}~>~100~GeV, \\
&M_{jj}~>~1200~GeV, \\
&\phi _{jj}~<~1
\end{eqnarray}
A mini-jet veto (no jet with $E_{T} >$ 20 GeV in the $\eta$ gap between
two tagging jets) and lepton veto (no lepton with $p_{T}>p_{T}^{cut}$) 
have also been used. The full set of these cuts we shall refer to, 
hereafter, as WBF cuts.

\section{Trigger on invisible Higgs}

A purely multi-jets plus $\sla{p}_{T}$ final state in the invisible Higgs
search requires a dedicated calorimeter trigger both at Level 1 and at 
High Level. Off line analysis exploits a specific feature of the two tagging 
jets accompanying Higgs production via weak boson fusion, in particular a big 
gap in rapidity between them (1). Such a requirement on the two jet topology 
could be applied already in on-line selections - at Level 1 and in High Level 
trigger (HLT). Together with a cut on calorimeter $\sla{p}_{T}$, these cuts 
allow to suppress the QCD background rate down to an acceptable level of a few 
Hz, as will be shown later.

The Forward Calorimeters of the ATLAS/CMS detectors will play a crucial 
role in the
on-line and off-line selections of invisible Higgs due to the presence of two
forward-backward tagging jets. The acceptance of the CMS Hadron Forward (HF) 
calorimeter ( 3.0 $<|\eta|<$ 5.0) for these jets with $E_{T}>$ 30 GeV is 
shown in Tab.~\ref{tab:eta_tag_jets}, before and after cut on the rapidity gap 
between jets. One can see that with the rapidity gap constraint, 
almost 80 \%  of the Higgs events will have at least 
one tagging jet in the pseudorapidity 
region covered by HF.
\begin{table}[htb]
  \begin{center}
  \caption{Acceptance (in \%) of the CMS Forward Hadron Calorimeter for 
   tagging jets ($E_{T}>$ 30 GeV) in $qq \rightarrow qqH$.}
  \label{tab:eta_tag_jets}
    \begin{tabular}{|c|c|c|c|} \hline
         & no jets in HF& one jet in HF & 2 jets in HF \\ \hline
\hline
no cut on $\mid \eta _{j1} - \eta_{j2} \mid$ & 49  & 45 &  6 \\ \hline
$\mid \eta _{j1} - \eta_{j2} \mid > $4.4     & 22  & 65 & 13 \\ \hline
\hline
    \end{tabular}
  \end{center}
\vspace*{-.6cm}
\end{table}

\subsection{Level 1 trigger}

\noindent

{\bf ATLAS}: The implementation of a specific trigger for Weak Boson Fusion
processes is still under discussion in ATLAS. Up to now the strategy
at LVL1 is to not include a trigger on jets with $|\eta| >
3.2$. However all the information on jet energy and $\sla{p}_{T}$ is
potentially available at LVL1 for all the covered regions ($|\eta| <
4.9$). 

A discussion on the implementation of a dedicated trigger for tagging
jets for the studies of the WBF channels is in progress. The invisible
Higgs channel has been used to demonstrate how such a trigger is
fundamental for such a search and, once it is implemented, 
how all the WBF channels
benefit of the trigger redundancy that will be important for precise
cross section measurements for these channels.

Using the results obtained with the off line analysis of the invisible
Higgs channel, the significance $S/\sqrt{B}$ at
an integrated luminosity $L = 10~\mathrm{fb}^{-1}$ has been evaluated
for two different regions of the jet rapidity acceptance, finding:
(a) $S/ \sqrt{B} \approx 10$,~~if~$|\eta| < 4.9$ and  
(b) $S/ \sqrt{B} \approx 4$,~~if~$|\eta| < 3.2$.  

The LVL1 hardware in ATLAS offers the capability to recognise the jet
hemisphere, therefore the data produced for the off line analysis have
been analysed with the cuts enumerated in the following to evaluate
the expected rate at LVL1 for the background processes:
(a)  2 tagging jets;  
(b) $|\eta _{j}| < 4.9$;
(c) $p_{T}^{j} > 40$~GeV; 
(d) $\eta_{j_{1}} \cdot \eta_{j_{2}} < 0$;
(e)  $\sla{p}_{T} > 85$~GeV.

The signal efficiency has been  evaluated to be about 95 \%
with these cuts for the events which passed off-line selections (1)-(4).
Table~\ref{tab:results} gives the rates for all the background channels
and for the signal for a luminosity of $L=10^{33}$cm$^{-2}$sec$^{-1}$
and the number of events for an integrated
luminosity $\int L dt = 10$ fb$^{-1}$ (corresponding to the first year
of running at LHC).
These preliminary results look very promising. Further studies are
going on to evaluate the errors on these numbers with a more accurate
simulation of the forward region.
\begin{table}[h]
    \begin{center}
    \caption{Expected rate and number of events for signal and backgrounds
during the first year of LHC operation.}
    \label{tab:results}
        \begin{tabular}{|l|c|c|}
            \hline \hline
            Process & Rate (Hz) & Nr. of events\\
            \hline \hline
        $H \to inv$ (120 GeV)  & $6.1 \cdot 10^{-4}$ & $6.1 \cdot 10^{3}$ \\ \hline 
         Wjj & $4.0 \cdot 10^{-2}$ & $4.0 \cdot 10^{5}$ \\ \hline       
        Zjj & $2.1 \cdot 10^{-2}$ & $2.1 \cdot 10^{5}$ \\ \hline        
        QCD (100-150 GeV) & $1.9 \cdot 10^{-2}$ & $1.9 \cdot 10^{5}$ \\ \hline
        QCD (150-200 GeV) & $3.3 \cdot 10^{-2}$ & $3.3 \cdot 10^{5}$ \\ \hline
        QCD (200-250 GeV) & $3.5 \cdot 10^{-2}$ & $3.5 \cdot 10^{5}$ \\ \hline
        QCD (250-300 GeV) & $2.4 \cdot 10^{-2}$ & $2.4 \cdot 10^{5}$ \\ \hline
        QCD (300-x GeV) & $4.5 \cdot 10^{-2}$ & $4.5 \cdot 10^{5}$ \\ \hline
        \end{tabular}
\end{center}
\end{table}

The trigger on the tagging jets has to be extended as much as possible
in $\eta$ to preserve a good signal to noise ratio, as demonstrated by the 
following study.
The trigger efficiency versus $\eta$ has been evaluated by measuring
the ratio between $N_{\eta}$, the number of events at a given LVL1 acceptance
$|\eta| \leq \eta_{LVL1}$,  and $N_{4.9}$, the number of events when
$|\eta| \leq 4.9$:
$$
\epsilon(\eta) = N_{\eta}/N_{4.9}
$$

In figure~\ref{fig:etaeff} the behaviour of $\epsilon$ versus $\eta$
is shown. It is clear from the plot that, reducing the acceptance
region, the signal is strongly reduced, while a lower impact is
obtained on the QCD background as well as on the $Wjj$ and $Zjj$
backgrounds. 
\begin{figure}[h]
\vspace*{-0.3cm}
\begin{center}
\mbox{\epsfig{file=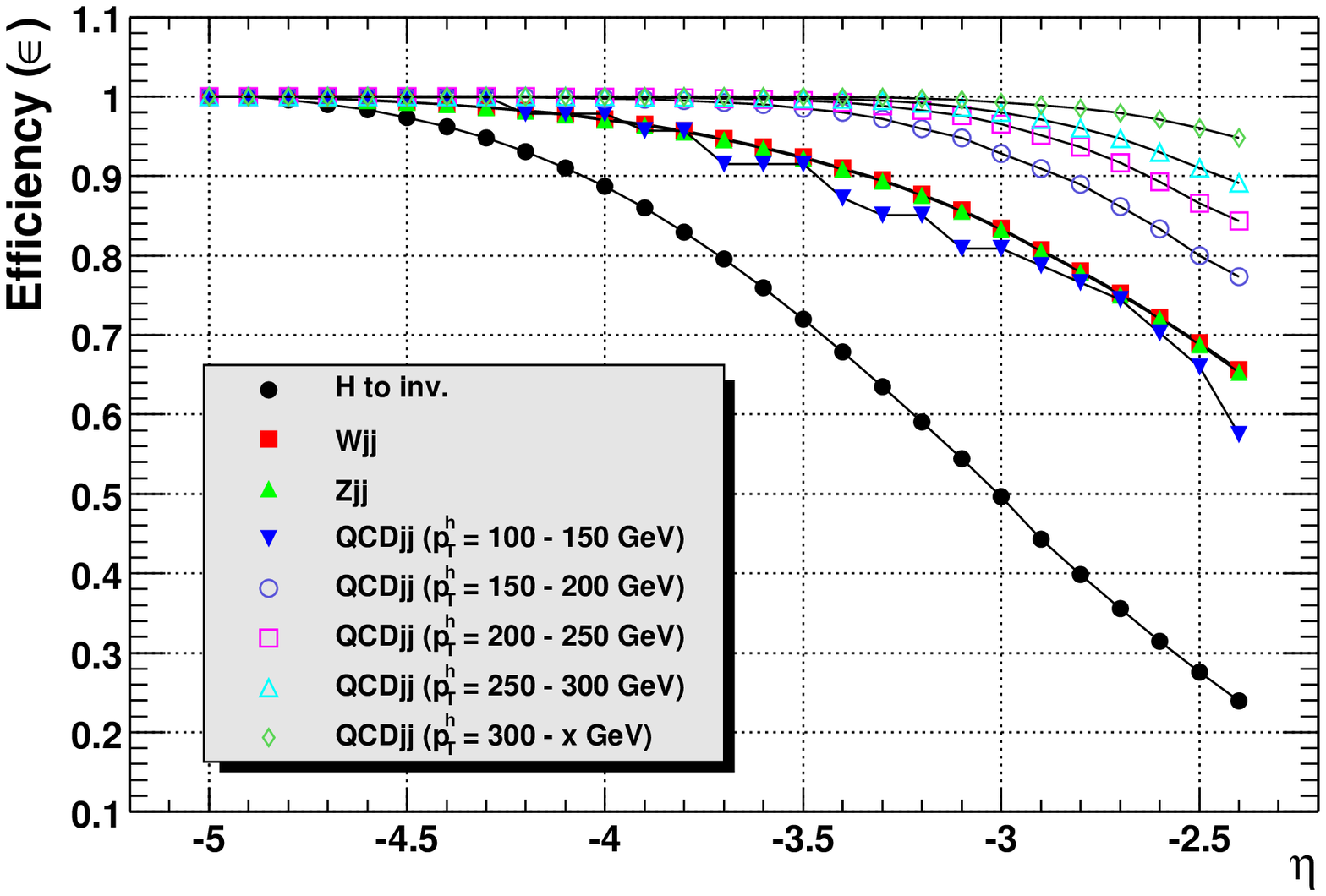,width=0.8\linewidth}}
\end{center}
\vspace*{-0.8cm}
\caption{Level 1 efficiency versus $\eta$ coverage of Level 1 calorimeter 
  trigger.}
\protect\label{fig:etaeff}
\vspace*{-0.3cm}
\end{figure}

\noindent
{\bf CMS}: CMS will have Jet triggers at Level 1 operating over the entire 
calorimeter
acceptance,
including Hadron Forward calorimeter \cite{trg_tdr}. The present Level 1
calorimeter trigger table includes a set of inclusive Jet triggers, missing
$E_{t}$ (MET) trigger as well as combined Jet+MET trigger. We have found that 
the combined Jet+MET trigger (with Jet trigger also implemented in the HF) 
is the most 
effective for the invisible Higgs selection. At low luminosity 
($L=2 \times 10^{33}$cm$^{-2}$sec$^{-1}$), with 
thresholds of 70 GeV on both Jet and MET, it provides $\simeq$ 96 \% 
efficiency for the events selected with off-line cuts (1)-(4) and 
at an acceptable background rate of 0.6 kHz \cite{sasha,dasu,pamela}. 
Such a high 
efficiency  can be understood from Fig.~\ref{fig:max_jet} and  
Fig.~\ref{fig:l2met} where transverse energy of the highest $E_{t}$ jet and 
calorimeter $\sla{p}_{T}$ reconstructed in off-line and at Level 1 is shown 
for the events which passed the off-line selections (1)-(4). 
In the off-line reconstruction both $E_{t}$ 
of jet and $\sla{p}_{T}$ are corrected for the effects of calorimeter 
non-linearity. Jet energy corrections are also applied at Level 1, 
while it is not foreseen to correct $\sla{p}_{T}$ at Level 1. Due to this
the Level 1 $\sla{p}_{T}$ spectrum shown in Fig.~\ref{fig:l2met} is shifted
in comparison with off line $\sla{p}_{T}$.
\begin{2figures}{hbtp}
\vspace*{-0.8cm}
  \resizebox{\linewidth}{!}{\includegraphics{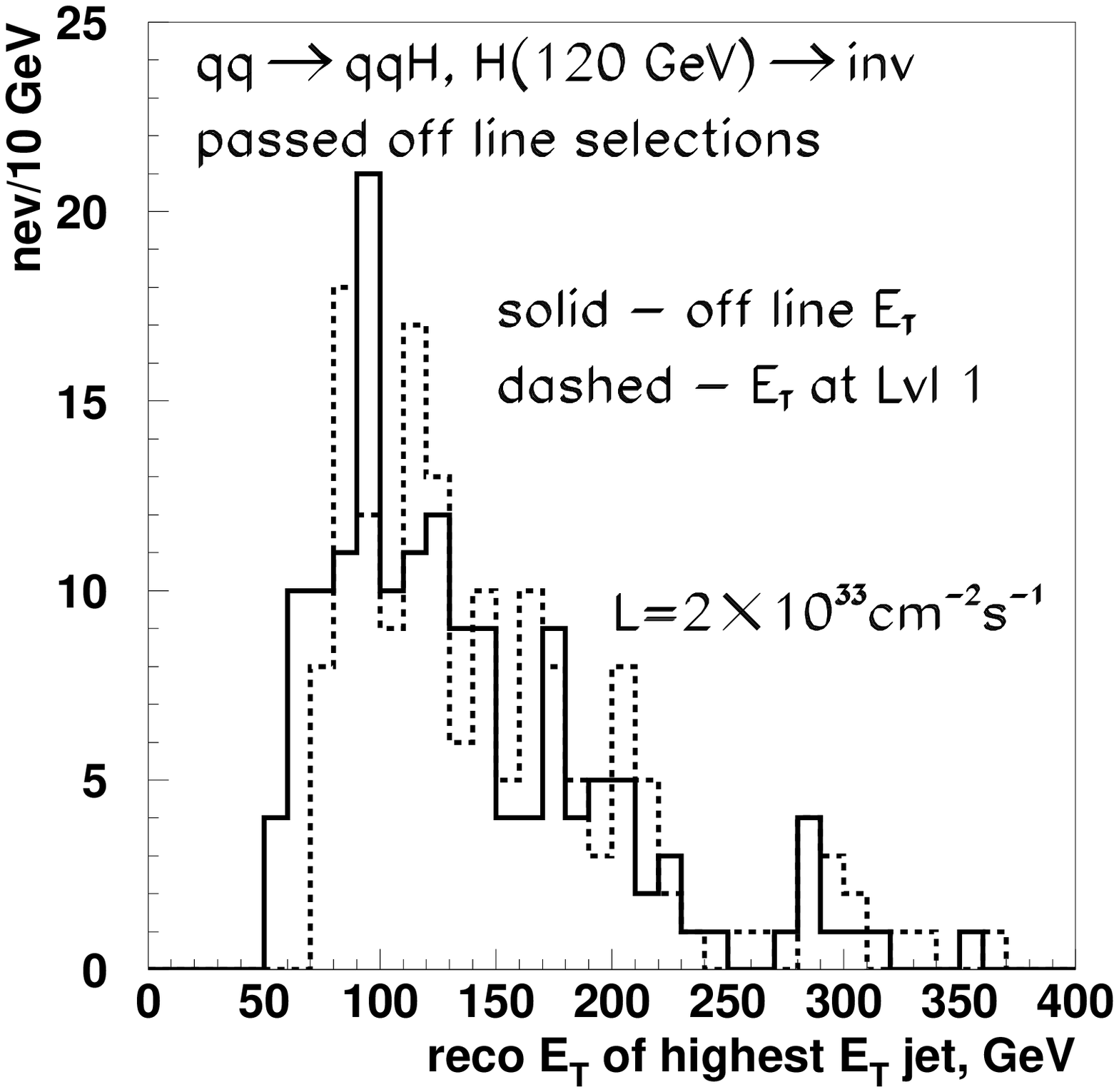}} & \vspace*{-0.8cm}
  \resizebox{\linewidth}{!}{\includegraphics{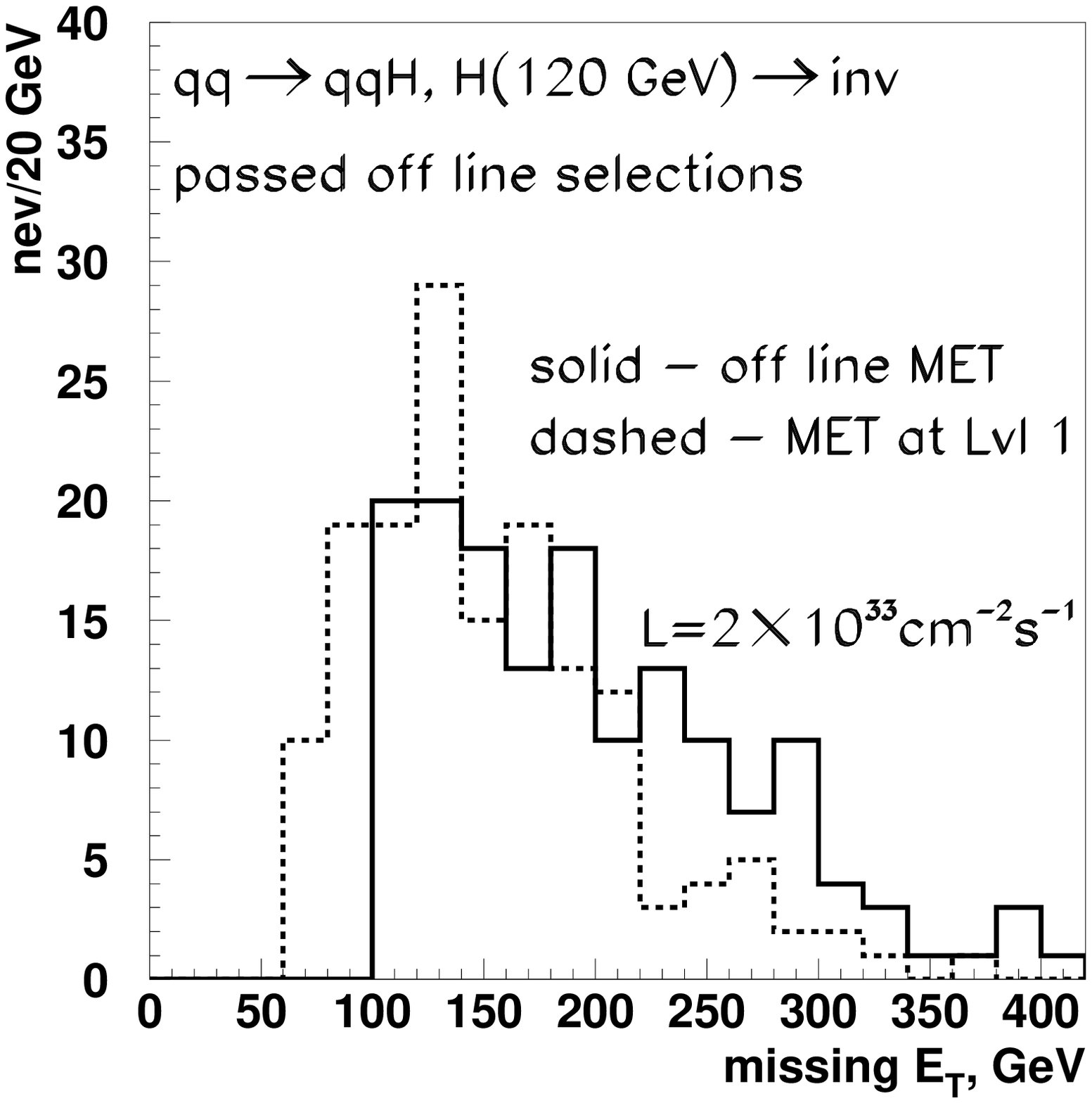}} \\[-0.6cm]
  \caption{Reconstructed $E_t$ of the highest $E_{t}$ 
           jet for the Higgs events passing cuts (1)-(4). Solid
           histogram - off line reconstruction, dashed histogram -
           reconstruction at Level 1.} 
  \label{fig:max_jet} &
  \caption{Calorimeter $\sla{p}_{T}$ reconstructed at Level 1 (dashed
           histogram) and off line (solid histogram) for
           the Higgs events passing cuts (1)-(4).} 
  \label{fig:l2met}
\vspace*{-1.9cm} 
\end{2figures}

\subsection{High Level Trigger}

{\bf CMS}: 
At High Level Trigger the off-line requirement (1) on the pseudorapidity 
gap between the two highest $E_{t}$ jets can be exploited together with the 
cut on 
$\sla{p}_{T}$. Full granularity calorimeter information is available at 
HLT and computer farms will perform jet and $\sla{p}_{T}$ 
reconstruction like in off-line analysis. In Fig.~\ref{fig:hltmet_rate} the 
rate of QCD multi-jet events for $L=2 \times 10^{33}cm^{-2}s^{-1}$ after 
cuts (1) is shown as a function of the cutoff on $\sla{p}_{T}$. This plot has 
been obtained with full detector \cite{cmsim} and reconstruction 
\cite{orca} simulations 
processing QCD multi-jet events with $\hat{p_{t}}$ bins from 15-20 GeV up to 
2600-3000 GeV (about 1M events). One can see that a cut on $\sla{p}_{T}$ above 
80 GeV will reduce the rate below 1 Hz. 
\begin{figure}[htp]
\vspace*{-0.3cm}
  \begin{center}
    \resizebox{10cm}{!}{\includegraphics{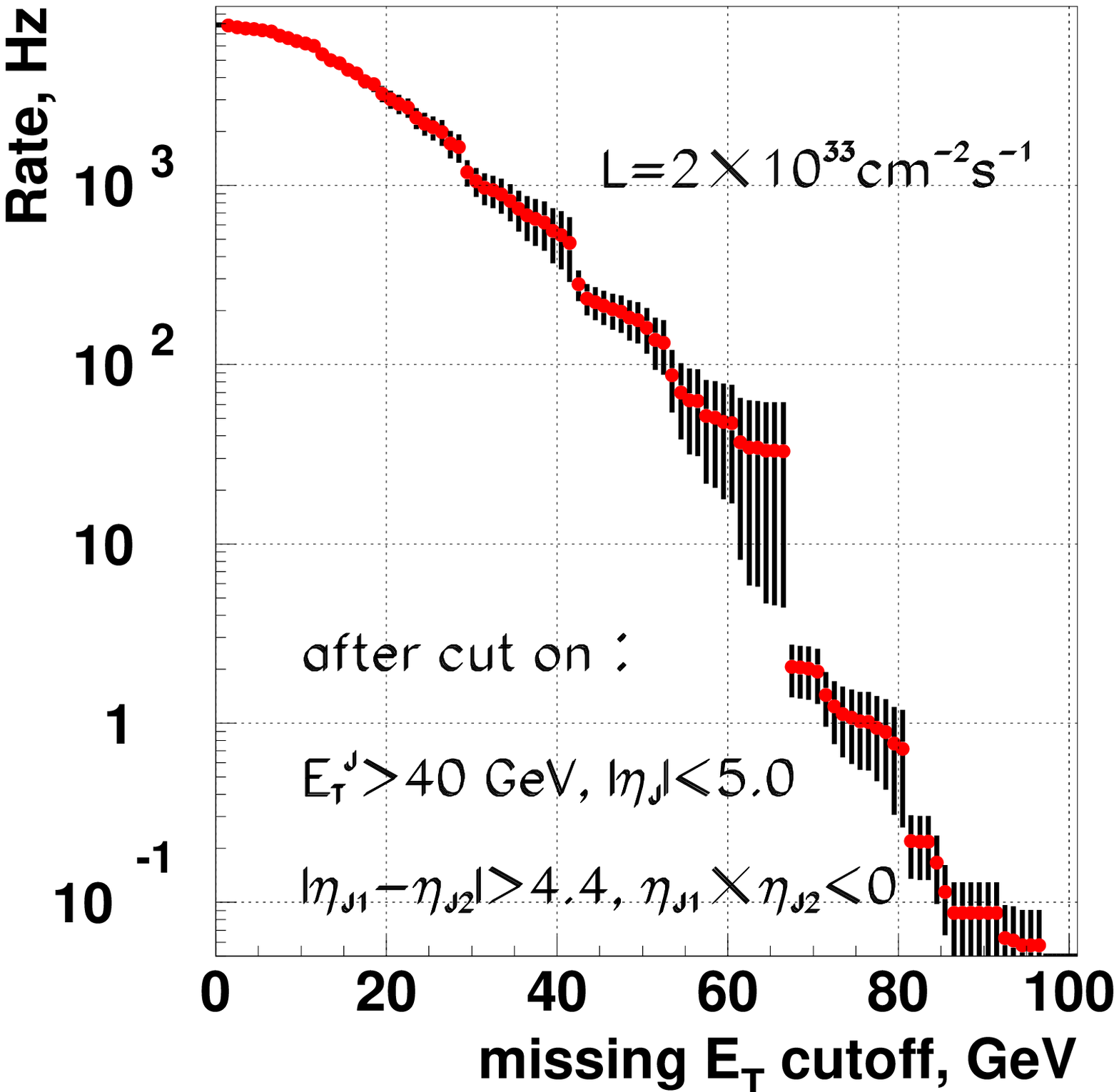}}
\vspace*{-0.3cm}
    \caption{QCD multi-jet background rate after cuts (1) as a function of
             the threshold on $\sla{p}_{T}$.}
    \label{fig:hltmet_rate}
  \end{center}
\vspace*{-0.8cm}
\end{figure}

\section{Results of off-line analysis with detector simulation}

We have performed off-line analysis, partially with full and partially with
fast
ATLAS/CMS detector simulation programs, using mainly the selection
criteria (1)-(4) together with mini-jet veto and lepton veto. As  
backgrounds, QCD multi-jet production and QCD and electroweak (EW) W and Z 
production with more than 1 reconstructed jet have been considered. 
Below, we separately provide two independent ATLAS and CMS analyses. 

\subsection{CMS analysis}

{\bf Kinematics simulation}: 
Signal events have been generated with PYTHIA6.158 (with CTEQ5L 
structure functions) \cite{pythia} for a SM Higgs of mass $M_{H}$=120 GeV,
produced via weak boson fusion.
All backgrounds (except QCD multi-jet production) have been simulated with LO 
matrix elements interfaced with PYTHIA for hadronization and additional 
initial and final state radiation. Colour and flavour information
at the parton level is passed to PYTHIA. EW Zjj and Wjj events were
generated with COMPHEP 
\cite{comphep} (with CTEQ5L); QCD Zjj and Wjj events (generated with CTEQ4L) 
were generated with MadCUP programs \cite{MadCUP} based on the work of
Ref.\cite{HZ}. Loose selection criteria have been used 
to produce events at the parton level with tree level matrix elements:
\begin{eqnarray}
&p_{T}^{j} > 20 GeV,~~|\eta _{j}| < 5.0,~~
|\eta _{j1} - \eta _{j2}| > 4.2,~~\eta _{j1} \eta _{j2} < 0, \\
&M_{jj} > 900 GeV,
\end{eqnarray}
These events are further processed through PYTHIA.
Initial cross sections (in pb) given by the matrix element calculations with 
these cuts are presented in Tab.~\ref{tab:init_cross}. Cross sections 
include $Br(Z \rightarrow \nu \nu)$ and $Br(W \rightarrow \ell\nu)$ for
three lepton generations.
\begin{table}[htb]
  \begin{center}
  \caption{Cross sections (in pb) for backgrounds as given by LO matrix
           element calculations with cuts (5), (6). 
           $Br(Z \rightarrow \nu \nu)$ and 
           $Br(W \rightarrow \ell \nu)$ is included.}
  \label{tab:init_cross}
    \begin{tabular}{|c|c|c|c|} \hline
 QCD W+jj & QCD Z+jj & EW W+jj & EW Z+jj \\ \hline
\hline
  76.0    &  15.7    &  4.7   & 0.644    \\ \hline
\hline
    \end{tabular}
  \end{center}
\vspace*{-.5cm}
\end{table}

\noindent {\bf Detector simulation:}
Full detector simulation has been performed for the Higgs and QCD multi-jet
events at $L=2 \times 10^{33}cm^{-2}s^{-1}$ (on average 3.4 minimum bias 
events of PYTHIA MSEL=1 have been superimposed). One of the crucial questions 
of this study is a proper simulation of 
the tails in the $\sla{p}_{T}$ distribution of the QCD multi-jet background. 
Such tails
could be due to real $\sla{p}_{T}$ from heavy quarks decays, but also
due to a number of detector effects. To make confident estimates of 
such effects we used about 1 million QCD events, fully simulated 
\cite{cmsim} and reconstructed \cite{orca}. As mentioned already, 
events have been generated in different $\hat{p_{t}}$ bins  
from 10-15 GeV ($\sigma=8.868 \times 10^{12}$ fb) up to 2600-3000 GeV
($\sigma=11.25 fb$). However, as will be shown later, this statistics is
still not enough to directly prove that the QCD background could be suppressed 
to an acceptable level after all cuts are applied. 

The other backgrounds, QCD and EW production of Wjj and Zjj have been 
simulated with CMSJET \cite{cmsjet} fast simulation with no minimum bias
events superimposed.  

Another key point of all searches for a light Higgs produced via weak boson 
fusion is the use of a mini-jet veto, namely a veto of events with 
additional soft 
($E_{T} >$ 20 GeV) jet(s) inside the rapidity gap between two tagging jets. 
The efficiency of the mini-jet veto is expected to be sensitive to
detector effects like calibration, electronic noise and readout thresholds, 
interaction of soft particles in the tracker in front of the calorimeter, 
magnetic field, or pile up activity. 
Since we did not expect that the fast CMSJET simulation can properly reproduce
some of these effects, we did not evaluate mini-jet veto 
efficiency from CMSJET simulation. Instead we multiply the background
efficiency by $P_{surv}$ as estimated 
in \cite{Zeppenfeld-h-invisible}. $P_{surv}$ calculated in
\cite{dr} is a probability to radiate a jet (parton) in the 
rapidity gap between two tagging jets. In the parton level study of \cite{Zeppenfeld-h-invisible} 
it has been assumed that such jets will be reconstructed with 100 \% 
efficiency. CMS full simulation study on soft jet reconstruction 
\cite{soft_jet} shows that with a dedicated window algorithm it is possible 
to reconstruct 20 GeV jets at low luminosity with reasonably good purity and 
about 100 \% efficiency. The question of the false jet suppression is still
under investigation. 

\noindent
{\bf Results on QCD background:}
Fig.~\ref{fig:met_100gev} shows the $\sla{p}_{T}$ distribution of the
QCD jet background (blue empty histogram) and of the Higgs signal, for
$M_{H}$=120 GeV, (red full histogram) after cuts (1),(3). With an additional 
cut (4), $\sla{p}_{T}$ for the signal events is shown as the light green 
histogram in Fig.~\ref{fig:met_100gev}. One observes that the tail in 
the background distribution goes well beyond 100 GeV. 
In Fig.~\ref{fig:met_100gev}
\begin{figure}[htp]
\vspace*{-0.3cm}
  \begin{center}
    \resizebox{10cm}{!}{\includegraphics{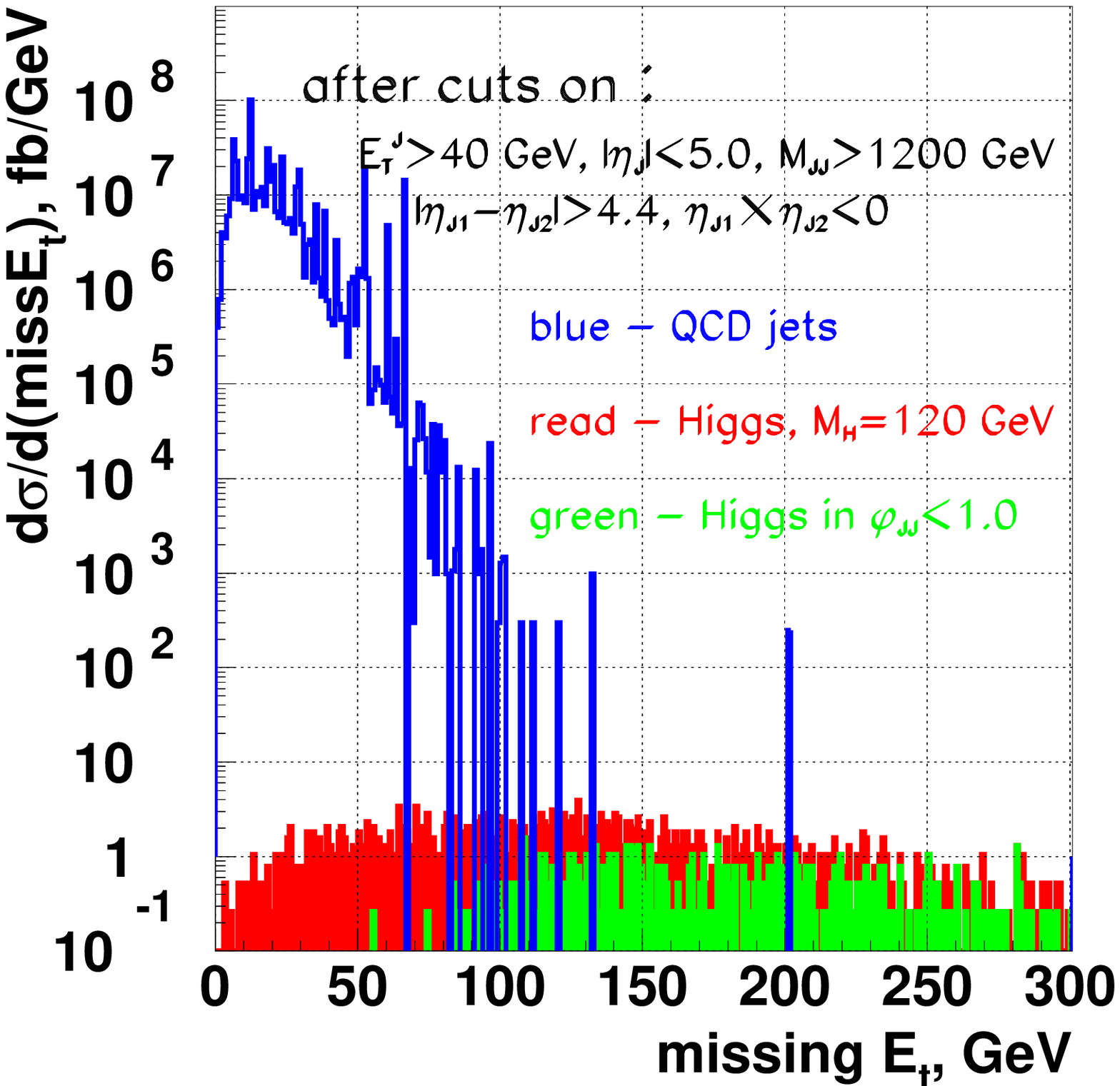}}
\vspace*{-0.3cm}
    \caption{$\sla{p}_{T}$ for Higgs of $M_{H}$=120 GeV and QCD multi-jet 
             background after selections (1),(3).}
    \label{fig:met_100gev}
  \end{center}
\vspace*{-0.5cm}
\end{figure}
QCD events in the tail come from $\hat{p_{t}}$ bins between 300 and 600
GeV. Once the cut (4) on $\phi _{jj}$ is applied, no background event 
with $\sla{p}_{T} >$ 100 GeV is left. 
With the statistics used in the analysis, this 
leads to an upper limit of $\simeq$ 1 pb on the QCD
background contribution which is about of 10 times higher than the signal
expected after the same selections (1)-(4). The ATLAS fast simulation study 
(see below), which uses much higher statistics, shows that the QCD background
can be suppressed to a negligible level with cuts (1)-(3) plus a cut on
the minimal angle in the transverse plane between $\sla{p}_{T}$ and a jet, or
with cuts (1)-(4).

\noindent
{\bf Results on Higgs signal and QCD and EW Z+jj, W+jj backgrounds:}

Estimated cross-sections (in fb) for the Higgs and backgrounds at different
steps of the event selection are shown in Tab.~\ref{tab:bkg_sign}. Numbers in
parentheses are the estimates obtained in \cite{Zeppenfeld-h-invisible}. Standard Model production
cross-sections and Br($H \rightarrow$ invisible) = 1 are assumed. \\
\begin{table}[htb]
  \begin{center}
  \caption{Cross sections in fb for the background and Higgs of 
           $M_{H}$=120 GeV assuming Br($H \rightarrow$ invisible)=1
           and Standard Model production cross-section for the Higgs.
           Numbers in parentheses are results from~\cite{Zeppenfeld-h-invisible}.} 
  \label{tab:bkg_sign}
    \begin{tabular}{|c|c|c|c|c|c|} \hline
cross section, fb & Higgs& QCD $Zjj$& QCD $Wjj$& EW $Zjj$& EW $Wjj$ \\ \hline
\hline
after cuts (1)-(3) and e($\mu$) veto for Wjj 
    &238(274) & 857(1254)& 1165 (1284) &141.5(151)& 145.1(101)\\ \hline
 + mini-jet veto &180(238) &240(351)& 237 (360)&116(124)& 84.5 (83) \\ \hline
 + $\phi _{jj}\le$ 1&74.7(96.7)&48.0(71.8)& 40.0 (70.2)&12.8(14.8)& 8.7(9.9) \\ \hline
\hline
    \end{tabular}
  \end{center}
\vspace*{-.5cm}
\end{table}
The first row of 
Tab.~\ref{tab:bkg_sign} presents cross sections after cuts (1)-(3) and
a veto on identified electrons (muons) for the Wjj backgrounds 
(including e($\mu$) from
$\tau$ decay in $W \rightarrow \tau + \nu$) with $p_{T}^{e(\mu)}>$10 (5) GeV
and $|\eta ^{e(\mu)}|<$2.5. The lepton veto in \cite{Zeppenfeld-h-invisible} includes a veto on
$\tau$ leptons with  $p_{T}^{\tau}>$20 GeV and $|\eta ^{e, \mu, \tau}|<$2.5. 
Here we discuss the veto on taus separately as a lepton veto or a jet veto,
dependent on whether the tau decays leptonically or hadronically.
Fig.~\ref{fig:inv_plot1} and Fig~.\ref{fig:bkg_plot1} show $\sla{p}_{T}$ 
distributions for the signal and background events after cuts (1) and (3) 
and e, $\mu$ veto for the $Wjj$ backgrounds. 
\begin{2figures}{hbtp}
\vspace*{-0.3cm}
  \resizebox{\linewidth}{!}{\includegraphics{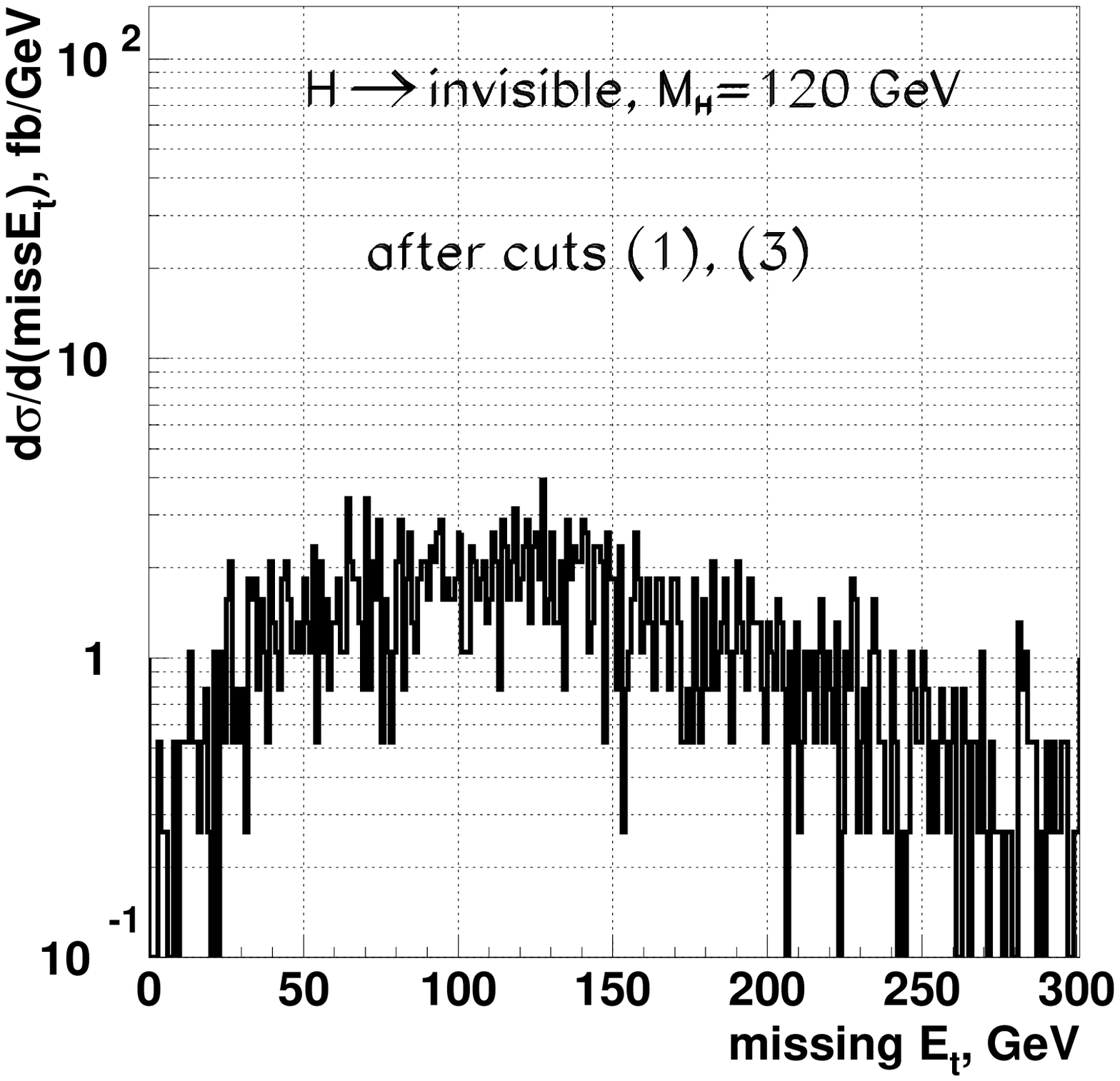}} &
\vspace*{-0.3cm}
  \resizebox{\linewidth}{!}{\includegraphics{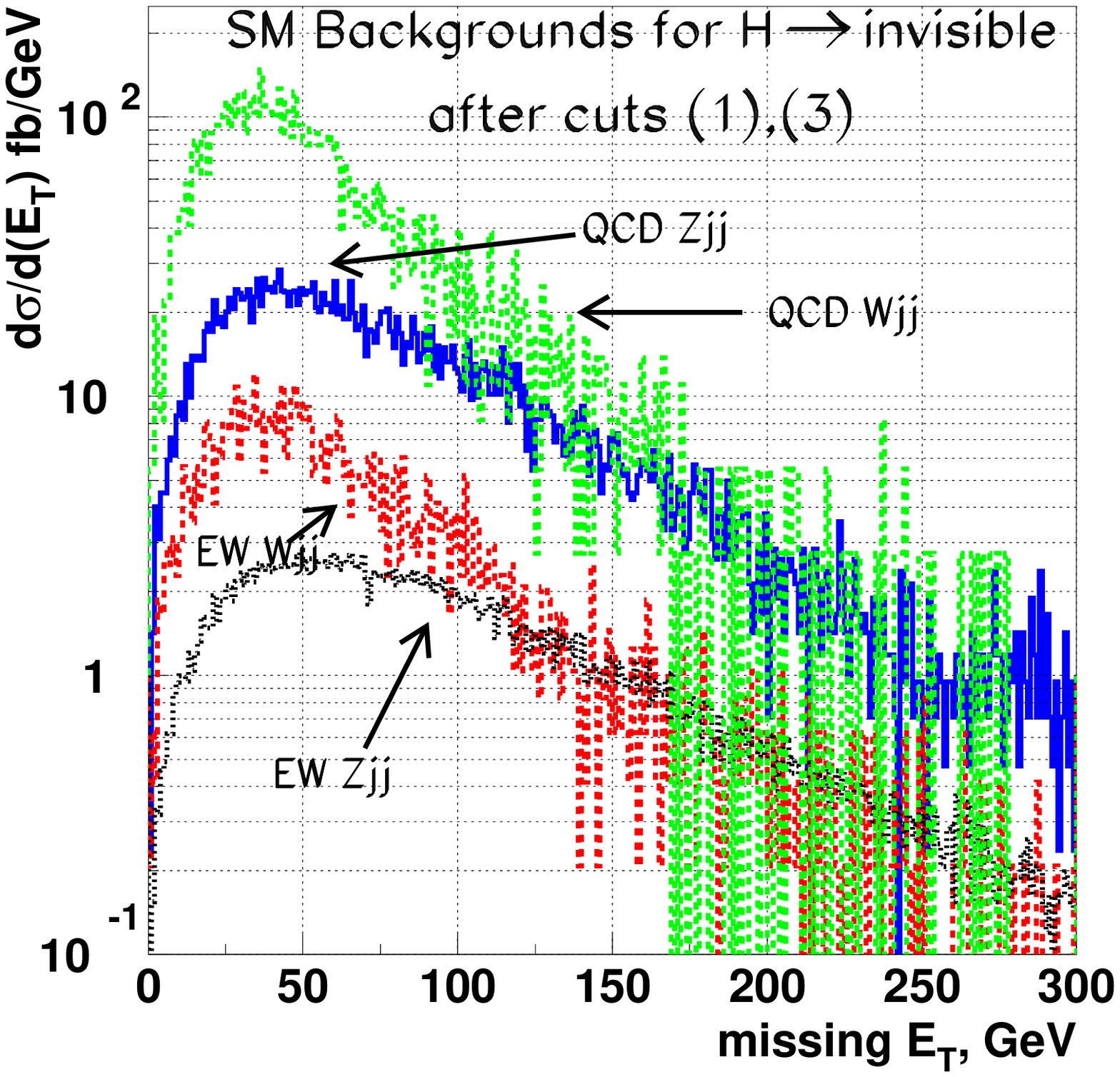}} \\[-0.6cm]
  \caption{$\sla{p}_{T}$ for Higgs of $M_{H}$=120 GeV after cuts (1), (3)} 
  \label{fig:inv_plot1} &
  \caption{$\sla{p}_{T}$ for backgrounds after cuts (1), (3) and e($\mu$) 
           veto for $Wjj$ background} 
  \label{fig:bkg_plot1}
\vspace*{-.7cm}
\end{2figures}


The second row of Tab.~\ref{tab:bkg_sign} presents cross-sections after 
mini-jet veto. As has been mentioned, the efficiency of the mini-jet veto 
for the backgrounds is taken from \cite{Zeppenfeld-h-invisible}. For the Wjj background with
$W\to \tau\nu$ and hadronic tau-decay, the veto on the $\tau$ jet 
is included in this second row.
The efficiency of the $\tau$ jet veto is estimated with fast CMSJET 
simulation, counting events with a reconstructed energy of the 
$\tau$ jet greater than 20 GeV. A mini-jet veto efficiency of 0.76 for the 
signal 
is obtained from the full simulation. It is lower than the efficiency of
0.87 used in \cite{Zeppenfeld-h-invisible}. This may be due to the reconstruction of additional 
soft jets from minimum bias events. Since we plan to use tracker 
information to suppress such contributions, our estimate of the mini-jet 
veto efficiency for the Higgs events is conservative. 

The last row of the Tab.~\ref{tab:bkg_sign} presents cross-sections after all 
selection cuts. Fig.~\ref{fig:inv_plot3} and Fig.~\ref{fig:bkg_plot3} show 
\begin{2figures}{hbtp}
\vspace*{-0.3cm}
  \resizebox{\linewidth}{!}{\includegraphics{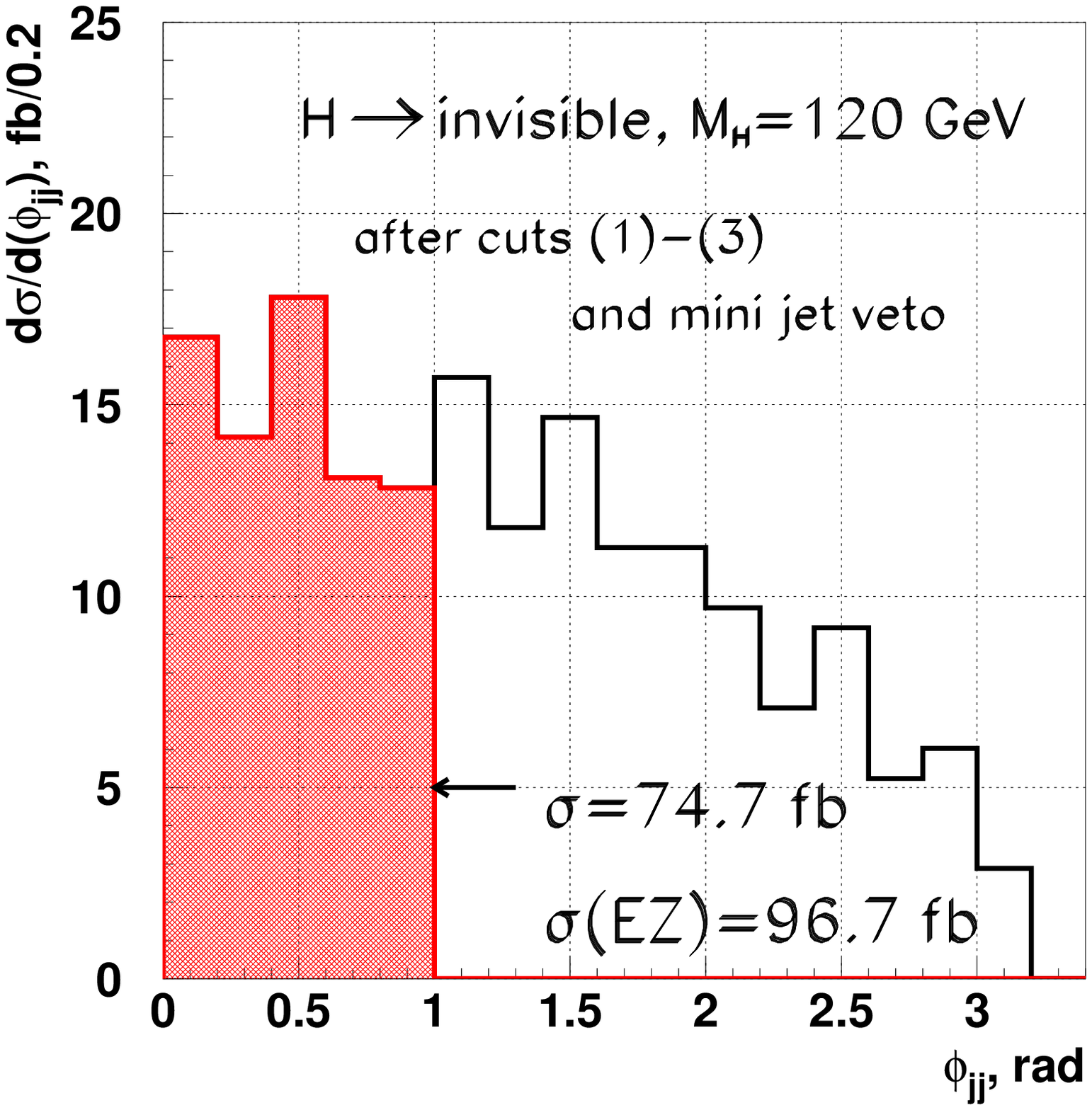}} &
\vspace*{-0.3cm}
  \resizebox{\linewidth}{!}{\includegraphics{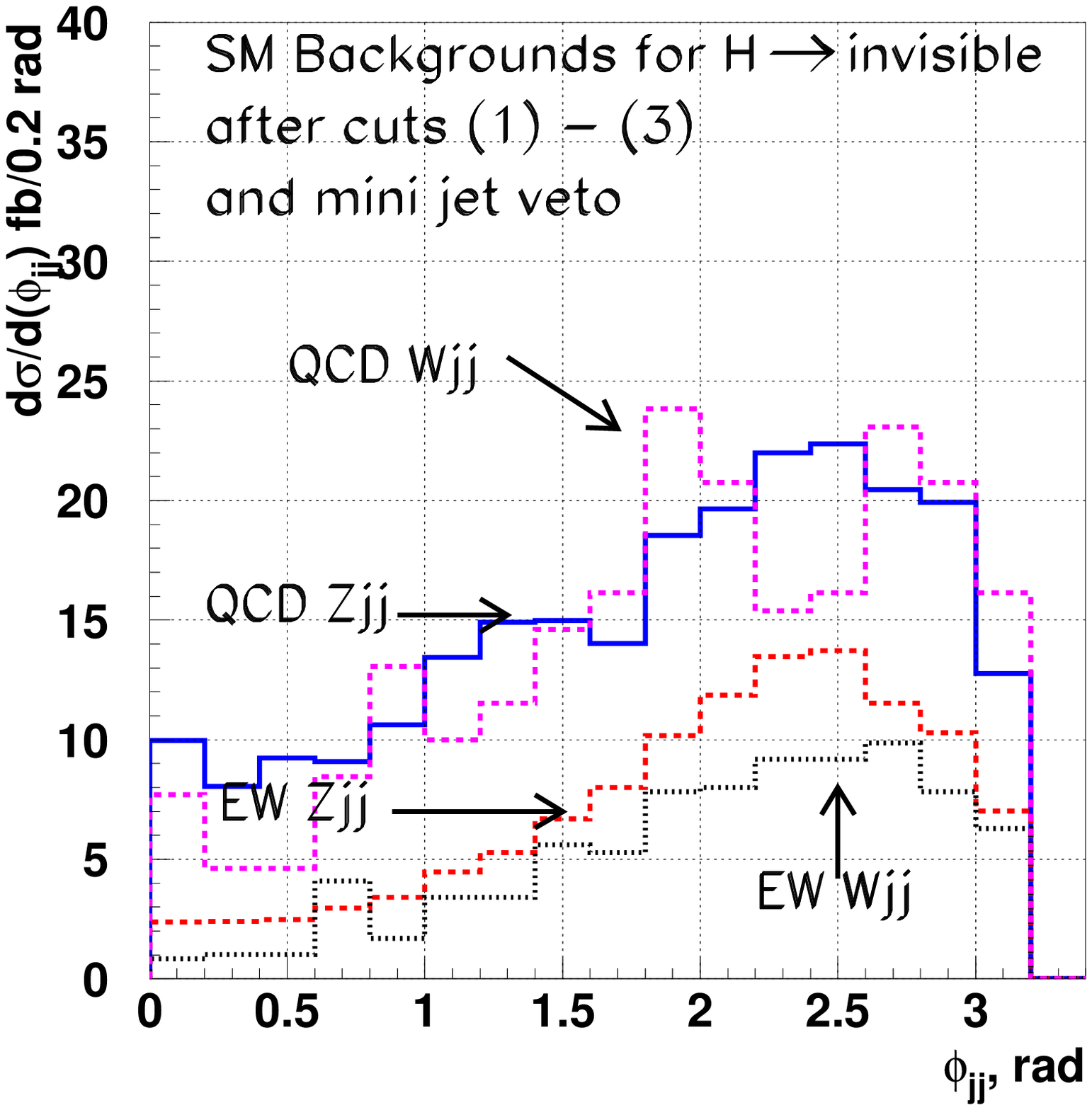}} \\[-0.6cm]
  \caption{$\phi _{jj}$ for Higgs of $M_{H}$=120 GeV after cuts (1)-(3) and
           mini-jet veto} 
  \label{fig:inv_plot3} &
  \caption{$\phi _{jj}$ for backgrounds after cuts (1)-(3), e, $\mu$ veto 
           for Wjj background and mini-jet veto} 
  \label{fig:bkg_plot3}
\vspace*{-.7cm}
\end{2figures}
$\phi _{jj}$ distributions for the Higgs and background events after 
cuts (1)-(3), e($\mu$) veto for Wjj background and mini-jet veto.
After all selections are applied, Tab.~\ref{tab:bkg_sign} shows good agreement
between our estimates and the ones obtained in \cite{Zeppenfeld-h-invisible} (except for a 
40 \% lower Wjj background).  Our simulations therefore confirm the 
conclusion reached in \cite{Zeppenfeld-h-invisible}: the LHC potential in the search 
for an invisibly decaying Higgs boson can be considerably extended by studying 
the weak boson fusion channel.

\subsection{ATLAS analysis}

\noindent
{\bf Kinematics and detector simulation:}
Higgs production, QCD multi-jet and QCD Zjj and Wjj production have been
generated with PYTHIA6.158 \cite{pythia}. QCD multi-jet 
events were generated with PYTHIA MSEL=1 subprocesses and for $\hat{p_{t}}$ 
in the interval 50-300 GeV, divided in bins of 50 GeV, and for 
$\hat{p_{t}}>$ 300 GeV. QCD Zjj (Wjj) backgrounds have been produced 
switching on the processes 15, 30 (16, 31) with $\hat{p_{t}}>$ 30 GeV. 
These processes generate Z(W)+1 parton only and additional jets are produced
due to initial and final state radiation (ISR, FSR) and fragmentation.
Processes 123 and 124 have been used to generate Higgs production via 
weak boson fusion.

Fast detector simulation with the ATLFast \cite{atlfast} package was 
performed both for the signal and all backgrounds.

\noindent
{\bf QCD background rejection:}
After cut (1) on the tagging jet topology, the QCD multi-jet background is 
about of factor $10^{4}$ larger than the other backgrounds.
Cuts (1)-(3), mini-jet 
and lepton veto suppress this background to the level of QCD Wjj and Zjj 
backgrounds. Considerable contributions of QCD multi-jet events are still 
expected in the region of $\sla{p}_{T}>$ 100 GeV, as shown in 
Fig.~\ref{fig:Ptmiss_NoIsolation}.
%
%
\begin{figure}[htp]
\vspace*{-0.3cm}
\begin{center}
\includegraphics[width=12.cm]{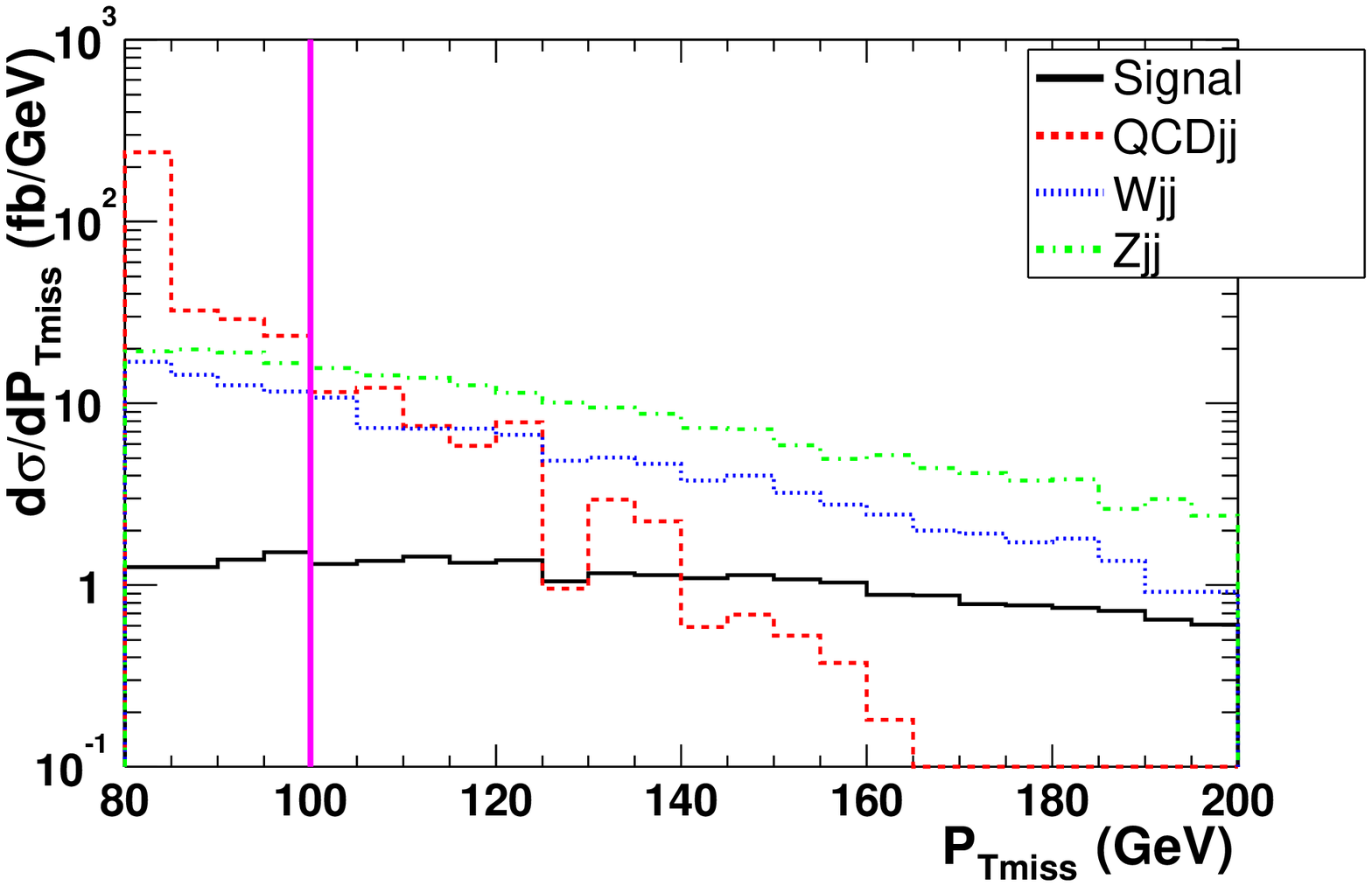}
\vspace*{-0.3cm}
\caption{Missing energy distribution after cuts (1), (3), mini-jet and lepton
veto.
\label{fig:Ptmiss_NoIsolation}}
\end{center}
\vspace*{-0.7cm}
\end{figure}
The tail of the $\sla{p}_{T}$ distribution in QCD multi-jet events is mainly 
due to 
semi-leptonic decays within jets ($>70\%$ of b-jets).  Therefore the missing 
energy is carried by the jet whereas it is carried by the Higgs for the 
signal. 

\begin{figure}[htp]
\vspace*{-0.3cm}
\begin{center}
\includegraphics[width=12.cm]{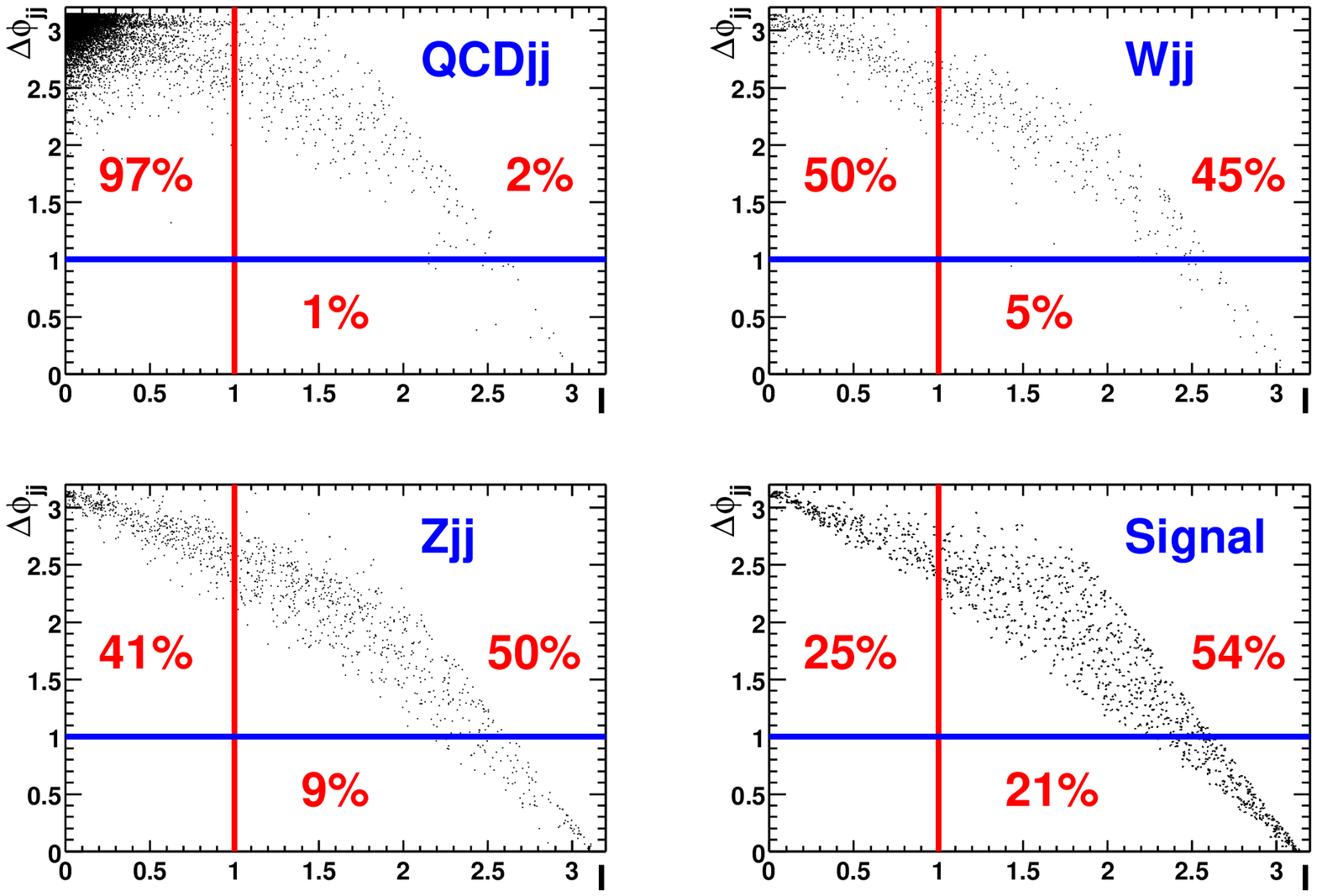}
\caption{Event distribution in $\phi_{jj}$ versus Isolation variable 
for backgrounds and 
signal.\label{fig:Isolation}}
\end{center}
\vspace*{-0.7cm}
\end{figure}
An additional cut is used to ensure that the QCD multi-jet background
is well under control. We define an isolation variable $I$  as the
minimal angle in the transverse plane between $\sla{p}_{T}$ and the tagging
jets: $I=min(|\phi(\sla{p}_{T})-\phi(j_{1,2})|)$.  The scatter plots in 
Fig.~\ref{fig:Isolation} show the correlation between $\phi_{jj}$ and
$I$ for the 
signal and the different backgrounds. A cut on $I>1$ removes 97\% of 
the QCD multi-jet background at the price of a 25\% loss of the signal, but it 
does not affect the $\phi_{jj}$ region which is important for the 
final counting.  
It may affect however, the accuracy in the prediction of the Zjj and Wjj 
backgrounds from the experimental data since it reduces the number of useful 
Zjj ($Z \rightarrow 2e,~2\mu$) and Wjj ($W \rightarrow e(\mu)+ \nu$) events by 
about of 40-50 \% as one can see in Fig.~\ref{fig:Isolation}. 
After selection of events with $I>1$ the QCD mini-jet contribution
in the region of  $\sla{p}_{T}>$ 100 GeV becomes negligible as shown
in Fig.~\ref{fig:Ptmiss_Isolation}.
\begin{figure}[htp]
\begin{center}
\includegraphics[width=12.cm]{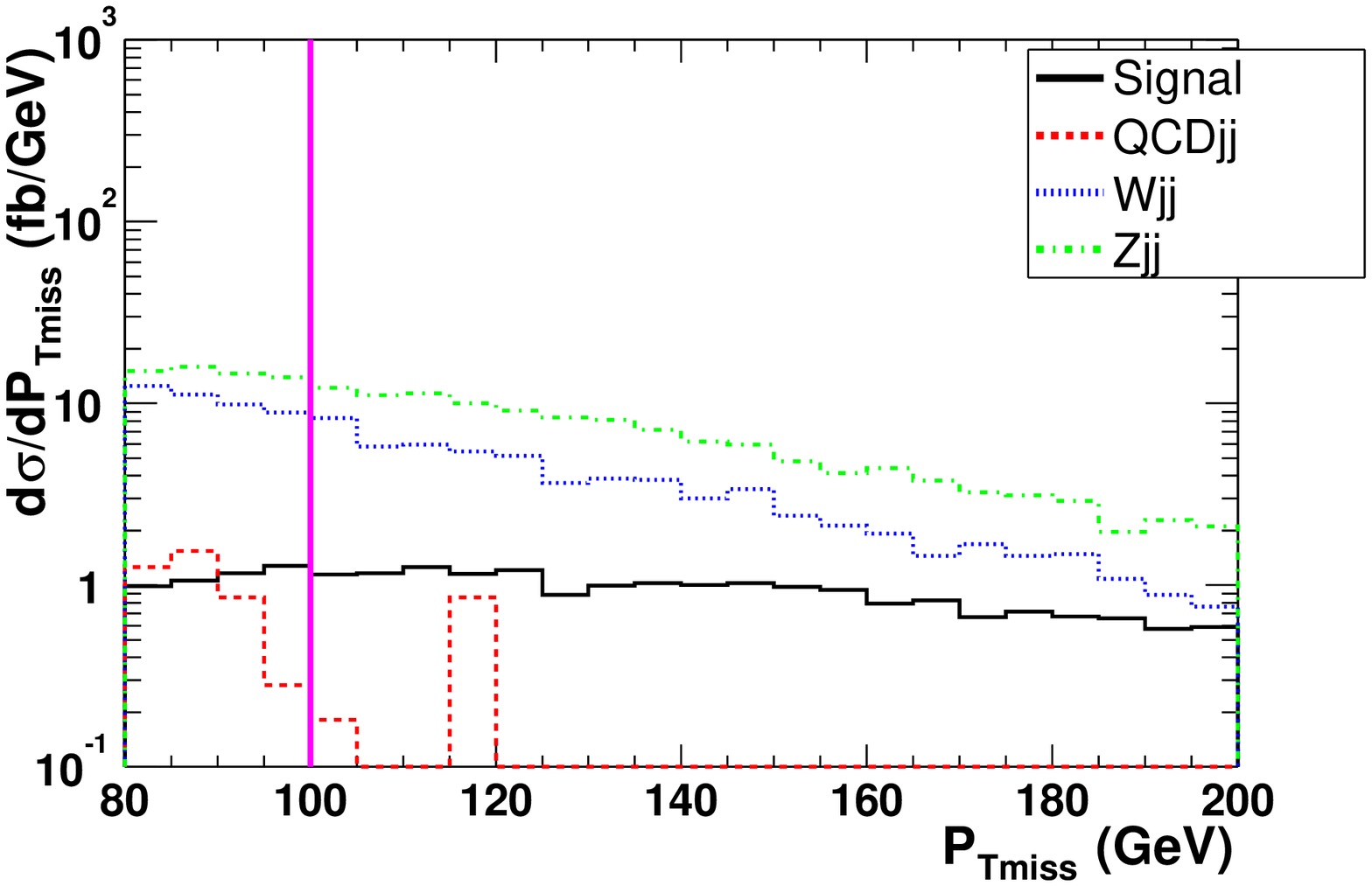}
\vspace*{-0.3cm}
\caption{Missing energy distribution after cuts (1), (3), mini-jet and
lepton veto and Isolation cut. 
\label{fig:Ptmiss_Isolation}}
\end{center}
\vspace*{-0.7cm}
\end{figure}

\noindent
{\bf Higgs signal and Wjj and Zjj background estimates:}
Tab.~\ref{tab:results1} presents the estimated cross-sections in fb for a
Higgs of $M_{H}$=130 GeV and QCD Wjj and Zjj backgrounds for the 
different selection criteria. Standard Model Higgs production cross-section 
and Br($H \rightarrow$ invisible) = 1 is assumed.   
\begin{table}[h]
    \begin{center}
  \caption{Cross sections in fb for the background and Higgs of 
           $M_{H}$=130 GeV assuming Br($H \rightarrow$ invisible)=1
           and Standard Model production cross-section for the Higgs.
           Numbers in parentheses are results from~\cite{Zeppenfeld-h-invisible}.
           \label{tab:results1}} 
        \begin{tabular}{|c|c|c|c|}
            \hline 
cross section, fb & Higgs (130 GeV) & QCD $Zjj$ & QCD $Wjj$ \\
            \hline \hline
after cuts (1)-(3) and lepton veto
                   & 187 (266) & 817 (1254)   & 899 (1284)  \\ \hline
+ mini-jet veto      & 146 (232) & 457 (351)    & 451 (360)   \\ \hline
+ $\phi_{jj} < 1$    & 60.1 (94.3) & 132.3 (71.8) & 125.6 (70.2) \\ 
            \hline  \hline
        \end{tabular}
\end{center}
\end{table}
Lepton veto in the first row of Tab.~\ref{tab:results1} includes veto of 
Wjj events with lepton ($\mu$, e, $\tau$) of 
$p_{T}^{e,~\mu ,~\tau}>$ 5, 6, 20 GeV and $|\eta ^{e,~\mu ,~\tau}|<$2.5. 
One can see that QCD backgrounds after all cuts are about of factor
1.8 larger than in \cite{Zeppenfeld-h-invisible}. It is mainly due to the fact that the rejection
factor due to the mini-jet veto obtained with this simulation is about a
factor 2 
smaller than the one used in \cite{Zeppenfeld-h-invisible}. The discrepancy in the signal is
understood as FSR in PYTHIA. For the QCD backgrounds it is known that PYTHIA 
predicts a smaller V+3jet cross section than matrix element calculations, 
when the hard process is simulated as V+1parton events. This might lead to an 
underestimate of mini-jet activity in PYTHIA. This point requires further 
study.
 
Since EW Wjj and Zjj backgrounds have not been simulated, we use the background
fractions of \cite{Zeppenfeld-h-invisible} and assume that  EW Wjj and Zjj events 
contribute $\simeq$ 20\% to the total background. 
Tab.~\ref{tab:results2} presents cross sections in fb for the total background 
and for a Higgs of $M_{H}$=130 GeV, after all cuts including the Isolation cut.
Fig.~\ref{fig:phijj} shows $\phi_{jj}$ distributions for the signal and
backgrounds after cuts (1)-(3), mini-jet and lepton veto and Isolation cut.
\begin{table}[h]
    \begin{center}
    \caption{Cross section in fb for background and Higgs of $M_{H}$= 
             130 GeV after all cuts including the Isolation cut. The 
             EW $Vjj$ background has been estimated to 
             contribute as 20 $\%$ of the total background and has been 
             added.\label{tab:results2}}
        \begin{tabular}{|c|c|c|c|}
            \hline 
cross section, fb & H (130 GeV) & $Zjj$ & $Wjj$ \\
            \hline \hline
all cuts not including cut on $\phi_{jj} < 1$ 
                            & 130  & 446   & 428   \\  \hline
with cut on $\phi_{jj} < 1$    & 60.1 & 158.  & 150.7  \\ \hline  \hline
        \end{tabular}
\end{center}
\end{table}
\begin{figure}[htp]
\vspace*{-0.3cm}
\begin{center}
\includegraphics[width=12.cm]{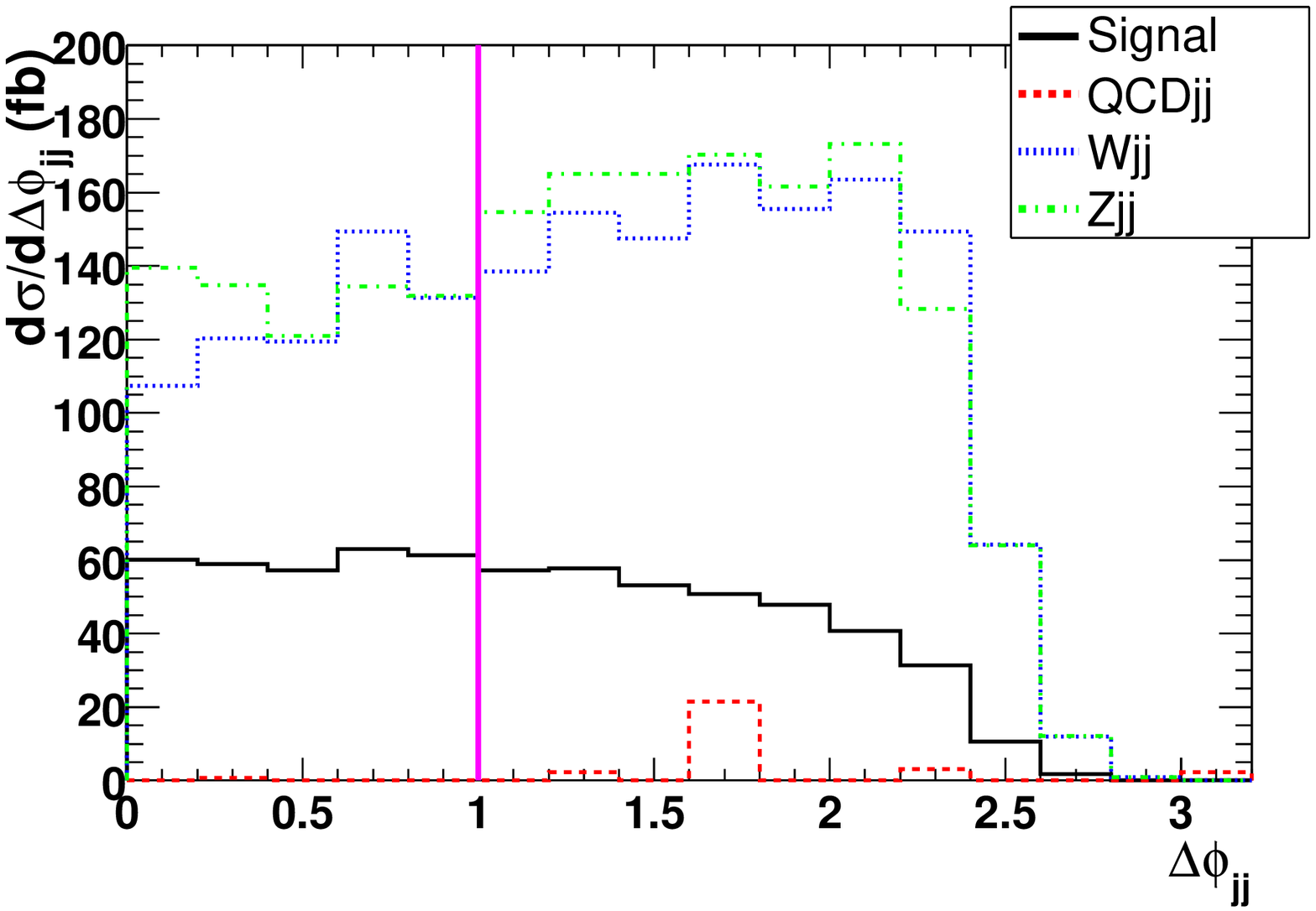}
\vspace*{-0.3cm}
\caption{$\phi _{jj}$ after cuts (1)-(3), mini-jet and lepton veto and 
Isolation cut\label{fig:phijj}}
\end{center}
\vspace*{-0.7cm}
\end{figure}

\noindent
{\bf Discovery potential:}
The observation of the invisible Higgs is fully determined by the knowledge of
background cross sections in the search area. 
At present, the leading order calculations 
for QCD Zjj and Wjj backgrounds lead to uncertainties of a factor of 3 to 4, 
depending on the renormalization scale \cite{Zeppenfeld-h-invisible}. 
However, these backgrounds could be directly predicted at LHC using 
$Z\rightarrow ll$ or $W\rightarrow l\nu$ data samples as proposed in 
\cite{Zeppenfeld-h-invisible}. Fig.~\ref{fig:Zjjprediction} shows the predicted 
\begin{figure}
\begin{center}
\includegraphics[width=12.cm]{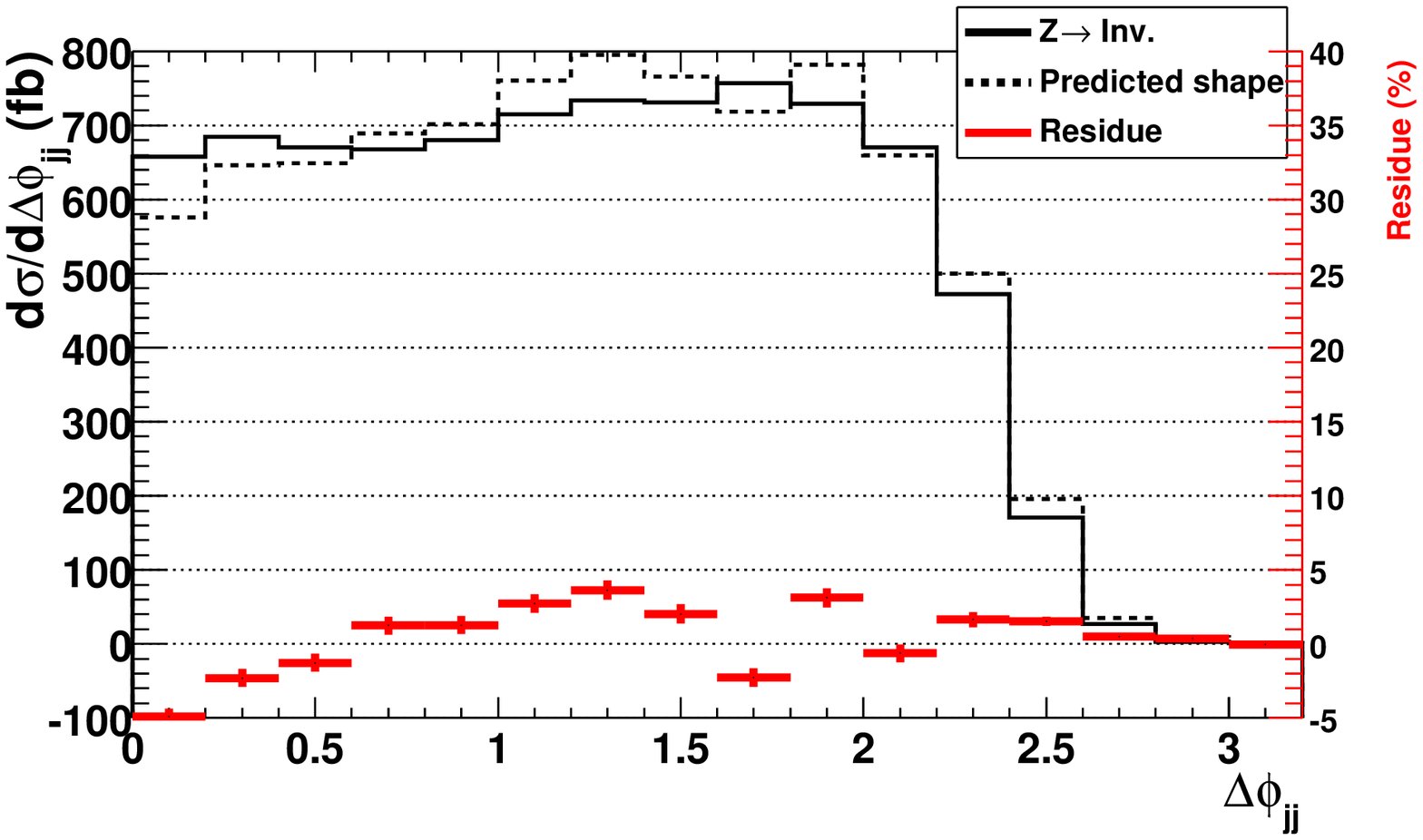}
\vspace*{-0.5cm}
  \caption{$\phi _{jj}$ distribution for the 
QCD Zjj background (solid line) and 
predicted from $Z \rightarrow e^{+}e^{-}$ events (dashed line). Cross points 
and the right y-axis are for the residue of the two distributions, defined as
$\frac{N_{Bkg} - N_{Pred}}{N_{bkg}^{tot}}$.} 
  \label{fig:Zjjprediction} 
\end{center}
\vspace*{-0.5cm}
\end{figure}
$\phi_{jj}$ distribution for the QCD Zjj background using $Z\rightarrow ee$ 
events. Residuals shown in this plot are defined as  
$(N_{Bkg} - N_{Pred})/{N_{bkg}^{tot}}$. Residuals indicate some pattern 
which is understood as the different acceptances in $\eta$ for 
$Z\rightarrow ee$  and $Z\rightarrow \nu\nu$ events. Moreover, the minimal 
lepton $p_{T}$ threshold to insure its observability induces a bias in 
the $Zjj$ prediction. Then the systematic error on the normalisation factor 
is $\sim\;2.4\%$. At NLO these uncertainties should be negligible and they are 
not considered in the following.

We define the sensitivity to invisible Higgs as 1.96 standard deviations 
(95 \% CL) from the background error which includes statistical error and 
the accuracy of the background prediction from the  ($Z \rightarrow 2l$)jj 
and ($W \rightarrow l+\nu$)jj data. The systematic error on the 
background prediction
is still under investigation, therefore we use the predicted accuracy of 3 \%
as evaluated in \cite{Zeppenfeld-h-invisible}. One should keep in mind however, that if the 
PYTHIA estimates of QCD Vjj cross sections with mini-jet veto are correct, 
then the
statistics for $lljj$ and $l \nu jj$ events, which are used for the 
background measurement, would also be a factor 2 higher. Hence the 3\% error 
would go down by a factor $\sqrt{2}$.  Tab.~\ref{tab:ksi2} and 
Fig.~\ref{fig:ksi2} show the parameter 
\begin{equation}
\xi ^{2}=Br(H\rightarrow invisible)\times\frac{\sigma _{(qq\rightarrow
qqH)}}{\sigma _{(qq\rightarrow qqH)SM}} 
\end{equation}
that can be probed at 95 \% CL as a function of $M_{H}$.
\begin{table}[h]
\begin{center}
\caption{Sensitivity to the $H \rightarrow invisible$ signal for different 
Higgs masses. The first line is the cross section after all cuts. 
The two last lines give values of $\xi^2$
which can be probed at 95$\%$ CL for an 
integrated luminosity of 10 $fb^{-1}$ and an expected background of $310$ fb 
without and with a $3\%$ uncertainty on the total background. \label{tab:ksi2}}
\begin{tabular}{|l||c|c|c|c|c|c|c|c|c|}
\hline
$M_H$ (GeV)                 & 110& 120& 130& 140& 150& 200& 250& 300& 400 
\\ \hline \hline
$\epsilon_{surv}\sigma(\phi_{jj}<1)$ (fb)&
56.2&61.1&60.1&64.6&64.2&58.8&51.2&42.5&31.0\\ \hline
$\xi^2$ ($\%$) (only stat.) &20.7&17.8&18.2&16.9&17.0&18.6&21.4&25.7&35.1 
\\ \hline
$\xi^2$ ($\%$) stat.+ 3\%   &40.4&34.8&35.4&32.9&33.1&36.2&41.7&50.1&68.5 
\\  \hline
\end{tabular}
\end{center}
\end{table}
\begin{figure}[htp]
\begin{center}
\includegraphics[width=12.cm]{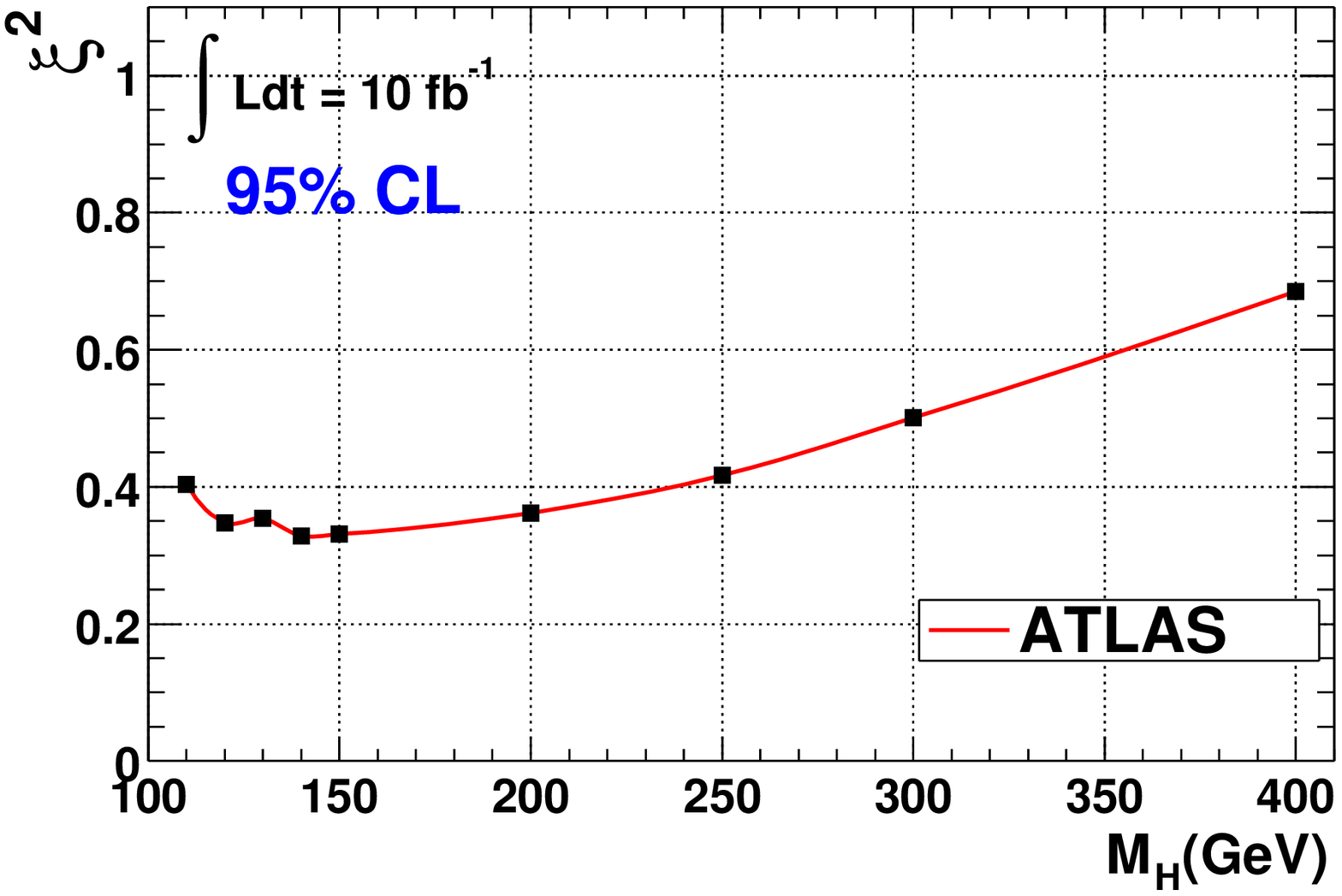}
\vspace*{-0.5cm}
\caption{ 95$\%$ CL sensitivity to $H \rightarrow invisible$ signal for an 
integrated luminosity of 10 $fb^{-1}$.\label{fig:ksi2}}
\end{center}
\vspace*{-0.9cm}
\end{figure}

We have investigated the possibility of the invisible Higgs observation in the
MSSM sector.  In this case, Higgs (h and H) production via weak boson fusion 
is suppressed by the factor 
\begin{equation}
\frac{\sigma_ {(qq\rightarrow
qqH)MSSM}}{\sigma _{(qq\rightarrow qqH)SM}}=(g _{V}^{h,H})^{2}
\end{equation}
with
$(g_V^{h})^2=sin^2(\alpha-\beta)$ and $(g_V^{H})^2=cos^2(\alpha-\beta)$.  
It means a Standard Model like production for h and a strongly suppressed 
production for H for large $M_{A}$ and $tan( \beta )$. 
Fig.~\ref{fig:MSSMproduction} shows Higgs production cross sections via 
weak boson fusion for different $tan(\beta)$ values. The black line is the 
Standard Model cross section. The marked line is the ATLAS sensitivity 
assuming that $Br(H\rightarrow invisible)=1$. The lines of different colours 
are production cross sections for MSSM h and H for different values of 
$tan(\beta )$.  
\begin{figure}
\begin{center}  
\includegraphics[width=12.cm]{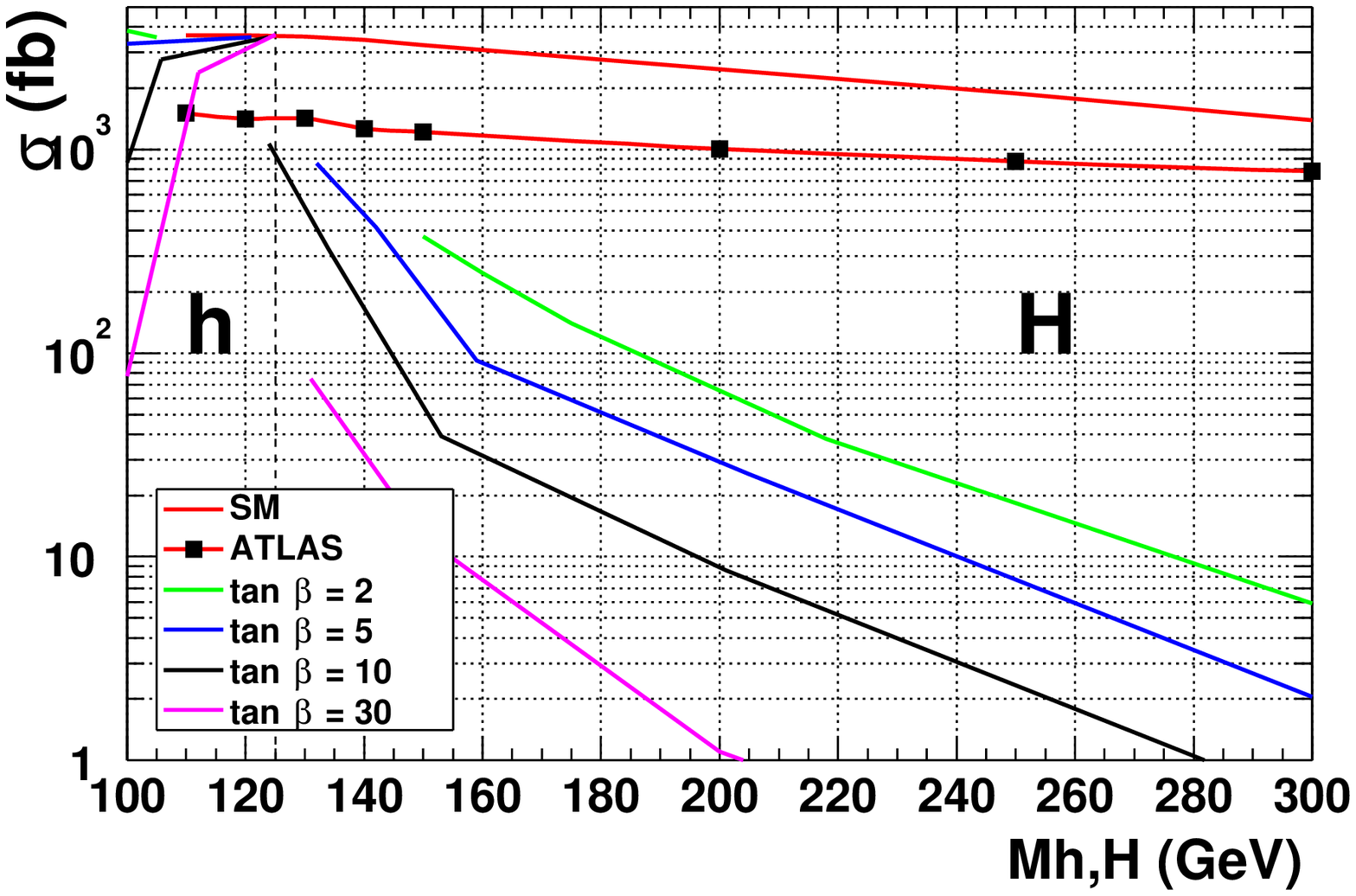}
\vspace*{-0.5cm}
\caption{ Higgs production cross sections via the WBF process for different 
$tan(\beta)$ values. The black line is the Standard Model cross section, the 
marked line is the ATLAS sensitivity for $Br(H\rightarrow invisible)=1$.
\label{fig:MSSMproduction}}
\end{center}
\vspace*{-0.5cm}
\end{figure}

In the case that gaugino mass unification is ruled
out~\cite{nous_hinvisible_lhc} and ${M1}/{M2}$=5 we have for medium
$tan\beta = 5$ a large Br($H\to\chi\chi$) which could cause
a dangerous situation for Higgs discovery at the LHC.  
At the decoupling limit ($M_{A}$=1TeV) and
for large stop mixing $A_{t}$=2.4 TeV and large stop mass (1 TeV) the
lightest Higgs mass is 120 GeV and it is produced with Standard Model 
cross section. Fig.~\ref{fig:M2mu} shows the region where the Higgs does not
escape detection. For the region which is not covered by the invisible Higgs 
search, the Higgs boson will be detected by other decay channels.
\\
\begin{figure}
\begin{center}
\includegraphics[width=12.cm]{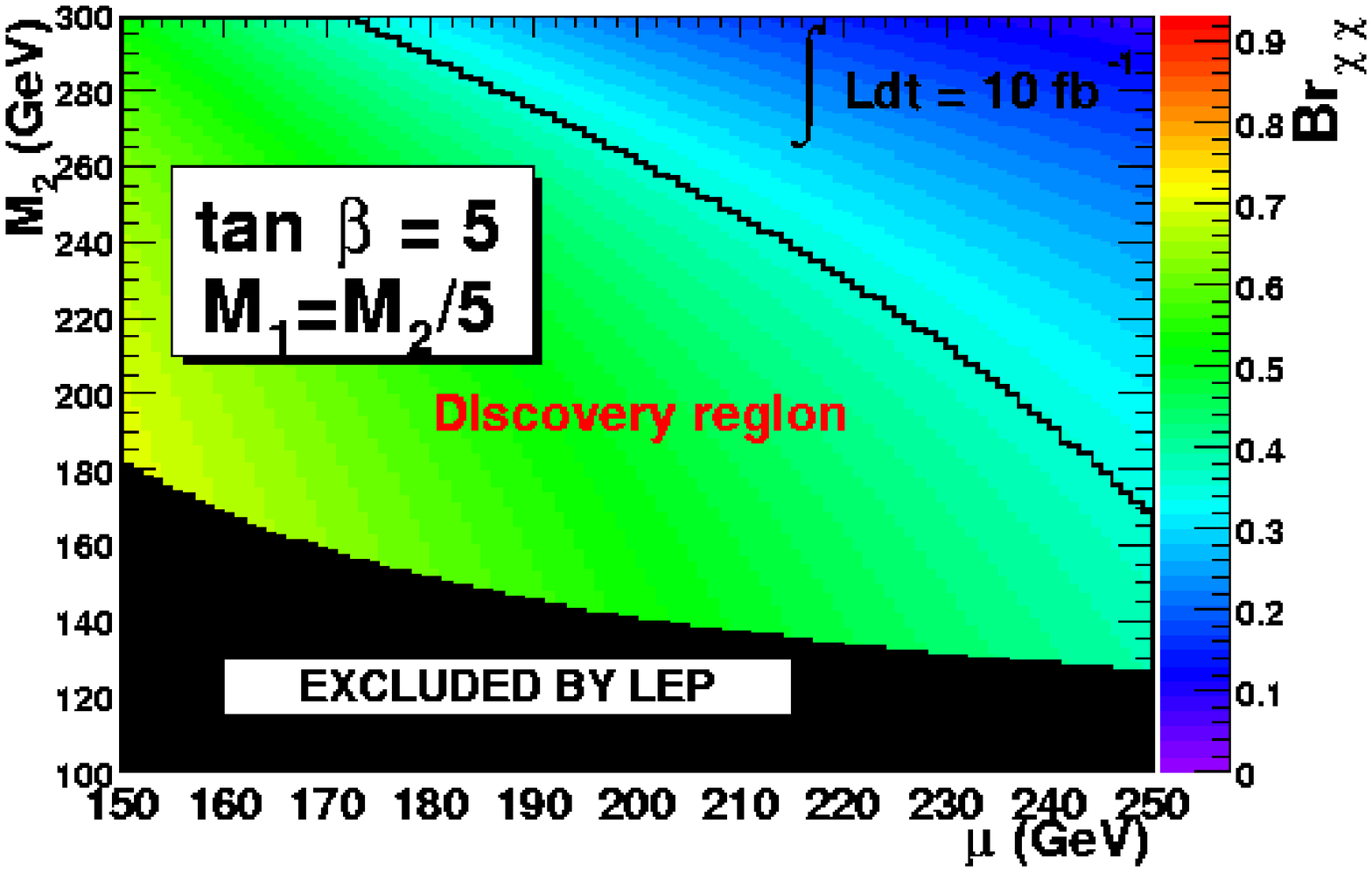}
\vspace*{-0.3cm}
\caption{\label{fig:M2mu}Branching ratio of $H\to\chi\chi$ in the 
(M2,$\mu$) plane. The red line is the ATLAS limit. The black area is 
already excluded by LEP direct chargino searches.}
\end{center}
\vspace*{-0.5cm}
\end{figure}

\section{Summary}

We have presented results of two independent studies of ATLAS and CMS on the
experimental observability of an invisible Higgs produced via 
weak boson fusion at the LHC. Background estimates in the two analyses differ
substantially even though
the performance of the two detectors is similar for this particular study. 
One of the reasons is that different MC samples are used in the analysis: 
purely PYTHIA ($2 \rightarrow 2$ processes) generation of W,Z plus
multi-jet backgrounds vs. generation with full LO matrix elements
implemented as an external process into PYTHIA. The most important reason 
however, which makes a difference of more than a factor of 2 between CMS and 
ATLAS estimates of QCD Wjj and Zjj backgrounds, is different rejection 
factors for the mini-jet (central jet) veto. In the CMS study, the survival 
probability of the mini-jet veto has been taken from analytical calculations 
as a multiplicative factor, while in the ATLAS study it comes from PYTHIA
generation of additional soft central jets between two tagging jets
and the further reconstruction of these jets with the calorimeter using the 
fast detector simulation. The expected performance of mini-jet veto in weak
boson fusion events requires further study, for all Higgs decay modes, not
just the invisible decay considered here. 

Even with the more pessimistic background estimates of the ATLAS simulations it
has been shown that an invisible branching ratio of $\simeq$ 30-40 \% can be
probed at 95 \% CL up to Higgs mass of 250 GeV with the first 10 fb $^{-1}$ 
of data. Fortunately, for the invisible Higgs search in the real 
experiment there will be the possibility to estimate Wjj and Zjj backgrounds
directly from the data. In addition one can directly measure the 
mini-jet veto efficiency with these events.

A detailed study of the possible trigger strategy at both Level 1 and High
Level trigger has now been performed. It was shown that very high trigger
efficiency ($\simeq$ 95 \%) for invisible Higgs can be achieved with 
an acceptable background rate, by making use of topological selections of the 
tagging jets in addition to a missing $E_{T}$ cutoff. 

\noindent
{\bf Acknowledgements} \\
A.N. would like to thank S. Ilyin for the generation of the EW Z+jj 
background with COMPHEP and explanation of the usage of this package. 
L.N. acknowledges partial support from Polish-French Collaboration
within IN2P3 during completing this study.
L.N. and B.D.G. would like to thank E. Richter-Was for fruitful discussions.
}

\setcounter{figure}{0}
\setcounter{table}{0}
\setcounter{section}{0}
\setcounter{equation}{0}
\newpage

{
%
%
%
\newcommand{\tc}{\textcolor}
\newcommand{\tcr}{\textcolor{red}}
\newcommand{\tcg}{\textcolor{green}}
\newcommand{\tcb}{\textcolor{blue}}
\newcommand{\tcm}{\textcolor{magenta}}
%
%
\newcommand{\ppbar}{\mbox{$p\overline{p}$}}
\newcommand{\rts}{\mbox{$\sqrt{s}$}}
\newcommand{\alphas}{\mbox{$\alpha_s$}}
\newcommand{\PT}{\mbox{$P_T$}}
\newcommand{\etmiss}{\mbox{$E_T^{miss}$}}
\newcommand{\ptmiss}{\mbox{$P_T^{miss}$}}
\newcommand{\ptw}{\mbox{$P_T(W)$}}
\newcommand{\ptz}{\mbox{$P_T(Z)$}}
\newcommand{\deleta}{\mbox{$\Delta \eta \times \Delta \phi$}}
\newcommand{\linteg}{\mbox{$\int{{\cal L} dt}$}}
\newcommand{\abseta}{\mbox{$\mid \eta \mid$}}
\newcommand{\lqcd}{\mbox{$\Lambda_{QCD}$}}
\newcommand{\QSQ}{\mbox{$Q^{2}$}}
\newcommand{\mevp}{MeV/$c$}
\newcommand{\meve}{MeV/$c^2$}
\newcommand{\gevp}{GeV/$c$}
\newcommand{\geve}{GeV/$c^2$}
\newcommand{\sw}{\sin\theta_W}
\newcommand{\cw}{\cos\theta_W}
\newcommand{\stbr}{\sigma \cdot BR}
%

%
%
\newcommand{\epem}{\mbox{$e^+e^-$}}
\newcommand{\mpmm}{\mbox{$\mu^+\mu^-$}}
\newcommand{\zzero}{\mbox{Z$^0$}}
%
%
\newcommand{\bs}{\mbox{$B_s^0$}}
\newcommand{\bd}{\mbox{$B_d^0$}}
\newcommand{\bsb}{\mbox{$\bar{B}_s^0$}}
\newcommand{\bdb}{\mbox{$\bar{B}_d^0$}}
\newcommand{\dms}{\mbox{$\Delta m_s$}}
\newcommand{\cl}{\mbox{$95 \% \ CL$}}
\newcommand{\pizero}{\mbox{$\pi^0$}}
\newcommand{\xrad}{\mbox{$X_0$}}
\newcommand{\degr}{\mbox{$^{\circ}$}}
\newcommand{\dsdpt}{\mbox{$frac{d\sigma}{dP_T}$}}
\newcommand{\jpsi}{\mbox{$J/\psi K_s^0$}}
\newcommand{\btojpsi}{\mbox{$B^0_d \rightarrow  J/\psi K_s^0$}}
\newcommand{\btopipi}{\mbox{$B^0_d \rightarrow  \pi \pi$}}
\newcommand{\sbeta}{\mbox{$\sin 2 \beta$}}
\newcommand{\salpha}{\mbox{$\sin 2 \alpha$}}
\newcommand{\dsbeta}{\mbox{$\Delta \sin 2 \beta$}}
\newcommand{\dsalpha}{\mbox{$\Delta \sin 2 \alpha$}}
%
%
\newcommand{\hgg}{\mbox{$H \rightarrow  \gamma \gamma$}}
\newcommand{\hbb}{\mbox{$H \rightarrow  b \bar{b}$}}
\newcommand{\hzz}{\mbox{$H \rightarrow  Z Z$}}
\newcommand{\hfourl}{\mbox{$H \rightarrow Z Z^{*} \rightarrow 4l$}}
\newcommand{\hzzfourl}{\mbox{$H \rightarrow Z Z \rightarrow 4l$}}
\newcommand{\hww}{\mbox{$H \rightarrow  W W $}}
\newcommand{\hwws}{\mbox{$H \rightarrow  W W^{(*)}$}}
\newcommand{\hwwsll}{\mbox{$H \rightarrow  W W^{(*)} \rightarrow l \nu l \nu $}}
\newcommand{\hzzs}{\mbox{$H \rightarrow  Z Z^*$}}
\newcommand{\gamgam}{\mbox{$\gamma \gamma$}}
\newcommand{\htautau}{\mbox{$H \rightarrow \tau \tau$}}

%
%
\newcommand{\vbfprocess}{\mbox{$qq \rightarrow qq H $}}

%
%
\newcommand{\wlnu}{\mbox{$W \rightarrow l \nu $}}
\newcommand{\wenu}{\mbox{$W \rightarrow e \nu $}}
\newcommand{\zll}{\mbox{$Z \rightarrow l l $}}
\newcommand{\zmumu}{\mbox{$Z \rightarrow \mu \mu $}}
%
%
%
\newcommand{\ttbar}{\mbox{$t \overline{t} $}}
\newcommand{\bbbar}{\mbox{$b \overline{b} $}}
\newcommand{\qqbar}{\mbox{$q \overline{q} $}}
\newcommand{\Zbbbar}{\mbox{$Z b \overline{b} $}}
\newcommand{\ttjets}{\mbox{$t \overline{t} + jets$}}

%
%
%
\newcommand{\fbs}{\mbox{$\rm{fb}^{-1}$}}
\newcommand{\pbs}{\mbox{$\rm{pb}^{-1}$}}
\newcommand{\lhigh}{\mbox{${\cal L} = 10^{34} \ \rm{cm}^{-2} \rm{sec}^{-1}$}}
\newcommand{\llow}{\mbox{${\cal L} =  10^{33} \ \rm{cm}^{-2} \rm{sec}^{-1}$}}
\newcommand{\lintyear}{\mbox{$\int {\cal L} dt \ = \ 10 \ fb^{-1}$}}
\newcommand{\lintlow}{\mbox{$\int {\cal L} dt \ = \ 30 \ fb^{-1}$}}
\newcommand{\linthigh}{\mbox{$\int {\cal L} dt \ = \  100 \ fb^{-1}$}}
%
%
%
\newcommand{\chiplus}{\mbox{$\chi^+$}}
\newcommand{\chiminus}{\mbox{$\chi^-$}}
\newcommand{\chizero}{\mbox{$\chi^0$}}
\newcommand{\chipm}{\mbox{$\chi^{\pm}$}}
\newcommand{\chione}{\mbox{$\chi^0_1$}}
\newcommand{\chitwo}{\mbox{$\chi^0_2$}}
\newcommand{\chithree}{\mbox{$\chi^0_3$}}
\newcommand{\sel}{\mbox{$\tilde{e}$}}
\newcommand{\smu}{\mbox{$\tilde{\mu}$}}
\newcommand{\stau}{\mbox{$\tilde{\tau}$}}
\newcommand{\slep}{\mbox{$\tilde{l}$}}
\newcommand{\slepton}{\mbox{$\tilde{l}$}}
\newcommand{\sfer}{\mbox{$\tilde{f}$}}
\newcommand{\sneu}{\mbox{$\tilde{\nu}$}}
\newcommand{\stauone}{\ensuremath{\tilde{\tau}_1 }}
\newcommand{\stautwo}{\ensuremath{\tilde{\tau}_2 }}
\newcommand{\sq}{\mbox{$\tilde{q}$}}
\newcommand{\sgl}{\mbox{$\tilde{g}$}}
\newcommand{\sbot}{\mbox{$\tilde{b}$}}
%
%
\newcommand{\msquark}{\mbox{$m_{\tilde{q}}$}}
\newcommand{\mzero}{\mbox{$m_{0}$}}
\newcommand{\msneu}{\mbox{$m_{\tilde{\nu}}$}}
\newcommand{\mchiplus}{\mbox{$m_{\tilde{\chi}^+}$}}
\newcommand{\mchi}{\mbox{$m_{\tilde{\chi^0}}$}}
%
%
\newcommand{\tanb}{\mbox{$\tan \beta$}}
\newcommand{\matb}{\mbox{$(m_A,\tan \beta )$}}
\newcommand{\atau}{\ensuremath{A_{\tau} }}
\newcommand{\phitau}{\ensuremath{\varphi_{\tau} }}
\newcommand{\delm}{\ensuremath{\Delta M}}
\newcommand{\mumtwo}{\ensuremath{(\mu - M_2)}}


\newcommand{\rh}{\rule[-3mm]{0mm}{8mm}}
\newcommand{\ru}{\rule[-0mm]{0mm}{5mm}}
\newcommand{\rl}{\rule[-3mm]{0mm}{5mm}}


%
%
\newcommand{\qcdtau}{QCD \mbox{$\tau \tau + jets$}}
\newcommand{\ewtau}{EW \mbox{$\tau \tau + jets$}}
\newcommand{\tautau}{\mbox{$\tau \tau + jets$}}
\newcommand{\qcdww}{QCD \mbox{$W W + jets$}}
\newcommand{\ewww}{EW \mbox{$W W + jets$}}
\newcommand{\wwjets}{\mbox{$W W + jets$}}
\newcommand{\ifb}{\mbox{$\rm fb^{-1}$}}
\newcommand{\ipb}{\mbox{$\rm pb^{-1}$}}
\newcommand{\sig}{\mbox{$\sigma$}}
\newcommand{\ttb}{\mbox{$t \overline{t}$}}
\newcommand{\bbb}{\mbox{$b \overline{b}$}}
\newcommand{\qqb}{\mbox{$q \overline{q}$}}
\newcommand{\hwwll}{\mbox{$H \rightarrow W W^{(*)} \rightarrow l \nu l \nu$}}

\noindent
{\Large\bf D. Search for the Standard Model Higgs Boson
using Vector Boson Fusion at the LHC} \\[0.5cm]
{\it G.\,Azuelos, C.\,Buttar, V.\,Cavasinni, D.\,Costanzo, 
T.\,Figy, R.\,Harper, K.\,Jakobs, M.\,Klute, R.\,Mazini, A.\,Nikitenko, 
E.\,Richter--Was, I.\,Vivarelli and D.\,Zeppenfeld}

\begin{abstract}
The weak boson fusion process has been suggested and discussed
recently as a discovery process for a Standard Model Higgs boson in
the intermediate mass range $m_H < 2 m_Z$ at the LHC. The additional 
jets in the 
forward region of the detector and the requirement of low jet activity 
in the central region allow for a significant background rejection. 
In the present paper 
the analyses for the $\hwws$ and the 
$H \rightarrow \tau\tau$ decay modes have been performed 
using a more realistic simulation of the expected performance of  
the LHC detectors. 
The results obtained confirm 
both the large discovery potential in the \hwws\ decay
channel and the sensitivity to Higgs boson decays into $\tau$-pairs, 
which is important for the determination of the Higgs boson coupling to 
fermions. 
\end{abstract}

\section{Introduction}

The search for the Higgs boson is one of the primary tasks of the 
experiments at the {\em Large Hadron Collider} (LHC). It has been established 
in many studies over the past years \cite{ATLASTDR,CMSTDR}
that a Standard Model Higgs boson can be identified with a high significance 
over the full mass range of interest, from the lower limit set by the 
LEP experiments of 114.1~GeV \cite{LEP-limit} up to about 1 TeV.

At the LHC the production cross section for a Standard Model Higgs boson 
is dominated by gluon-gluon fusion. 
The fusion of vector bosons radiated from initial state quarks represents 
the second most important contribution to the production cross section.
The relative contribution depends on the Higgs boson mass. In the 
intermediate mass range vector boson fusion amounts to about 20\% of the 
total production cross section and becomes more important
with increasing mass. 
However, for this production mode additional event characteristics can 
be exploited to suppress the large backgrounds. 
In these events the Higgs boson is accompanied by 
two jets in the forward region of the detector originating from the 
initial quarks from 
which the vector bosons are emitted. 
Another feature of the vector boson fusion process
is the lack of color exchange between the initial state quarks, which 
leads to suppressed jet production in the central region.  
This is in contrast to most background processes, where color flow in 
the t-channel appears.
Jet tagging in the forward region of the detector together with a veto 
of jet activity in the central region are therefore useful tools to enhance the signal to 
background ratio. These techniques have so far been applied in the 
search for heavy Higgs bosons \cite{ATLASTDR}. 

The observation of the Standard model Higgs boson at the LHC 
in the vector boson fusion channels in the intermediate mass range 
has first been discussed in Refs.~\cite{zeppenfeld} and \cite{zeppenfeld-ww} 
for the 
$H \rightarrow \gamma \gamma $ and 
$\hwws$  decay modes and in Ref.~\cite{zeppenfeld-tau} for the 
$\htautau$ decay mode. 
In the 
framework of the {\em Les Houches workshop} 
the analyses for the $WW^{(*)}$ and 
$\tau\tau$ decay modes have been repeated
using more realistic simulations of the performance of the LHC
detectors, including forward jet tagging and jet veto efficiencies.
In the present study the performance at low LHC 
luminosity, i.e. \llow\ , is addressed, and the discovery potential 
is evaluated 
for integrated luminosity values up to 30 $\fbs$, which are expected to be 
reached during the first years of operation.

\section{Signal and Backgrounds}
The cross sections for the vector boson fusion process have been calculated 
using the programme {\em VV2H} \cite{spira-hqq}. Although 
next-to-leading order calculations are available
\cite{fusion-nlo}, leading order cross sections have been used. The size 
of the QCD corrections amounts to about 10\% and is thus small. 
Another reason for this approach is the consistency with the background
estimates, for which NLO cross section calculations are not available 
for all relevant processes. The Higgs branching ratios have been calculated 
using the programme {\em HDECAY} \cite{hdecay}. 
The values for the total cross section for the vector boson fusion process as well as the cross sections times branching ratios for the \hwws\ and \htautau\ decay 
mode are given in Table \ref{t:sig_br} as a function of the Higgs boson mass. 
They have been computed using the 
CTEQ5L structure function parametrization \cite{cteq-sf}. 

\begin{table}[h]
\footnotesize
\begin{center}
\caption{\footnotesize \em Total vector boson fusion production cross sections
$\sigma (qqH)$ 
and 
$\sigma \cdot BR (H \rightarrow W W^{(*)})$ and 
$\sigma \cdot BR (H \rightarrow \tau \tau )$ as a function of the Higgs boson mass. 
}\label{t:sig_br}
\begin{tabular}{l r || c | c | c | c | c | c | c | c }
\hline
\hline
$m_H$ & (GeV) & 120 & 130 & 140 & 150 & 160 & 170 & 180& 190 \\
\hline
\hline
$\sigma (qq H)$ & (pb) & 4.36 & 4.04 & 3.72 & 3.46 & 3.22 & 3.06 & 2.82 & 2.64 \\
\hline
$\sigma \cdot BR (H \rightarrow WW^{(*)})$ & (fb) & 
531 & 1127 & 1785 & 2370 & 2955 & 2959 & 2620 & 2054 \\
$\sigma \cdot BR (H \rightarrow \tau \tau)$ & (fb) & 
 304 & 223 & 135 & 64.4 & 11.9 & 2.8 & 1.6 & 1.0 \\
\hline
\hline
\end{tabular}
\end{center}
\end{table}

The following background processes are common to all channels considered, as 
described in more detail in Ref.~\cite{zeppenfeld-ww}: 

\begin{itemize}
\item {\em \ttbar\ production:} due to the appearence of two b-jets, 
$\ttbar$ events contribute  
already at leading order to the background, if the two 
b-jets fulfill the identification criteria of the two tag jets. 

\item {\em QCD WW background:} the continuum production of W-pairs, 
where two or more tag jet candidates 
arise from parton emission. 

\item  {\em Electroweak WW background:} 
pair production of $W$ bosons 
via t-channel vector boson exchange. 
Due to the similarity to the signal process the rejection of this particular
background is expected to be much harder than for the QCD type backgrounds. 

\item {\em QCD Drell-Yan $Z / \gamma^* + jet$ production}, with 
$Z / \gamma^* \rightarrow ee, \mu\mu $ and $\tau \tau$.

\item {\em Electroweak $\tau \tau $ production:}
tau pair production via a t-channel weak boson exchange.

\end{itemize}

\begin{table}[h]
\footnotesize
\begin{center}
\caption{\footnotesize \em Cross sections times leptonic branching ratios 
($W \rightarrow l \nu$, 
$l = e, \mu $ and $\tau$) for the major background processes.}
\label{t:backgr}
\begin{tabular}{l|l|r}
\hline
\hline
process & $p_T$-cutoff& cross-section  \\
\hline
\hline
\ttbar   & & 55.0 pb \\
\qcdww   & & 16.7 pb \\
$Z / \gamma^* + jets$, $Z / \gamma^* \rightarrow \tau \tau$ & $ > 10$ GeV& 1742.0  pb \\
\ewww    & & 81.6 fb \\
\ewtau   & & 170.8 fb \\
\hline
$Z / \gamma^* + jets$, $Z / \gamma^* \rightarrow ee /\mu \mu$ & $ > 10$ GeV & 3485.0  pb \\
$ZZ$ & & 37.8 pb \\
$H \rightarrow ZZ$ & & 0.26 - 2.5 pb \\
\hline
\hline
\end{tabular}
\vspace{0.5cm}
\end{center}
\end{table}

The signal processes and all background processes except the electroweak 
WW and $\tau \tau $ background have been generated 
using the PYTHIA 6.1 Monte Carlo 
event generator \cite{pythia}. The Drell-Yan $Z / \gamma^* + jet$ background
has been generated using matrix element calculations for $q \bar{q} \rightarrow Z g$ 
and $q g \rightarrow Z q$ with 
a \PT\ cutoff of the outgoing quark or gluon of 10 GeV. A summary of the major 
background processes and the relevant cross sections 
multiplied by the branching ratio $BR(W \rightarrow l \nu)$, where 
$l=e,\mu$ and $\tau$ are listed in Table \ref{t:backgr}. 
In the PYTHIA simulation initial and final 
state radiation (ISR and FSR) and fragmentation have been switched on, thereby
allowing for a study of the 
jet activity in the central detector region due to radiation.
The CTEQ5L parametrization \cite{cteq-sf} of the parton distribution functions 
has been used in the generation of all signal and background processes.
To take the spin correlations in tau decays properly into account, tau decays
have been modelled using the TAUOLA $\tau$ decay library \cite{tauola}.
The two electroweak processes, which are not included in PYTHIA, have been 
generated by interfacing the matrix element calculation of Ref.~\cite{zeppenfeld-ww}
to PYTHIA, which was then used to perform the parton showering, including 
initial and final state radiation\cite{rachid}. The $W + jet$ 
background which is relevant for the 
$ H \rightarrow WW^{(*)} \rightarrow \ l \nu \ jj$ decay channel has been
generated using the matrix elements from the VECBOS Monte Carlo \cite{vecbos},
interfaced to PYTHIA. The fast simulation packages 
ATLFAST \cite{atlfast} and CMSJET \cite{cmsjet} of the ATLAS and CMS
detectors have been used to perform the detector simulation.

\section{Experimental Issues in the Search for the Vector Boson Fusion Process}

\noindent
{\bf Trigger aspects:}
all channels considered in the following have leptons ($e$ or $\mu$) in the final state and can be
triggered by either the single or the di-lepton trigger. 
It has been assumed that full trigger efficiency can be reached for a single
electron or muon for \PT\ values above 25 GeV or 20 GeV respectively. The corresponding 
threshold values for the lepton pair triggers are 15 GeV (for $e$) and 
10 GeV (for $\mu$). 

\noindent
{\bf Lepton Identification:}
it has been assumed that leptons ($e$ and $\mu$) can be identified
in the pseudorapidity range, $|\eta | < 2.5$,
with an efficiency of 90\%.
Hadronically decaying taus can be identified over the
same range of pseudorapidity. 
The tau reconstruction efficiency
is correlated with the rejection against QCD jets and the results
obtained in detailed simulation studies  
\cite{ATLASTDR,CMSTDR} have been used.

\noindent
{\bf Jet Tagging:}
from the signal production process it is expected that the two tag jets are 
reconstructed with a sizeable \PT\ in opposite hemispheres and have a 
large separation in pseudorapidty. In case where there is no further hard 
initial or final state radiation the transverse momentum of the tagging jets 
should be balanced by the transverse momentum of the Higgs boson. 

In the present study the two tag jets are searched over the full 
calorimeter coverage of the detectors $(| \eta | < 4.9)$. For all
jets a calibration has been applied which corrects the jet energy on average 
back to the original parton energy.  After calibration  the jet with
the highest \PT\ in the positive and negative region of pseudorapidity
is considered as the tag jet candidate. 
Detailed studies have shown \cite{jakobs01} that this choice of the 
tag jets has a high efficiency for a correct tag jet identifiction. 
Since the tag jets originate from quarks in the incoming proton it
is unlikely that they are b-jets. Consequently a b-jet veto has been 
applied in the pseudorapidity range of the detectors, where 
b-jet tagging is available, i.e. $| \eta | < 2.5$. In this  
procedure a b-tagging efficiency of 60\% has 
been assumed with a corresponding efficiency of about 99\% for a 
light quark or gluon jet not to be b-tagged \cite{ATLASTDR}. 

An important question is how well the tag jets can be identified at the LHC 
in the presence of pile-up. To answer this question a full GEANT 
simulation of the performance of the ATLAS detector in which also pile-up 
effects have been considered, has been performed \cite{pisa-tagging}. In this 
study it has been demonstrated that tag jets can be reliably reconstructed 
in the ATLAS detector and that the fast simulation package of the ATLAS 
detector provides a sufficiently good description of the tagging efficiency. 
Differences between the fast and full simulation have been found in the 
transition regions between different calorimeters and at very forward 
rapidities. The ratio between the efficiency for reconstructing a jet with 
\PT\ above 20 GeV as determined in the full and fast simulation 
has
been parametrized as a function of \PT\ and $\eta$ and has been used
to correct the fast simulation results accordingly \cite{pisa-tagging}. 

\noindent
{\bf Jet Veto Efficiencies:}
at the LHC, jets in the 
central region can also be produced by pile-up events. In the full simulation 
study \cite{pisa-tagging} it has been found that after applying a
threshold cut on the calorimeter
cell energies of 0.2 GeV at low and 1.0 GeV at high luminosity, that fake 
jets from pile-up events can be kept at a low level, provided that \PT\
thresholds of 20 GeV at low and 30 GeV at high luminosity are used for the 
jet definition.

\section{The \hwws\ decay mode}
In this Section the analyses of the  \hwws\ channels is briefly described. 
The acceptance cuts proposed in Ref.~\cite{zeppenfeld-ww} have been 
used as a starting point. Finally a multi-variate optimisation has been 
performed to find the best combination of values for the cuts \cite{jakobs01}
for Higgs boson masses in the range between 150 and 170 GeV. 
The cuts found in this optimization have also been used to get a first estimate
of the discovery significance outside this mass range. The signal significance 
may still be improved, if the cut optimization is done as a function of mass, 
in particular for lower Higgs boson masses.

\subsection{Di-lepton final states: $\hwwll$ }
As discussed already previously \cite{ATLASTDR, zeppenfeld-ww} a large 
rejection against the \ttbar\ and the $WW$ backgrounds is obtained by 
exploiting 
the anti-correlation
of the W spins from the decay of the scalar Higgs boson \cite{Dittmar:1997ss}.
Background from real taus from $Z + jet$ production with 
$Z \rightarrow \tau \tau$ can be rejected if the tau momenta and thereby the 
$\tau \tau$ invariant mass can be reconstruced in the collinear 
approximation \cite{zeppenfeld-ww}. Due to the high \PT\ of the $Z$ boson 
in $Z + jet$ events it can be assumed that the neutrinos in the tau decays 
are emitted in the direction of the visible charged leptons. From the 
lepton momenta and the $\ptmiss$ vector the 
fractions $x_{\tau_1}$
and $x_{\tau_2}$ of the $\tau$ energy carried by each lepton and thereby the 
$\tau \tau$ invariant mass $m_{\tau \tau}$ can be reconstructed. 
For decays of real $\tau$'s values of $x_{\tau_{1,2}}$  in the range  
$ 0 < x_{\tau_{1,2}} < 1$ are expected. 
The background from $Z / \gamma^*$ Drell-Yan production in association with jets 
can be efficiently rejected by a cut on the reconstructed 
transverse mass $m_T (ll \nu) $ of the di-lepton and neutrino
system, defined as $m_T (ll \nu) = \sqrt{2 P_T^{ll} 
\ptmiss \cdot ( 1 - cos \Delta \phi )}$,
where $\Delta \phi$ is the angle between the di-lepton vector and the \ptmiss\ 
vector in the transverse plane.

In the event 
selection the following cuts have been applied: 
\begin{itemize}
\item Two isolated leptons with
$ P_{T} > 20$ GeV and $| \eta| < 2.5 $;  

\item Two tag jets with $P_T^1 > 40 $ GeV, $P_T^2 > 20 $ GeV and 
$ \Delta \eta_{tags} = |\eta_{tag}^{1} - \eta_{tag}^{2} | > 3.8 $;  \\
in addition it has been required that the leptons are 
reconstructed within the pseudorapidity gap spanned
by the two tag jets: 
$\eta_{tag}^{min} <  \eta_{l_{1,2}}  < \eta_{tag}^{max}$;
\item Lepton Angular Cuts:
$ \Delta \phi_{ll} \le 1.05, \ \, \ \ \ \Delta R_{ll} \le 1.8, \ \ \ 
\ \cos \theta_{ll} \ge 0.2$ \\
\hspace*{4.0cm}
$M_{ll} < 85 \  {\rm GeV}  ,  \;  \; \; P_{T} (l_{1,2}) < 120 \; {\rm GeV}$,

where $\Delta \phi_{ll}$ is the azimuthal separation between the leptons, 
$\cos \theta_{ll}$ is the cosine of the polar opening angle, 
$\Delta R_{ll}$ is the separation
in $\eta - \phi$ space, and $M_{ll}$ is the invariant mass of the di-lepton 
system. 

\item Real tau rejection: events are rejected, if 
$ x_{\tau_{1}}, x_{\tau_{2}} > 0.0$ and \\ 
$ M_{Z} - 25 \; {\rm GeV} \; < M_{\tau \tau} <  M_{Z} + 25 \; {\rm GeV}$;

\item Invariant mass of the two tag jets: $M_{jj} > 550$ GeV; 

\item Transverse momentum balance: $ | \vec{P}_T^{tot} | < 30$ GeV. \\
If no hard initial or final state gluons are radiated, it is 
expected that the transverse momentum of the Higgs boson is balanced 
by the transverse momentum of the two tag jets, such that an upper
cut on the modulus of the vector 
\[
\vec{P}_T^{tot} = \vec{P}_T^{l,1} + \vec{P}_T^{l,2} + \vec{P}_T^{miss} 
                + \vec{P}_T^{j,1} + \vec{P}_T^{j,2} 
\]
can be used to reject background.

\item Jet veto: no jets with $\PT > 20$ GeV in the pseudorapidity range 
$| \eta | < 3.2$;

\item $Z / \gamma^{*}, Z / \gamma^* \rightarrow \tau \tau$ rejection: $m_T (ll \nu) > 30$ GeV.
\end{itemize}

The additional background contributions 
for the signal from same-flavour leptons, of which the 
the $ee$- and $\mu \mu$-Drell-Yan backgrounds are the dominant ones, 
can be efficiently rejected by tightening the di-lepton mass cut and by introducing 
a \ptmiss\ cut:

\begin{itemize}
\item $ M_{ll} < 75$ GeV   and  $\ptmiss > 30$ GeV. 
\end{itemize}

The acceptance for a Higgs boson with a mass of 160 GeV and for the 
backgrounds after the application of successive cuts 
is summarized for the $e \mu$ final state in detail in Table \ref{t:ww-acc}. In addition
to the signal from the vector boson fusion also contributions from
the gluon gluon fusion process $gg \rightarrow 
\hwws $ where the two tag jets are produced from initial and final 
state radiation, have been found to contribute to the final signal rate.

\begin{table}
\footnotesize
\begin{center}
\caption{\footnotesize \em Accepted signal (for $m_H = 160 \; {\rm GeV}$) and background cross sections 
 in fb for the $H \rightarrow WW \rightarrow e\mu$ channel after the application of 
successive cuts. For the signal the contributions via the vector boson fusion and the 
gluon fusion channel are given separately. The last two lines give the
 final numbers if the contributions from $W\rightarrow \tau \nu
 \rightarrow l \nu \nu \ \nu$ are added for both the $e \mu$ and the
 $ee / \mu\mu$ final states.}
\label{t:ww-acc}
\begin{tabular}{l||r r||r|r|r|r|r|r}
\hline
\hline
 & \multicolumn{2}{c||}{signal (fb)} & \multicolumn{6}{c}{background (fb)} \\
 &VV & gg & \ttbar\ \hspace*{0.4cm}& \multicolumn{2}{c|}{\wwjets} & \multicolumn{2}{c|}{
$Z / \gamma^* + jets$} & total \\
 & \multicolumn{2}{c||}{$m_H$=160 GeV} &  & EW & QCD & EW & QCD & \\
\hline
Lepton acceptance       & 25.3 & 107.4 & 5360 & 12.9 &513.7 & 3.56 &12589& 18479 \\
+ Forward Tagging       & 10.7 &  2.35 &186.4 & 7.79 & 1.37 & 1.04 &125.8& 322.4 \\
+ Lepton angular cuts   & 6.99 &  1.46 & 22.0 & 0.47 & 0.12 & 0.40 & 22.7&  45.7 \\
+ Real $\tau$ rejection & 6.69 &  1.44 & 21.0 & 0.42 & 0.12 & 0.06 & 3.54&  25.1 \\
+ Inv. mass $M_{jj}$    & 5.30 &  0.89 & 12.5 & 0.42 & 0.05 & 0.06 & 2.54&  15.6 \\
+ $P_T^{tot}$           & 4.56 &  0.63 & 2.71 & 0.33 & 0.04 & 0.05 & 1.77&  4.90 \\
+ Jet veto              & 3.82 &  0.45 & 0.72 & 0.31 & 0.03 & 0.04 & 1.16&  2.26 \\
+ $M_T$-cut             & 3.71 &  0.42 & 0.69 & 0.30 & 0.03 & 0.01 & 0.03&  1.06 \\
\hline 
\hline
$\hwws \rightarrow e \mu + X $&    &       &      &      &      &      &     &      \\
incl. $\tau \rightarrow e, \mu$ contribution
                        & 4.14 &  0.46 & 0.71 & 0.33 & 0.03 & 0.01 & 0.03& 1.11 \\
\hline
$\hwws \rightarrow ee / \mu \mu + X$  &    &       &      &      &      &      &     &      \\                              
incl. $\tau \rightarrow e, \mu$ contribution
                        & 3.89 &  0.43 & 0.64 & 0.33 & 0.02 & 0.01 & 0.15& 1.15 \\
\hline
\hline
\end{tabular}
\end{center}
\end{table}

All numbers given in the upper part of Table~\ref{t:ww-acc}
come from direct decays into 
electron and muon final states. Di-leptons can, however, also be produced 
via cascade decays of tau leptons, for example, 
$ W \rightarrow \tau \nu \rightarrow \ l \nu \bar{\nu} \ \nu$.  
These contributions have also been 
calculated and have been added to the accepted signal and background 
cross sections. An increase of about 10\% for the cross sections 
has been found.
Due to the softer \PT\ spectra of leptons
from tau decays this contribution is smaller than the one expected 
from a scaling of branching ratios. 
The final acceptance including the contributions from $\tau$ cascade
decays, is also given for the sum of 
the $ee$ and $\mu \mu$ final states. 
Due to the additional cuts the signal acceptance is  slightly
lower than in the $e \mu$ case. 

It has to be pointed out that the numbers for the dominant \ttbar\ background
given in Table~\ref{t:ww-acc} have been obtained from the PYTHIA parton shower
simulation. An independent estimate of that background has been made by using
tree level matrix element calculations for $\ttbar + \ 0, 1, $ and $2-jets$.  
In order to avoid double 
counting when adding the three contributions the procedure proposed in 
Ref.\cite{zeppenfeld-ww}, to define three distinct final state jet topologies, 
has been adopted. For \ttbar\ + 0 \ jets, only the two b-jets are considered
as tag jet candidates. Initial and final state radiation in these events may 
lead to a rejection of the event due to the jet veto. A distinctively different
class is defined by those \ttbar + 1 \ jet\ events where the final state 
light quark or 
gluon gives rise to one tag jet and one of the two b-jets is identified as the
other tag jet. Finally, a third class is defined where in \ttbar\ + 2 jet events 
the final state light quarks or gluons are identified as tag jets. 

Using this procedure the total \ttbar\ background in the $e \mu$ channel has
been estimated to be 1.65 fb, which is about a factor of 2.3 higher than the 
PYTHIA prediction. The largest contribution has been found to arise from 
events where one tag jet originates from a b-jet and the second one from 
an emitted parton. In the following estimate of the signal significance, a
conservative approach is taken and this number is assumed for the \ttbar\ 
background. 

After all cuts 
a large signal to background ratio can be reached, which leads to 
an impressive discovery potential for a Higgs boson with a mass 
around 160 GeV in this channel. It has to be pointed out that even if the 
larger \ttbar\ background estimate is taken, 
the signal to background ratio is much better than in the 
$gg \rightarrow WW^{(*)}$ channel considered so
far \cite{ATLASTDR, jakobs-ww}. 
Therefore, the final signal significance is much less affected by 
systematic uncertainties on the background. Similar to the situation 
in the $gg \rightarrow WW^{(*)}$ channel no mass peak can be
reconstructed. Evidence for a signal has to be extracted from an 
excess of events above the sum of the backgrounds, for example, in the 
transverse mass spectrum. 

Following the discussion in Ref.~\cite{zeppenfeld-ww} the transverse
mass of the Higgs boson has been calculated as

\[ 
M_T = \sqrt{(E_T^{ll} + E_T^{\nu \nu})^2 - ( \vec{p}_T^{\ l l} + 
\vec{p}_T^{\ miss} )^2 }.
\]
where 
\[ 
E_T^{ll} = \sqrt{(P_T^{ll})^2 + m^2_{ll}} \ \, \ \ \ \ \ 
E_T^{\nu \nu} = \sqrt{(\ptmiss)^2  + m^2_{ll}} . 
\]

The corresponding distribution is shown in Fig.~\ref{f:mt-plot}
for Higgs boson signals of 140 GeV and 160 GeV above the total background.  
\begin{figure}[hbtn]
\begin{center}
\begin{minipage}{7.7cm}
 \mbox{\epsfig{file=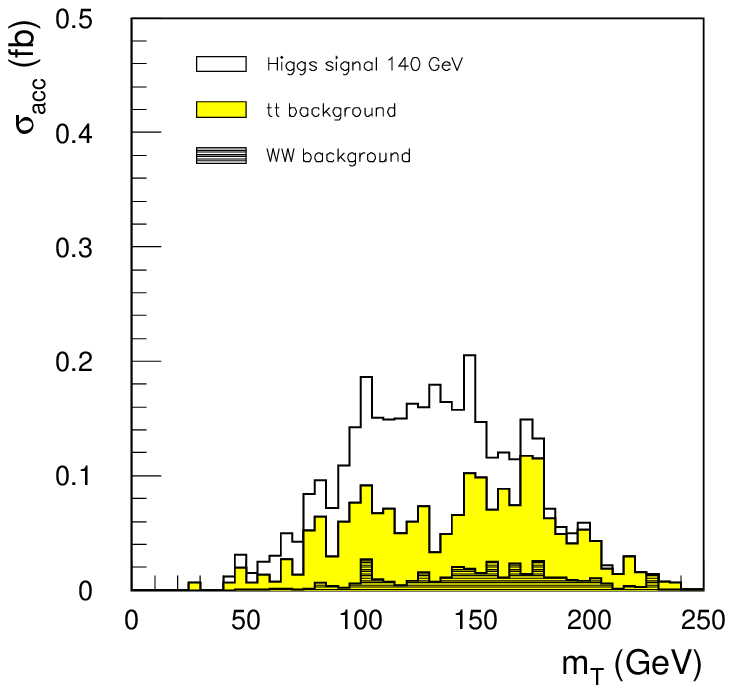,height=7.5cm}}
\end{minipage}
\begin{minipage}{7.7cm}
\mbox{\epsfig{file=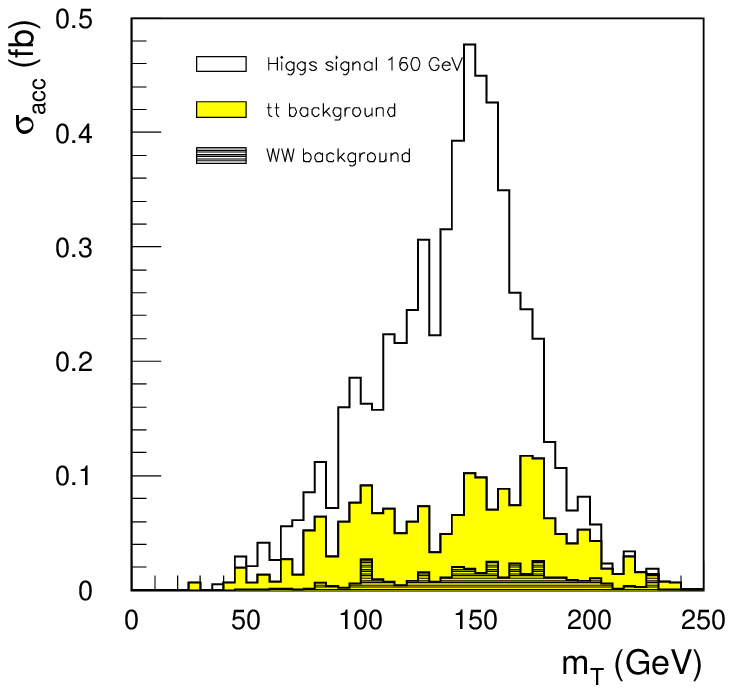,height=7.5cm}}
\end{minipage}
\vspace*{-0.5cm}
\caption{\it 
Distributions of the transverse mass $M_T$ for Higgs boson signals of 140 GeV (left) 
and 160 GeV (right) above the total background after all cuts are applied. 
The accepted cross sections  
$\sigma_{acc}$ (in fb/5 GeV) including all efficiency and acceptance factors 
are shown in both cases. The number of events observed in 
the detector is obtained by multiplying with the integrated luminosity.}
\label{f:mt-plot}
\end{center}
\vspace*{-0.5cm}
\end{figure}

It should be noted that in the present study numbers for signal and background have
been found which are somewhat different from the numbers quoted in the original 
parton level study of Ref.~\cite{zeppenfeld-ww}. A detailed comparison 
between both 
simulations has been performed and the main differences have been understood. 
One reason for a reduced signal efficiency observed in the present study is 
a lower 
lepton acceptance. In addition, the efficiency for reconstructing
the tag jets is found to be lower. Both   
are related to effects from initial and final state gluon radiation. They lead to 
a degraded lepton isolation as well as to non-Gaussian tails in the jet response
which can not be fully corrected in jet calibration procedures. However, the 
main conclusions of Ref.~\cite{zeppenfeld-ww}, that the search 
for vector boson fusion 
in the intermediate mass range at the LHC has a large discovery potential for 
a Standard Model Higgs boson in the \hwws\ decay channel are confirmed.

\subsection{The $l \nu$-jet-jet channel}

It has also been investigated whether the larger branching ratio of
the W-bosons into quark pairs can be used and the process
$ qq \rightarrow qqH \rightarrow qqWW^{(*)} \rightarrow qq \ l \nu \ jj$ can be 
identified above the larger backgrounds, in particular the $W + jet$
background.
This process has already been
established as a discovery channel for a heavy Higgs boson
\cite{ATLASTDR} in the vector boson fusion process, but has so far
not been considered in the intermediate mass region. 

The final signal rate in this channel is expected to be 
much lower than the corresponding numbers in the 
di-lepton channel \cite{pisa-ww}. However, a possible 
observation of a Higgs boson with a mass 
around 160 GeV can be
confirmed in this channel for higher integrated luminosities around 30 \fbs . 
It must be stressed that very hard cuts on the \PT\ and on the 
invariant mass of the forward tag jets, as well as on the separation 
$\Delta R$ between the lepton and the jets from the W-decay are 
necessary to extract the signal above the large backgrounds. 
These extreme cuts might also lead to larger systematic uncertainties on the 
background prediction.

\subsection{Discovery potential as a function of mass}

The analyses outlined above have been performed in the full range of Higgs boson 
masses from 110 to 190 GeV. The expected numbers of signal and
background events expected 
in the transverse mass interval $ 50 < M_T < m_H +
40$ GeV are given in Table~\ref{t:ww_sig} for integrated luminosities
of 5 and 30 $\fbs$, respectively, for the three $WW^{(*)}$ 
channels considered. The interval of transverse mass has been chosen to maximize
the signal to background ratio. For signal events about 98\% of all events 
are contained in that interval. For the estimate of the signal significance the
more conservative matrix element estimate of the \ttbar\ background has been used. 

\begin{tiny}
\begin{table}[h]
\footnotesize
\begin{center}
\caption{\footnotesize \em Expected signal and background rates for the three $WW^{(*)}$
decay channels as a function of $m_H$ assuming an
integrated luminosity of 5 $\fbs$ and 30 $\fbs$ (for the $l \nu jj$ channel).
In addition, the signal significances are given for an integrated
luminosity of 5 and 30 $\fbs$.  They have been 
computed using Poisson statistics and assuming a systematic uncertainty 
of 5\% on the background.}\label{t:ww_sig}
\begin{tabular}{l r || c| c | c | c | c | c | c | c | c }
\hline
\hline
$m_H$ & (GeV) & 110 & 120 & 130 & 140 & 150 & 160 & 170 & 180& 190 \\
\hline
\hline
\multicolumn{2}{c ||}{$ \hwws \rightarrow e \mu + X$}  & & & & & & & & & \\
Signal     & (5 \fbs\ )  & 0.4  & 1.8  & 4.8  & 8.4  & 13.9 & 22.7 & 21.6 & 16.9 & 12.7 \\
Background & (5 \fbs\ )  & 5.2  & 6.0  & 7.0  & 8.0  & 8.3  &  9.1 &  9.3 &  9.4 &  9.8 \\
Stat. significance  & (5 \fbs\ )  
                         &  -   &  -   & 1.5  & 2.5  & 3.9  & 5.8  & 5.5  & 4.4  & 3.4  \\
Stat. significance  & (30 \fbs\ )  
                         &  -   & 1.5  & 3.9  & 6.9  &11.1  & 17.3 & 16.3 & 12.6 & 9.3  \\
\hline
\hline
\multicolumn{2}{c||}{$ \hwws \rightarrow ee/\mu\mu + X$} & & & & & & & & &  \\
Signal     & (5 \fbs\ )  & 0.3  & 1.6  & 4.4  & 7.9  & 13.2 & 21.5 & 20.4 & 16.6 & 11.6 \\
Background & (5 \fbs\ )  & 5.2  & 5.9  & 6.9  & 7.6  & 8.0  &  8.6 &  8.8 &  8.9 &  9.2 \\
Stat. significance  & (5 \fbs\ )  
                         &  -   &  -   & 1.4  & 2.4  & 3.8  &  5.6 &  5.3 &  4.4 & 3.2  \\
Stat. significance & (30 \fbs\ )  
                         &  -   & 1.6  & 3.9  & 6.7  & 10.8 & 16.9 & 15.8 & 12.8 & 8.8  \\
\hline
\hline
\multicolumn{2}{c||}{$ \hwws \rightarrow l \nu \ jj  + X$} & & & & & & & & & \\
Signal     & (30 \fbs\ )  &  -   &  -   & 4.5  & 7.5  & 10.5 & 24.0 &24.0  &18.0 & 15.0 \\
Background & (30 \fbs\ )  &  -   &  -   & 6.0  & 6.0  &  6.0 & 18.0 &18.0  &18.0 & 18.0 \\
Stat. significance & (30 \fbs\ )  
                   &  -   &  -   & 1.5  & 2.4  &  3.3 &  4.6 &  4.6 & 3.5 & 3.0 \\
\hline
\hline
\end{tabular}
\end{center}
\end{table}
\end{tiny}

\normalsize
The signal significances, expressed in the 
equivalent number of Gaussian standard deviations, have been calculated 
using Poisson statistics and assuming an integrated luminosities of 5
and 30 \fbs\ and a systematic 
uncertainty of 5\% on the background. 
A 5\sig\ discovery can be claimed for 5 \fbs\ 
for $m_H = 150 - 185$ GeV if the $e \mu$ and $ee / \mu \mu$ channels are combined. 
For an integrated luminosity 
of 30 \fbs\ the discovery range increase to $m_H = 130 - 190$ GeV.

\section{The $ H \rightarrow \tau \tau$ decay mode}

In the following searches for $H \rightarrow \tau \tau$ decays  
using the  double leptonic decay mode, 
$ qqH  \rightarrow qq \ \tau \tau \rightarrow qq \ l^+ \nu \bar{\nu} \  
l^- \bar{\nu} \nu$ and the lepton-hadron decay mode
$ qqH  \rightarrow qq \ \tau \tau \rightarrow qq \ l^{\pm} \nu \nu \ had \ \nu$,
are described. 
Due to the $\tau \tau$ 
final state the $Z + jet, \ Z \rightarrow \tau \tau$ background contributes 
to the irreducible background and constitues the principal 
background for $H \rightarrow \tau \tau$ decays at low Higgs boson masses. 
The main points of the analyses are briefly 
summarized in the following subsections. For details the reader is referred to 
Refs.~\cite{klute} and \cite{rachid-tau}. 

\subsection{Di-lepton final states: $H \rightarrow \tau \tau \rightarrow l^+ l^- \ptmiss + X$ }
The di-lepton final state is characterized by 
two tag jets in the forward regions of the detector, two leptons in the 
central region and missing transverse momentum. 
The following cuts have been applied to select $ e \mu$ final states: 

\begin{itemize}
\item Two isolated leptons with
$ P_{T}(e) > 15 $  GeV  and $| \eta_{e}| \le 2.5 $ and \\
$ P_{T}(\mu) > 10$  GeV and $| \eta_{\mu}| \le 2.5 $;

\item Two tag jets with $P_T^1 > 50$ GeV,  $P_T^2 > 20$ GeV  and 
$ \Delta \eta_{tags} = |\eta_{tag}^{1} - \eta_{tag}^{2} | \ge 4.4 $. \\  
In addition, it has been required that the leptons are 
reconstructed within the pseudorapidity gap spanned
by the two tagging jets: 
$\eta_{tag}^{min} <  \eta_{l_{1,2}}  < \eta_{tag}^{max}$;

\item Missing transverse momentum: $ \ptmiss > 50$ GeV; 

\item Invariant mass of the two tag jets: $M_{jj} > 700$ GeV; 

\item Jet veto: no jets with $\PT > 20$ GeV in the pseudorapidity range 
defined by the two tag jets $\eta_{tag}^{min} < \eta_j^{veto} < 
\eta_{tag}^{max}$;

\item Azimuthal separation $\Delta \phi_{jj}$ between the tag jets: 
 $\Delta \phi_{jj} < 2.2$.\\
This cut is applied to reduce the electroweak $Zjj$ background, for which 
back-to-back jets are preferred \cite{Zeppenfeld-h-invisible}.
\item Separation $\Delta R_{e \mu}$ in $\eta-\phi$ space
between the two leptons: $\Delta R_{e \mu} < 2.6$;

\item Real tau reconstruction: 
$ x_{\tau_{1}}, x_{\tau_{2}} > 0$ and 
$ x_{\tau_{1}}^2 + x_{\tau_{2}}^2 < 1$; 

\item Mass window around the Higgs boson mass: 
$m_H - 10$ GeV $< m_{\tau \tau} < m_H + 15$ GeV. 

\end{itemize}

For $ee$ and $\mu \mu$ final states the additional background 
from Z decaying into 
$ee$ or $\mu \mu$ is efficiently rejected by requiring in addition: 
$ m_{ll} < m_Z - 15$ GeV.

The results are summarized for both the $e \mu$ and the 
sum of the $ee$ and $\mu \mu$ channel in Table \ref{t:tau01}, where the 
accepted cross sections for the signal with $m_H = 120$ GeV 
and the background contributions 
are given after the application of all cuts. 
\begin{table}
\footnotesize
\begin{center}
\caption{\footnotesize \em Accepted signal (for $m_H = 120 \; {\rm GeV}$) and background cross sections 
 in fb for the $H \rightarrow \tau \tau \rightarrow e\mu$ channel after the 
application of all cuts for both the $e \mu$ and the sum of the $ee$ and $\mu \mu$ 
channels. For the signal the contributions via the vector boson 
fusion and the 
gluon fusion channel are given separately.}
\label{t:tau01}
\begin{tabular}{l||r r||r|r|r|r|r|r}
\hline
\hline
 & \multicolumn{2}{c||}{signal (fb)} & \multicolumn{6}{c}{background (fb)} \\
 &VV & gg & \ttbar\ & \multicolumn{2}{c|}{\wwjets} & \multicolumn{2}{c|}{\tautau} & Total \\
 & \multicolumn{2}{c||}{$m_H$=120 GeV}& & EW & QCD & EW & QCD & \\
\hline
$H \rightarrow \tau \tau \rightarrow e \mu$
                        & 0.23 & 0.01  & 0.02 & 0.01 & 0.0  & 0.02 & 0.04 &  0.09 \\
\hline
\hline
$H \rightarrow \tau \tau \rightarrow ee / \mu \mu$  
                        & 0.24 & 0.02  & 0.05 & 0.01 & 0.0  & 0.04 & 0.08 &  0.17\\
\hline
\hline
\end{tabular}
\end{center}
\end{table}
After $\tau$ reconstruction the signal to background 
ratio is still much smaller than 1. This situation is drastically changed after the 
application of the mass cut around the Higgs boson mass. 
Due to the reconstructed Higgs
boson mass the sidebands can be used for the determination of the absolute 
level of the background. 

The distribution of the reconstructed $\tau \tau$ invariant mass is shown 
in Fig.~\ref{f:mtautau} (left) for the sum of the $e \mu , ee$ and $\mu \mu$ channels 
for a Higgs boson signal of 120 GeV above the background assuming an
integrated luminosity of 30 \fbs\ . 
\begin{figure}[hbtn]
\begin{center}
\begin{minipage}{7.5cm}
\mbox{\epsfig{file=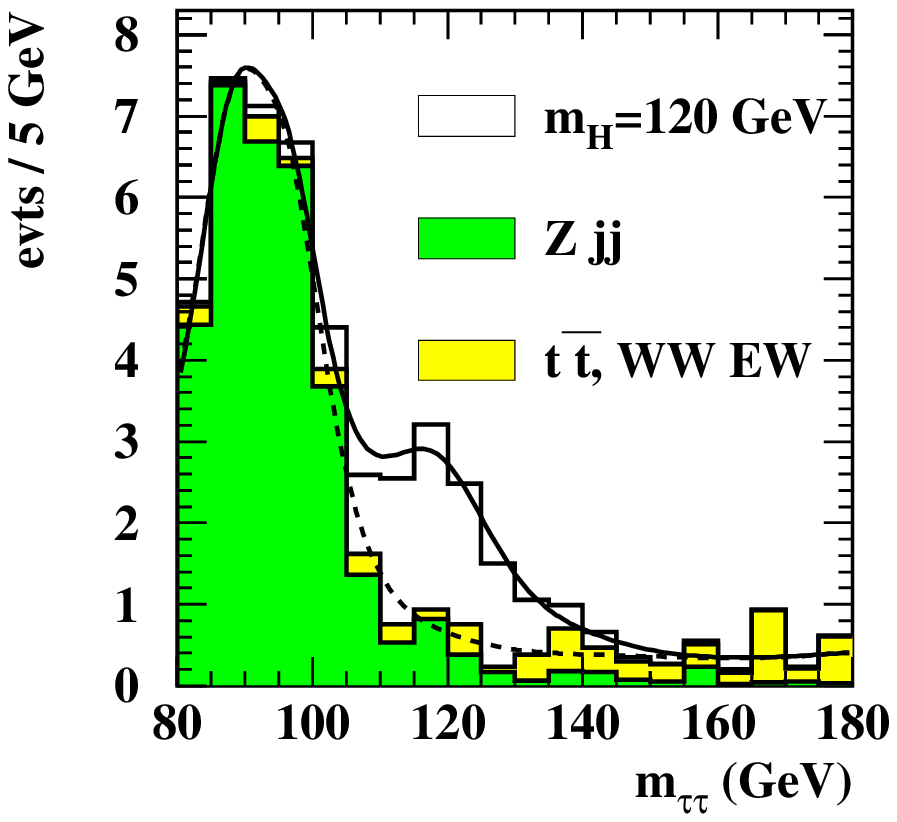,height=7.0cm}}
\end{minipage}
\begin{minipage}{7.5cm}
\vspace*{-1.0cm}
\mbox{\epsfig{file=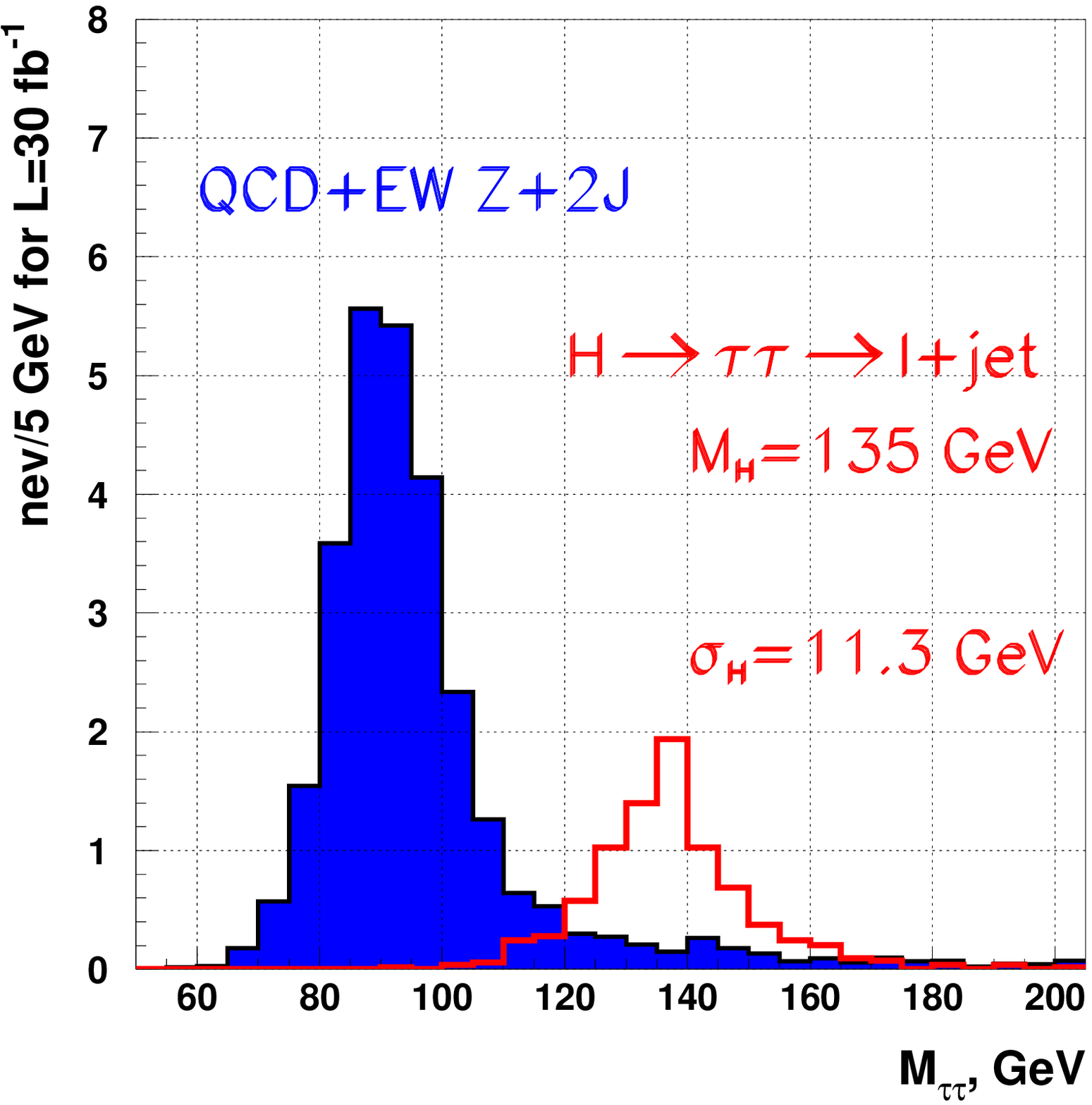,height=7.5cm}}
\end{minipage}
\vspace*{-0.3cm}
\caption{\it 
left: The reconstructed $\tau \tau $ invariant mass for a Higgs boson signal 
of 120 GeV in the $ll$-channel  
above all backgrounds after application of all cuts except the mass window cut. 
right: 
The reconstructed $\tau \tau $ invariant mass for a Higgs boson signal 
of 135 GeV in the $(l-had)$-channel compared to 
the QCD plus electroweak $Z+jj~(Z \rightarrow \tau \tau \rightarrow 
l+ had$
backgrounds after application of all cuts except the mass window cut. 
In both cases the signal and background numbers are shown for an integrated
luminosity of 30 $\fbs$.
}
\label{f:mtautau}
\end{center}
\vspace*{-0.5cm}
\end{figure}

The analysis has been performed for Higgs boson 
masses in the range from 110 to 150 GeV. The expected numbers of signal and
background events and the statistical significance for a Higgs boson discovery 
expressed in terms of Gaussian standard deviations are given in Table~\ref{t:tt_sig} for 
an integrated luminosity
of 30 \fbs\ .
\begin{table}[h]
\footnotesize
\begin{center}
\caption{\footnotesize \em Expected signal and background rates and statistical significance 
for the three $\tau \tau$
decay channels as a function of $m_H$ assuming an
integrated luminosity of 30 \fbs\ .}\label{t:tt_sig}
\begin{tabular}{l r || c| c | c | c | c }
\hline
\hline
$m_H$ & (GeV) & 110 & 120 & 130 & 140 & 150 \\
\hline
\hline
\multicolumn{2}{c ||}{$H \rightarrow \tau \tau \rightarrow e \mu \ \ptmiss $}  & & & & &  \\
Signal     & (30 \fbs\ )  & 7.7  & 7.0  & 5.1  & 3.3  & 1.5  \\
Background & (30 \fbs\ )  & 7.0  & 2.6  & 2.3  & 1.9  & 1.5  \\
Stat. significance  & (30 \fbs\ )  
                    & 2.4  & 3.2  & 2.5 & 1.8  &  -  \\
\hline
\hline
\multicolumn{2}{c ||}{$H \rightarrow \tau \tau \rightarrow ee / \mu\mu \ \ptmiss$}  & & & & &  \\
Signal     & (30 \fbs\ )  & 9.2  & 7.2  & 5.7  & 3.1  & 1.5 \\
Background & (30 \fbs\ )  & 10.5 & 5.2  & 3.8  & 3.1  & 2.3 \\
Stat. significance  & (30 \fbs\ )  
                    &  2.4  &  2.6 & 2.3  & 1.4  & - \\
\hline
\hline
\end{tabular}
\end{center}
\end{table}

\subsection{The lepton-hadron decay mode:
$H  \rightarrow \  \tau \tau \rightarrow  \ l^{\pm} \nu \nu \ had \ \nu$}

The $l-had$ decay mode of the $\tau \tau$ final state has also been studied using a fast 
simulation of the CMS detector \cite{cmsjet}. The 
backgrounds considered in this study are QCD and electroweak production of
$Z+jj~(Z \rightarrow \tau \tau \rightarrow l+\tau$-jet) and 
$W+3j~(W \rightarrow e(\mu)+ \nu)$. As in the previous studies, the 
$b \bar{b}+jj$ background it expected to be small \cite{zeppenfeld-tau}. 
QCD $Z+jj$ production has been generated using the lowest order matrix
element provided by the authors of Ref.~\cite{zeppenfeld-tau}
interfaced to PYTHIA. The electroweak $Z+jj$ production has been simulated 
with COMPHEP \cite{comphep}, again interfaced to PYTHIA. The $W+3j$
events have been produced with PYTHIA which may lead to an 
underestimate of this background cross-section.

For the present study the jet veto efficiency has not yet been evaluated in a
full detector simulation and the survival probability as determined in 
Ref.~\cite{zeppenfeld-ww} has been used to account for the acceptance
of the jet veto cut. In Ref.~\cite{zeppenfeld-ww} the 
jet veto efficiency was found to be 0.87 for signal events and
background from electroweak production and 
0.28 for QCD type backgrounds. In addition, it has been assumed that
the jets are reconstructed with full efficiency. Results based on a full
simulation of the CMS detector have shown that a reconstruction 
of low \PT\ jets around 20 GeV is possible with a high efficiency in a
low luminosity scenario. For the identification of the hadronic tau with the 
calorimeter and the tracker information an 
efficiency of 0.32 has been used. Using these criteria a probability of 
0.0019 is obtained for mis-identifying jets as hadronic
$\tau$'s.

In the event selection the following cuts are applied: 
\begin{itemize}
\item One isolated lepton with
$ P_{T} > 15 $  GeV  and $| \eta| \le 2.4 $; 
\item One hadronic tau jet with 
$ P_{T} > 30 $  GeV  and $| \eta| \le 2.4 $;

\item Two tag jets with 
$ \Delta \eta_{tags} = |\eta_{tag}^{1} - \eta_{tag}^{2} | \ge 4.4 $. \\  
In addition, it has been required that the lepton and the tau-jet are 
reconstructed within the pseudorapidity gap
$\eta_{tag}^{min} +0.7 <  \eta_{l,\tau -jet}  < \eta_{tag}^{max} -0.7$;
\item Invariant mass of the tag jets: $M_{jj}~>~1000$ GeV;
\item Transverse mass $m_{t}(l,\ptmiss):= \sqrt{2 P_T^l \ptmiss (1 - cos \Delta \phi)}~
<~30$ GeV; 
\item Tau reconstruction: $0~<x_{\tau_l}~<0.75,~~0~<x_{\tau_h}~<1$;
\item Mass window: $ |m_{\tau \tau}~-m_{H}|~<~15~GeV$.
\end{itemize}

The number of signal events for $m_H$=135 GeV and the 
number of different background events 
expected after all selections for an integrated luminosity $30~fb^{-1}$ are 
given in Table.~\ref{tab:cms_h_2tau_lj_1}. The errors quoted result from the 
statistical uncertainty of the Monte Carlo data samples. 
For comparison the number of events estimated in Ref.~\cite{zeppenfeld-tau} 
are also shown in the second row of the table.
\begin{table}[h]
\footnotesize
\begin{center}
\caption{\footnotesize \em Number of signal ($m_H = 135$ GeV) and background events 
in the $l-had$ channel
expected after all selections for an integrated luminosity $30~fb^{-1}$
\label{tab:cms_h_2tau_lj_1}}
\begin{tabular}{c|c|c|c}
\hline
\hline
Higgs, $M_{H}$=135 GeV & QCD Z+jj        & EW Z+jj       & W+3j    \\ \hline
\hline
  6.7$\pm$0.3  & 0.63$\pm$0.10   & 0.74$\pm$0.08 & 0.14$\pm$0.05 \\ \hline
  6.2 (from Ref.~\cite{zeppenfeld-tau}) & 
 \multicolumn{3}{c}{total background from Ref.~\cite{zeppenfeld-tau} is 1.1} 
 \\ \hline 
\hline
\end{tabular}
\end{center}
\end{table}

The number of signal and the total number of background events expected 
after all selections for different Higgs boson masses in the range between
115 and 145 GeV and assuming 
an integrated luminosity $30~fb^{-1}$ are shown in 
Tab.~\ref{tab:cms_h_2tau_lj_2}. 
\begin{table}[h]
\footnotesize
\begin{center}
\caption{\footnotesize \em Expected signal and background rates and statistical 
significance 
for the l-had  $\tau \tau$
decay channels as a function of $m_H$ assuming an
integrated luminosity of 30 $\fbs$.}
\label{tab:cms_h_2tau_lj_2}
\begin{tabular}{l r || c| c | c | c  }
\hline
\hline
$m_H$ & (GeV) & 115 & 125 & 135 & 145 \\
\hline
\hline
\multicolumn{2}{c ||}{$H \rightarrow \tau \tau \rightarrow l had \  \ptmiss $} & & & &  \\
Signal     & (30 \fbs\ )  & 12.6  & 9.9 & 6.7  & 3.6   \\
Background & (30 \fbs\ )  & 5.5  & 2.3  & 1.5  & 1.1   \\
Stat. significance  & (30 \fbs\ )  
                    & 4.1 & 4.5 & 3.7 & 2.4    \\
\hline
\hline
\end{tabular}
\end{center}
\end{table}

The reconstructed $\tau \tau$ invariant mass  
for the QCD and EW 
$Z+jj~(Z \rightarrow \tau \tau \rightarrow l+\tau$-jet) backgrounds 
and for a Higgs boson with 
$m_{H}$=135 GeV is shown in Fig.~\ref{f:mtautau} (right). 
The distributions 
are normalised to the expected number of events after all cuts, except the 
mass window cut, for an integrated luminosity of $30~fb^{-1}$.

\section{Conclusions}

The discovery potential for the Standard Model Higgs boson in the
intermediate mass range has been studied using the
vector boson fusion process. It has been demonstrated that the 
LHC experiments have a large discovery potential in the 
$ \hwws \rightarrow l^+ l^- \ptmiss $ channel. The additional signature
of tag jets in the forward and of a low jet activity in the
central region of the detector allow for a significant background 
rejection, such that a better signal to background ratio than in 
the inclusive $H \rightarrow WW^{(*)}$, which is dominated by gluon
gluon fusion process,  is obtained. As in the inclusive case, only 
the transverse mass of the Higgs boson can be reconstructed and a
signal has to be claimed from an excess of events above the
background. Due to the larger signal to background ratio in the
search for the fusion process the signal sensitivity is less affected
by systematic uncertainties on the background prediction. The present 
study shows that the ATLAS and CMS
experiments at the LHC would be sensitive to a Standard Model Higgs
boson in this decay channel in the mass range between 150 and 185 GeV 
with data corresponding to an integrated luminosity of 5 \fbs\
only. 

In addition, it has been shown that in the low mass region for 
$m_H < 140$ GeV the LHC experiments are also sensitive to 
the $\tau \tau$ decay mode of the Standard Model Higgs 
boson, if the characteristics of the 
vector boson fusion are exploited. However, a discovery in this
final state would require an integrated luminosity of about 
30 $\fbs$ and a combination of the $l-l$ and $l-had$ decay modes.
The measurement of the $\tau$ decay mode is 
particularly important for a measurement of the Higgs boson coupling
to fermions. 

The present study confirms the results
published earlier \cite{zeppenfeld,zeppenfeld-ww,zeppenfeld-tau},
that the search for vector boson fusion 
in the intermediate mass range at the LHC has a large discovery
potential over the full range from the lower limit set by the LEP 
experiments up to  $2 \ m_Z$, where the high sensitivity 
$H \rightarrow  Z Z \ \rightarrow 4 \ l$ channel takes over. 


\noindent
{\bf Acknowledgments} \\
The authors from the ATLAS and CMS collaborations would like to deeply 
thank D.\,Zeppen\-feld and D.\,Rainwater for very useful discussions and for 
providing the matrix element calculations for the various background 
processes. A.N.\,is grateful to S.\,Ilyin for the generation of the 
electroweak $Z+jj$ background with the COMPHEP programme. The work 
of E.R.--W. has been partially supported by the Polish Government grant KBN
2P03B11819 and by the Polish-French Collaboration within IN2P3.
Work supported in part by the European Community's Human Potential
Programme under contract HPRN--CT--2000--00149 Physics at Colliders.
All of us would like to thank the organizers of the {\em Les Houches
workshop} for the stimulating atmosphere and the fruitful workshop in the 
French alps. 
}

\setcounter{figure}{0}
\setcounter{table}{0}
\setcounter{section}{0}
\setcounter{equation}{0}
\newpage

{
%
\newcommand{\freccia}{\mbox{$\rightarrow$}}
\newcommand{\hilumi}{\mbox{$\rm {10^{34}~cm^{-2} s^{-1}}$}}
\newcommand{\medlumi}{\mbox{$\rm {10^{33}~cm^{-2} s^{-1}}$}}
\newcommand{\lowlumi}{\mbox{$\rm {10^{32}~cm^{-2} s^{-1}}$}}
\def\lumi#1#2{\mbox{$\rm{{#1}\cdot 10^{#2}~cm^{-2} s^{-1}}$}}
\newcommand{\fbm}{\mbox{$\rm {fb^{-1}}$}}
 
\newcommand{\degr}{\mbox{$^\circ$}}
\newcommand{\detaphi}{\mbox{$\Delta\eta\times\Delta\phi$}}
\newcommand{\dRf}%
{\mbox{$\Delta R=\sqrt{\Delta ^{2} \eta  + \Delta ^{2} \phi }$}}
\newcommand{\dR}{\mbox{$\Delta R$}}
\newcommand{\EM}{\mbox{e.m.\ }}
\newcommand{\ET}{\mbox{$E_T$}}
\newcommand{\ETjet}{\mbox{$E_T^{jet}$}}
\newcommand{\ETtau}{\mbox{$E_T^{\tau}$}}
\newcommand{\ETmiss}{\mbox{$E^{miss}_T$}}
\newcommand{\pT}{\mbox{$p_T$}}
\newcommand{\pTmiss}{\mbox{$p_T^{miss}$}}
\newcommand{\pTjet}{\mbox{$p_T^{jet}$}}
\newcommand{\pTtau}{\mbox{$p_T^{\tau}$}}
\newcommand{\abseta}{\mbox{$|\eta|$}}
\newcommand{\deta}{\mbox{$\Delta\eta$}}
\newcommand{\dphi}{\mbox{$\Delta\phi$}}
\newcommand{\qqbar}{\mbox{$\rm {q\overline{q}}$}}
\newcommand{\ppbar}{\mbox{$\rm {p\overline{p}}$}}
\newcommand{\ccbar}{\mbox{$\rm {c\overline{c}}$ }}
\newcommand{\bbbar}{\mbox{$\rm {b\overline{b}}$ }}
\newcommand{\ttbar}{\mbox{$\rm {t\overline{t}}$ }}
\newcommand{\gaga}{\mbox{$\rm {\gamma\gamma}$ }}
\newcommand{\A}{\mbox{$A$}}
\newcommand{\Z}{\mbox{$Z$}}
\newcommand{\Aev}{\mbox{$\rm{A/H \rightarrow \tau \tau}$}}
\newcommand{\Atau}{\mbox{$\rm{A \rightarrow \tau \tau}$}}
\newcommand{\Ataulh}{\mbox{$\rm{A \rightarrow \tau \tau 
  \rightarrow lepton-hadron }$}} 
\newcommand{\Atauhh}{\mbox{$\rm{A \rightarrow \tau \tau 
  \rightarrow hadron-hadron }$}} 
\newcommand{\Ataull}{\mbox{$\rm{A \rightarrow \tau \tau 
  \rightarrow lepton-lepton }$}} 

\newcommand{\nut}{\mbox{$\rm{\nu_{\tau}}$}}
\newcommand{\anut}{\mbox{$\rm{\overline{\nu_{\tau}}}$}}
\newcommand{\anul}{\mbox{$\rm{\overline{\nu_{l}}}$}}

\newcommand{\Htau}{\mbox{$\rm{H \rightarrow \tau \tau}$}}
\newcommand{\Hcs}{\mbox{$\rm{H^{\pm} \rightarrow cs}$}}
\newcommand{\Ztau}{\mbox{$\rm{Z \rightarrow \tau \tau}$}}
\newcommand{\Zgatau}{\mbox{$\rm{Z/\gamma^* \rightarrow \tau \tau}$}}
 
\newcommand{\mt}{\mbox{$m_{\rm{t}}$}}
\newcommand{\mH}{\mbox{$m_{\rm{H}}$}}
\newcommand{\mh}{\mbox{$m_{\rm{h}}$}}
\newcommand{\mA}{\mbox{$m_{\rm{A}}$}}
\newcommand{\mZ}{\mbox{$m_{\rm{Z}}$}}
\newcommand{\mHplus}{\mbox{$m_{\rm{H}^+}$}}
\newcommand{\Hplus}{\mbox{$\rm{H}^+$}}
\newcommand{\Hminu}{\mbox{$\rm{H}^-$}}
\newcommand{\Htaunu}{\mbox{$\rm{H^{\pm} \rightarrow \tau
\nu}$}}
\newcommand{\tbar}{\overline{t}}
\newcommand{\ptmx}{\mbox{$p_x^{miss}$}}
\newcommand{\ptmy}{\mbox{$p_y^{miss}$}}
\newcommand{\tanbeta}{\mbox{$tan \beta$}}
\newcommand{\mtt}{\mbox{$m_{\rm{\tau \tau}}$}}

\def\np{\newpage}                               
\def\bl{\vskip16pt\relax}                       

\newenvironment{2figures}[1]{\begin{figure}[#1] 
  \begin{center}
    \begin{tabular}{p{.47\textwidth}p{.47\textwidth}} }
 {  \end{tabular}
  \end{center} 
 \end{figure}
}

\noindent
{\Large\bf{E. Study of  the MSSM channel \Aev~ at the LHC}} \\[0.5cm]
{\it D.\,Cavalli, R.\,Kinnunen, G.\,Negri,
 A.\,Nikitenko and J.\,Thomas}

\begin{abstract}
Sudies both from ATLAS and CMS with fast and full detector
simulation have shown that the discovery potential of the \Aev~ channel
in the MSSM is large in the \mA~ range from $\sim$100 GeV to $\sim$1 TeV 
already with 30 \fbm~ collected at low LHC 
luminosity (\medlumi). 
The results of these studies, in particular for  the lepton-hadron and 
the hadron-hadron final decay channels, are presented here. 
 The question of the  trigger   for 
 the hadron-hadron final state  that is a very important issue
for this purely hadronic final state process is also discussed here.
\end{abstract}

\section{Introduction}

   In the minimal supersymmetric extension of the Standard Model~(MSSM),
two Higgs doublets are required, resulting in 5 physical
states,  referred to as
\Hplus, \Hminu, h (neutral lighter scalar), H (neutral heavier scalar)
and~A (neutral pseudoscalar). At tree level their masses can be
computed in terms of only two parameters, typically \mA\  and
\tanbeta~ (the ratio of the vacuum expectation values of the two
doublets). 
\par
The MSSM \Htau~ and \Atau~ rates are strongly enhanced with respect to the SM
case over a large region of the parameter space.
\par
A/H can be produced via  two different mechanisms.
For low values of \tanbeta, the  
$gg\rightarrow \A/H $ 
(direct production mode - Fig. \ref{mec1})
\begin{figure}[b]
\begin{center}
  \mbox{\epsfig{figure=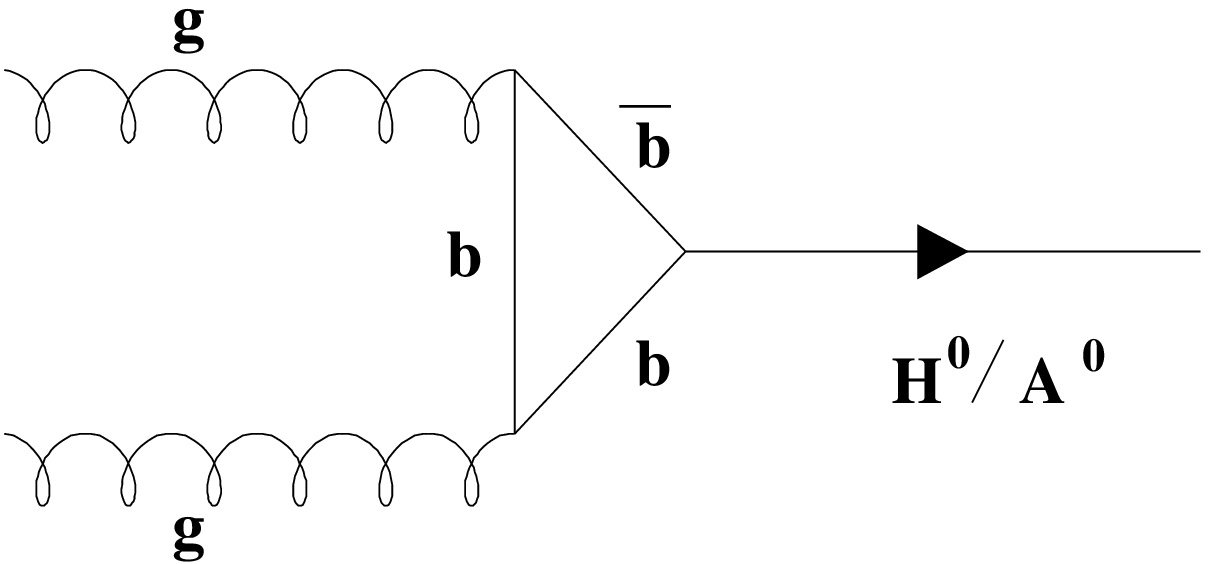,width=0.4\textwidth}}
\vspace*{-0.2cm}
 \caption{Direct A/H production  mode. } 
 \label{mec1}
\end{center}
\vspace*{-0.5cm}
\end{figure}
 rates are dominant and
significantly larger than in the SM case. For large values of  \tanbeta,
the production is dominated by 
$gg, q\overline{q} \freccia \bbbar A/H $, 
$gg \freccia \bbbar A/H$ is largely dominant between the two at the LHC,
  (associated \bbbar production mode- Fig. \ref{mec2}).
\begin{2figures}{hbtp}
  \resizebox{\linewidth}{40 mm}{\includegraphics[width=3.0cm]
{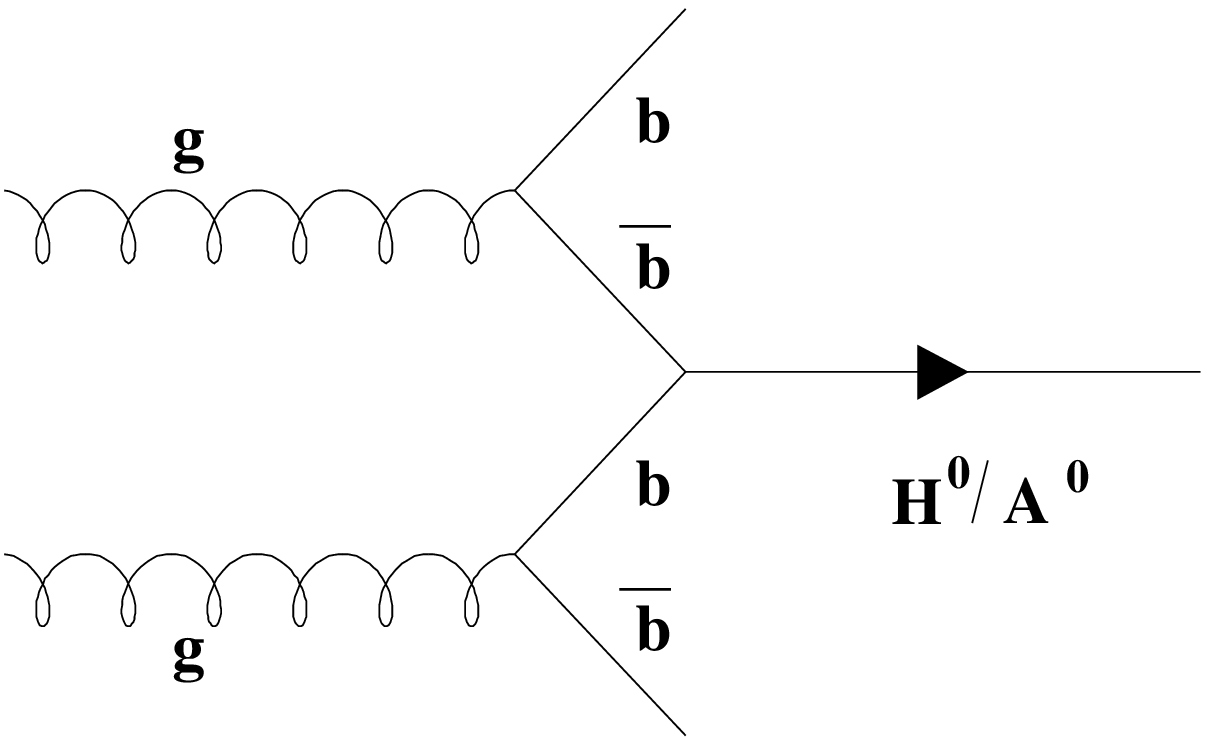}} &
  \resizebox{\linewidth}{35 mm}{\includegraphics[width=3.0cm]
{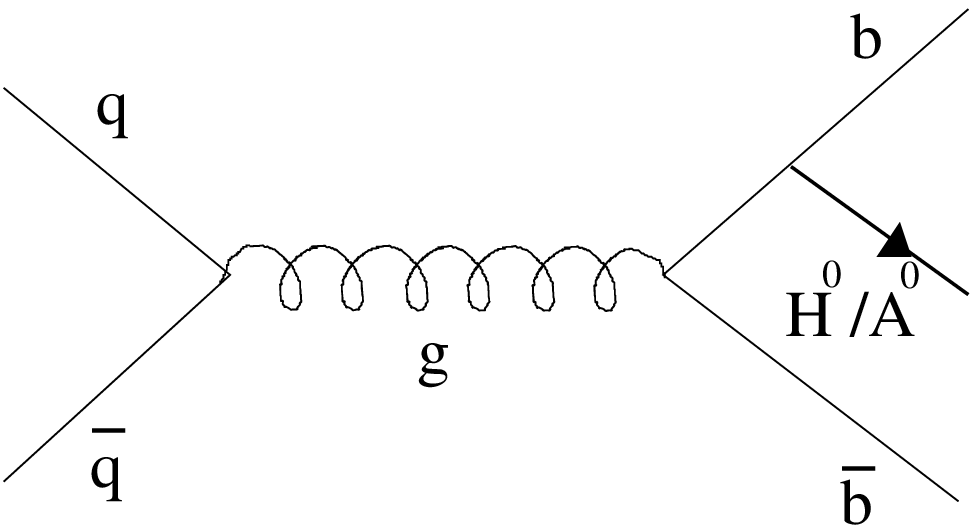}} \\
\vspace*{-0.2cm}
 \caption{Associated \bbbar A/H production modes. } 
 \label{mec2} & \\[-1.0cm]
\end{2figures}
\par
For \mA$>150$~GeV, the H and A bosons are degenerated in mass, so the
signal rates in the $\tau \tau$-channel can be added, whereas a more
complicated procedure depending on the experimental resolution and on the
mass difference \mH-\mA~ has to be applied for \mA$<150$~GeV \cite{NOI}.
 Higgs-boson
masses below 120 GeV have not been considered in this channel because of
the large resonant background from \Ztau~ decays.
\par
For high \tanbeta~ values A and H couple dominantly to the heaviest lepton 
and to the heaviest down-type quark;  the branching ratio of A/H is 
$\sim 90 \%$ into \bbbar and $\sim 10 \%$ into $\tau \tau$.
\par
Including the decay of the $\tau$ leptons, the three possible 
final states are: 
\begin{itemize}
\item
the lepton-lepton ($e \mu$) channel, with a branching ratio (BR) of $6.3 \%$;
\item
the lepton-hadron channel, with $BR=46 \%$;
\item
the hadron-hadron channel, with  $BR=41 \%$. 
\end{itemize}
\par
The lepton-lepton channel has been studied in the low \mA~ region
($<\sim$500 GeV)  
both in ATLAS \cite{NOI}  and CMS \cite{sami}: compared to
the lepton-hadron channel it turns out to
provide a worse sensitivity to a possible signal, due to both
its lower rate and  the less favourable kinematics of the
$\tau$-decay. CMS is studying the possibility to use the impact information
to reduce the backgrounds in this channel. 
\par
The lepton-hadron channel  has been studied in the low  \mA~ region
both in ATLAS \cite{ATLASTDR} \cite{noinuova} and CMS \cite{hljet}.
In this channel  the application of $\tau$-jet identification 
strongly reduces  the jet-background from various sources.
In ATLAS the \mA~ region studied has been recently extended to $\sim$1 TeV
\cite{noinuova} with promising results.
\par
 The hadron-hadron channel has been
 studied in  CMS \cite{h2jet} and recently in ATLAS \cite{juergenhh} 
in the higher \mA~
region ($> \sim$ 500 GeV). The leptonic decay channels include a trigger
lepton which allows for an efficient background reduction; in the
hadron-hadron channel the
purely hadronic final states compete with QCD jets, so it is difficult to
maintain the trigger rates acceptable and it is also difficult to find
criteria to reduce the huge QCD background. 
To exploit fully the
$2~\tau$-jet final states,  especially in the very low ($\sim$ 200~GeV)
 mass range, 
 an efficient hadronic $\tau$ trigger has been developed in CMS  
based on Level-1 calorimeter selection, Level-2 electromagnetic
calorimeter isolation \cite{trigger1} and a Level-3 tracking 
(isolation) \cite{trigger3}.
\par
The search strategy for all channels is based on kinematical cuts,
$\tau$-jet identification  (for the channels where at least one $\tau$
decays hadronically) and the recontruction of the $\tau \tau$ invariant
mass  \mtt, so it relies on two very
important detector performance  requirements.
One important feature of the \Aev~ analysis is in fact the
possibility to reconstruct the invariant $\tau \tau$ mass.
The energies of the two $\tau$'s 
 are evaluated from the energies of the $\tau$ decay products,
assuming that they have the same direction of the $\tau$-parent; the neutrino
energies are obtained solving a system containing the two \pTmiss~ components.
Therefore,   it is crucial to have a
very good \pTmiss~ resolution. 
  For the channels where at least one $\tau$ decays to
hadrons, a very good
$\tau$-jet identication is also crucial, 
to have the possibility to reject the huge
 jet-background from different sources.

\section{ATLAS Results}
\subsection{Event Generation,
 A/H Production Cross-Sections and Branching Ratios to $\tau \tau$}
\label{crosse}
The signal and background events were generated with the PYTHIA 6.152 
Monte Carlo event generator.
 The CTEQ5L parametrisation of the structure functions was used.
The fast ATLAS detector simulation was used \cite{atlfast}.
\par
The direct A/H production (from  
$gg\rightarrow \A/H \rightarrow\tau\tau$)  
cross-section 
is calculated using the program HIGLU \cite{spiraprog}, 
based on the results of \cite{glufusnlo}.
The associated \bbbar A/H cross-section 
is calculated using the program 
HQQ, which  calculates the production cross-section of Higgs
 bosons via  
$gg, q\overline{q} \freccia \bbbar A/H $
according to the results presented in \cite{gunion}.
The MSSM Higgs
  sector is implemented in the approximate two-loop RGE approach of
\cite{mhiggsRG1}.
\begin{2figures}{hbtp}
  \resizebox{\linewidth}{80 mm}{\includegraphics{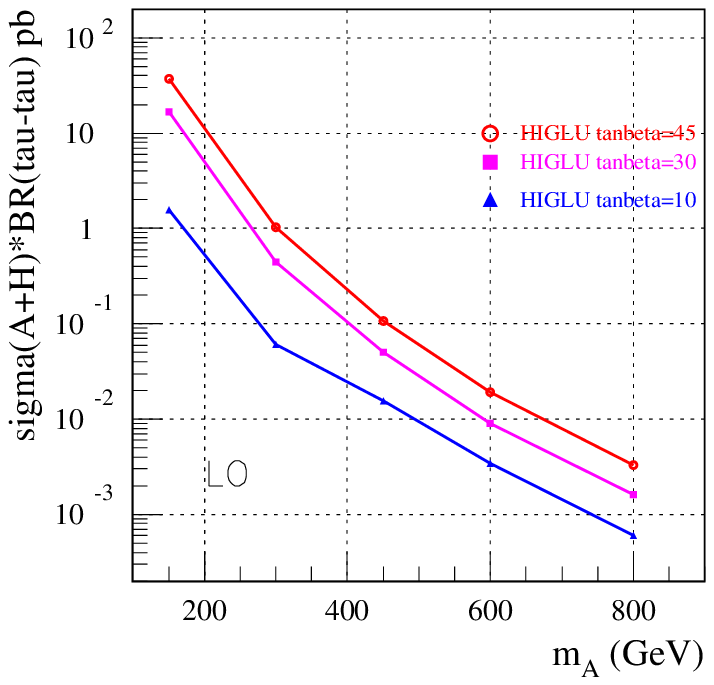}} &
  \resizebox{\linewidth}{80 mm}{\includegraphics{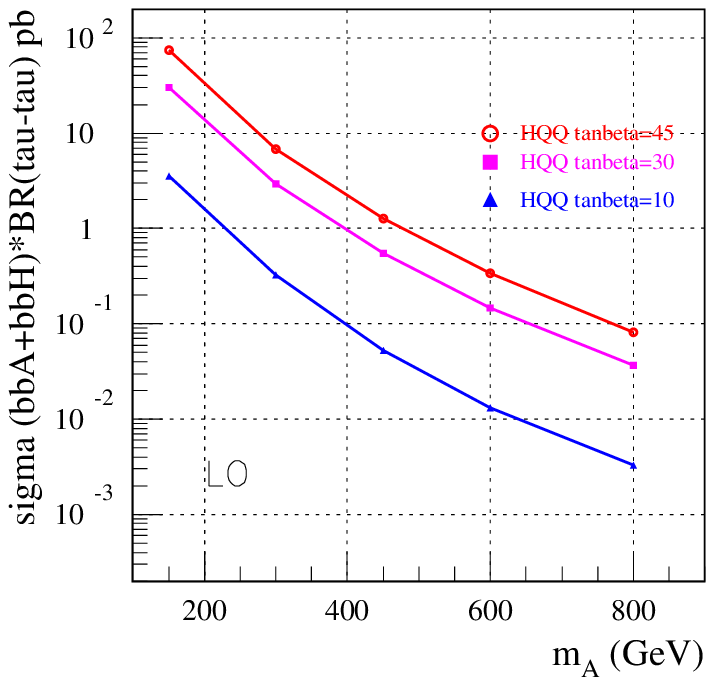}} \\[-1.2cm]
 \caption{Direct A/H production cross-section 
 times BR($A/H \freccia \tau \tau$)~ as a function of
 \mA~ for different \tanbeta~ values. }
  \label{sigmadir} &
 \caption{Associated \bbbar A/H production cross-section 
  times BR($A/H \freccia \tau \tau$)~ as a function of \mA~ 
  for different \tanbeta~ values. }
  \label{sigmass} \\[-1.0cm]
\end{2figures}
The  program  HDECAY is used to calculate the total decay widths 
and the branching ratios \cite{hdecay}.
The cross-sections  calculated at leading order (LO) are used here.
  The direct A+H production cross-sections 
 and  the associated \bbbar A+\bbbar H cross-sections,
both  multiplied by the  BR($A/H \freccia \tau \tau$), 
  are shown in Fig. \ref{sigmadir}
 and in  Fig. \ref{sigmass} respectively, for three different \tanbeta~
values.
 For large values of  \tanbeta,
the production is dominated by the associated production mode, moreover,
for a fixed \tanbeta~ value, the ratio between the associated and direct
production increases as \mA~ increases.
\begin{figure}[hbtp]
 \begin{center}
  \mbox{\epsfig{figure=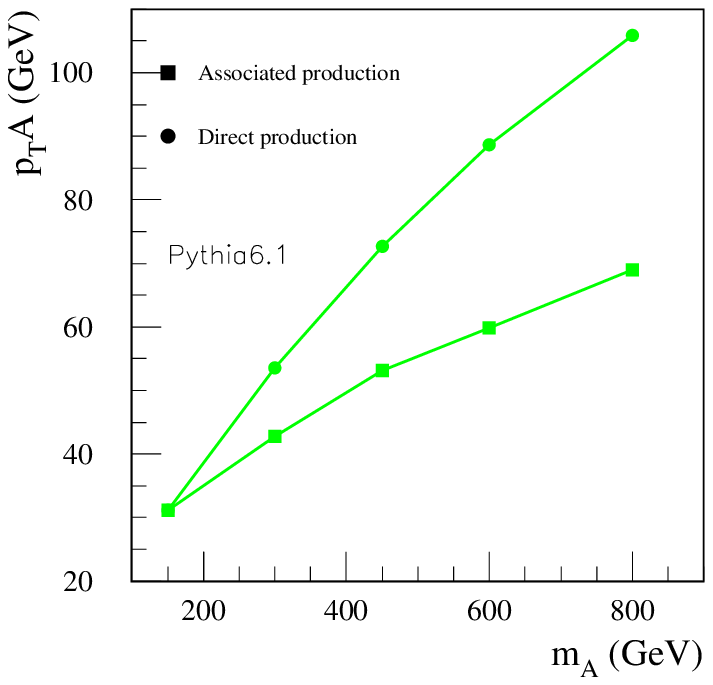,width=0.5\textwidth}}
 \end{center}
\vspace*{-1.5cm}
 \caption{ Average \pT(A/H) as a function of \mA~ in 
direct and associated production in  PYTHIA6.1}
 \label{pta}
\vspace*{-0.2cm}
\end{figure}
\par
There are  differences in the event topology and kinematics 
between the events from the two different production processes:
\begin{itemize}
\item
there are b-jets in \bbbar A/H events
\item
the \pT~ distribution of the generated A/H is different.
As  can be seen from Fig. \ref{pta}, the average \pT A is larger in direct
production for masses larger than 150 GeV.
\end{itemize}
\par
It must be underlined that the theoretical uncertainty for both the
computation of cross-section and simulation of events for the
\bbbar associated production is still large \cite{eduardo}. 

\subsection{$\tau$-jet identification, \pTmiss~ resolution 
and the reconstructed \mtt~ resolution in ATLAS}
\label{requ}
An excellent 
{\it $\tau$-jet identification performance} to suppress the  huge 
jet-background from various sources is necessary for the \Aev~ study for
the channels where at least one $\tau$ decays hadronically. In ATLAS the
$\tau$-jet identification/jet-rejection
 has been studied with  full detector simulations 
of signal and background events \cite{TAU}.
The criteria to identify 
a hadronic jet  as a $\tau$-jet are based on both calorimeter and tracker
information.
\par
In the \pT~ region 30-150 GeV, the requests are (here  called TDR criteria
because they were used to obtain the results reported in
in the ATLAS Physics Performance Technical Design Report 
(TDR)\cite{ATLASTDR}):\\
$\bullet$ $R_{em}~<~0.07$, where
$R_{em}$ is the jet radius computed using only the \EM~cells contained
in the jet; \\
$\bullet$ $\Delta E_T^{12}~<~0.1$,
where $\Delta E_{T}^{12}$ is the difference between the transverse
energies contained in cones of size \dR~=~0.2 and~0.1,
normalised to the total jet transverse energy~\ET; \\
$\bullet$ $N_{tr}~=~1$, where $N_{tr}$ is the number of reconstructed
charged tracks with \pT~$>$~2~GeV pointing to the cluster.
\par
With these cuts 
the {\it $\tau$-jet identification efficiency} ($\epsilon_{\tau}$)
 is $\sim 25 \% $ for  $\tau$'s from \mA=150 GeV
($<\pT^{\tau-jet}> \sim 50$ GeV)
and the
$jet-rejection$ goes from  $\sim 170$ to $\sim1700$ 
for jets in 30$<$\pT$<$150 GeV, 
depending on the \pT~ and on the jet type (light quark, gluon, b-jet).
With the same criteria 
 an $\epsilon_{\tau} \sim 40 \% $ for \mA=800 GeV 
($<\pT^{\tau-jet}> \sim 200$ GeV) and a 
$jet-rejection \sim 2500$ against QCD jets with \pT$>$150 GeV can be achieved.
In Fig. \ref{taueff} the
  $\tau$-jet identification efficiency as a function of \mA~ is shown,
while in Fig. \ref{effjet} the jet-efficiency is shown 
as a function of the jet \pT.
\begin{2figures}{hbtp}
  \resizebox{\linewidth}{80 mm}{\includegraphics{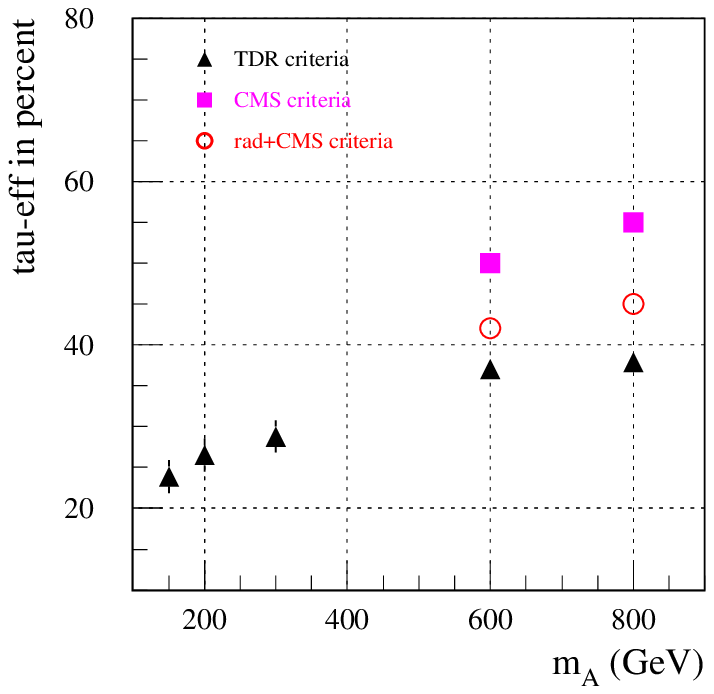}} &
  \resizebox{\linewidth}{80 mm}{\includegraphics{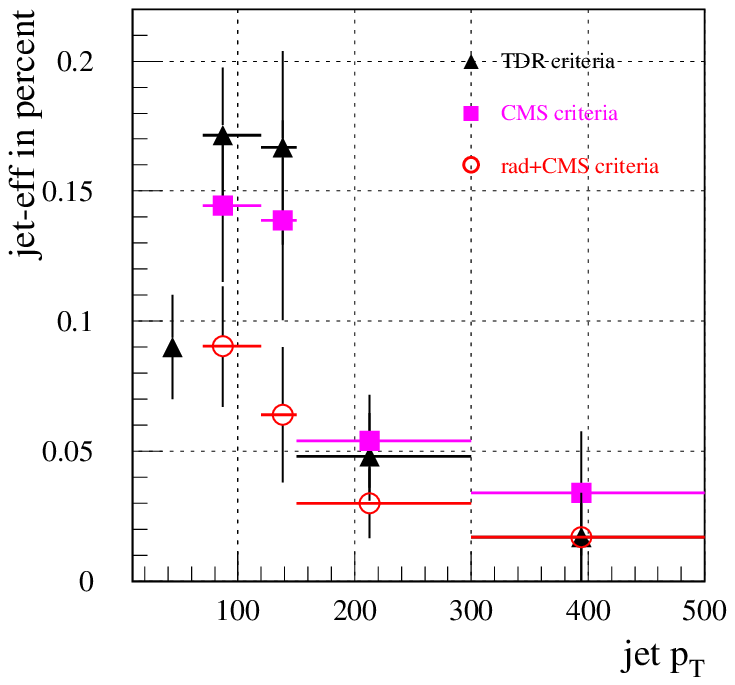}} \\[-1.2cm]
  \caption{ Efficiency of the $\tau$-identification criteria for $\tau$'s
   from \Atau~ decays as a function of  \mA.} 
  \label{taueff} &
  \caption{ Efficiency of the $\tau$-identification criteria in 
    QCD jet events  as a function of  \pTjet.}
  \label{effjet} \\[-1.2cm]
\end{2figures}
The $\tau$-jet identification criteria were optimized 
 for the high \mA~ and \pT ($>$150 GeV) \cite{noinuova}
combining the Atlas TDR criteria and the criteria used by CMS in \cite{h2jet}.
 Asking for:\\
$\bullet$ $R_{em} <0.12$\\
$\bullet$  CMS cuts
\par
- 1 isolated track with \pT$>$40~GeV within \dR$<$0.1 from the jet axis
\par
- track isolation: no other track with \pT$>$1 GeV in a cone of \dR=0.4\\
 an $\epsilon_{\tau} \sim 45 \%$ ($\epsilon_{\tau} \sim 55 \%$ - only CMS
criteria) for  $\tau$'s from \mA=800 GeV can be achieved and   
 the jet-rejection can be significantly improved to  
$ \sim 3500$ ( $\sim 2000$ - only CMS criteria)
 for QCD jets with \pT$>$150 GeV. 
\par
However, the jet-efficiencies are still determined with large errors:
 a larger statistics of fully  
simulated events containing jets  is still needed 
to complete this study.
\vskip 0.5cm
\par
Moreover a very good 
{\it \ETmiss -resolution performance }
for the reconstruction of the $\tau \tau$ mass is required 
 for the \Atau~ channel study.
Crucial for a good \pTmiss~ resolution \cite{PTMISS} are the
calorimeter coverage until \abseta$<$5,
\ptmx~and \ptmy~ have to be reconstructed from all calorimeters cells 
(in clusters and outside the clusters),
a careful calorimeter calibration and intercalibration is necessary and
finally a careful choice of electronic noise cutoff has to be made 
(only cells with \ET$>1.5\sigma(noise)$ are kept).
\par
In this way, the \pTmiss~ resolution in ATLAS is found to be:
$$\sigma(\pTmiss)=0.46*\sqrt{\Sigma\ET}$$
where $\Sigma\ET$ is the total transverse energy in the calorimeters
expressed in GeV.
\par
This formula is valid at low luminosity and it takes into account 
both the coverage effect and the  energy resolution.
At high luminosity, there is a strong  degradation of the \pTmiss~
resolution (about a factor of 2 worse) 
due to the pile-up as described in the TDR \cite{ATLASTDR}.
\vskip 0.5cm
\par
The invariant mass of the $\tau$-pair in \Aev~ 
can be reconstructed  in the collinear approximation  that
 the directions of the two neutrino systems
from each $\tau$-decay 
coincide with the ones of the measured $\tau$-decay products
 and under the condition that the $\tau$-decay products
are not back-to-back:\\
$$m_{\tau \tau}~=~\sqrt{2(E_1+E_{\nu_1})(E_2+E_{\nu_2})(1-cos\theta)}$$
where
 $E_1$, $E_2$ are the energies 
of the measured $\tau$-decay products,
$E_{\nu_{1}}$, $E_{\nu_{2}}$ are the energies of the two neutrino 
systems  and $\theta$ is the angle between the directions
of the measured $\tau$-decay products.
$E_{\nu_1}$ and $E_{\nu_2}$ are obtained by solving a system
containing the two \pTmiss~ components.
The measurement uncertainties on \ptmx~, \ptmy~ combined with
the assumption on the  directions of the decay-products often result in
unphysical negative solutions for the  neutrino energies, in that case the
\A~ mass cannot be reconstructed.
\par
The mass resolution $\sigma(\mtt)$ is  proportional to 
$\sigma(\pTmiss)/|sin(\Delta\phi(p_1 p_2))|$, 
 therefore both the
$\pTmiss$ resolution and the $\Delta\phi$ separation between the charged
$\tau$-decay products are important in the
$\tau\tau$ mass reconstruction \cite{NOICOMB}.
\par
The reconstructed \mtt~  resolution has been compared for the three
different final states (after having applied a cut on the lepton
\pT$>$24 GeV and on the $\tau$-jet \pT$>$40 GeV - the dependence of
the mass resolution on the \pT~ cutoffs is weak - and the cut  
$\Delta \phi < 165^{\circ}$)  (see Fig. \ref{resolm}) 
\begin{figure}[hbtp]
 \begin{center}
  \mbox{\epsfig{figure=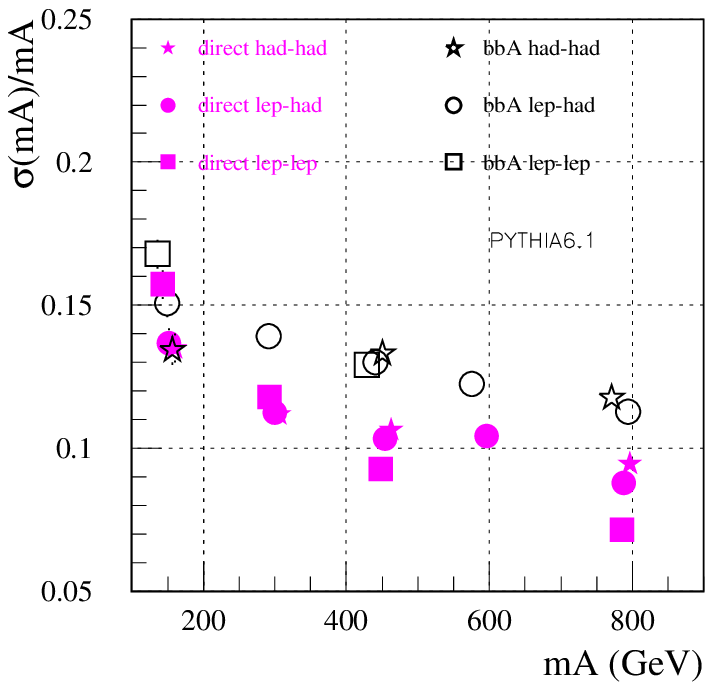,width=0.5\textwidth}}
 \end{center}
\vspace*{-1.5cm}
 \caption{Relative reconstructed \mtt~ resolution as a function of  A mass.} 
 \label{resolm}
\vspace*{-0.2cm}
\end{figure}
and it is found to be comparable, 
as it is expected on the base of  the \pTmiss~ resolution,
shown in Fig. \ref{resolptm}, and the 
 $\Sigma \ET$  in calorimeters, shown in  Fig. \ref{sumet}.  

\begin{2figures}{hbtp}
  \resizebox{\linewidth}{80 mm}{\includegraphics{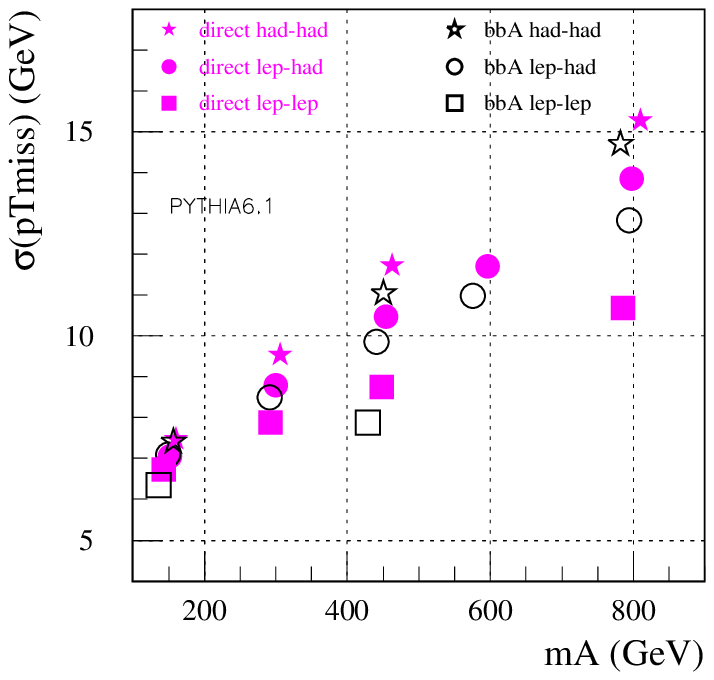}} &
  \resizebox{\linewidth}{80 mm}{\includegraphics{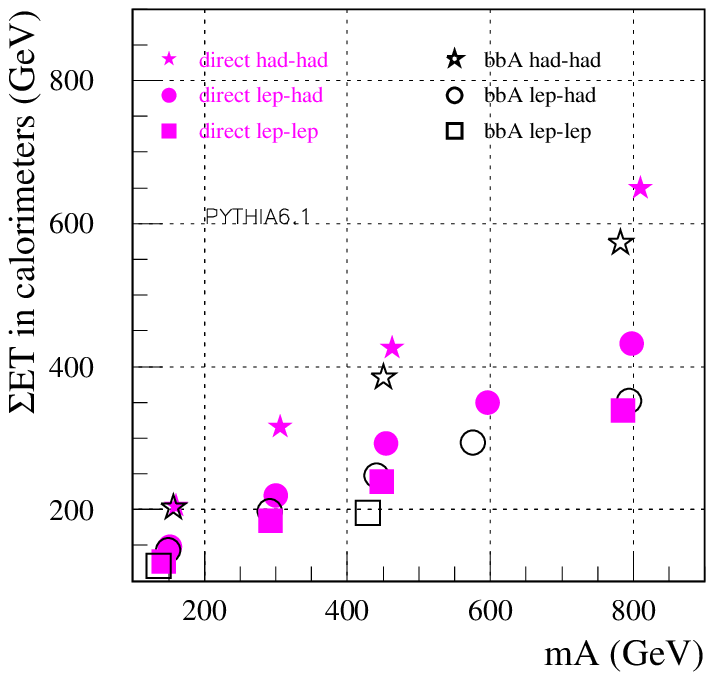}} \\[-1.2cm]
  \caption{\pTmiss~ resolution in A events as a function of A mass.} 
  \label{resolptm} &
  \caption{$\Sigma$ \ET~ in calorimeters in A events as a function of A mass.}
  \label{sumet} \\[-1.5cm]
\end{2figures}
The reconstructed mass resolution is systematically worse for the associated
production events.
In the associated production events in fact
the  average \pT~  of the generated A  is lower with 
 respect to the direct A production and the difference increases with \mA~
(see Fig. \ref{pta}).
 This implies that
in $\bbbar\A$ events the two $\tau$'s from the 
\A~ tend to be more back to back, with two important consequences  that are
a lower acceptance of the $\Delta\phi$ cut  and
 a final worse solution of the system giving the neutrino energies 
and consequently
a lower efficiency in the mass reconstruction and a worse resolution.
At \mA=450 GeV the relative $\sigma(\mtt)$  is $\sim 10 \%$ and 
$\sim 13 \%$ respectively for direct and associated production with an
overall efficiency  of the $\Delta \phi$ cut and of the request of positive
solutions for the neutrino energies 
 of $\sim 30 \%$ and $\sim 20 \%$.
 
\subsection{The lepton-hadron channel analysis in ATLAS}
\label{clh}

The irreducible backgrounds for this channel are 
$\ttbar \freccia bW^+ \overline{b} W^- \freccia \bbbar lep~ \tau$ 
and \Zgatau~, 
the reducible backgrounds
are $\ttbar \freccia bW^+ \overline{b} W^- \freccia \bbbar lep~ had$,
 $W \freccia lep$ +jets and $\bbbar \freccia lep~ had$.
\par
The analysis criteria  
({\it Standard A analysis}) are:\\
$\bullet$ \pT$^{lepton}~>$~24~GeV (40 GeV for \mA$>$500 GeV) 
and \abseta$^{lepton}~<$~2.5;\\
$\bullet$ Isolation of the trigger lepton
(which rejects leptons
from \bbbar~by a factor~100 for a 90\% efficiency for isolated leptons);\\
$\bullet$   m$_T(lepton-\pTmiss)~<~25~GeV$ 
(against the backgrounds containing W), 
where m$_T(lepton-\pTmiss)$ is the transverse mass of the lepton-neutrino
system;\\
$\bullet$ \pTmiss~$>$~18~GeV (40 GeV for \mA$>$500 GeV);\\
$\bullet$ \ET$^{jet}~>$~40~GeV (80 GeV for \mA$>$500 GeV), 
\abseta$^{jet}~<$~2.5 ~~~~~~~~~($\tau$-Candidate);\\
$\bullet$  $\Delta \phi(jet-lepton)$ in $100^{\circ}-165^{\circ}$;\\
$\bullet$ m$_{\tau \tau}$~in the window \mA~$\pm~\Delta$M~ ($\Delta$M =
1.5$\sigma_{m_{\tau \tau}}$).\\
\\
Each event is weighted using the $\tau$-jet identification factor for
the  $\tau$-jet candidate  (see section \ref{requ}).
\par
Due to the topological and kinematical differences in the direct and
associated events,
 two different analyses, one optimized for the direct
production process, the other one optimized for the associated production
are performed with the following criteria:\\
- {\it Direct analysis:}\\
$\bullet$ zero b-jet tagged (against \ttbar~ 
  and \bbbar~ backgrounds);\\
$\bullet$ all cuts of the {\it Standard \A~ analysis} 
($\tau$-jet identification,    kinematic and mass cuts).\\ 
- {\it Associated analysis:}\\
$\bullet$  1 b-jet tagged (against \Z~ and W+jets backgrounds);\\
$\bullet$  number of non b-jets $<$ 3 (against \ttbar~
         backgrounds); \\
$\bullet$  cuts of {\it Standard \A~ analysis}
         ($\tau$-jet identification, kinematic
         and mass cuts) {\it except the cut on}  $\Delta\phi(jet-lepton)$, 
to not reduce too much the signal acceptance.\\
To choose the b-jets, a b-tagging efficiency of 60$\%$, with a
corresponding rejection of 100 against other jets and of 10 against c-jets
has been randomly applied.
\par
The dominant background selected by the direct analysis arises from
W+jets, which have the largest production cross-section and from
 the \Ztau~  at the lower masses. The analysis 
optimized for the associated production rejects much better W+jets and 
\Ztau~ backgrounds and the \ttbar background becomes dominant.
\par
Having the opposite request to have or not to have a b-jet tagged,
the two analyses are not correlated, so, after having applied them
separately  to both signal
samples (direct and associated A production) and to background events,
 the significances can be combined.
\par
The results at lower masses have been compared to the results reported in
the TDR and they have been found to be in reasonable agreement \cite{noinuova}.
The differences observed are due to the different PYTHIA version, to the use
of fast instead full simulation and to the use of different cross-sections
values for signals and backgrounds.
\vskip 0.5cm
\par
At the higher masses ($\mA >$ 500 GeV), the analysis is performed only
on the events from the \bbbar A production channel, due to the complete
dominance of that production mode (see section \ref{crosse}).
\par
Despite the low production cross-section 
and  the low acceptances of the analysis  ($\sim 0.5 \%$ for \mA=800 GeV)
  the
  backgrounds are strongly reduced  (the total background for \mA=800 GeV
is  $\sim$4.6 events, dominated by \ttbar~ background, in 30 \fbm).
\par
Figure \ref{sigbac} shows the
distribution of \mtt~ after the  analysis cuts (except the mass cut) 
 for associated signal events at \mA=800 GeV for \tanbeta=45
and for the main backgrounds normalized to the expected event number
for an integrated luminosity of 30 \fbm.
\begin{figure}[htbp]
\vspace*{-0.5cm}
 \begin{center}
  \mbox{\epsfig{figure=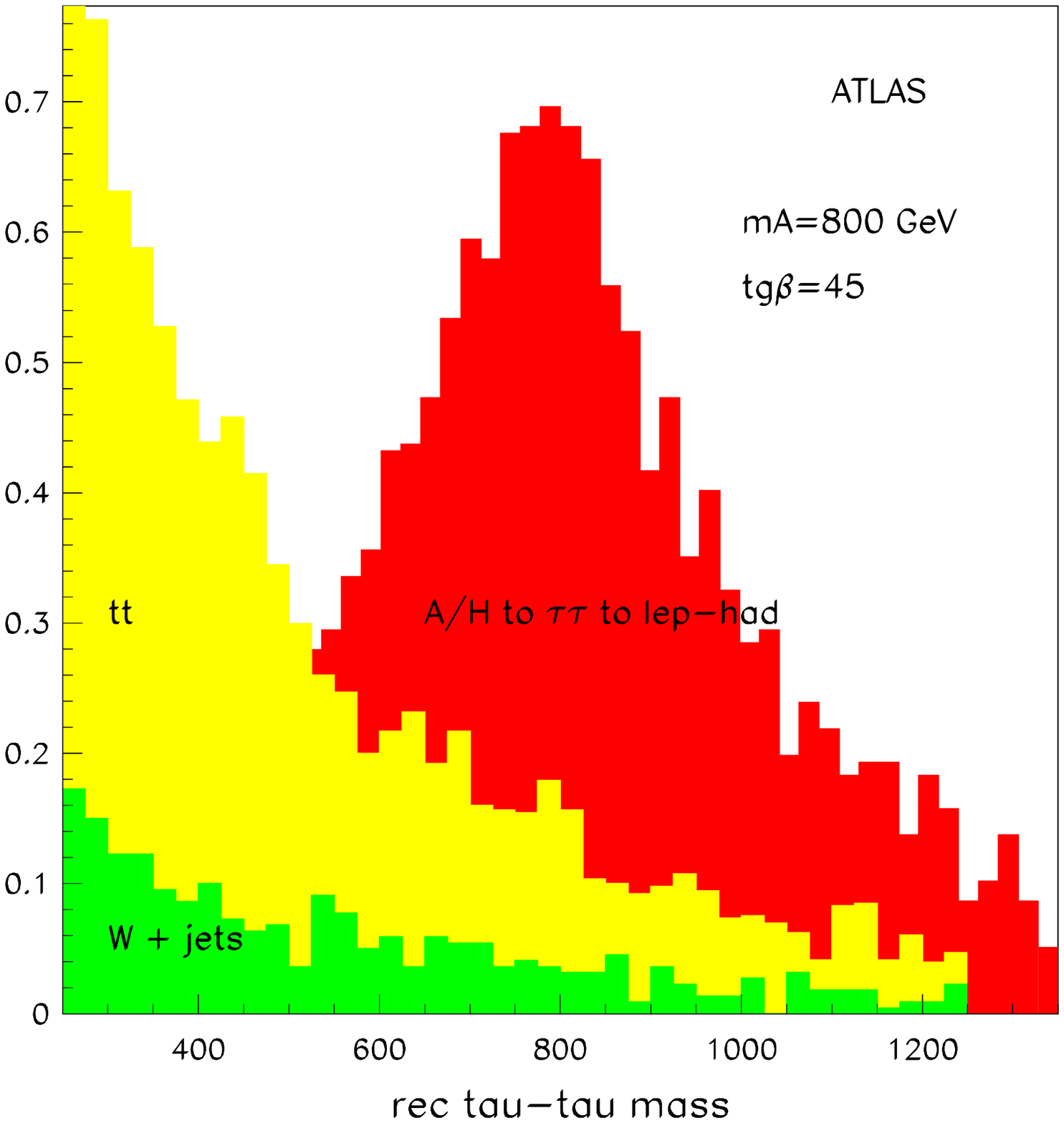,width=0.5\textwidth}}
 \end{center}
\vspace*{-0.8cm}
 \caption{Reconstructed \mtt~ in the lepton-hadron channel 
after the  analysis cuts for  \mA=800 GeV for \tanbeta=45
and for the \ttbar and W+jets backgrounds (plotted separately) 
assuming an integrated luminosity of 30 \fbm.}
 \label{sigbac}
\vspace*{-0.2cm}
\end{figure}
\subsection{The hadron-hadron channel in ATLAS: trigger study and analysis}
\label{hadhadatl}
Due to the very low production cross-section of the $A/H$ especially in the
higher mass region between 0.5 and 1 TeV, the
trigger acceptance for the signal needs to be excellent; on the other hand,
 the signature of the signal with only hadronic decays is similar to
QCD-background.\\
In the first level trigger, the total rate of Jet and $\tau$ triggers is
dominated by 2-jet QCD-background,
which has to be controlled by appropriately high settings of the 
\ET~ thresholds for jet triggers \cite{RolfNote}.
However, these settings also reduce the acceptance of the
trigger for the hadron/hadron channel of the $A/H$ decay. 
A high input acceptance of the  
trigger for \Atauhh~ signal events can be achieved by
using combined Jet+\ETmiss~ and $\tau$+\ETmiss~ triggers.
\par
The fast trigger simulation ATL1CT \cite{ATL1CT} \cite{Kambiz}
interfaced to the fast ATLAS simulation program has been
used for the study.
The following trigger types are of 
relevance for the hadron-hadron channel:
\begin{itemize} 
 \item { Jet + \ETmiss}~(Threshold: \ETjet $ > 50{\rm GeV}$, 
\ETmiss $> 50{\rm GeV}$)
 \item { $\tau$ + \ETmiss}~(Threshold: \ETtau$ > 20{\rm GeV}$,
 \ETmiss$ > 30{\rm GeV}$)
 \item { Single Jet} (\ETjet$ > 180{\rm GeV}$)
 \item { Three Jet} (\ETjet$ > 75{\rm GeV}$)
 \item { Four Jet} (\ETjet$ > 55{\rm GeV}$)
\end{itemize}
The threshold settings on the cluster $E_{T}$ 
of the jets and $\tau$'s 
are set to accept 90 \% of the jets with the $E_T$ value given in the
trigger menu, which is identified with the $E_T$ value in the reconstruction 
\cite{Kambiz}. The isolation criteria of the $\tau$ trigger are set to
fixed values of 2 GeV for the electromagnetic and 4 GeV for
hadronic trigger towers \cite{TriggerTDR}. \\
The total acceptance in the Level-1 trigger is determined by
the number of events accepted by at least one of the trigger
types. This is equivalent to a logical {\it OR}. Many events are
accepted by more than one trigger.
\par
The input acceptance of the Level-1 trigger for signal events 
$m_{A/H} = 450, 600~{\rm and}~800~{\rm GeV}$ 
for the individual trigger types listed above and 
their combination ({\it OR}) are shown in Fig. \ref{acc_masses}. For 
$m_{A/H} = 800~{\rm GeV}$, an input acceptance of 76.6 \% is reached 
using fixed isolation thresholds in the $\tau$ trigger. Using an alternative
scheme of a dynamic isolation in the $\tau$ trigger (electromagnetic trigger
towers: 4 \% of cluster energy, hadronic: 8 \% of cluster energy), this value
can be improved to 80.6 \% \cite{juergenhh}. 
The total trigger rate was evaluated being $\sim 1.4~{\rm kHz}$, 
which fits well within the limitations of the Level-1 trigger menu \cite{TriggerTDR}.
\\
The influence of the trigger acceptance on the discovery contour, however, 
is given by the combination of the trigger acceptance and the offline
analysis, discussed after.
\begin{figure}[htbp]
 \begin{center}
   \begin{minipage}{11cm}
    \includegraphics[width=11cm]{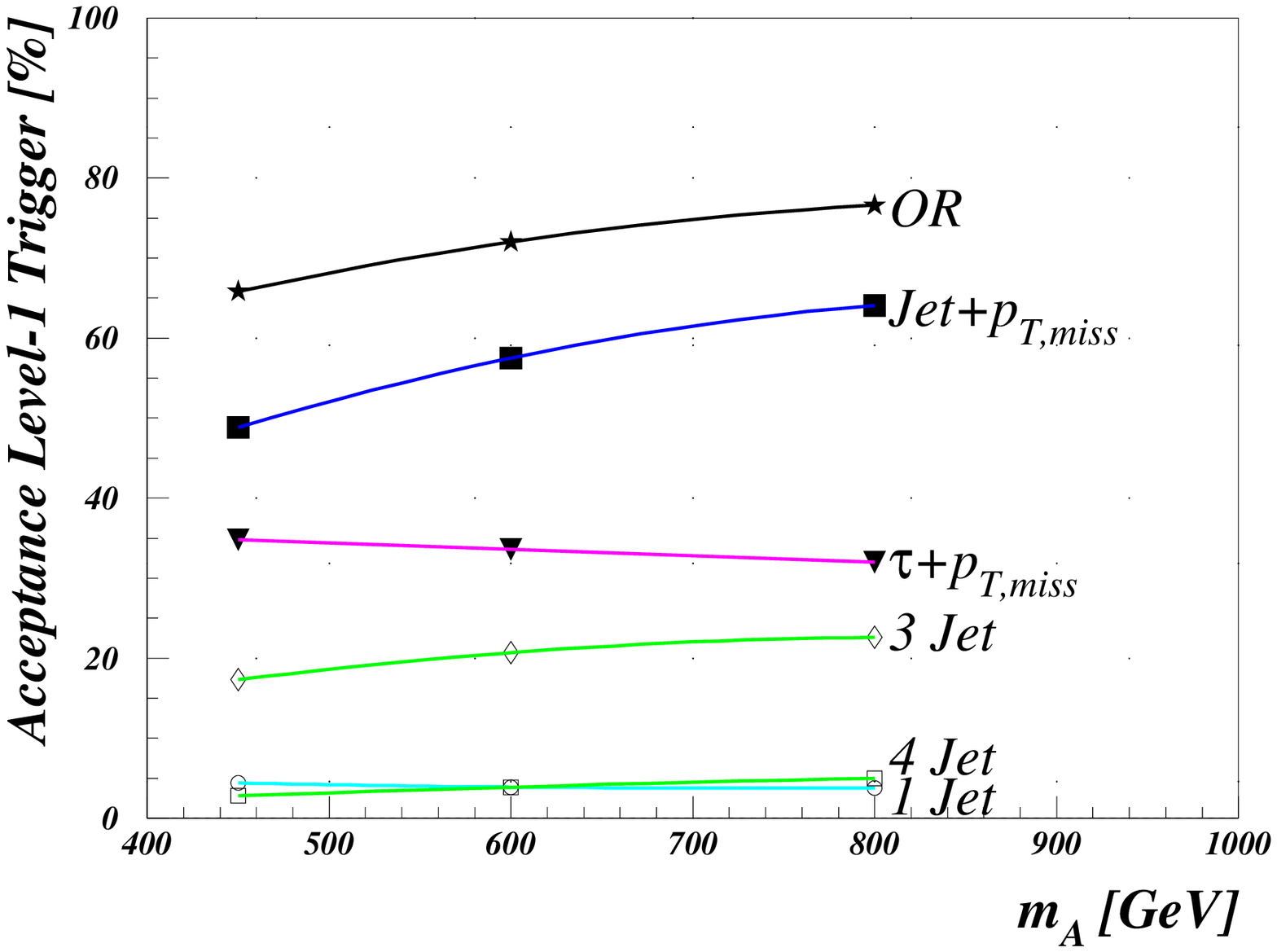}
   \end{minipage}
 \end{center}
\vspace*{-0.8cm}
 \caption
{Level-1 trigger input acceptance for Signal  $m_{A/H}$ = 450, 600 and 800 GeV:
total acceptance by all Jet and $\tau$-triggers ({\it OR}), and acceptances
of each separate trigger. }
\label{acc_masses}
\vspace*{-0.2cm}
\end{figure}
\par
Since the associated production is dominant 
for high masses, the expected signal event signature of the hadron-hadron
channel consists of two 
high-$p_T$ $\tau$'s in hadronic decay with two b-jets.
Backgrounds for this channel are 2-jet QCD, \ttbar, W+jets and Z+jets events.
The analysis uses event weighting, where the
two jets with the highest $p_T$ are considered as '$\tau$ candidates', while
b-tagging is used for all other jets, where one of the two expected
 b-jets is requested to be tagged.
Using this method, 
background events are not rejected by the requests on the number
of $\tau$'s and b-jets, but weighted accordingly, therefore background rate
estimates can also be given for channels with extremely low acceptance,
especially 2-jet background from QCD and also \ttbar. 
The $\tau$
identification described in Sec. \ref{requ} is used to derive the 
$\tau$ acceptance factors for each $\tau$ candidate. The $\tau$ identification
is assumed to be $\epsilon_{\tau} = 55 \%$ and the corresponding jet
rejection is used for the other jets.
The b-jet tagging efficiency is here assumed being $\epsilon_b = 70 \%$.
\par
The following cuts have been applied:\\
  $\bullet$ { Two jets}  in the event with $p_T > 100~{\rm GeV}$ and 
  $|\eta| < 2.5$ ($\tau$ candidates);\\
  $\bullet$ { No lepton} (e, $\mu$) with $p_T > 10~{\rm GeV}$;\\
  $\bullet$ { Not more than 4 jets} in $|\eta|<3.2$ with 
  $p_{T} > 20~{\rm GeV}$;\\
  $\bullet$  at least   { one b-jet } tagged ;\\
  $\bullet$ { \pTmiss $>$ 65 GeV};\\
  $\bullet$ { $\Delta\phi$ between $\tau$ candidates in } 
    $145^{\circ}-175^{\circ}$;\\
  $\bullet$ { transverse mass cut:} $m_{T}$ $<$ 50 GeV (minimum of
    $m_{T} (\tau_{1},\pTmiss)$ and $m_T (\tau_{2},\pTmiss)$);\\
$\bullet$ m$_{\tau \tau}$~ in the window \mA~$\pm~\Delta$M~ ($\Delta$M =
1.5$\sigma_{m_{\tau \tau}}$).\\
Some cuts lower
the acceptance of signal events significantly, especially the \pTmiss~ cut and
b-tagging, however, those  cuts are necessary to suppress the
background channels efficiently.
The acceptance of the analysis cuts for \mA=800 GeV is $\sim 0.6 \%$ 
and the total background is $\sim$ 5.4 events, with dominance of
\ttbar (2.2 events) followed by Z (0.8 events) and 2-jet QCD background 
 in 30 \fbm \cite{juergenhh}.
\par
Figure \ref{sigbg_800} shows the
distribution of \mtt~ after the cuts (except the mass cut) 
for the events at \mA=800 GeV for \tanbeta=50
and for the main backgrounds normalized to the expected event number
for an integrated luminosity of 30 \fbm.  
\begin{2figures}{hbtp}
\vspace*{-0.6cm}
  \resizebox{\linewidth}{80 mm}{\includegraphics{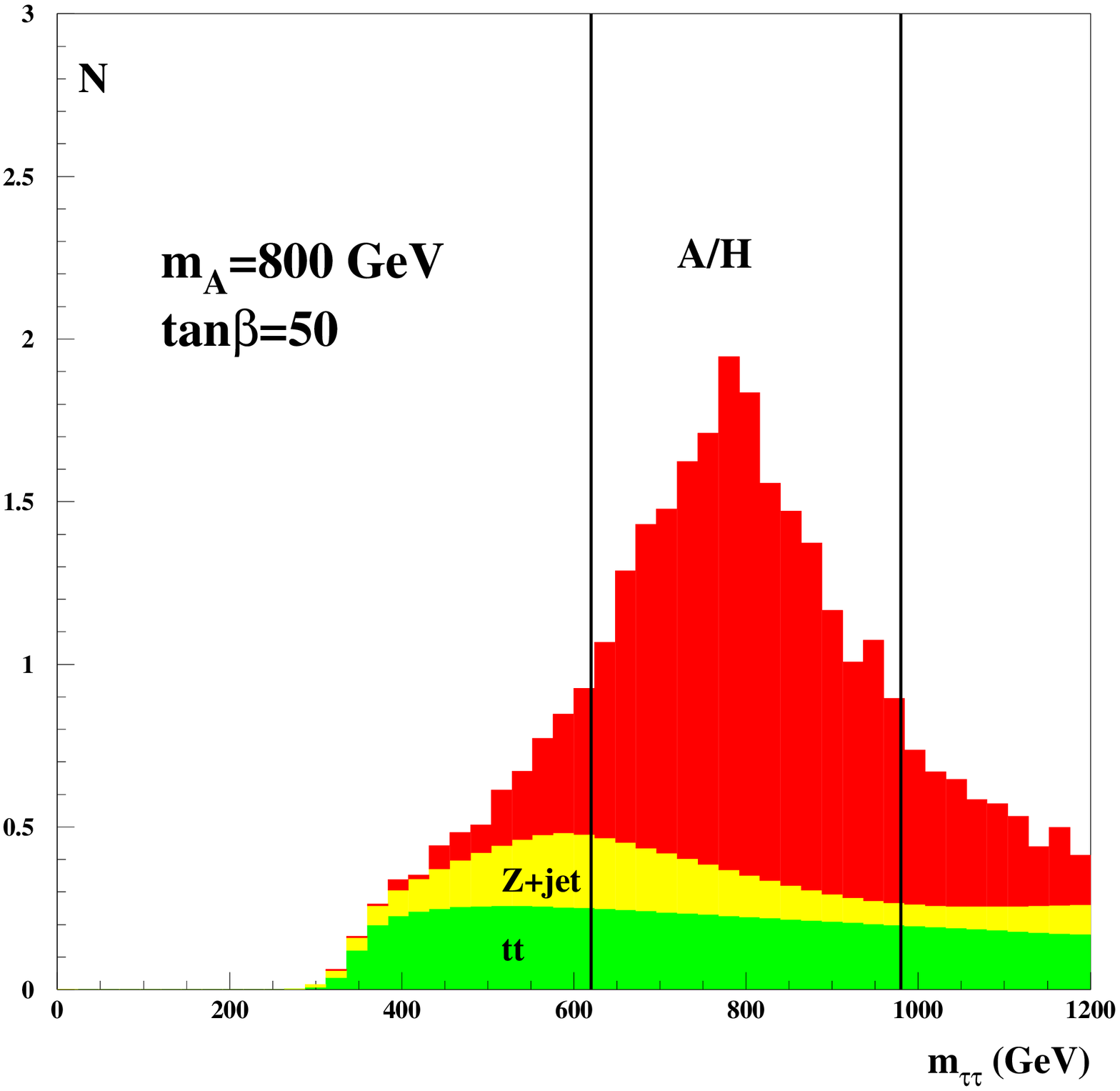}} &
\vspace*{-0.6cm}
  \resizebox{\linewidth}{80 mm}{\includegraphics{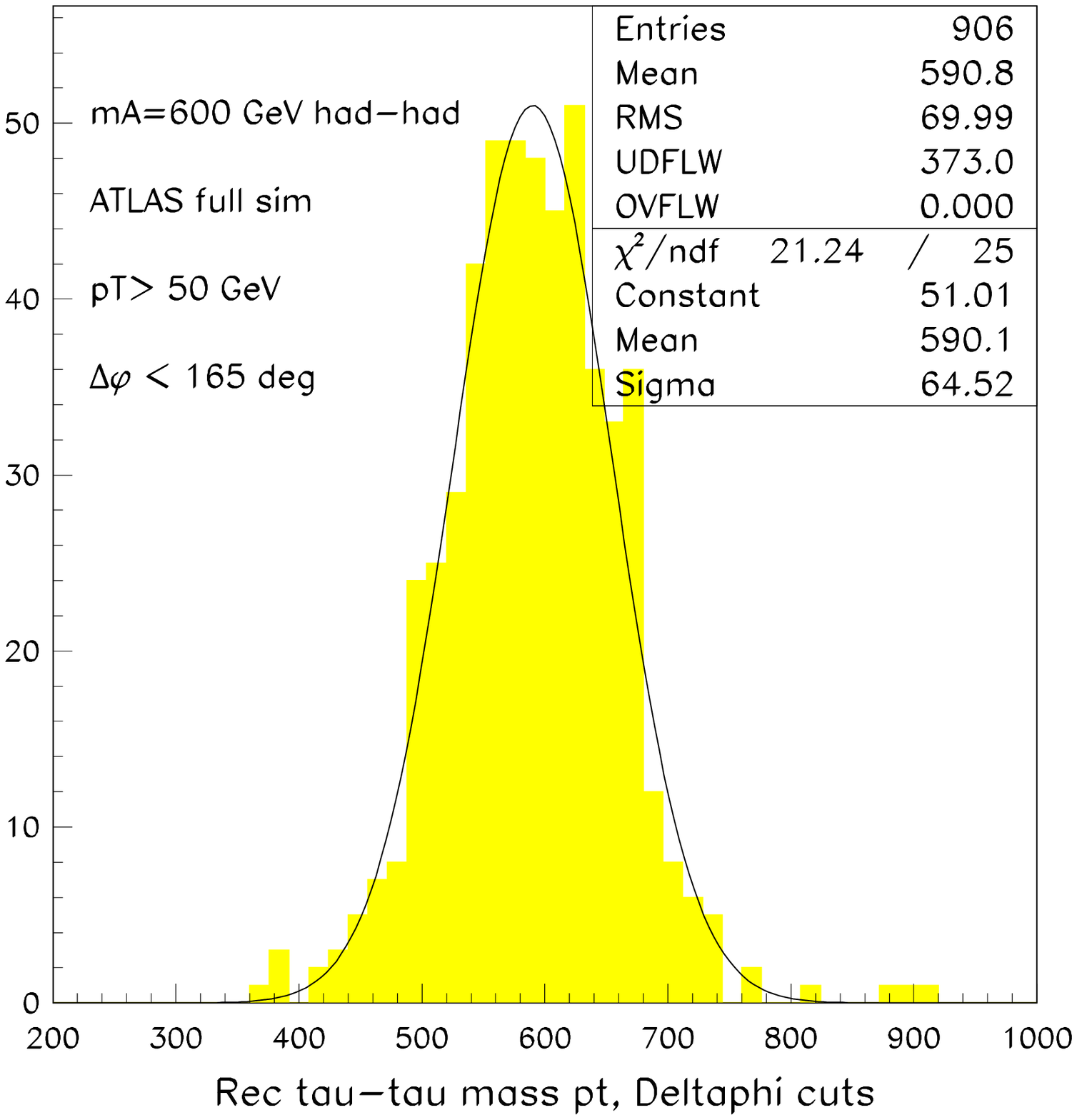}} \\[-0.6cm]
 \caption{Reconstructed \mtt~ in the hadron-hadron channel after the  
analysis cuts for  \mA=800 GeV for \tanbeta=50
and for the  \ttbar and Z backgrounds assuming an integrated luminosity 
of 30 \fbm.} 
 \label{sigbg_800} &
  \caption{Reconstructed \mtt~ for signal events with \mA=600 GeV
   in the hadron-hadron channel in full ATLAS detector simulation.}
 \label{massfull600} \\[-1.2cm]
\end{2figures}
\par
In Fig. \ref{massfull600}  
the reconstructed \mtt~ is shown for \mA=600 GeV 
 in the \Atauhh~ full simulated events,
 after the $\pT^{jet}>50$ GeV  and $\Delta\phi < 165^{\circ}$ cutoffs.
\vskip 0.5cm
\par
The combined acceptance of the Level-1 trigger  
and offline analysis for the hadron-hadron channel have been studied.
The acceptance after all cuts for events passing the trigger conditions 
are evaluated to be $ 92.5 \%, 95.6 \% $ and $ 95.3 \%$
respectively for \mA=450, 600 and 800 GeV.
Especially for high $m_A$, the acceptance is
very good due to the similarity of the kinematic cuts 
used in the Level-1 trigger and offline analysis; both require
large values of \ETmiss~ and jets/$\tau$'s with high $p_{T}$.

\subsection{Combinations of the lepton-hadron and the hadron-hadron channels 
 in ATLAS}
\label{fine}
For  higher mass values, the results from the lepton-hadron and from
the hadron-hadron channels can be combined to improve the signal significance.
\par
In Table \ref{disclimits}, the discovery \tanbeta~ values
(giving a 5$\sigma$ significance) are reported for the
two channels separately and combined.
\begin{table}
\begin{center}
 \caption{Discovery \tanbeta~ values in ATLAS (5$\sigma$ confidence).}
\label{disclimits}
\begin{tabular}{|l|c|c|c|}
\hline
$m_A$ (GeV) & lepton-hadron & hadron-hadron & combined \\
\hline
450 & 20.7 & 22.1 & 19.0 \\
600 & 32.9 & 30.0 & 25.2 \\
800 & 50.0 & 45.0 & 41.4 \\
\hline
\end{tabular}
\end{center}
\vspace*{-0.5cm}
\end{table}
The extended discovery contour plot is given in  Fig. \ref{contourco} in 
the \tanbeta~ logarithmic scale 
and with a linear  \tanbeta~ scale in  Fig. \ref{contourcl}.
\begin{2figures}{hbtp}
  \resizebox{\linewidth}{80 mm}{\includegraphics{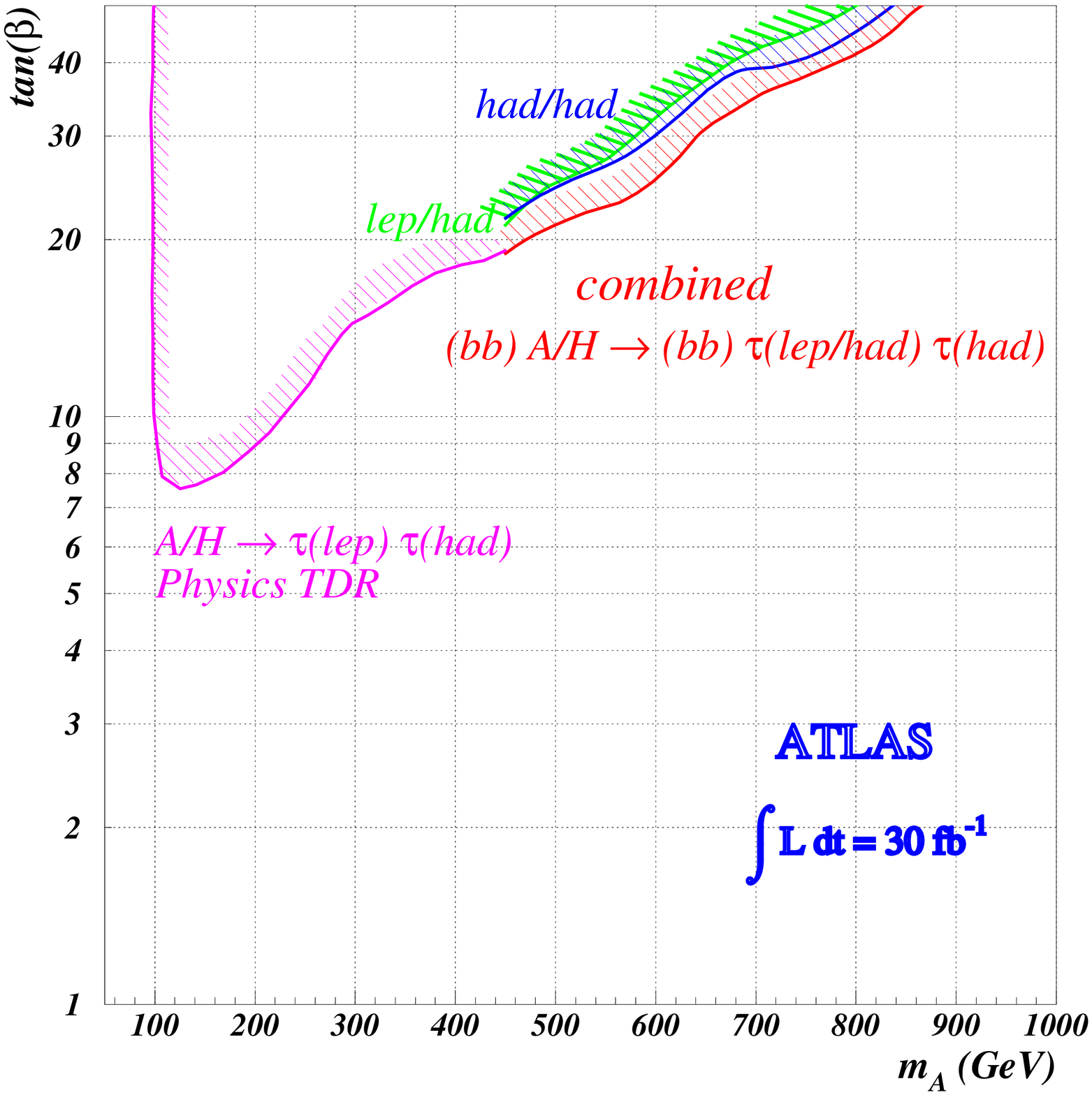}} &
  \resizebox{\linewidth}{80 mm}{\includegraphics{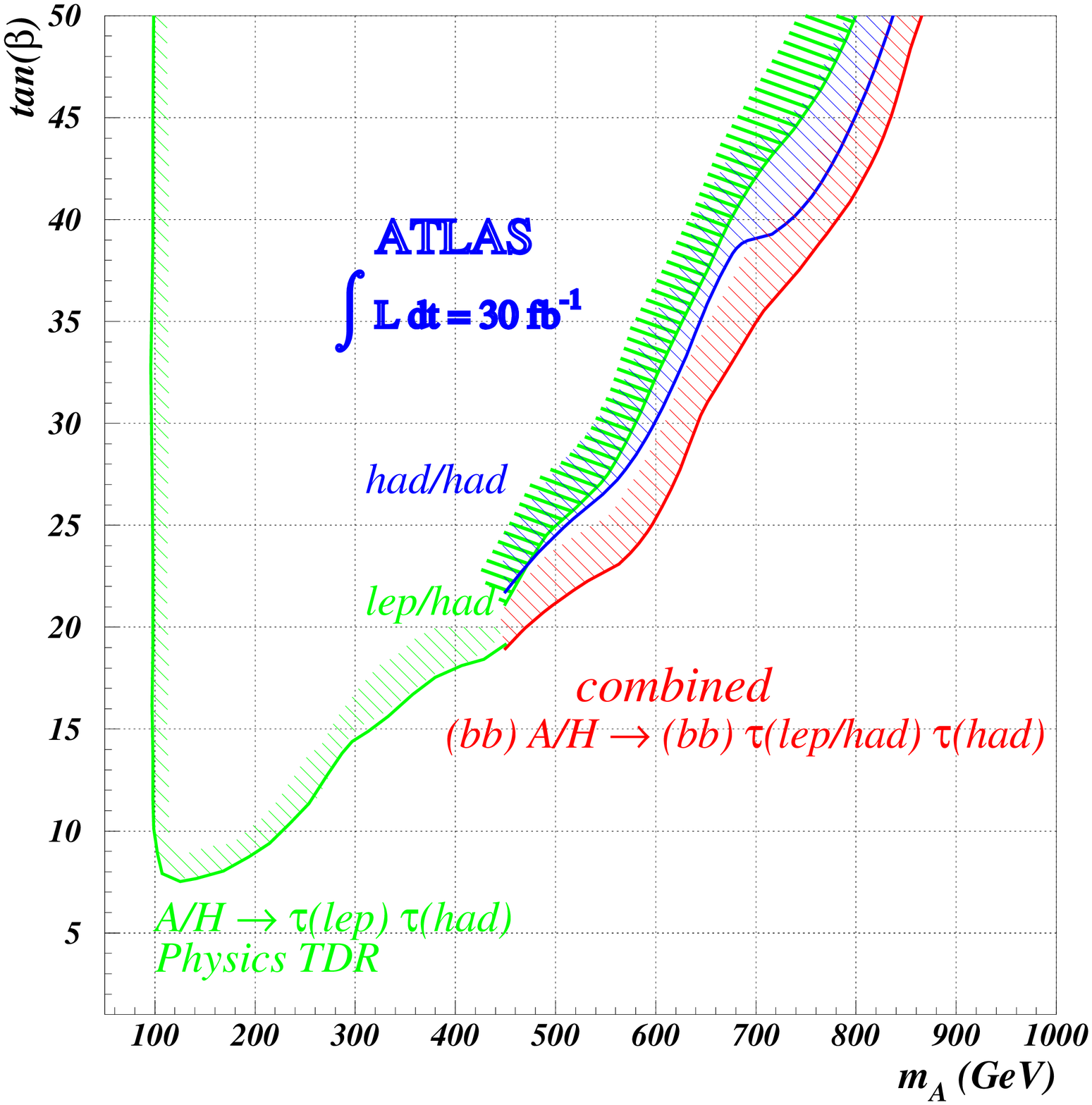}} \\[-0.8cm]
  \caption{Discovery contour curves for the \Aev~ channel in the
  (\mA,\tanbeta) plane in ATLAS for an integrated luminosity of 30 \fbm.}
  \label{contourco} &
  \caption{Discovery contour curves for the \Aev~ channel in the
  (\mA,\tanbeta) plane in ATLAS for an integrated luminosity of 30 \fbm
 shown using a linear scale for \tanbeta.}
  \label{contourcl} \\[-1.2cm]
\end{2figures}

\section{CMS Results}

\newcommand{\nc}{\newcommand}
\nc{\gsim}{\mbox{\raisebox{-.6ex}{~$\stackrel{>}{\sim}$~}}}
\nc{\esim}{\mbox{\raisebox{-.6ex}{~$\stackrel{-}{\sim}$~}}}

\newcommand{\ra}{\rightarrow}

\subsection{$A, H \ra \tau\tau \ra 2~\tau~jets$ in CMS}

$A, H \ra \tau\tau$ with 2~$\tau~jet$ hadronic final states have been shown 
to extend significantly the SUSY Higgs discovery reach into the large mass 
(600 - 800~GeV) range \cite{h2jet}. To exploit fully the
$2~\tau~jet$ final states - especially in the low ($\sim$ 200~GeV) mass range -
 an efficient hadronic $\tau$ trigger has been developed 
based on Level-1 calorimeter selection, Level-2 electromagnetic
calorimeter isolation \cite{trigger1} and a Level-3 tracking 
(isolation) using only the pixel detector information \cite{trigger3}.

Level-1 calorimeter single or double Tau trigger with thresholds of 80 and
65 GeV for $L=2 \times 10^{33}cm^{-2}s^{-1}$ selects 
$A, H \ra \tau\tau \ra 2 \tau$-jet events useful for off-line analysis with 
an efficiency of about 0.9 while giving an output QCD background rate of about 6 kHz.
A further reduction of the QCD background rate by a factor $\sim$ 10$^3$ is 
possible at the High Level trigger path (Level-2 calorimeter and Level-3 
Pixels) with an efficiency of $\sim$ 40\% for the signal at $m_H$ = 200 and 
500~GeV \cite{sasha4} as one can see in Figure \ref{fig:path_l2calol3pxl}. 
Even better performance is expected using the regional tracking option
of the CMS High Level trigger once the CPU performance is proven to be satisfactory.
\begin{figure}
\vspace*{-0.5cm}
\begin{center}
\includegraphics[width=10.cm]{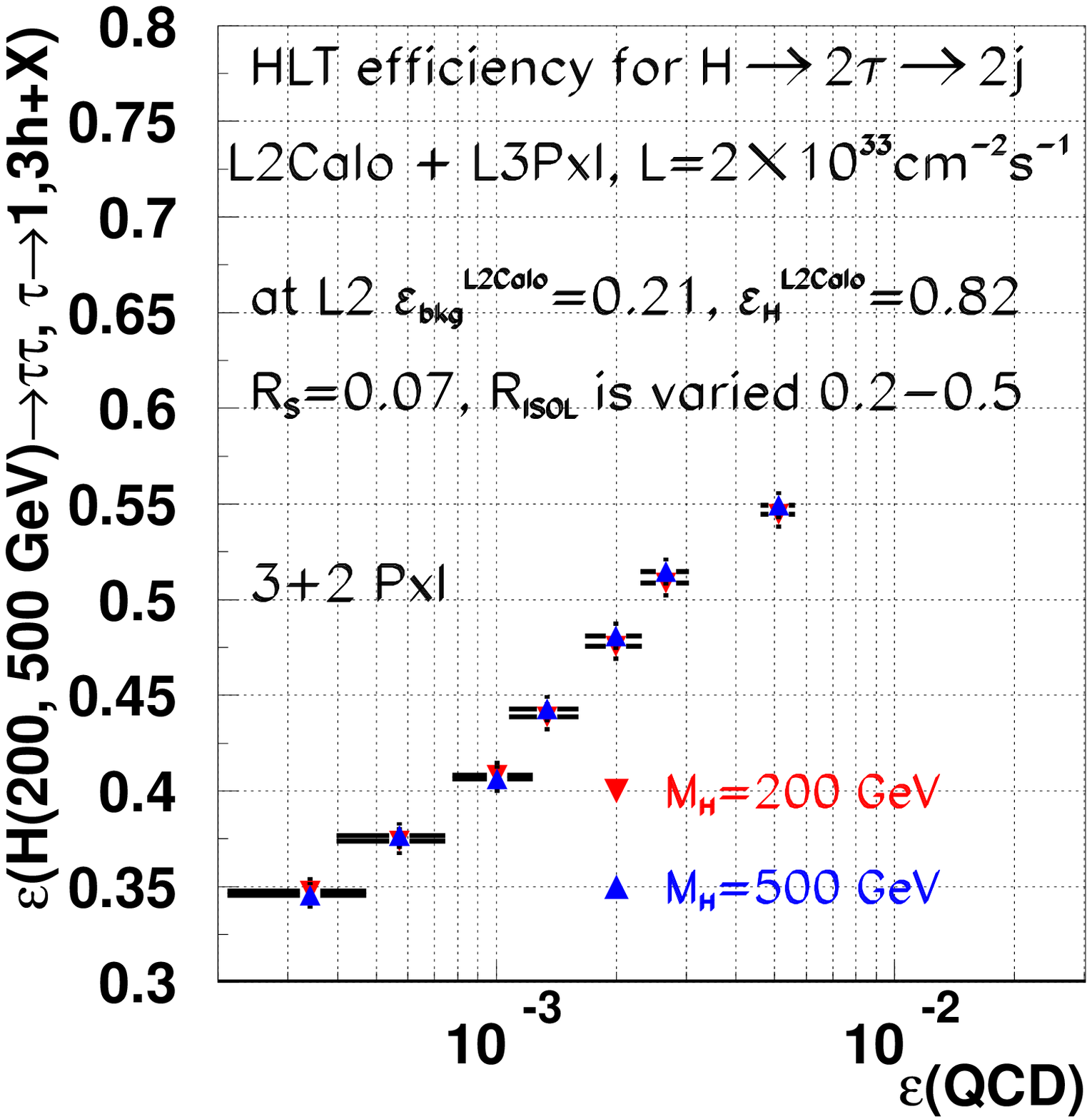}
\vspace*{-0.5cm}
\caption{Efficiency of QCD background and Higgs signal at High Level trigger
when the size of the pixel isolation cone is varied. 
\label{fig:path_l2calol3pxl}}
\end{center}
\vspace*{-0.5cm}
\end{figure}

For the off-line $\tau$ identification the tracker information is used.
The fast simulation of the CMS detector \cite{cmsjet} is used to study 
the signal to background ratios. The track reconstruction efficiency evaluated with full simulation of CMS tracker is included as a function of $p_t$ and $\eta$ for the track. The $\tau$
jet candidate ($E_t >$ 60 GeV) is required to contain 
a hard ($p_t >$40 GeV) charged track within $\Delta R <$ 0.1 around 
the calorimeter jet axis. Around this leading track in a cone of
 $\Delta R <$ 0.03 two other tracks with $p_t >$ 1 GeV are 
accepted to include the 3-prong $\tau$ decays. 
This narrow cone with one or three hard tracks is required to be 
isolated demanding that no track with  $p_t >$ 1 GeV is found in the 
surrounding larger cone of $\Delta R <$ 0.4. The efficiency for this $\tau$
selection is 7.2\% for $m_A$ = 200 GeV and 34\% for $m_A$ = 500 GeV.
 Accepting the 3-prong decays in the narrow cone of $\Delta R <$ 0.03
increases the event rate for $A, H \ra \tau\tau \ra 2~\tau~jets$ 
in the high mass range ( $m_A$ = 500 GeV) by $\sim$ 1.7 but also degrade significantly 
the QCD rejection factor for hard QCD jets. Figure \ref{fig:qcd_rej} shows the rejection factor against the QCD jets for the 1/3 prong selection a function of $E_t$ jet
compared to the one prong selection with one hard ($p_t >$40 GeV) charged
track within $\Delta R(jet,track)<$ 0.1. Optimization is still needed for the 
low mass range for more efficient selection mainly by increasing the size of 
the narrow cone.
   
A further suppression can be obtained exploiting the $\tau$ lifetime using
a $\tau$ vertex reconstruction or impact parameter measurement or a combination
of them. 
A full simulation study indicates that an additional
rejection factor of $\sim$ 5 against the 3-prong QCD jets and an efficiency of $\sim$ 70\% for the $\tau$ jets can be obtained with $\tau$ vertex reconstruction \cite{tau_vertex}.    
Promising results are also obtained from the impact parameter method in the 
channel $A, H \ra \tau\tau \ra \ell^+\ell^- + X$ using full
 simulation combining the 
impact parameter measurements for the two leptons from $\tau$ decays 
to reduce the backgrounds 
with $W \ra \ell\nu$ and $Z \ra \ell\ell$ decays \cite{sami}.
\begin{2figures}{hbtp}
\vspace*{-1.3cm}
  \resizebox{\linewidth}{80 mm}{\includegraphics{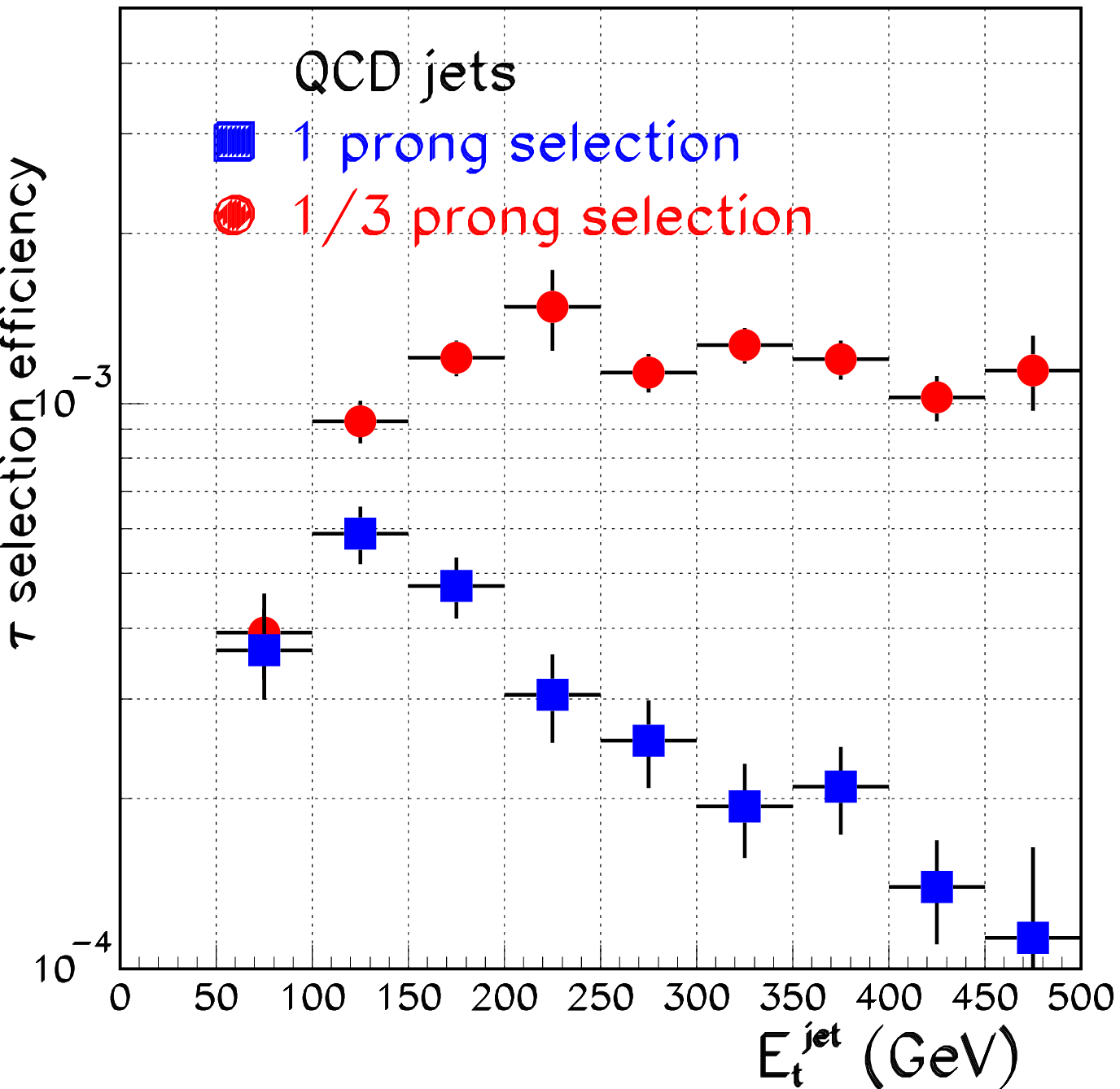}} &
\vspace*{-1.3cm}
  \resizebox{\linewidth}{80 mm}{\includegraphics{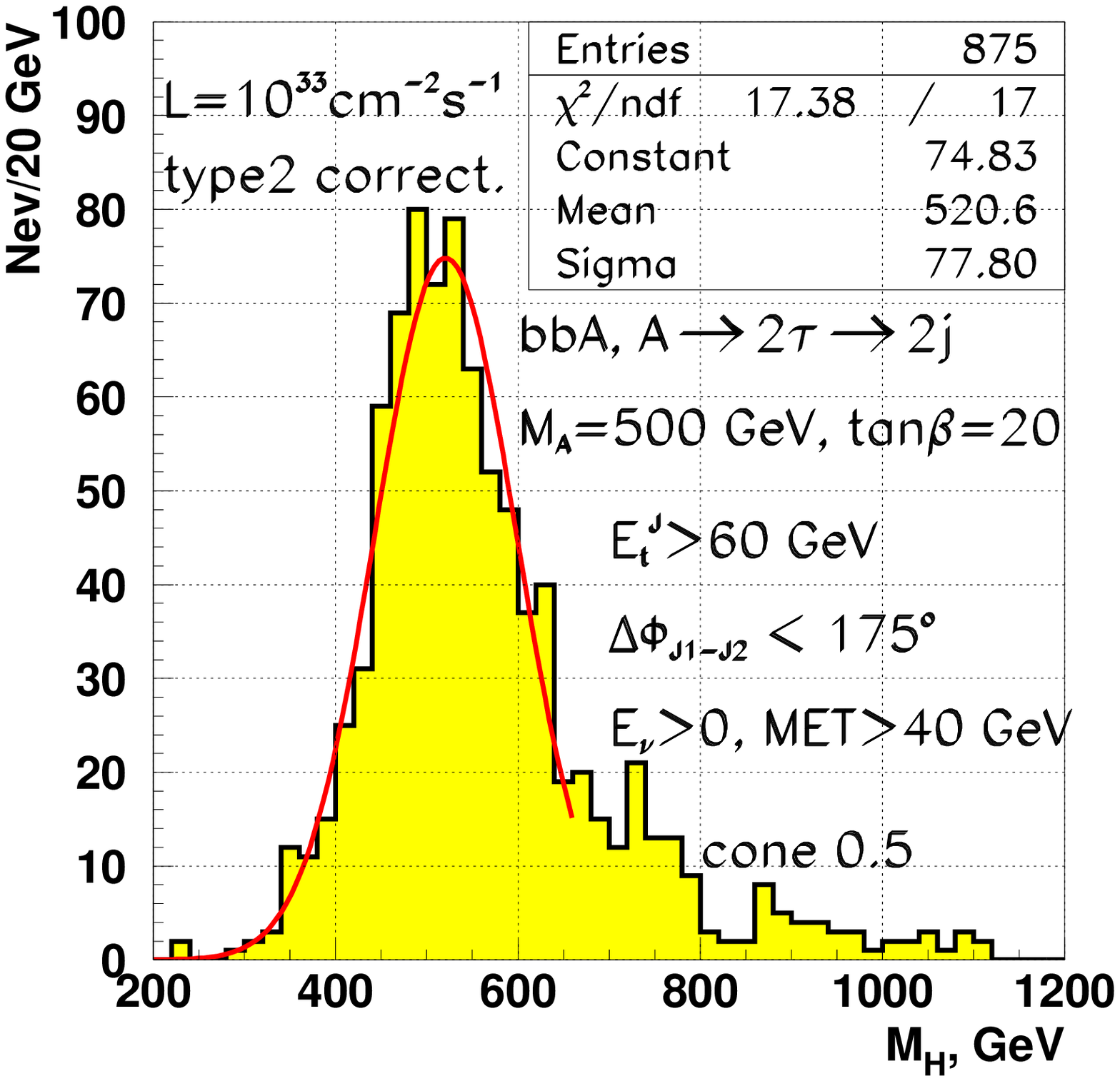}}
\\[-0.7cm]
  \caption{Rejection factor against the QCD jets as a function 
  of $E_t^{jet}$ for the 1 and 1/3 prong $\tau$ selection.} 
  \label{fig:qcd_rej} &
  \caption{Reconstructed Higgs mass for $b\overline{b}$, 
  $H \ra \tau\tau \ra 2~\tau~jets$ with $m_H$ = 500 GeV. }
  \label{fig:hmass} \\[-1.2cm]
\end{2figures}

The resolution of the reconstructed Higgs mass and even more so the mass reconstruction
efficiency in $A, H \ra \tau\tau$ events is very sensitive 
to the $E_t^{miss}$ measurement. The absolute value of $E_t^{miss}$  is relatively
small in these events making the mass reconstruction 
and background reduction with 
a cut in $E_t^{miss}$ a difficult task. Figure \ref{fig:hmass} shows the 
Higgs mass reconstructed with full simulation for $b\overline{b}A$,
$A \ra \tau\tau \ra 2~\tau~jets$ with $m_A$ = 500 GeV and $tan\beta$= 20
\cite{sasha0}. 
The resolution of the reconstructed Higgs mass is 14.5\% for $m_A$ = 200 GeV  and 
14.9\% for $m_A$ = 500 GeV and the corresponding reconstruction efficiencies are
 37\% and 36\%, respectively (including $\Delta \phi < 175^{\circ}$ cut and requiring positive
neutrino energies). This confirms the earlier results of the fast 
simulation study \cite{h2jet}. 

The large $Z, \gamma^* \ra \tau\tau$ background can be reduced efficiently only 
with b-tagging in the associated production processes $b\overline{b}H_{SUSY}$. 
The associated b-jets are soft and uniformly distributed over $|\eta|<$ 2.5.
Nevertheless, a study with full simulation shows that a b-tagging efficiency 
of $\sim$ 34\% per jet can be obtained for the signal events with a mistagging 
rate less than 1\% for $Z+jets$ events \cite{sami}. Requiring
one tagged b-jet reduces efficiently also the QCD background thus
improving significantly the signal visibility. Figure \ref{fig:h500} shows
the signal for $m_H$ = 500 GeV and $tan\beta$ = 25 superimposed on the 
total background with 
b-tagging. The missing transverse energy $E_t^{miss}$ plays a major role 
in the Higgs mass reconstruction as discussed above. However, a cut in $E_t^{miss}$ does not improve significantly the mass resolution and therefore,
 in order to retain the signal statistics, is not used in this study.
Figure \ref{fig:maxmix_30fb} shows the expected discovery reach for 
 for 30~$fb^{-1}$ assuming a maximal 
stop mixing scenario \cite{benchmark, benchmarkhad}. 
The expectations for other important
MSSM Higgs discovery channels in CMS \cite{summary} and the exclusion region from 
LEPII \cite{mhLEP2001} are also shown in the figure. The $A, H \ra \tau\tau$
channels are found to be insensitive for stop mixing, the SUSY scale
and for the sign of the Higgsino mass parameter $\mu$ for high enough 
$tan\beta$ ($\gsim$10). A systematic study 
of the $A, H \ra \tau\tau$ with $\ell^+\ell^-$, $lepton+\tau~jet$ and $2~\tau~jet$
final states is presently in progress in CMS including full simulation
of the hadronic $\tau$ trigger, $\tau$ identification, $\tau$ tagging with 
impact parameter and vertex reconstruction, Higgs mass reconstruction and 
b-tagging in the associated production channels.
\begin{2figures}{hbtp}
\vspace*{-1.2cm}
  \resizebox{\linewidth}{80 mm}{\includegraphics{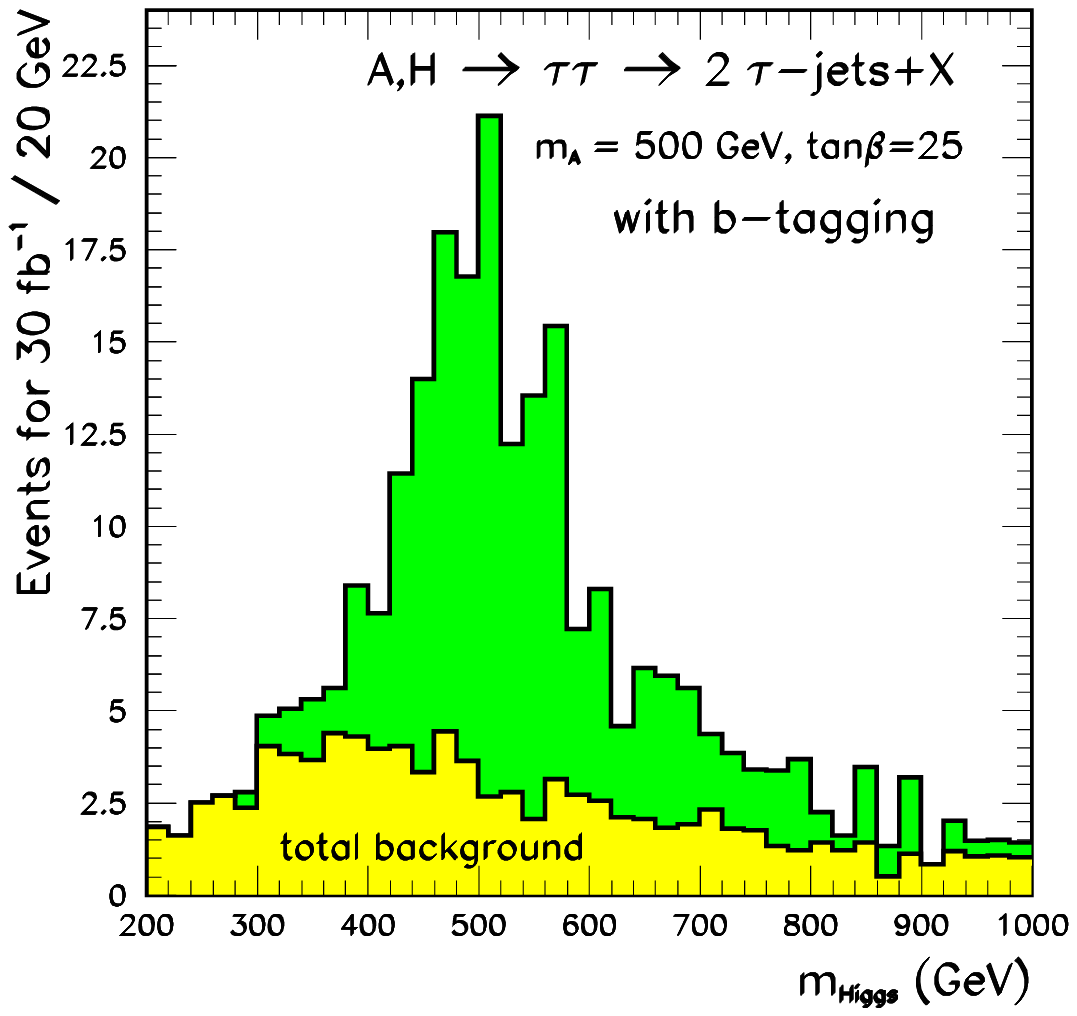}} &
\vspace*{-1.2cm}
  \resizebox{\linewidth}{80 mm}{\includegraphics{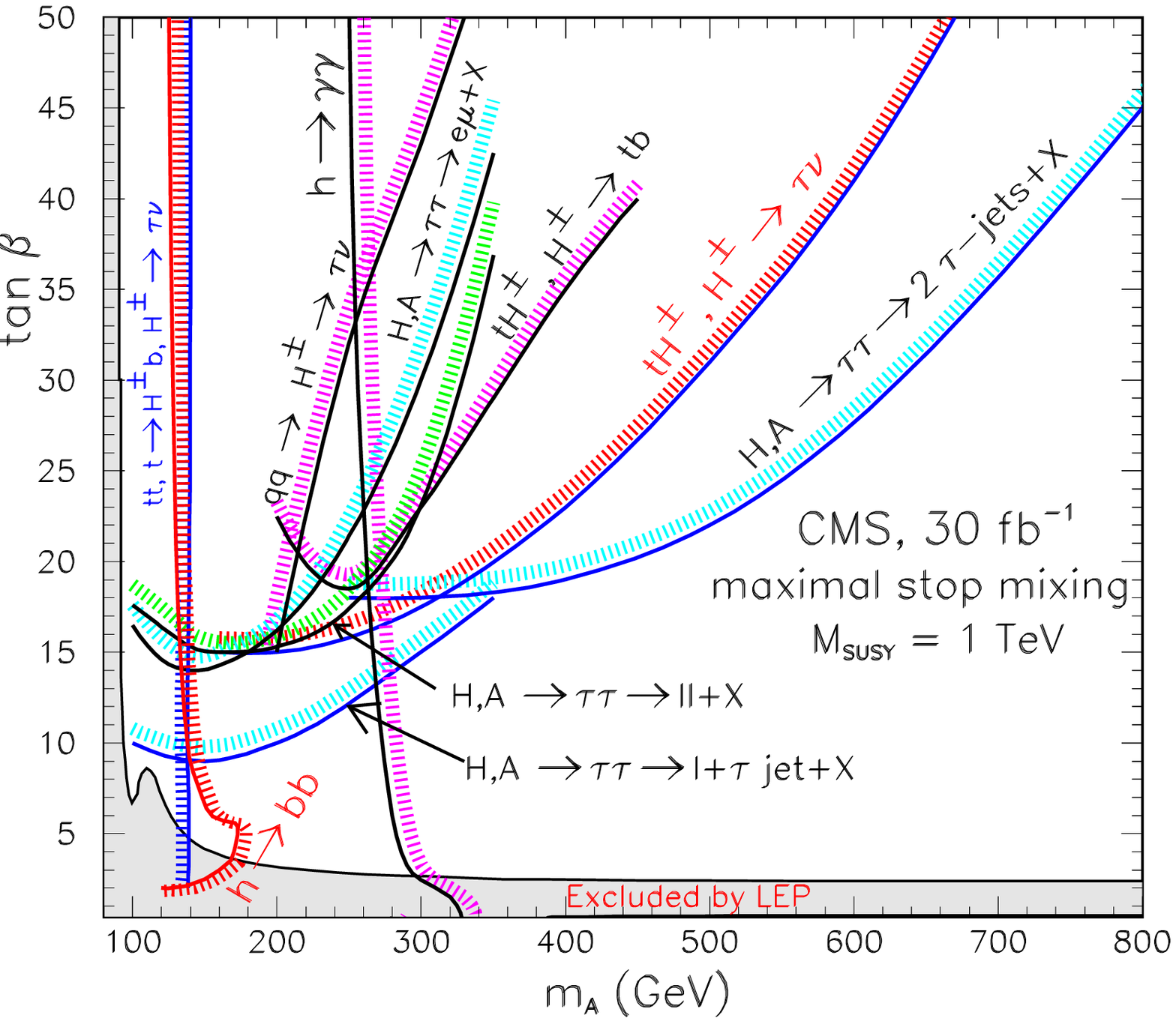}}
\\[-0.7cm]
  \caption{Higgs mass for $H \ra \tau\tau \ra 2~\tau~jets$ with $m_H$ = 500 GeV 
   and $tan\beta$=25 superimposed on the total background for 30~$fb^{-1}.$
    One tagged b-jet is required.}
  \label{fig:h500} &
  \caption{Expected 5$\sigma$ discovery reach for 
            the MSSM Higgs bosons in 
            CMS in the maximal mixing scenario for 30$fb^{-1}$
            as a function of $m_A$ and $tan\beta$. 
            The shaded area is excluded by LEP \cite{mhLEP2001}.}
  \label{fig:maxmix_30fb} \\[-1.8cm]
\end{2figures}

\section{Conclusions}

The LHC discovery potential for  \Aev~ has been studied in ATLAS and CMS
 in the three different final decay
channels at low luminosity (\medlumi) in the  \mA~ range from $\sim$100 GeV
 until $\sim$1 TeV.
\begin{itemize}
\item
at lower masses ($< \sim$500 GeV), the lepton-lepton and lepton-hadron
channels have been studied in both experiments; 
 the lepton-hadron gives the best sensitivity. To study the hadron-hadron
channel also in the lower mass range, CMS is developing 
 an efficient hadronic $\tau$ trigger (Level-1 + High Level trigger).
\item 
at higher masses, for large \tanbeta~ values the \A~ production is dominated
by \bbbar A.
In this mass range the hadron-hadron channel can be  studied because it is 
possible to reject
the huge QCD background with kinematical cuts and $\tau$-jet identification.
The trigger is a very important point for this purely hadronic channel
and it has been studied in the two experiments.
The lepton-hadron channel has been also studied in ATLAS in the high \mA~ 
region giving promising results; the combination of the hadron-hadron and
lepton-hadron channels improves the discovery potential.
\end{itemize}
 The expected 5$\sigma$-discovery contour curves for the combined 
$A/H \rightarrow \tau \tau$ signal show that a signal should be 
observed over a large region of the (\mA,\tanbeta) plane,
with \mA~ up to $\sim$ 1 TeV 
  assuming an integrated luminosity of 30 \fbm.\\
At high \mA~  ($\sim$0.5-$\sim$1 TeV) the \Aev~ channel should be observable 
for \tanbeta~ values greater than $\sim$ 25.
  This is an important result
because the \Aev~ channel is up to now the only one giving
access to this high  \mA~ region.
\par
Studies are still in progress, both on the experimental and on the
theoretical sides, to improve the results presented here.

\section*{Acknowledgments}
The authors would like to thank E. Richter-Was for helpful discussions.
They would also like to thank the organisers of the LesHouches
workshop for the warm atmosphere and the fruitful work.
}

\setcounter{figure}{0}
\setcounter{table}{0}
\setcounter{section}{0}
\setcounter{equation}{0}
\newpage

{
\noindent
{\Large \bf F. Searching for Higgs Bosons in $t\bar t H$ Production} \\[0.5cm]
{\it V.\,Drollinger}

\begin{abstract}
Higgs boson production in association with $t\bar t$ pairs with the
subsequent decay into $l^\pm \nu q\bar{q} b\bar{b} b\bar{b}$ is analyzed
including all relevant background processes. Excellent $b$-tagging
performance and a good mass resolution turn out to be the most important
components for a successful analysis. The top Yukawa coupling can be
determined with an accuracy of about 17\% in this process, provided the
branching ratio of $H\to b\bar b$ is known with a sufficient accuracy.
Finally, a first estimate of the potential size of higher order
corrections to the $t\bar tb\bar b$ background is given.
\end{abstract}

\section{Introduction}

If the Higgs boson is lighter than 130 $GeV/c^2$, it decays mainly to a $b\bar{b}$ pair. To observe the Higgs boson at the LHC, the $t\bar{t}H^0$ channel turns out to be the most promising channel among the Higgs production channels with $H^0 \rightarrow b\bar{b}$ decay \cite{THESIS}. In this study, we discuss the channel $t\bar{t} H^0 \rightarrow l^\pm \nu q\bar{q} b\bar{b} b\bar{b}$ (Figure~\ref{fig:fey_tth}), where the Higgs Boson decays to $b\bar{b}$, one top quark decays hadronically and the second one leptonically. The relevant signal and background cross sections at the LHC ($\sqrt{s_{pp}} = $ 14 $TeV$) and particle masses used in the simulation are listed in Table~\ref{tab:crossections}.
\vspace*{-3mm}
\begin{table}[ht]
 \begin{center}
  \caption{\sl CompHEP \cite{CompHEP} cross sections for signal and background relevant for the $t\bar{t} H^0 \rightarrow l^\pm \nu q\bar{q} b\bar{b} b\bar{b}$ channel, calculated with parton density function CTEQ4l \cite{PDFLIB}. The branching ratio of the semileptonic decay mode (one $W^\pm$ decays to quarks the other $W^\pm$ decays leptonically, where only decays to electrons or muons are taken into account) is 29\%  (not included in the cross sections of this table) and $m_{W^\pm} =$ 80.3427 $GeV/c^2$.\rm}
 \begin{tabular}{|lcr|lcr|}
 \hline
 \multicolumn{3}{|c|}{LO cross sections} & \multicolumn{3}{|c|}{masses}\\
 \hline
  $\sigma_{t\bar{t}H^0} \times BR_{H^0 \rightarrow b\bar{b}}$ & = 
 & 1.09 - 0.32 $pb$ &  $m_{H^0}$ & = & 100 - 130 $GeV/c^2$ \\
    $\sigma_{t\bar{t}Z^0}$ & = & 0.65 $pb$ & $m_{Z^0}$ & = & 91.187 $GeV/c^2$\\
 $\sigma_{t\bar{t}b\bar{b}}$ & = & 3.28 $pb$ &   $m_{b}$ & = & 4.62 $GeV/c^2$\\
       $\sigma_{t\bar{t}jj}$ & = & 507  $pb$ &   $m_{t}$ & = &  175 $GeV/c^2$\\
 \hline
 \end{tabular}
 \end{center}
\label{tab:crossections}
\vspace*{-0.5cm}
\end{table}
This is the first set of signal and background processes completely calculated at LO for the $t\bar{t} H^0$ channel. The hard processes are generated with CompHEP and then interfaced to PYTHIA, where the fragmentation and hadronisation are performed \cite{CompHEP}-\cite{pythia}. The combined package CompHEP-interface-PYTHIA includes all features of a pure PYTHIA simulation, such as initial sate radiation, final state radiation, multiple interactions and underlying event. After the final state has been obtained, the CMS detector response is simulated, with track and jet reconstruction with parametrisations FATSIM \cite{FATSIM} and CMSJET \cite{cmsjet}, obtaining in this way tracks, jets, leptons (the electron or muon reconstruction efficiency is assumed to be 90\%; taus are not considered here) and missing transverse energy. These parametrisations have been obtained from detailed simulations based on GEANT.
\begin{figure}[ht]
\begin{center}
 \includegraphics[width=0.44\textwidth,angle=+0]{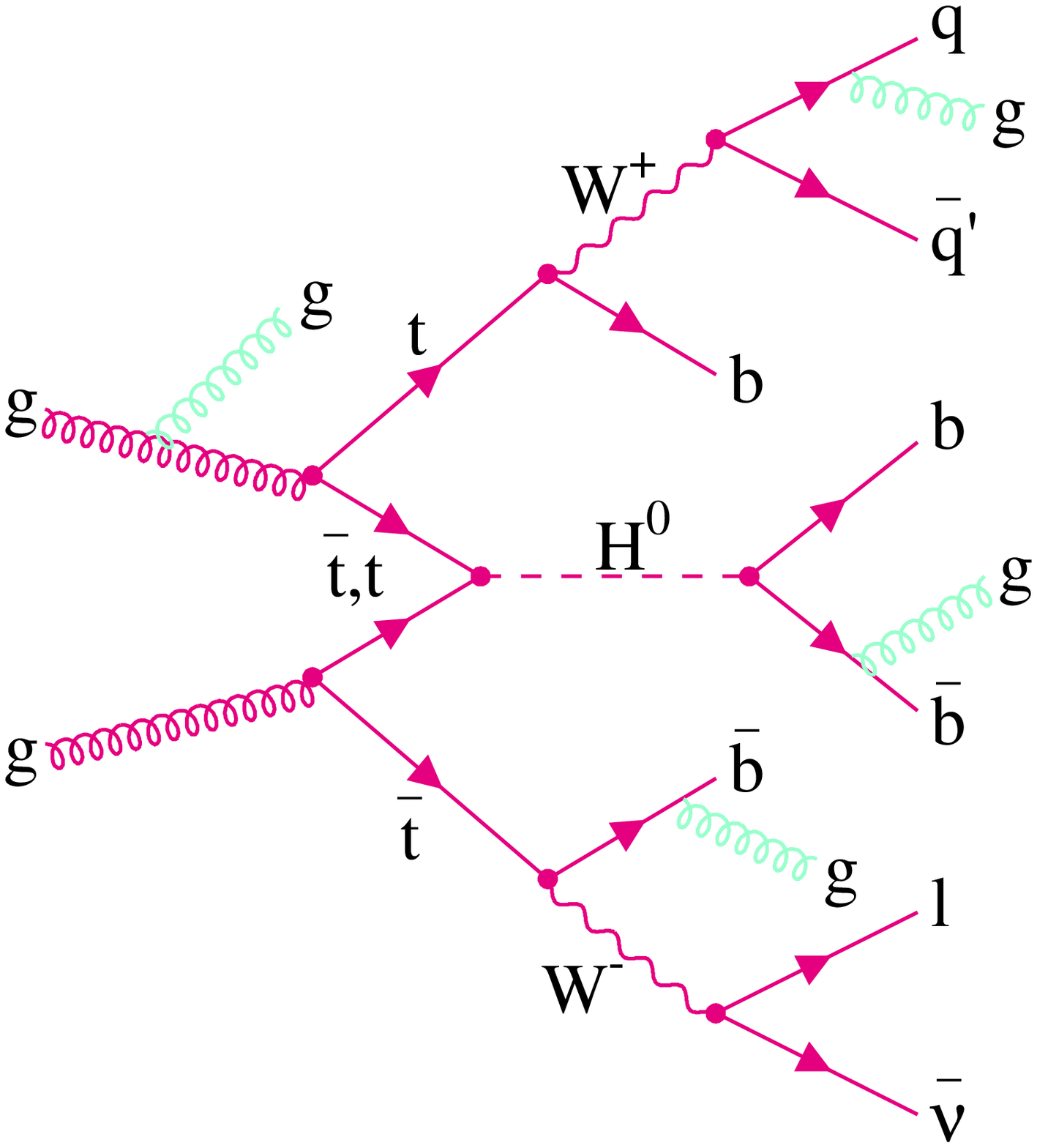}
\vspace*{-0.3cm}
 \caption{\sl Example of a $pp \rightarrow t\bar{t} H^0 \rightarrow l^\pm \nu q\bar{q} b\bar{b} b\bar{b}$ signal event. The LO process is drawn in red. HO events include gluon radiation (light green) in addition. The expected final state consists of one isolated lepton, missing transverse energy, four $b$-jets and two (or more) non-$b$-jets.\rm}
 \label{fig:fey_tth}
\end{center}
\vspace*{-9mm}
\end{figure}



\section{Reconstruction}

From Figure~\ref{fig:fey_tth} we expect to find events with one isolated lepton, missing transverse energy $E_T^m$ and six jets (four $b$-jets and two non-$b$-jets), but initial and final state radiation are sources of additional jets. So the number of jets per event is typically higher than six. On the other hand, not all six quarks of the hard process can be always recognised as individual jets in the detector, in which case it is impossible to reconstruct the event correctly - even if there are six or more jets.

For the reconstruction of resonances it is necessary to assign the $n$ jets of an event to the corresponding quarks of the hard process. In general, and ignoring information on $b$-jets, the number of possible combinations $N$ is given in Table~\ref{tab:combinations} as a function of the number of jets per event. We obtain $N$ for the case, when the masses of the Higgs boson, both top quarks and the hadronically decaying W boson are reconstructed. The nominal mass of the leptonically decaying W boson, together with $E_T^m$ and the lepton four momentum, is used to calculate two solutions of the longitudinal momentum of the neutrino $p_Z(\nu)$ which is needed for the mass reconstruction of the leptonically decaying top.
\vspace*{-3mm}
\begin{table}[ht]
 \begin{center}
  \caption{\sl Number of jets per event $n$ and the corresponding number of possible combinations $N$. If there are more than a dozen jets, only the twelve with highest $E_T$ are considered.\rm}
\label{tab:combinations}
 \begin{tabular}{|ccccccccc|}
 \hline
 \multicolumn{9}{|c|}{\rule[-3mm]{0mm}{8mm} $N = \binom{n}{6} \times 6! \times \frac{1}{2} \times \frac{1}{2} \times 2\ = \binom{n}{6} \times 360$}\\
 \hline
 $n$ & = &   6 &    7 &     8 &     9 &    10 &     11 &     12 \\
 $N$ & = & 360 & 2520 & 10080 & 30240 & 75600 & 166320 & 332640 \\
 \hline
 \end{tabular}
 \end{center}
\vspace*{-0.5cm}
\end{table}

Good mass resolution and the identification of $b$-jets is essential to reduce the number of wrong combinations in the event reconstruction. A good mass resolution can be obtained when the energy and direction of each reconstructed jet agree as closely as possible with the quantities of the corresponding parent quark. This can be achieved with jet corrections as described in \cite{JETCOR} and \cite{JETRAD}. For $b$-tagging we use the $b$-probability functions which depend on impact parameters of tracks and leptons inside the jets. They are determined using $t\bar{t}$ six jet events, as described in \cite{THESIS}. The identification of $b$-jets is even more important for efficient background suppression.
\begin{figure}[ht]
\begin{center}
 \includegraphics[width=0.60\textwidth,angle=+0]{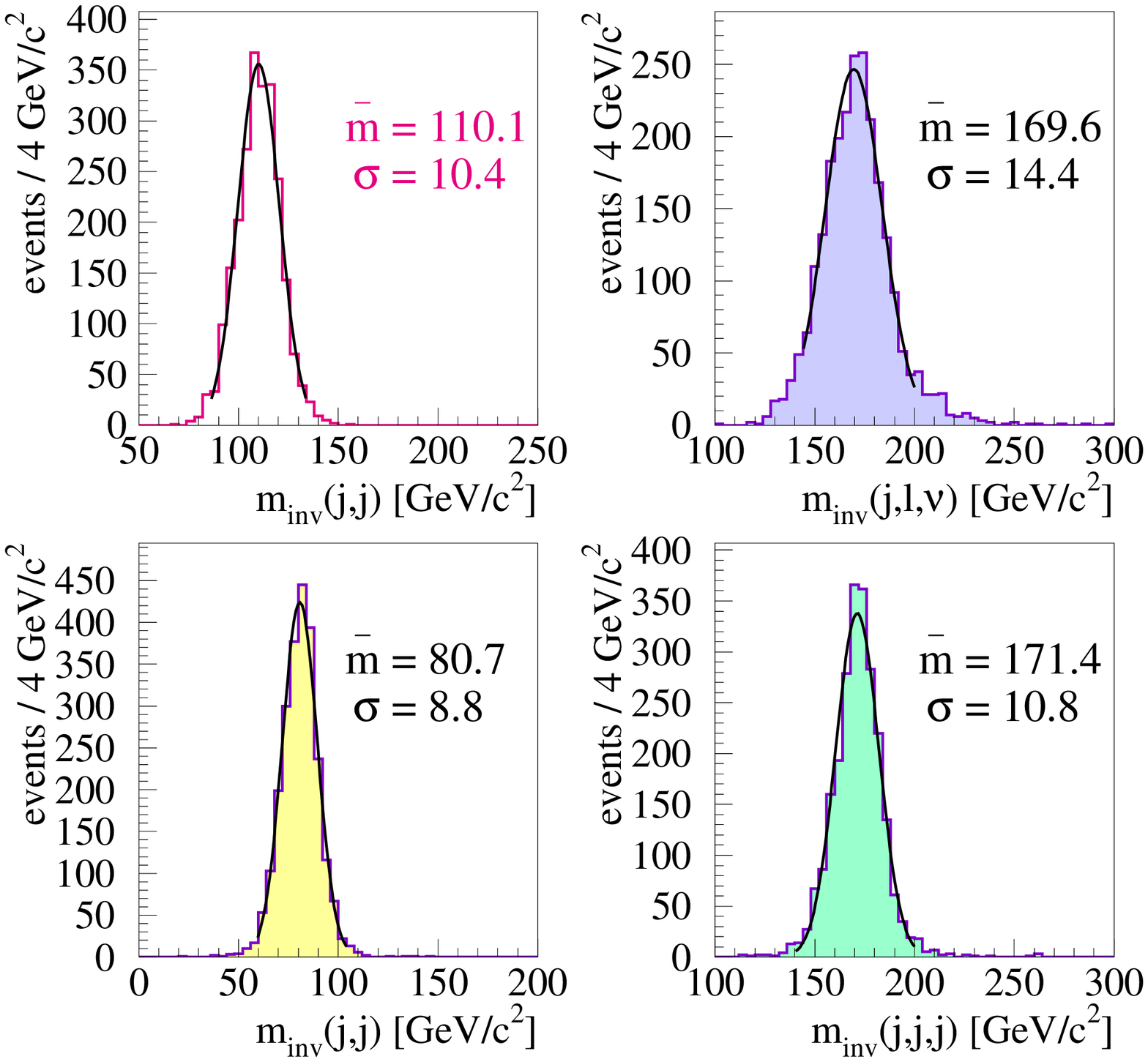}
\vspace*{-0.5cm}
 \caption{\sl Invariant resonance masses of the $t\bar{t} H^0 \rightarrow l^\pm \nu q\bar{q} b\bar{b} b\bar{b}$ signal: Higgs boson, leptonic top, hadronic top and hadronic $W^\pm$. The leptonic $W^\pm$ is not reconstructed but its nominal mass is used to calculate $p_Z(\nu)$. The generated masses are: $m_{H^0} =$ 115 $GeV/c^2$, $m_{t} =$ 175 $GeV/c^2$ and $m_{W^\pm} =$ 80.3427 $GeV/c^2$.\rm}
 \label{fig:resonances}
\end{center}
\vspace*{-7mm}
\end{figure}

Figure~\ref{fig:resonances} shows the invariant mass distributions of the reconstructed resonances of $t\bar{t} H^0 \rightarrow l^\pm \nu q\bar{q} b\bar{b} b\bar{b}$ events in the case of an ideal reconstruction: after the ``preselection'' and the calculation of $p_Z(\nu)$ (see later on) each quark of the hard process is matched with exactly one jet, the closest one in $R = \sqrt{\phi^2 + \eta^2}$ if $\Delta R(q,j) <$ 0.3 and if the jet energy is closer than $\pm$ 30 \% to the parent quark energy. The mean values and widths of the top and W mass distributions are used to define likelihood functions used in the selection procedure described in the following.

{\bf\boldmath$\diamond$ Preselection}\\
Events are selected if there is an isolated lepton ($e^\pm$ or $\mu^\pm$ with $p_T >$ 10 $GeV/c$ within the tracker acceptance; no other track with $p_T >$ 1 $GeV/c$ in a cone of 0.2 around the lepton) and at least six jets ($E_T >$ 20 $GeV$ , $|\eta| <$ 2.5).

{\bf\boldmath$\diamond$ Event Configuration}\\
In order to be able to reconstruct the Higgs mass, we have to find the correct event configuration among all possible combinations listed in Table~\ref{tab:combinations}. The best configuration is defined as the one which gives the highest value of an event likelihood function (\ref{EV-L-SHORT}) which takes into account $b$-tagging of four jets, anti-$b$-tagging of the two jets supposed to come from the hadronic $W^\pm$, mass reconstruction of $W^\pm$ and the two top quarks, and sorting of the $b$-jet energies.
\begin{align}
 {\it\bf L\_EVNT} & = \prod_{i=1,4}P_b(b_i) 
               \times \prod_{i=1,2}[1 - P_b(q_i)]
               \times \prod_{i=W^\pm ,t,\bar{t}}e^{-0.5 \times [\frac{m_i - \bar{m_i}}{\sigma_i}]^2}
               \times f[E_b(t,\bar{t})-E_b(H^0)]
\label{EV-L-SHORT}
\end{align}

{\bf\boldmath$\diamond$ Jet Combinations}\\
Events with more than six jets can contain gluon jets from final state radiation, which are not yet used in the analysis. The combination of these jets with the correct quark jets can improve the event reconstruction further. The additional jets are combined with the decay products of both top quarks if they are closer than $\Delta R(j,j) <$ 1.7, if the corresponding mass is closer to the expected value of Figure~\ref{fig:resonances}. If there are still jets left, they are considered as Higgs decay products and are combined with the closest of the corresponding two $b$-jets, if $\Delta R(j,j) <$ 0.4. 

{\bf\boldmath$\diamond$ Event Selection}\\
Three likelihood functions: for resonances ($L\_RESO >$ 0.05), $b$-tagging ($L\_BTAG >$ 0.50), and kinematics ($L\_KINE >$ 0.2) are used to reduce the fraction of background events. Finally, the events are counted in a mass window around the expected Higgs mass peak ($m_{inv}(j,j)$ in $\bar{m}\ \pm$ 1.9 $\sigma$ ; $\bar{m}$ and $\sigma$ are obtained from mass distributions as shown in Figure~\ref{fig:resonances} with various generated Higgs masses).

The overall efficiency for a triggered event to be finally selected is 1.3\% for $t\bar{t}H^0$ ($m_{H^0} =$ 115 $GeV/c^2$), 0.2\% for $t\bar{t}Z^0$, 0.4\% for $t\bar{t}b\bar{b}$ and 0.003\% for $t\bar{t}jj$ events. This shows that the reducible background is reduced very effectively. In addition, there is little combinatorial background left (see Figure~\ref{fig:tth_sum115}) with this reconstruction method.
\vspace*{-3mm}
\begin{figure}[ht]
\begin{center}
 \includegraphics[width=0.53\textwidth,angle=+0]{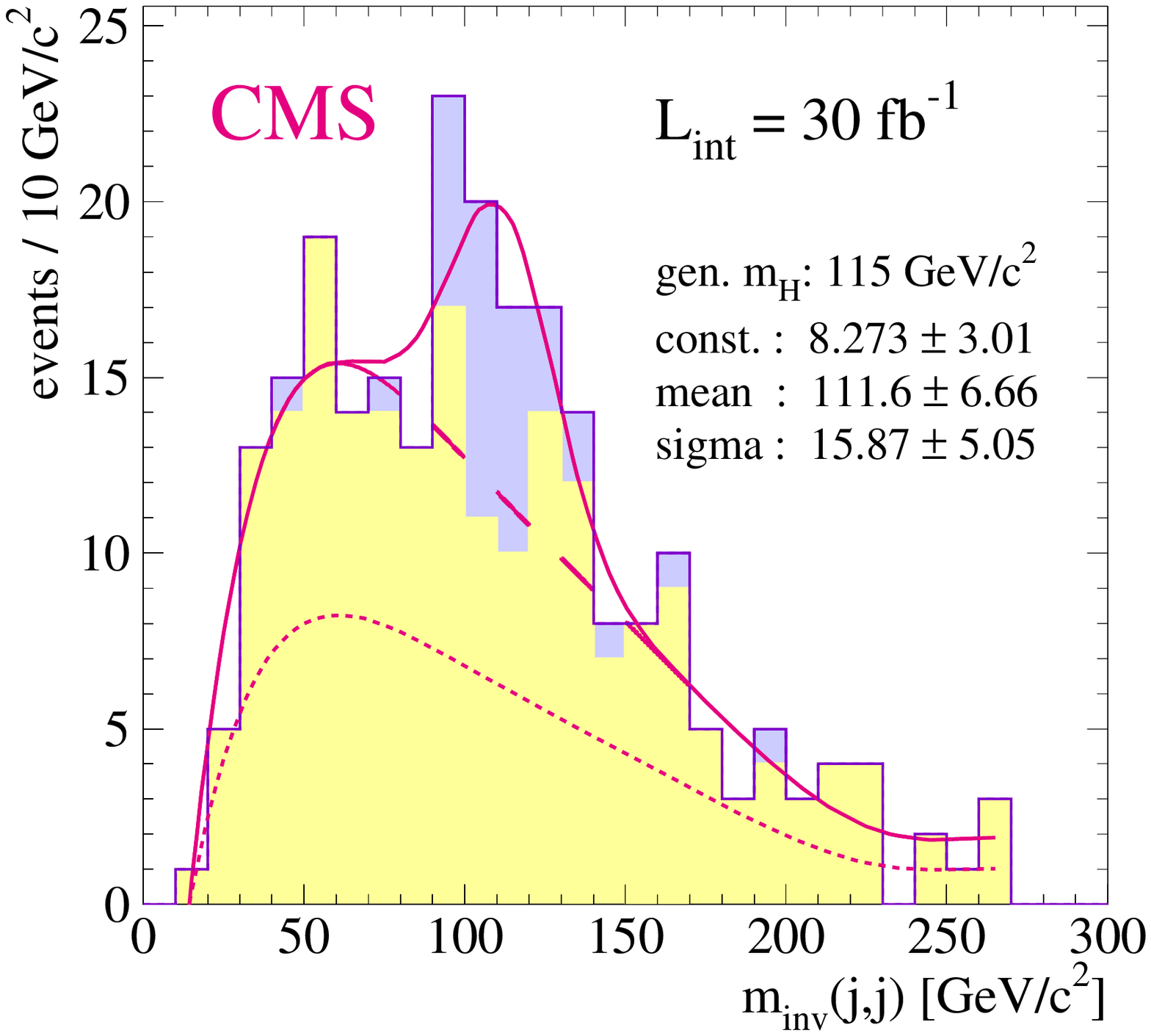}
\vspace*{-0.5cm}
 \caption{\sl Simulated invariant mass distribution of signal (dark shaded, $m_{H^0} =$ 115 $GeV/c^2$) plus background for $L_{int} = $ 30 $fb^{-1}$. The dashed curve is obtained from the fit of the background without signal, the solid line describes the fit of signal plus background. The small dashed line corresponds to the LO background contribution.\rm}
 \label{fig:tth_sum115}
\end{center}
\vspace*{-9mm}
\end{figure}



\section{SM Results}

After the whole reconstruction and event selection procedure, it turns out that the irreducible background (with four real $b$-jets) is dominant. Even the $t\bar{t}jj$ background, where only two $b$-jets from the top decays are generated in the hard process, is dominated by events with four real $b$-jets. This is possible after the fragmentation of PYTHIA: e.g. $gg \rightarrow t\bar{t} gg \rightarrow l^\pm \nu q\bar{q} b\bar{b} g b\bar{b}$ with one $b\bar{b}$ pair coming from $g \rightarrow b\bar{b}$ (gluon splitting). In this case the final state consists of nine partons or leptons which is one more than expected at LO and is therefore considered as HO (in this case NLO) process. Together with the number of $t\bar{t}b\bar{b}$ events (considered as LO) we obtain an intrinsic k-factor k$_{t\bar{t} q\bar{q}} =$ 1.9 for all $t\bar{t} q\bar{q}$ events as indicated in Figure~\ref{fig:tth_sum115}. (For the full $m_{H^0} =$ 115 $GeV/c^2$ selection we get 23 events from ``$t\bar{t} b\bar{b}$'' and 20 events from ``$t\bar{t} jj$'' with four real $b$-jets plus 6 events with two real and two false $b$-tags. The total number for the non resonant background amounts to 49 events, whereas the corresponding number from the PYTHIA $t\bar{t}$ process with additional jets from fragmentation is only 24 events.) In case of the $t\bar{t}H^0$ signal additional $b$-jets from fragmentation cannot enhance the signal, but rather complicate the reconstruction of the correct invariant mass, the Higgs mass, in the end. 

The signal to background ratio $S / B$, the significance $S / \sqrt{B}$ for $L_{int} = $ 30 $fb^{-1}$, the integrated luminosity $L_{int}$ required for a significance of five or more and the precision on the top Higgs Yukawa coupling $y_t$ for $L_{int} = $ 30 $fb^{-1}$ are shown in Figure~\ref{fig:tth_stat4} as a function of the generated Higgs mass: $S / B$ is around 50\% and an relatively low integrated luminosity is sufficient to discover the Higgs boson in this channel with a significance above five. An integrated luminosity $L_{int} = $ 100 $fb^{-1}$ would be enough to explore all points considered in Figure~\ref{fig:tth_stat4} up to a Higgs mass of 130 $GeV/c^2$. If we assume a known branching fraction of the decay $H^0 \rightarrow b\bar{b}$, it is possible to determine the precision of $y_t$ with accuracy of about 17\%. Apart from these results, the Higgs mass can be determined from the Gaussian fit of the final mass distribution (see Figure~\ref{fig:tth_sum115}) with a precision of better than 6\% for $L_{int} = $ 30 $fb^{-1}$.
\begin{figure}[ht]
\begin{center}
 \includegraphics[width=0.55\textwidth,angle=+0]{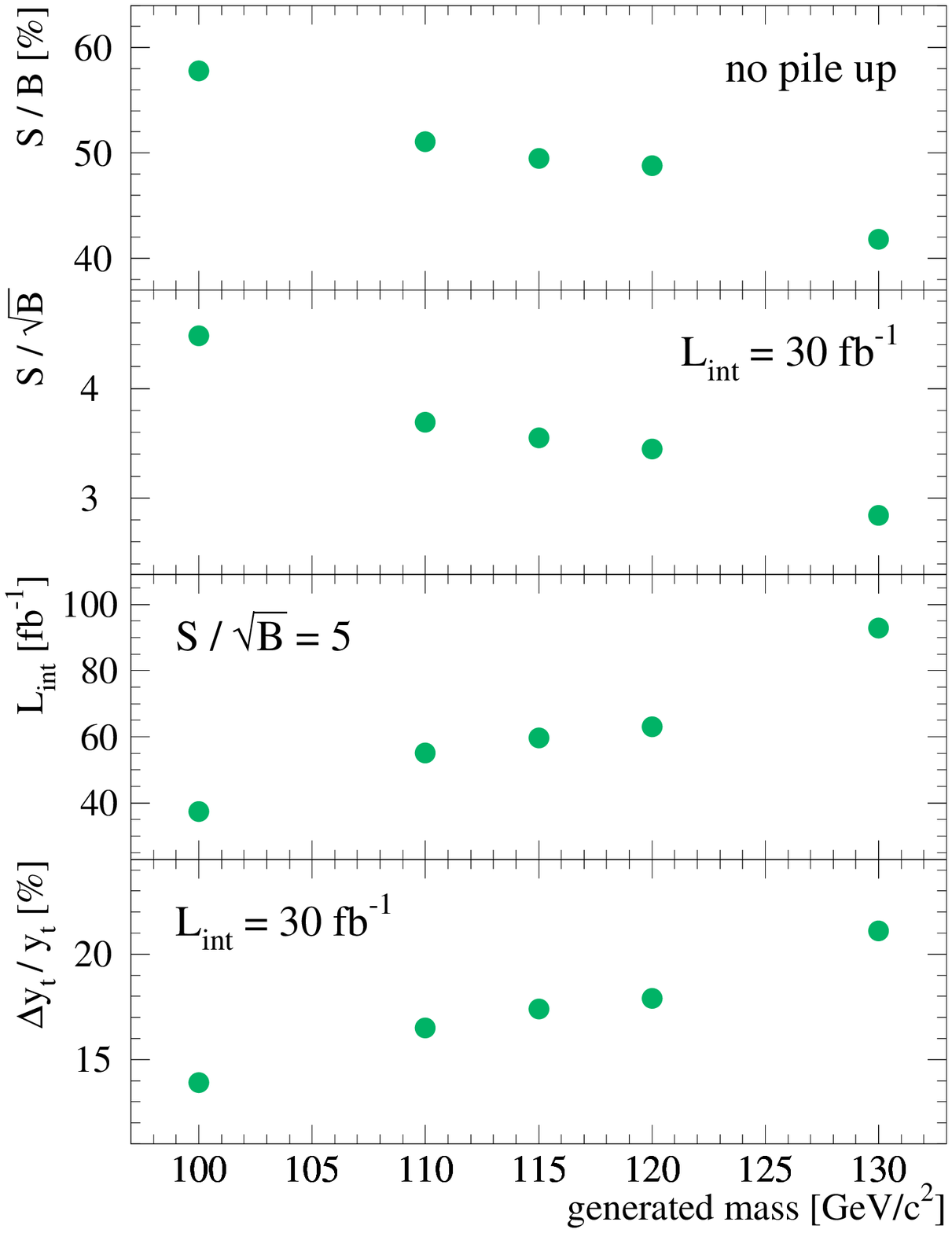}
\vspace*{-5mm}
 \caption{\sl $S / B$, $S / \sqrt{B}$, $L_{int}$ (required for $S / \sqrt{B} =$ 5) and $\Delta y_t / y_t$ versus generated Higgs mass in the SM. All results are based on CompHEP cross sections calculated at LO; from the fragmentation with PYTHIA an intrinsic k-factor k$_{t\bar{t} q\bar{q}} =$ 1.9 for $t\bar{t} q\bar{q}$ background events is included.\rm}
 \label{fig:tth_stat4}
\end{center}
\vspace*{-8mm}
\end{figure}


\section{Conclusions}

From our present understanding, it is experimentally possible to observe the $t\bar{t} H^0 \rightarrow l^\pm \nu q\bar{q} b\bar{b} b\bar{b}$ channel. Most important for a successful analysis are excellent $b$-tagging performance and a good mass resolution. The effects of event pile up still have to be investigated.

From theoretical point of view, the first complete LO simulation has
been performed for signal and background. After the K-factor for the
signal has been calculated (see Ref.\,\cite{tthnlo}), the main uncertainty in this channel is background cross section at HO, because our first estimate gives a factor of almost two.


\section*{Acknowledgments}
We would like to thank Viacheslav Ilyin and Nikita Stepanov for many fruitful discussions about LO and HO aspects of this study before and during this workshop.


\begin{appendix}

\section{\boldmath Comparison of $t\bar{t}b\bar{b}$ Events}

\vspace*{-3mm}
\begin{table}[htb]
\begin{center}
 \caption{\label{tab:comppyt1} CompHEP (ISR and FSR included) PYTHIA (default) comparison of $t\bar{t}b\bar{b}$ background: the cross sections do not include any branching fractions (all decays are allowed). All four $b$-quarks are required to be within $|\eta| < $ 2.5 and additional the $b$-quarks of the top decays have to satisfy $p_T(b_{top}) > $ 15 $GeV/c$. The $p_T(b_{glu})$ cut for both $b$-quarks not coming from the top decay is varied.}
\begin{tabular}{ |l|c|c|c|}
\hline
 $p_T$ Cuts & CompHEP & PYTHIA & CompHEP / PYTHIA \\\hline
\hline
 $p_T(b_{glu}) > $  15 $GeV/c$ & $\sigma =$ 2407 $fb$ & $\sigma =$ 2927 $fb$ & 0.82 \\\hline
 $p_T(b_{glu}) > $  30 $GeV/c$ & $\sigma =$ 1123 $fb$ & $\sigma =$ 1189 $fb$ & 0.94 \\\hline
 $p_T(b_{glu}) > $  50 $GeV/c$ & $\sigma =$  512 $fb$ & $\sigma =$  431 $fb$ & 1.19 \\\hline
 $p_T(b_{glu}) > $ 100 $GeV/c$ & $\sigma =$  116 $fb$ & $\sigma =$   53 $fb$ & 2.19 \\\hline
 $p_T(b_{glu}) > $ 200 $GeV/c$ & $\sigma =$   13 $fb$ & $\sigma =$    2 $fb$ & 6.73 \\\hline
 $p_T(b_{glu}) > $ 300 $GeV/c$ & $\sigma =$    3 $fb$ & $\sigma =$    0 $fb$ & ---- \\\hline
 \end{tabular}
\end{center}
\vspace*{-5mm}
 \end{table}

\end{appendix}
}

\setcounter{figure}{0}
\setcounter{table}{0}
\setcounter{section}{0}
\setcounter{equation}{0}
\newpage

{
\renewcommand{\thesection}{\arabic{section}.}

\def\cO#1{{\cal{O}}\left(#1\right)}
\newcommand{\be}{\begin{equation}}
\newcommand{\ee}{\end{equation}}
\newcommand{\br}{\begin{eqnarray}}
\newcommand{\er}{\end{eqnarray}}
\newcommand{\ba}{\begin{array}}
\newcommand{\ea}{\end{array}}
\newcommand{\bi}{\begin{itemize}}
\newcommand{\ei}{\end{itemize}}
\newcommand{\bn}{\begin{enumerate}}
\newcommand{\en}{\end{enumerate}}
\newcommand{\bc}{\begin{center}}
\newcommand{\ec}{\end{center}}
\newcommand{\ul}{\underline}
\newcommand{\ol}{\overline}
\newcommand{\ar}{\rightarrow}
\newcommand{\sm}{${\cal {SM}}$}
\newcommand{\as}{\alpha_s}
\newcommand{\aem}{\alpha_{em}}
\newcommand{\ycut}{y_{\mathrm{cut}}}
\newcommand{\susy}{{{SUSY}}}
\newcommand{\Dir}{\kern -6.4pt\Big{/}}
\newcommand{\Dirin}{\kern -10.4pt\Big{/}\kern 4.4pt}
\newcommand{\DDir}{\kern -10.6pt\Big{/}}
\newcommand{\DGir}{\kern -6.0pt\Big{/}}
\def\Ecm{\ifmmode{E_{\mathrm{cm}}}\else{$E_{\mathrm{cm}}$}\fi}
\def\gluino{\ifmmode{\mathaccent"7E g}\else{$\mathaccent"7E g$}\fi}
\def\photino{\ifmmode{\mathaccent"7E \gamma}\else{$\mathaccent"7E \gamma$}\fi}
\def\mgluino{\ifmmode{m_{\mathaccent"7E g}}
             \else{$m_{\mathaccent"7E g}$}\fi}
\def\taugluino{\ifmmode{\tau_{\mathaccent"7E g}}
             \else{$\tau_{\mathaccent"7E g}$}\fi}
\def\mphotino{\ifmmode{m_{\mathaccent"7E \gamma}}
             \else{$m_{\mathaccent"7E \gamma}$}\fi}
\def\ML{\ifmmode{{\mathaccent"7E M}_L}
             \else{${\mathaccent"7E M}_L$}\fi}
\def\MR{\ifmmode{{\mathaccent"7E M}_R}
             \else{${\mathaccent"7E M}_R$}\fi}

\def\lsim{\buildrel{\scriptscriptstyle <}\over{\scriptscriptstyle\sim}}
\def\gsim{\buildrel{\scriptscriptstyle >}\over{\scriptscriptstyle\sim}}
\def\jp #1 #2 #3 {{J.~Phys.} {#1} (#2) #3}
\def\pl #1 #2 #3 {{Phys.~Lett.} {#1} (#2) #3}
\def\np #1 #2 #3 {{Nucl.~Phys.} {#1} (#2) #3}
\def\zp #1 #2 #3 {{Z.~Phys.} {#1} (#2) #3}
\def\pr #1 #2 #3 {{Phys.~Rev.} {#1} (#2) #3}
\def\prep #1 #2 #3 {{Phys.~Rep.} {#1} (#2) #3}
\def\prl #1 #2 #3 {{Phys.~Rev.~Lett.} {#1} (#2) #3}
\def\mpl #1 #2 #3 {{Mod.~Phys.~Lett.} {#1} (#2) #3}
\def\rmp #1 #2 #3 {{Rev. Mod. Phys.} {#1} (#2) #3}
\def\sjnp #1 #2 #3 {{Sov. J. Nucl. Phys.} {#1} (#2) #3}
\def\cpc #1 #2 #3 {{Comp. Phys. Comm.} {#1} (#2) #3}
\def\xx #1 #2 #3 {{#1}, (#2) #3}
\def\preprint{{preprint}}
\def\Ord{\lower .7ex\hbox{$\;\stackrel{\textstyle <}{\sim}\;$}}
\def\OOrd{\lower .7ex\hbox{$\;\stackrel{\textstyle >}{\sim}\;$}}

\newcommand{\dx}{\mbox{\rm d}}
\newcommand{\ra}{\rightarrow}
\newcommand{\tb}{\tan \beta}
\newcommand{\s}{\smallskip}
\newcommand{\nn}{\noindent}
\newcommand{\non}{\nonumber}
\newcommand{\beq}{\begin{eqnarray}}
\newcommand{\eeq}{\end{eqnarray}}
\newcommand{\miss}{\not\hspace*{-1.8mm}E}
\newcommand{\ct}[1]{c_{\theta_#1}}
\newcommand{\st}[1]{s_{\theta_#1}}

\noindent
{\Large \bf G. Studies of Charged Higgs Boson Signals for the Tevatron
and the LHC} \\[0.5cm]
{\it K.A.\,Assamagan, M.\,Bisset, Y.\,Coadou,
A.K.\,Datta, A.\,Deandrea, A.\,Djouadi,
M.\,Guchait, Y.\,Mambrini, F.\,Moortgat and S.\,Moretti}

\begin{abstract}
Two Higgs doublet models are a viable extension to the Standard Model (SM) and can be incorporated 
into supersymmetry (SUSY). In such models, electroweak symmetry breaking leads to five Higgs 
particles, three neutral and a charged pair. We discuss various analyzes of the charged Higgs boson, 
carried out in the context of the Minimal Supersymmetric extension of the Standard Model (MSSM) and 
also in models with singlet neutrinos in large extra dimensions. Specific studies for the Large 
Hadron Collider (LHC) and the Tevatron are presented. 
\end{abstract}

\section{Introduction}
\label{sec:intro}

The Higgs sector of the MSSM contains five physical states, two of which are charged, $H^\pm$, and the other three are neutral ($h^0$, $H^0$, and $A^0$)~\cite{Higgs,hhg}. Searches for the charged Higgs boson have been carried out at LEP and at the Tevatron: at LEP 2, a lower bound of 78.6~GeV has been set on the charged Higgs boson mass independent of the $H^\pm\rightarrow\tau^\pm\nu_\tau$ branching ratio (BR)~\cite{LEP}. At the Tevatron, CDF and D{\O} performed direct and indirect searches for the charged Higgs boson, and excluded the low and high $\tan\beta$ regions up to $\sim 160$~GeV~\cite{Teva}. 
\par
The sensitivity of the ATLAS and CMS detectors at the LHC to the discovery of the charged Higgs boson has been investigated in detail~\cite{ATLASTDR,CMSTDR}. Some of these studies have been carried out as particle-level event generation in PYTHIA, HERWIG and ISAJET~\cite{pythia,herwig,isajet}, at $\sqrt{s}=14$~TeV, with the detector resolutions and efficiencies parameterized in ATLFAST~\cite{atlfast} and in CMSJET~\cite{cmsjet} from the full detector simulations. Some of the LHC studies assume that the mass scale of supersymmetric partners of ordinary matter is above the charged Higgs boson mass so that charged Higgs boson decays into supersymmetric partners are forbidden. The main production processes considered in these studies are the gluon fusion mechanism, $gg\rightarrow tbH^\pm$ and the $2\rightarrow 2$ process $gb\rightarrow tH^\pm$ shown in Figure~\ref{fig:figure1}. A central value of 175~GeV is used for the top-quark mass. 
\begin{figure}[!htb] 
\begin{center} 
\epsfig{file=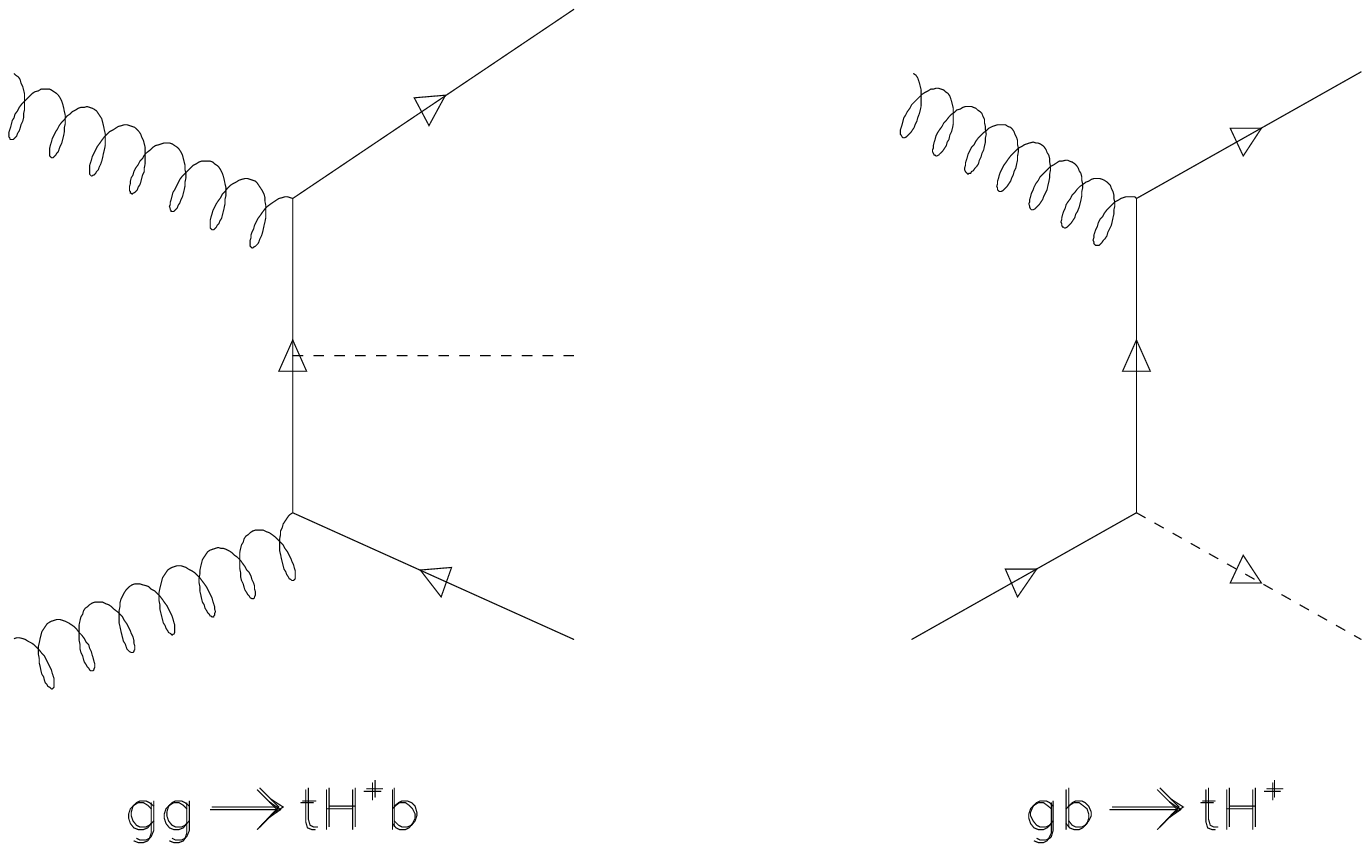,width=4in} 
\caption{\small The charged Higgs boson production at the LHC through the $2\rightarrow 3$ process,  
$gg\rightarrow tbH^\pm$ and the $2\rightarrow 2$ process, $gb\rightarrow 
tH^\pm$. The inclusive cross section is the sum of both contributions after 
the subtraction of the common terms.} 
\label{fig:figure1}  
\end{center}  
\end{figure} 
The decay channel $H^\pm\rightarrow\tau^\pm\nu_\tau$ has been studied extensively for ATLAS  for 
$m_{H^\pm} < m_t$, and the signal appears as an excess of $\tau$ leptons~\cite{cava}. The channel $H^\pm\rightarrow Wh^0$ is only relevant in a tiny range of MSSM parameter space although it constitutes a unique test for MSSM and may be sensitive to the singlet extension to MSSM, i.e., NMSSM~\cite{dress,assa1}. $H^\pm\rightarrow tb$ and $H^\pm\rightarrow\tau^\pm\nu_\tau$ are the dominant decay channels of the charged Higgs boson in most of the parameter space. In the $H^\pm\rightarrow tb$ channel, upwards of 5-$\sigma$ discovery can be achieved above the top-quark mass in the low and high $\tan\beta$ regions up to $\sim$400~GeV~\cite{assa2}. $H^\pm\rightarrow\tau^\pm\nu_\tau$ extends the discovery reach to high Higgs boson masses and to lower $\tan\beta$ values in the high $\tan\beta$ region as seen in Figure~\ref{fig:figure2}. 
\begin{figure}[!htb]
\begin{center}
\epsfig{file=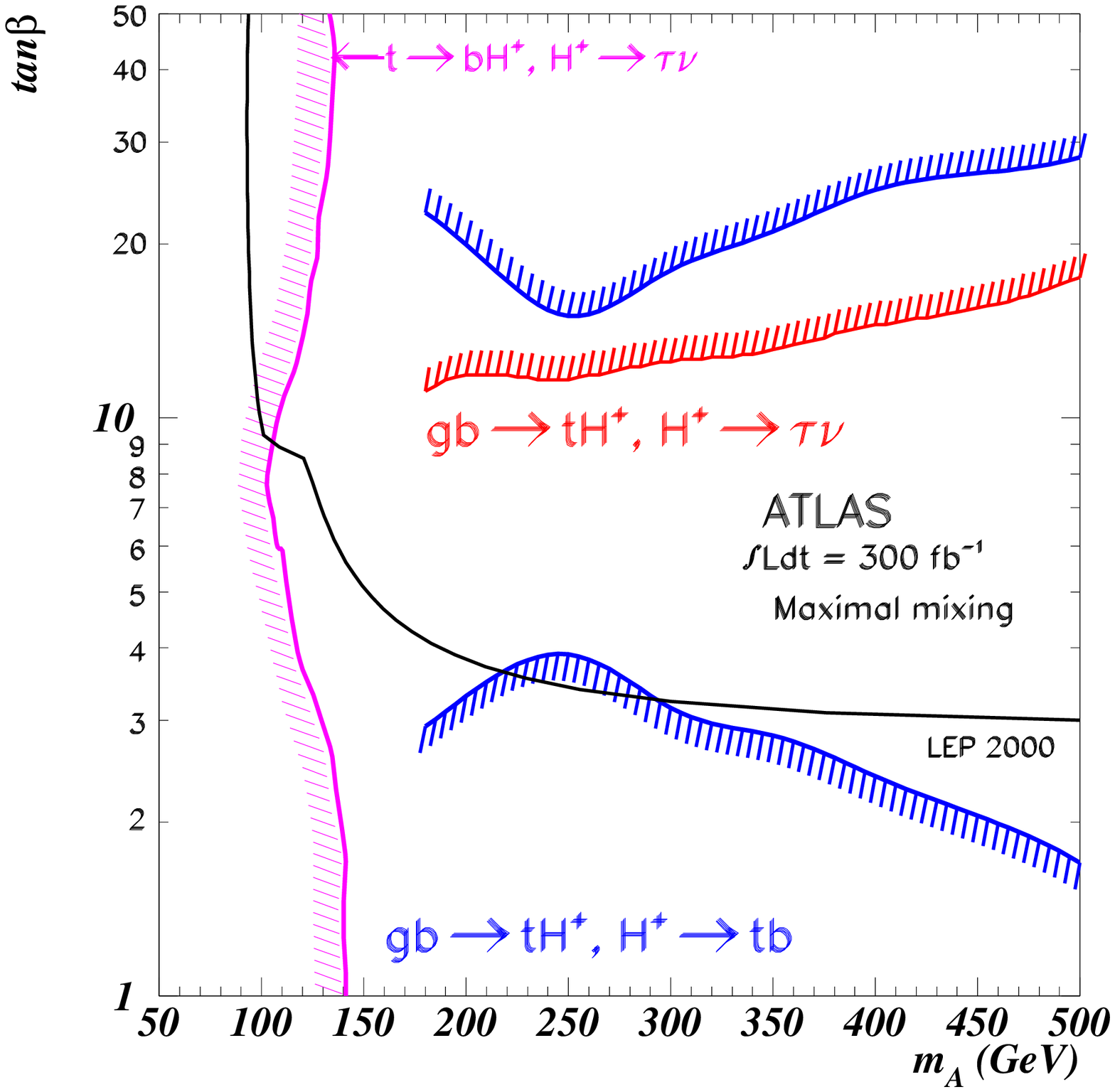,width=7cm,height=5.3cm}
\epsfig{file=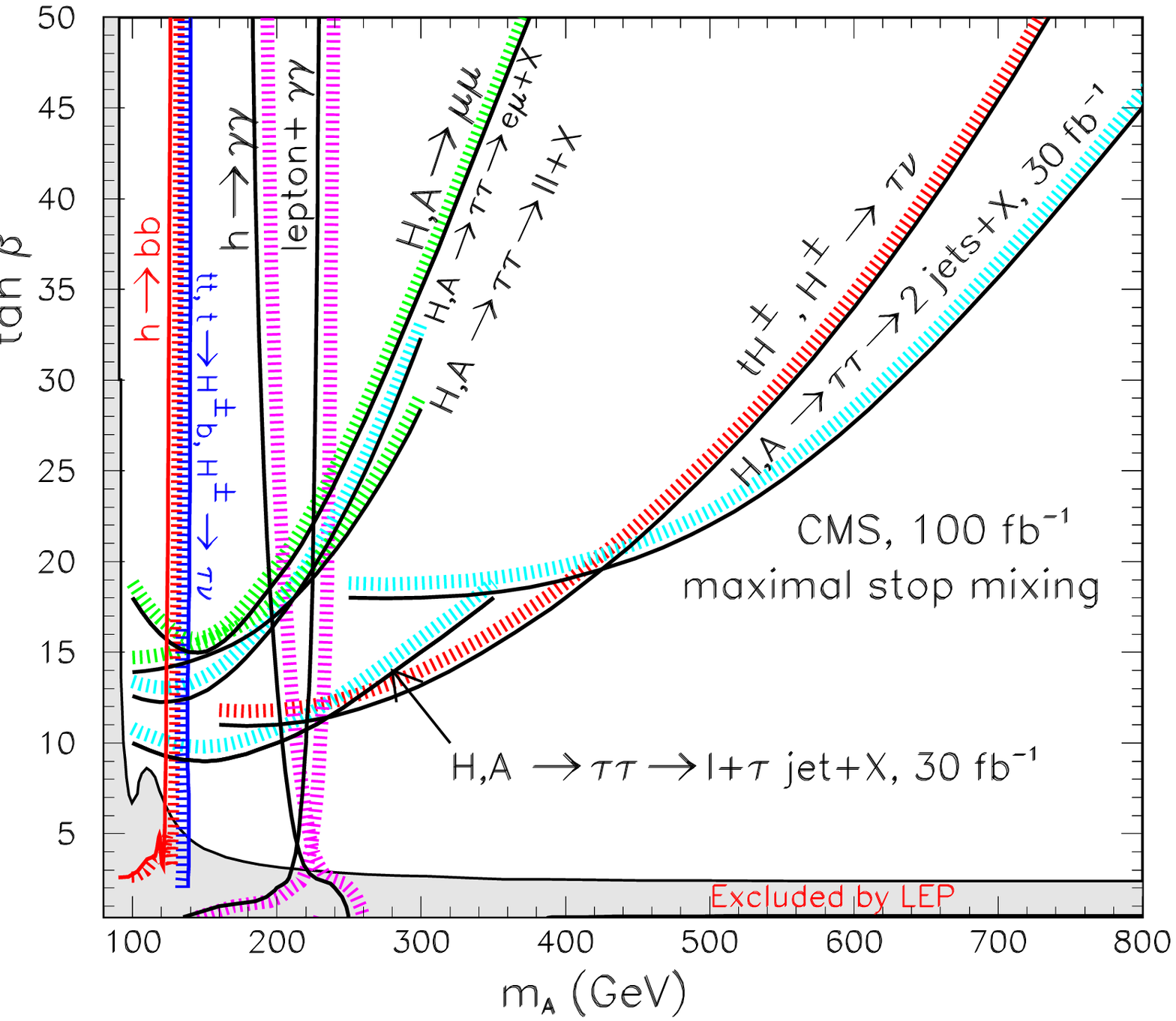,width=6cm}
\caption{\small The ATLAS 5-$\sigma$ discovery contour of the charged Higgs boson for an integrated luminosity of 300~fb$^{-1}$ (left plot); the CMS 5-$\sigma$ discovery contour of the MSSM Higgs bosons for 100~fb$^{-1}$ of luminosity (right plot). Below the top-quark mass, the charged Higgs boson is produced from top decay and the $\tau^\pm\nu_\tau$ channel provides coverage for most $\tan\beta$ below $\sim$160~GeV. Above the top-quark mass, the $tb$ channel covers the low and the high $\tan\beta$ regions while the $\tau^\pm\nu_\tau$ channel extends the discovery reach to high Higgs boson mass and to lower $\tan\beta$ in the high $\tan\beta$ region.}
\label{fig:figure2}
\end{center}
\end{figure}
However, in the low $\tan\beta$ region, the $\tau^\pm\nu_\tau$ channel offers no sensitivity for the charged Higgs boson discovery as the $H^\pm\rightarrow\tau^\pm\nu_\tau$ branching vanishes~\cite{assa3}. Further studies are needed to cover the remaining areas of the 5-$\sigma$ discovery contour of Figure~\ref{fig:figure2}: 
\begin{itemize}
\item The lack of sensitivity in the intermediate $\tan\beta$ region is due to the fact that the charged Higgs boson coupling to SM fermions is proportional to
\begin{equation}
\label{eq:H+couplings}
H^+(m_t\cot\beta\bar{t}b_L+m_b\tan\beta\bar{t}b_R),
\end{equation}
the square of which goes through a minimum at $\tan\beta=\sqrt{m_t/m_b}$. The studies of charged 
Higgs boson production from SUSY cascades and charged Higgs boson decays to SUSY particles might help cover this region. 
\item The gap in the $m_A$ axis around $m_A=160$~GeV corresponds to the transition 
region where, for the correct description of the charged Higgs boson production and decay mechanisms, it 
is mandatory to use the production process $gg\rightarrow tbH^\pm$ which includes not only 
$gg\rightarrow t\bar{t}$ with $t\rightarrow bH^\pm$, but also the Higgs-strahlung mechanism and the 
relative interferences~\cite{guch}. 
\item The discovery reach could be extended to high Higgs boson 
masses by studying the process $gg\rightarrow tbH^\pm$ with $H^\pm\rightarrow tb$ and tagging all 
the four b-jets in the spectrum~\cite{mor1}. 
\end{itemize}
Recent studies which attempt to cover these remaining regions of the parameter space are presented 
along with the observability of charged Higgs boson signals in models with singlet neutrinos in large 
extra dimensions and the prospects for the determination of the charged Higgs boson mass and $\tan\beta$ 
at the LHC.  

\section{$\mathbf{H^\pm}$ Mass and $\mathbf{\tan\beta}$ Determination at the 
LHC}
\label{sec:mass}

In this section, we discuss the expected precisions on the charged Higgs boson mass and $\tan\beta$ 
measurements at the LHC --- above the top-quark mass --- in the $H^\pm\rightarrow\tau^\pm\nu_\tau$ 
and $H^\pm\rightarrow tb$ channels. Details of this analysis can be found in~\cite{assa4}.

\subsection{Motivation}

The detection of a charged Higgs boson signal would constitute an irrefutable proof for physics beyond the 
SM. The subsequent determination of the charged Higgs boson parameters such as the mass, the decay width, 
the spin, the rates in the various decay channels and the couplings to SM and SUSY particles will be 
necessary not only to establish that the observed particle is indeed consistent with a charged 
scalar boson but also to identify the actual scenario that is realized. The measurements of the 
charged Higgs boson mass and $\tan\beta$ will be essential to the determination of the charged Higgs 
boson properties. 

\subsection{$\mathbf{H^\pm}$ Mass Determination in $\mathbf{H^\pm\rightarrow\tau^\pm\nu_\tau}$}

This channel does not offer the possibility for the observation of a resonance peak above the 
background, only the transverse Higgs boson mass can be reconstructed because of the neutrino in the final 
state. The background comes from single top $W^\pm t$, and $t\bar{t}$ productions with one 
$W^\pm\rightarrow\tau^\pm\nu_\tau$. Thus, the transverse mass is kinematically constrained to be 
less than the $W^\pm$ mass while in the signal the upper bound is the charged Higgs boson mass. Furthermore, 
the distributions of one-prong hadronic decays of $\tau^\pm$'s, 
\begin{eqnarray}  
\label{eq:pinu} 
\tau^\pm  \rightarrow & \pi^\pm\nu_\tau  & ~~~(11.1\%) \\
\tau^\pm  \rightarrow & \rho^\pm(\rightarrow\pi^\pm\pi^0)\nu_\tau  & ~~~(25.2\%) \nonumber \\
\tau^\pm  \rightarrow & a_1^\pm(\rightarrow\pi^\pm\pi^0\pi^0)\nu_\tau & ~~~(9.0\%), \nonumber  
\end{eqnarray} 
are sensitive to the polarization state of the $\tau$-lepton~\cite{roy,hagiwara}. In fact, it is to be noted that the spin state of $\tau^\pm$'s coming from $H^\pm$- and $W^\pm$-boson decays are opposite (neglecting leptonic mass effects, as we did here). This is true for the case of one-prong decays into both $\pi^\pm$'s and longitudinal vector mesons, while the transverse component of the latter dilutes the effect and must be somehow eliminated by requiring that 80\% of the $\tau$-jet (transverse) energy is carried away by the $\pi^\pm$'s, i.e.: 
\begin{equation}
\label{eq:pfrac}
R=\frac{p^{\pi^\pm}}{p_T^{\tau}}> 0.8.
\end{equation} 
Ultimately, the polarization effect leads to a significantly harder momentum distribution of charged pions from $\tau$-decays for the $H^\pm$-signal compared to the $W^\pm$-background, which can then be exploited to increase the signal-to-background ratios and the signal significances~\cite{assa3,ritva}. Indeed the background is relatively small as shown in Figure~\ref{fig:figure3} where the transverse mass 
\begin{equation}
\label{eq:transM}
m_T =\sqrt{2 p_T^{\tau} {p\!\!\!/}_T (1 - \cos\Delta\phi)},
\end{equation}
is reconstructed from the visible $\tau$-jet and the missing energy. 
\begin{figure}[!htb]
\begin{center}
   \epsfig{file=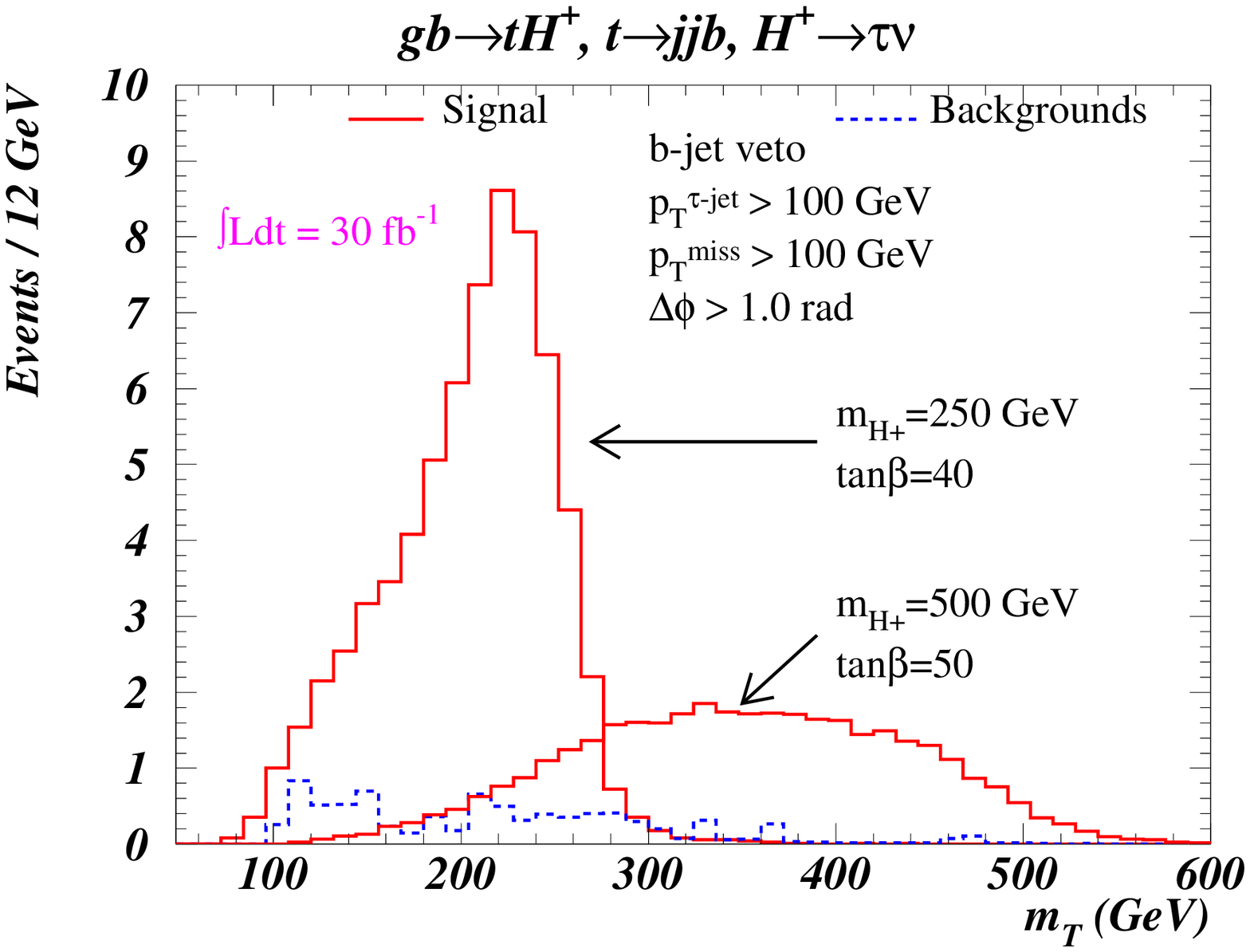,width=5in}
\caption{\small The reconstruction of the transverse charged Higgs boson mass in $H^\pm\rightarrow\tau^\pm\nu_\tau$ for $m_{H^\pm}=250$ and 500~GeV. The background is relatively small in this channel. The discovery reach is limited to high $\tan\beta$ but extended to higher mass compared to the $tb$ channel.}
\label{fig:figure3}
\end{center}
\end{figure}
As a result, although there is no resonance peak in this channel, the charged Higgs boson mass can be extracted from the transverse mass distribution with a relatively good precision. For the mass determination in this channel, we use the likelihood method described in~\cite{hohl}, which we summarize as follows:
\begin{itemize}
\item Suppose we wish to estimate the expected precision $\delta m_0$ on a Higgs boson reference mass $m_0$. We generate samples of events with charged Higgs boson masses $m_k=m_0+k\times\delta m$ and for each $m_k$ we calculate the probability density function $P_k(m)$ from the reconstructed transverse mass distribution of a charged Higgs boson with mass $m_k$. For example, for a charged Higgs boson reference mass $m_0=250$~GeV, we generate signal events at charged Higgs boson masses $m_k=230$, 235, 240, 245, 250, 255, 260, 265 and 270~GeV.  
\item Assuming $N_0$ is the expected number of events --- signal and background --- corresponding to the reference mass $m_0$, we draw randomly $N\equiv N_0+\delta N_0$ masses $m_j$ from each distribution $P_k(m)$ ($\delta N_0$ is the statistical error on $N_0$). For each $m_k$, we calculate the likelihood function ${\mathcal{L}}_k = \Sigma^N_{j=1}\log(P_k(m_j))$. The differences $\Delta{\mathcal{L}}_k = {\mathcal{L}}_0-{\mathcal{L}}_k$ show a minimum around $m_0$, where a parabolic fit is performed to get the actual expected value of $m_0$. This exercise can be repeated many times within the statistical error $\delta N_0$ and the distribution of the expected values, so obtained, of $m_0$ would be a Gaussian whose mean is the reconstructed mass and whose standard deviation is the statistical precision on the reconstructed mass. 
\item Three main sources of systematic uncertainties are included in the mass determination: the shape of the background, the background rate and the energy scale. The background shape becomes more significant at lower Higgs boson masses where there is more overlap between signal and background. To include this effect, we assumed a linear variation of the background shape, from $-10$\% to $+10$\% between the minimum and the maximum of the transverse mass distribution. Another source of systematic uncertainty is the rate of the backgrounds. It is expected that the background rate ($W^\pm t$ and $t\bar{t}$) could be known to 5\%~\cite{hohl}. Therefore, to take this effect into account, we increase the background rate by 5\% while at the same time we decrease the signal by 5\%. Finally, we also include the scale uncertainty: 1\% for jets and 0.1\% for photons, electrons and muons. The overall precisions on the charged Higgs boson mass determination, including the systematic uncertainties, are shown in Table~\ref{tab:mass} and Figure~\ref{fig:figure4}. 
\end{itemize}

\subsection{$\mathbf{H^\pm}$ Mass Determination in $\mathbf{H^\pm\rightarrow tb}$}

In the $tb$ channel, the full invariant mass can be reconstructed although this channel suffers from the irreducible $t\bar{t}b$ background and the signal combinatorial background~\cite{assa2}. The determination of the mass can be done using the likelihood method described above or by fitting the signal and the background. In the latter case, one assumes that the background shape and normalization can be determined by fitting outside the signal region, thus, the systematic uncertainties include only the scale uncertainty. We assume a Gaussian shape for the signal and an exponential for the background and fit signal$+$background including the statistical fluctuations and the scale uncertainty. Both methods are in agreement on the mass determination. The results are shown in Table~\ref{tab:mass} and Figure~\ref{fig:figure4}.
\begin{table}[!htb]
 \begin{center}
\caption{The overall precisions on the mass determination are better in the $\tau\nu$ channel than in the $tb$ channel. This is due to the fact that the latter suffers from large $t\bar{t}b$ and signal combinatorial backgrounds (${\mathcal{L}}=100$~fb$^{-1}$).}
\begin{tabular}{ccccc}\hline\hline
$m_{H^\pm}$~(GeV) & \multicolumn{2}{c}{$H^\pm\rightarrow\tau^\pm\nu_\tau$} & \multicolumn{2}{c}{$H^\pm\rightarrow tb$} \\ \hline
      & $<m>$   & $\delta m$ & $<m>$ & $\delta m$ \\
225.9 & 225.9 & 2.9 & 226.9 & 1.8 \\
271.1 & 271.0 & 3.9 & 270.1 & 10.1 \\
317.8 & 319.7 & 5.9 & 320.2 & 11.3 \\
365.4 & 364.9 & 8.1 & 365.4 & 12.1 \\
413.5 & 414.8 & 8.0 & 417.4 & 17.6 \\
462.1 & 460.7 & 10.6 & 465.9 & 24.1 \\
510.9 & 511.4 & 15.7 &             &           \\ 
\hline\hline
\end{tabular}
\label{tab:mass}
\end{center}
\end{table}

\subsection{Determination of $\mathbf{\tan\beta}$}

$\tan\beta$ can be obtained by measuring the signal rate in the $\tau^\pm\nu_\tau$ channel where 
the backgrounds are relatively low. The main systematic error would come from the knowledge of the 
luminosity, whose uncertainty is taken conservatively to be 10\%. The error in the rate measurement 
can be estimated as~\cite{s1996}
\begin{equation}
\label{eq:rate}
\frac{\Delta (\sigma\times BR)}{\sigma\times BR} = \sqrt{\frac{S+B}{S^2} + \left(\frac{\Delta \mathcal{L}}{\mathcal{L}}\right)^2},
\end{equation}
where $S$ and $B$ are the numbers of signal and background events respectively. The uncertainty on 
$\tan\beta$ is computed as
\begin{equation}
\label{eq:dtan}
\Delta\tan\beta \simeq \Delta(\sigma\times BR)\left[\frac{d(\sigma\times BR)}{d\tan\beta}\right]^{-1}.
\end{equation}
The production cross-section for $gb\rightarrow tH^\pm$ and the branching ratio of $H^\pm\rightarrow\tau^\pm\nu_\tau$ above the top-quark mass can be written respectively as~\cite{hhg}
\begin{equation}
\label{gbcross}
\sigma(gb\rightarrow tH^\pm) \propto m^2_t\cot^2\beta + m^2_b\tan^2\beta,
\end{equation}
and
\begin{equation}
\label{eq:ratio}
BR (H^\pm\rightarrow\tau^\pm\nu_\tau) \simeq \frac{m^2_\tau\tan^2\beta}{3(m^2_t\cot^2\beta+m^2_b\tan^2\beta) + m^2_\tau\tan^2\beta}.
\end{equation}
Using the relations~(\ref{gbcross}) and (\ref{eq:ratio}), the rate in the $\tau^\pm\nu_\tau$ channel at large $\tan\beta$ is obtained as:
\begin{equation}
\label{eq:sigbr}
\sigma\times BR \propto \tan^2\beta.
\end{equation}
From the Equations~(\ref{eq:dtan}) and~(\ref{eq:sigbr}), we get
\begin{equation}
\label{eq:errtan}
\frac{\Delta\tan\beta}{\tan\beta} = \frac{1}{2}\frac{\Delta(\sigma\times BR)}{\sigma\times BR}.
\end{equation}
The expected uncertainties on $\tan\beta$ determination from the measurement of the rate in the $H^\pm\rightarrow\tau^\pm\nu_\tau$ channel are shown in Table~\ref{tab:tanb} and Figure~\ref{fig:figure4}.
\begin{table}[!htb]
 \begin{center}
\caption{The overall precisions on $\tan\beta$ determination in the $H^\pm\rightarrow\tau^\pm\nu_\tau$ channel for ${\mathcal{L}}=30$, 100 and 300~fb$^{-1}$, and for $m_{H^\pm} = 250$~GeV.}
\begin{tabular}{cccc}\hline\hline
$\tan\beta$ & \multicolumn{3}{c}{$\Delta\tan\beta/\tan\beta$ (\%)} \\ \hline
                   & 30~fb$^{-1}$ & 100~fb$^{-1}$ & 300~fb$^{-1}$ \\
 20                & 15.4         & 10.6          & 7.4   \\
 25                & 12.2         & 8.7           & 6.5  \\
 30                & 10.5         & 7.7           & 6.1   \\
 35                & 9.1          & 7.0           & 5.7   \\
 40                & 8.4          & 6.6           & 5.6  \\
 45                & 7.7          & 6.6           & 5.5   \\
 50                & 7.3          & 6.1           & 5.4  \\
\hline\hline
\end{tabular}
\label{tab:tanb}
\end{center}
\end{table}

\begin{figure}[!htb]
\begin{center}
  \epsfig{file=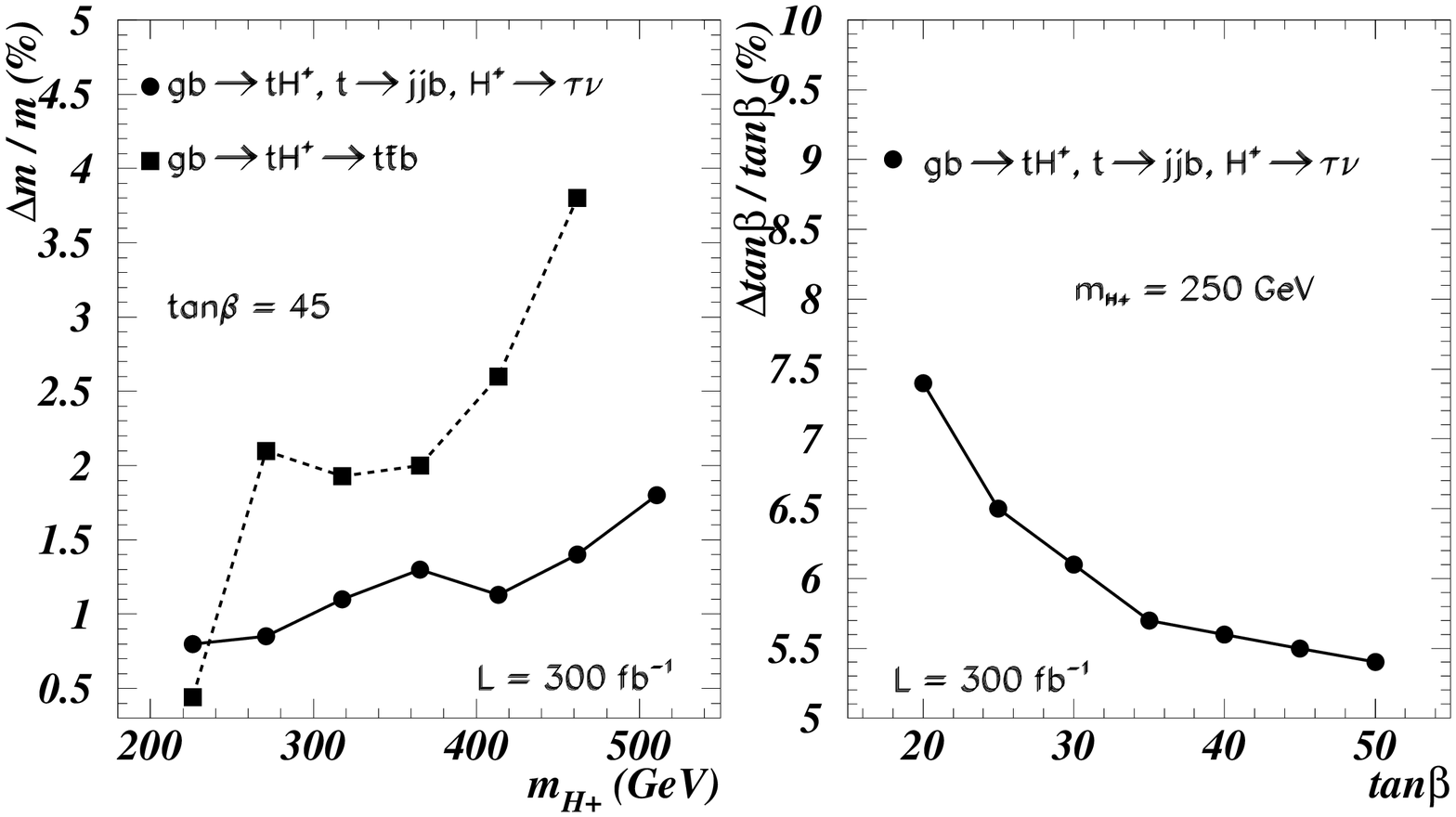,width=\textwidth}
\caption{\small The expected overall precision on the charged Higgs boson mass and on $\tan\beta$ measurements, as a function of the charged Higgs boson mass (left plot) and $\tan\beta$ (right plot) respectively. For the mass determination, the $H^\pm\rightarrow\tau^\pm\nu_\tau$ channel gives better precisions than $H^\pm\rightarrow tb$ except at low $m_{H^\pm}$. In addition, $H^\pm\rightarrow\tau^\pm\nu_\tau$ allows for the determination of $\tan\beta$ by measuring the rate in this channel.}
\label{fig:figure4}
\end{center}
\end{figure}

\subsection{Conclusions}

In the $\tau^\pm\nu_\tau$ channel, there is no resonance peak, only the transverse mass is reconstructed. A likelihood method is used to estimate the expected precisions on the mass measurements. The systematic effects include the background shape, the background rate and the energy scale. The overall relative precision in this channel ranges from 1.3\% at $m_{H^\pm}=226$~GeV to 3.1\% at $m_{H^\pm}=511$~GeV for an integrated luminosity of 100 fb$^{-1}$. At 300~fb$^{-1}$, the precision improves to 0.8\% at $m_{H^\pm}=226$~GeV and 1.8\% at $m_{H^\pm}=511$~GeV. 
\par
The $tb$ channel offers a resonance peak with a large background from $t\bar{t}b$ and the signal combinatorial. It is possible to use the likelihood method for the mass determination in this channel. Alternatively, a fit of the signal and background can be performed provided the background shape and normalization can be determined by fitting outside the signal region. Results from both methods are in agreement. The relative precision in this channel ranges from 0.8\% at $m_{H^\pm}=226$~GeV to 5.2\% at $m_{H^\pm}=462$~GeV for 100~fb$^{-1}$. For 300~fb$^{-1}$, the precision improves to: 0.5\% at 226~GeV and 3.5\% at 462~GeV. 
\par
In either channel, the overall uncertainties are dominated by the statistical errors. The $\tau\nu$ channel offers better precisions on the charged Higgs boson mass determination than the $tb$ channel, except at low $m_{H^\pm}$ where the $\tau^\pm\nu_\tau$ channel suffers from a much reduced cut efficiency. 
\par
$\tan\beta$ can be measured in the $H^\pm\rightarrow\tau^\pm\nu_\tau$ channel (by measuring the rate)  where the background is relatively low and the discovery reach is extended to high masses compared to $H^\pm\rightarrow tb$. Assuming a 10\% uncertainty on the luminosity, the relative precision on $\tan\beta$ ranges from 15.4 to 7.3\% for $\tan\beta=20$ to 50, at low luminosity. For an integrated luminosity of 300~fb$^{-1}$, the precision improves to: 7.4\% at $\tan\beta=20$ to 5.4\% at $\tan\beta=50$.

\section{$\mathbf{H^\pm}$ Boson in the Threshold Region}
\label{sec:thresh}

In this section, we discuss charged Higgs boson analyzes in the threshold region, i.e., for $m_{H^\pm}\sim m_t$, taking into account the correct description of the charged Higgs boson production and decay mechanism in this region.
  
\subsection{Motivation}

In the MSSM, the LEP 2 limits on the mass of the lightest Higgs boson
convert directly (see \cite{hhg}) into a lower bound
on the charged Higgs boson mass (at least, at low $\tan\beta$,
say, around 3): $m_{H^\pm}^2\approx m_{W^\pm}^2+m_{h^0}^2\OOrd 
(140~\mathrm{GeV})^2$ \cite{LepTre}. Whereas the charged
Higgs boson mass region just above this value is theoretically well 
understood, the description of the so-called (top) `threshold region',
$m_{H^\pm}\sim m_t$, requires careful considerations
when it comes to $H^\pm$ production
and decay in the context of a Monte Carlo (MC) 
simulation, as explained below. The main production mode of $H^\pm$ scalars with mass 
strictly below the top-quark mass,
$m_{H^\pm}<m_t$, is the decay of the top
(anti)quarks themselves, the latter being produced via QCD in the annihilation
of gluon-gluon and quark-antiquark pairs. So far, standard
MC programs, such as PYTHIA, HERWIG and ISAJET \cite{pythia,herwig,isajet}, have 
accounted for this process through the usual procedure of factorizing
the production mode, $gg,q\bar q\to t\bar t$, times the
decay one, $\bar t\to \bar b H^-$, in the so-called Narrow Width
Approximation (NWA). However, this description
fails to correctly account for the production and decay 
phenomenology of charged Higgs bosons when their mass approaches or
exceeds that of the top-quark, hence undermining the ability of experimental analyzes 
in pinning down the real nature of these particles (if not detecting them 
altogether). 
\begin{figure}[!htb]
\begin{center}
\epsfig{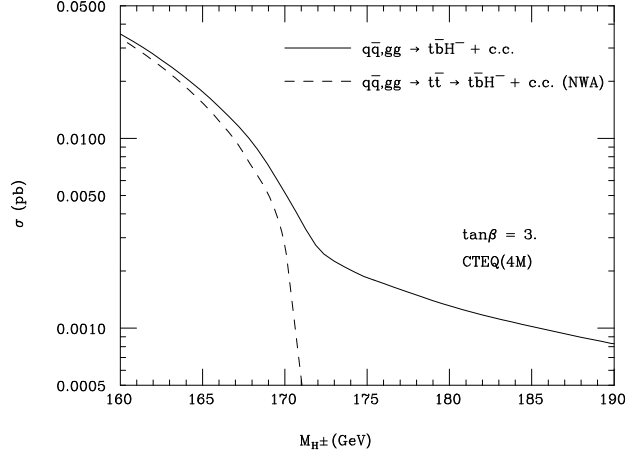}
\caption{\small Cross section for $gg,q\bar q\to t\bar b H^-$ and
$gg,q\bar q\to t\bar t\to t\bar b H^-$ in NWA, 
at the Tevatron with $\sqrt s=2$ TeV,
as a function of $m_{H^\pm}$ for a representative value of $\tan\beta$ (the kinematical effects discussed are the same irrespective of the latter). Hereafter, charge conjugated rates are always included. Besides, 
both top and bottom quark masses in the Higgs boson Yukawa couplings are
non-running and set to 175 and 4.25 GeV, respectively. CTEQ4M \cite{cteq}
is used
for the Parton Distribution Functions (PDFs), with scale $m_{H^\pm}$.}
\label{fig:fig_threshold}
\end{center}
\end{figure}

This is particularly a pressing issue at the Tevatron Run 2 \cite{guch},
as the collider reach in $m_{H^\pm}$ dips precisely into the threshold 
region \cite{reviewTEV}. Here, the use of the $2\to 3$ hard scattering
process $gg,q\bar q\to t\bar b H^-$  \cite{tbH}, in place of the
`factorization' procedure \cite{CDFD0}, is mandatory, as one can clearly see 
from  Figure~\ref{fig:fig_threshold} where the discrepancies in the shape and normalization 
come from the Higgs-strahlung mechanism and the relative interferences as mentioned earlier.

\begin{figure}[!htb]
\begin{center}
\epsfig{file=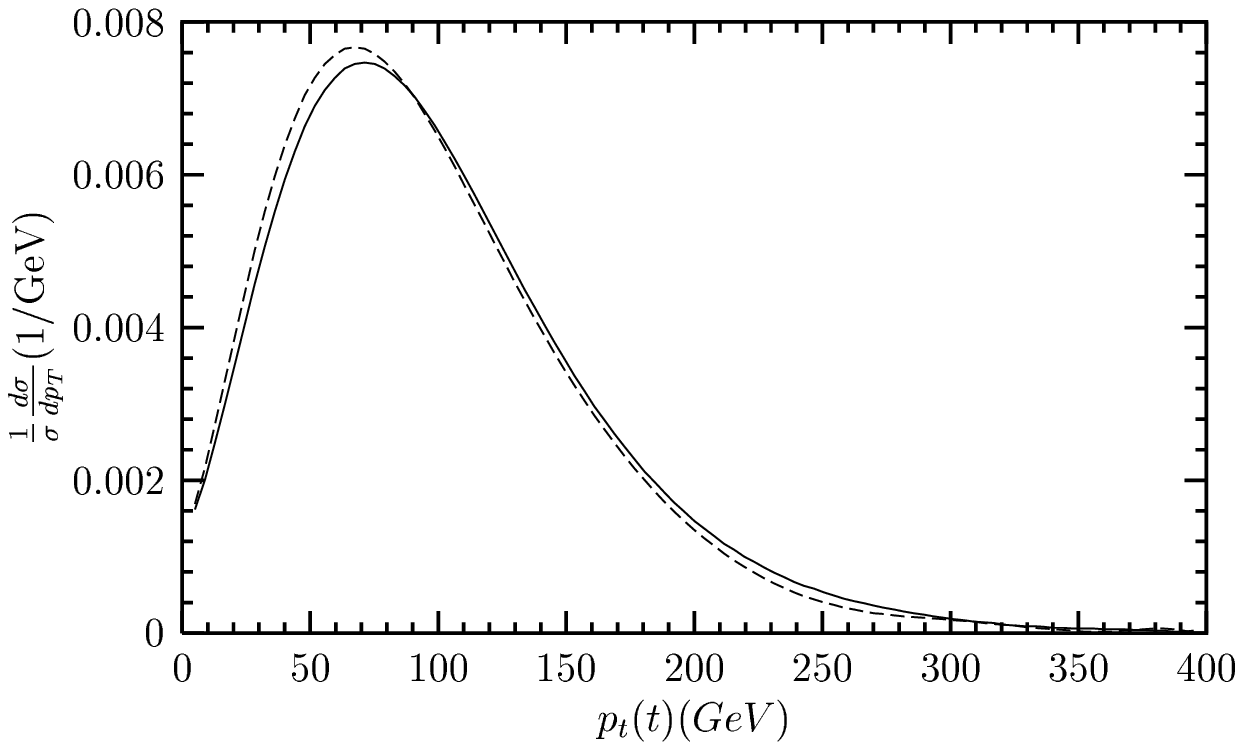, width=70mm, height=70mm}
\epsfig{file=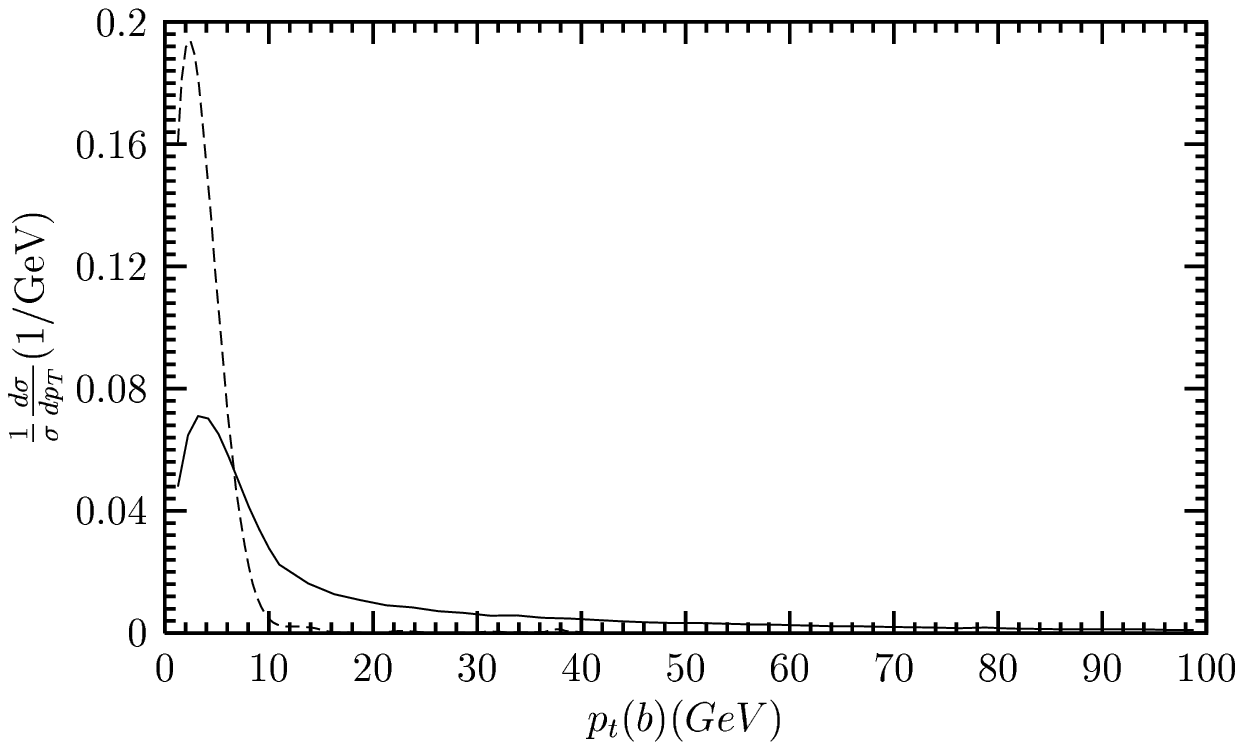, width=70mm, height=70mm}
\vspace*{-0.15in}
\caption{\small Transverse momentum distributions of the final state
quarks in $gg,q\bar q\to t\bar b H^-$ (solid) and
$gg,q\bar q\to t\bar t\to t\bar b H^-$ (dashed)
in NWA, at Tevatron with $\sqrt s=2$ TeV, for $m_{H^\pm}=170$ GeV.
Again, the actual $\tan\beta$ value is irrelevant.}
\label{fig:fig_pt}
\end{center}
\end{figure}

Also differential distributions can strongly be affected by an
approximated modeling of the production and decay process in the
threshold region, as one can appreciate from
Figure~\ref{fig:fig_pt}. Here, differences are
clearly sizeable also for the top quark. However, in this case
one should expect the impact to be marginal, as
this particle is actually unstable and since its three-body decay products are 
subject to the effect of usual detector resolution uncertainties.
In contrast, this
is no longer true for the bottom quark, which fragments directly
into hadrons. Besides,
the availability of the newly implanted silicon vertex detector  
may render the tagging of $b$-quarks a crucial ingredient in detection 
strategies of charged Higgs bosons at Run 2, pretty much along
the same lines as established at the LHC \cite{mor1}.

\subsection{Analysis}

If one looks at the most promising (and cleanest) charged Higgs boson decay 
channel, i.e., $H^\pm\to\tau^\pm\nu_\tau$~\cite{BRs}, while reconstructing the 
accompanying top quark hadronically, the prospects of detection are rather good. 
This is made clear in Table~\ref{tab:tab_Hpm}. Even if one neglects 
the tagging of the $b$-quarks in the final state, the final results are very 
different between the full process and the NWA. In Table~\ref{tab:tab_Hpm}, 
we have reported the signal and dominant (irreducible) background rates 
(that is, from $gg,q\bar q \to t\bar b W^-$ + c.c. events, yielding the same 
final state as the signal) after the following sequence of cuts:

\begin{enumerate}
\item
Tau-jets are selected if they satisfy the criteria:
$p_T^{\tau} >$15 GeV and $|\eta^\tau|<$ 2.5. 
\item
We require ${p\!\!\!/}_T>$ 20 GeV, since the 
presence of neutrinos from $H^-$ decays and invisible decay products 
of $\tau$'s (mainly $\pi^0$'s) implies that a significant fraction of the 
transverse momentum goes undetected.
\item
Quark-jets are selected by imposing $p_T^j >$ 20 GeV and $|\eta^j| <$ 2.5.
We require at least one of these to be tagged as a $b$-jet. 
\item 
We demand that two un-tagged jets have an invariant mass
around $m_{W^\pm}$, e.g., $|m_{q\bar q'} - m_{W^\pm}| <$ 10 GeV
and that the $b$-jet in combination with the other two un-tagged jets produces
an invariant mass close to $m_t$, e.g., $|m_{b q \bar q'} - m_t| <$
15 GeV. 
\item
We require that the reconstructed transverse mass, Equation~(\ref{eq:transM}), be above the 
$W^\pm$-boson mass: $m_T >m_{W^\pm}\approx$ 80 GeV.
\end{enumerate}

The $\tau$'s can be tagged as narrow jets in their `one-prong' hadronic decay 
modes --- see the relations~(\ref{eq:pinu}) --- which represent 90\% of the hadronic decay rate 
and about 50\% of the total. This distinguishing feature is in contrast to the typical appearance 
of quark- and gluon-jets, which yield `multi-prong' hadronic topologies in the detectors, typical 
of QCD backgrounds of the form $W^\pm + {\mathrm {jets}}$ and $Z^0 + {\mathrm {jets}}$.

\begin{table}[!htb]
\begin{center}
\caption{\small The signal rates (in fb) for the process 
$q\bar q,gg \rightarrow t \bar b  H^-(\rightarrow \tau^-
\bar\nu_{\tau})$, at Tevatron with $\sqrt s=2$ TeV,  for representative 
values of $m_{H^\pm}$ and $\tan\beta$, after all cuts described in the text. The corresponding rate 
of the background is 0.22~fb independent of $m_{H^\pm}$.} 
\begin{tabular}{lccc}
\hline\hline
$m_{H^\pm}$ (GeV) $\downarrow$ \, / $\tan\beta$ $\rightarrow $  & 3  & 6 & 40 \\
\hline
150 & 6   & 3   & 52  \\
160 & 2.8 & 1.5 & 22  \\
170 & .4  & 0.25 & 3.5 \\
175 & .13 & .08 & 1.42 \\ 
180 & .067 & .061 & 1.09 \\
\hline\hline
\end{tabular}
\label{tab:tab_Hpm}
\vspace*{-2mm}
\end{center}
\end{table}

\subsection{Conclusions}

In the end, despite the fact that
one should more realistically expect both signal 
and background rates to be further reduced by a factor of 4 or so
($\tau$-identification efficiencies are estimated to be of order 
 50\% \cite{tauid}, similarly for the tagging of any $b$-jet 
\cite{reviewTEV}), the final message that emerges is that 
the chances of extracting the $H^\pm\to\tau^\pm\nu_\tau$ signal 
after 15 fb$^{-1}$ of luminosity at the Tevatron Run 2 are rather good 
for $m_{H^\pm}$ up to 180 GeV or so at large $\tan\beta$, while being 
negligible at low to intermediate $\tan\beta$ values. Conclusions would 
obviously be drastically different in the NWA scenario, if one recalls 
Figure~\ref{fig:fig_threshold}.
 
\begin{figure}[!htb]
\begin{center}
\epsfig{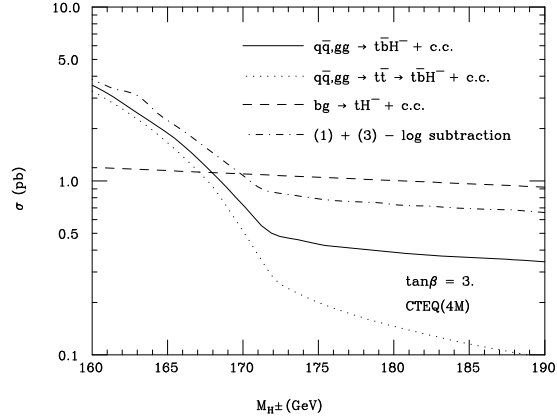}
\caption{\small Cross section for $gg,q\bar q\to t\bar b H^-$;
$gg,q\bar q \to t\bar t \to t\bar b H^-$ with finite top-quark width; $bg\to tH^-$ and
the combination of the first and the last, at the LHC with $\sqrt s=14$ TeV,
as a function of $m_{H^\pm}$ for a representative value of $\tan\beta$.}
\label{fig_LHC}
\end{center}
\end{figure}

The situation can be improved even further by taking advantage of the $\tau$ polarization effects as explained above. In this respect, the enforcement of the constraint~(\ref{eq:pfrac}) reduces the background by a factor of 5, while costing to the signal only a modest --- in comparison --- 50\% suppression (for any charged Higgs boson mass in the usual interval between 160 and 190 GeV). 

Although we have relied here on a parton-level analysis,
it is clear that its main features would remain valid even in presence
of fragmentation/hadronization effects. In fact, work is currently ongoing
in order to include the latter, as well as a more realistic detector
simulation, to emulate the real potential of the Tevatron experiments,
by exploiting the mentioned $2\to3$ description of the $H^\pm$ 
production dynamics and the spin correlations in $\tau$-decays, as
they are now both available in version 6.4 
\cite{HW64} of the HERWIG event generator (the latter also through
an interface to TAUOLA \cite{tauola}).

The problematic is very similar at the LHC, if anything more complicated.
In fact, at the CERN hadron collider, the above $2\to3$ reaction is
dominated by the $gg$-initiated subprocesses, rather than by $q\bar q$
annihilation, as is the case at the Tevatron. This means that a potential 
problem of double counting arises in the simulation of $t\bar b H^-$ + c.c. 
events at the LHC, if one considers that Higgs-strahlung can also be 
emulated through
the $2\to2$ process $bg\to t H^-$ + c.c. The difference between the
two descriptions is well understood, and a prescription exists for
combining the two, through the subtraction of a common logarithmic
term: see Refs~\cite{subtraction,mor-roy,reviewLHC,Jaume}. Figure~\ref{fig_LHC}
summarizes all the discussed issues in the context of the LHC.
The $2\to3$ process is available in HERWIG and detailed
simulations of the $\tau^\pm\nu_\tau$ channel 
at the CERN hadron collider are now possible
also for the threshold region,   as already done for
other mass intervals \cite{assa1,assa2,assa3,assa4,ritva,reviewLHC}.

\section{$\mathbf{H^\pm}$ Boson in Large Extra Dimensions}
\label{sec:large}

In this section, we discuss the LHC sensitivity to the charged Higgs boson discovery in  
the channel $H^-\rightarrow\tau_L^-\nu$ in models with singlet neutrinos in 
large extra dimensions. The observation of such a signal would provide a 
distinctive evidence for these models since in the standard two Higgs doublet 
model type II, $H^-\rightarrow\tau_L^-\nu$ is completely suppressed. Details of this
analysis can be found in~\cite{assa5}. 

\subsection{Motivation}  

In models where extra dimensions open up 
at the TeV scale, small neutrino masses can be generated without implementing 
the seesaw mechanism \cite{arka}. These models postulate the existence of 
$\delta$ additional spatial dimensions of size $R$ where gravity and perhaps 
other fields freely propagate while the SM degrees of freedom are confined to 
(3+1)-dimensional wall (4D) of the higher dimensional space. 
The true scale of gravity, or fundamental Planck scale $M_*$, of the ($4+\delta$)D 
space time is related to the reduced 4D Planck scale $M_{Pl}$, by
$M_{Pl}^2 = R^\delta M_*^{\delta+2}$,
where $M_{Pl}=2.4 \times 10^{18}$ GeV is related to the usual Planck mass
$1.2 \times 10^{19}$ GeV  $=\sqrt{8\pi} M_{Pl}$.
Since no experimental deviations from Newtonian gravity are 
observed at distances above 0.2 mm \cite{expgra}, the extra dimensions must be 
at the sub-millimeter level with $M_*$ as low as few TeV and $\delta \geq 2$.  
 
The right handed neutrino can be interpreted as a singlet with no quantum 
numbers to constrain it to the SM brane and thus, it can propagate into the 
extra dimensions just like gravity~\cite{Kane-invisible-lhc}. Such singlet states in the bulk 
couple to the SM states on the brane as right handed neutrinos with small 
couplings -- the Yukawa couplings of the bulk fields are suppressed by the 
volume of the extra dimensions. The interactions between the bulk neutrino  
and the wall fields generate Dirac mass $m_D$ terms between the wall fields and
all the Kaluza-Klein modes of the bulk neutrino:
\begin{equation}  
\label{eq:dm}  
m_D =\frac{\lambda}{\sqrt{2}}\frac{M_*}{M_{Pl}}v,  
\end{equation}
where $\lambda$ is a dimensionless constant. The mixing between the lightest neutrino 
with mass $m_D$ and the heavier neutrinos introduces a correction $N$ to the
Dirac mass such that the physical neutrino mass $m_\nu$ is
$m_\nu =m_D/N$, where  
\begin{equation}  
\label{eq:N}  
N \simeq 1 + \sum_{\vec{n}}^{|\vec{n}|< M_* R}  \left(\frac{m_D
R}{\vec{n}}\right)^2 \; ,   
\end{equation}  
$\vec{n}$ is a vector with $\delta$ integer components counting the number of
states and the summation is taken over the Kaluza-Klein states up to the
fundamental scale $M_*$. The sum over the different Kaluza-Klein states can be
approximately replaced by a continuous integration. 
As shown in Table~\ref{tab:table1}, small neutrino masses, $m_\nu$, can be obtained 
consistent with atmospheric neutrino oscillations~\cite{superK}.  
\begin{table}[!htbp] 
\begin{center}  
\caption{\small \label{tab:table1}The parameters used in the current analysis of the 
signal with the corresponding polarization asymmetry. In general, $H^-$ would 
decay to $\tau^-_L$ and $\tau^-_R$, $H^-\rightarrow\tau_R^-\bar{\nu} + 
\tau_L^-\psi$, depending on the asymmetry. For the decay 
$H^-\rightarrow\tau^-_R\bar{\nu}$ (as in MSSM), the asymmetry is $-1$. The 
signal to be studied is $H^-\rightarrow\tau^-_L\psi$.}  
\begin{tabular}{ccccccc}
\hline\hline    
 & $M_*$ (TeV) & $\delta_\nu$, $\delta$ & $m_{H^\pm}$ (GeV) & $\tan\beta$ & $A_{LR}$  & $m_\nu$ (eV) \\ 
\hline  
Sig.-1 & 2 & 4,4 &  219.9 &  30  & $\sim 1$ & 0.5 $10^{-3}$ \\  
Sig.-2 & 20 & 3,3 & 365.4 & 45 & $\sim 1$ & 0.05 \\
Sig.-3 & 1 & 5,6 & 506.2 & 4 & $\sim 1$ & 0.05 \\   
Sig.-4 & 100 & 6,6 & 250.2 & 35 & $\sim -1$ & 0.005 \\ 
Sig.-5 & 10 & 4,5 & 350.0 & 20 & $\sim -1$ & 0.04 \\  
Sig.-6 & 50 & 5,5 & 450.0 & 25 & $\sim -1$ & 0.04 \\  
\hline\hline
\end{tabular}
\end{center}  
\end{table} 
The spectrum of many extensions of the SM includes a charged Higgs boson state. 
We consider as a prototype of these models the 2-Higgs Doublet Model of type 
II (2HDM-II). $H^-$ decays to the right handed 
$\tau^-$ through the $\tau$ Yukawa coupling:
$H^-\rightarrow \tau_R^-\bar{\nu}$.
The $H^-$ decay to left handed $\tau^-$ is completely suppressed in MSSM. 
However, in the scenario of singlet neutrinos in large extra dimensions, $H^-$ 
can decay to both right handed and left handed $\tau^-$ depending on the 
parameters $M_*$, $m_D$, $\delta$, $m_{H^\pm}$ and $\tan\beta$, due to the large number of 
Kaluza-Klein states of the right handed bulk neutrino: 
$H^- \rightarrow \tau_R^-\bar{\nu} +\tau_L^-\psi$,   
where $\psi$ is a bulk neutrino and $\nu$ is 
dominantly a light neutrino with a small admixture of the Kaluza-Klein modes 
of the order $m_DR/|n|$. The measurement of the polarization asymmetry
\begin{equation}  
\label{eq:asym}  
A_{LR} =\frac{\Gamma(H^-\rightarrow\tau_L^-\psi)- 
\Gamma(H^-\rightarrow\tau_R^-\bar{\nu})} 
{\Gamma(H^-\rightarrow\tau_L^-\psi)+ 
\Gamma(H^-\rightarrow\tau_R^-\bar{\nu})}, 
\end{equation}    
can be used to distinguish between the ordinary 2HDM-II and the 
scenario of singlet neutrinos in large extra dimensions.  
 
The singlet neutrino may propagate into a subset $\delta_\nu$ ($\delta_\nu \leq \delta$) of 
the $\delta$ additional spatial dimensions, in which case the formalism for 
the generation of small Dirac neutrino masses is merely a generalization of 
the case $\delta_\nu=\delta$~\cite{dimo}. 
 
The charged Higgs boson 
decay to right handed $\tau$, $H^-\rightarrow\tau_R^-\bar{\nu}$ has been 
extensively studied for the LHC~\cite{assa4,ritva}. Here we discuss 
the possibility to observe $H^-\rightarrow\tau_L^-\psi$ at the LHC above
the  top-quark mass. Table~\ref{tab:table1} shows the parameters selected for
the  current analysis. The cases where the asymmetry is $+1$ are discussed in 
details. We assume a heavy SUSY spectrum with maximal stop mixing. The present 
analysis is conducted in the framework of PYTHIA6.1 and ATLFAST~\cite{pythia,atlfast}, 
and the Higgs boson masses and couplings are calculated to 1-loop with 
FeynHiggsFast~\cite{svenH}.
   
\subsection{Analysis} 

We consider the $2\rightarrow 2$ production 
process where the charged Higgs boson is produced with a top-quark, $gb\rightarrow 
tH^\pm$ as shown in Figure~\ref{fig:figure1}. Further, we require the hadronic decay of the 
top-quark, $t\rightarrow Wb\rightarrow jjb$ and the charged Higgs boson decay to $\tau$-leptons. 
\begin{table}[!htbp] 
\begin{center}  
\caption{\small \label{tab:table2}The expected rates ($\sigma\times$ BR), for the signal 
$gb\rightarrow t H^\pm$  with $H^-\rightarrow\tau_R^-\bar{\nu}+\tau_L^-\psi$ 
and $t\rightarrow jjb$, and for the backgrounds:  $W t$ and $t\bar{t}$ 
with $W^-\rightarrow\tau_L^-\bar{\nu}$ and $W^+\rightarrow jj$. We assume an 
inclusive $t\bar{t}$ production cross section of 590~pb. Other cross  sections 
are taken from PYTHIA~6.1 with CTEQ5L parton distribution function.  
See  Table~\protect{\ref{tab:table1}} for the 
parameters used for Sig.-1, Sig.-2 and Sig.-3. In the  last columns, we 
compare the $H^-\rightarrow\tau_R^-\bar{\nu}$ branching ratios in this model 
to the corresponding MSSM branching ratios from HDECAY~\cite{hdecay}.} 
\begin{tabular}{cccc}
\hline\hline 
Process & $\sigma\,\times\,$~BR (pb) & BR & BR(MSSM) \\  
\hline  
Sig.-1 & 1.56 & 0.73 &  0.37 \\ 
Sig.-2 & 0.15 & 1.0 & 0.15  \\  
Sig.-3 & 0.04 & 1.0 & 0.01  \\  
\hline 
$t\bar{t}$ & 84.11 &  &     \\  
$gb\rightarrow Wt$ ($p_T>30$~GeV) & 47.56 &  &   \\ 
\hline\hline
\end{tabular}
\end{center}  
\end{table}   
The major backgrounds are the single top production 
$gb\rightarrow Wt$, and $t\bar{t}$ production with one $W^+\rightarrow jj$ and 
the other $W^-\rightarrow\tau_L^-\bar{\nu}$\footnote{There is no enhancement
in the background rate from the contribution $W^- \to \tau^-_L\psi$.}. 
Depending on the polarization 
asymmetry, $H^-\rightarrow\tau_R^-\bar{\nu}$ will  contribute as an additional background. In 
Table~\ref{tab:table2}, we list the rates for the signal and for the backgrounds. The polarization 
of the $\tau^\pm$-lepton is included in this analysis through TAUOLA~\cite{tauola}. We consider 
the hadronic one-prong decays of the $\tau^\pm$-lepton --- see the relations~(\ref{eq:pinu}) --- which are believed to carry a better imprint of the $\tau^\pm$ 
polarization~\cite{roy}.   

For the signal in MSSM, right handed $\tau_R^-$'s 
come from the charged Higgs boson decay, $H^-\rightarrow\tau_R^-\bar{\nu}$, while in 
the backgrounds, left handed $\tau_L^-$'s come from the decay of the 
$W^-(\rightarrow\tau_L^-\bar{\nu})$. Because of the 
neutrino in the final state, only the transverse mass, Equation~(\ref{eq:transM}), can be 
reconstructed. In the framework of large extra dimensions, we are interested in 
$H^-\rightarrow\tau_L^-\psi$ where the polarization of the 
$\tau$-lepton would be identical to the background case but opposite to the 
MSSM case. Therefore, the polarization of the $\tau$-lepton would not help in 
suppressing the backgrounds. 
\begin{figure}[!htb] 
\epsfysize=11truecm 
\begin{center} 
\epsffile{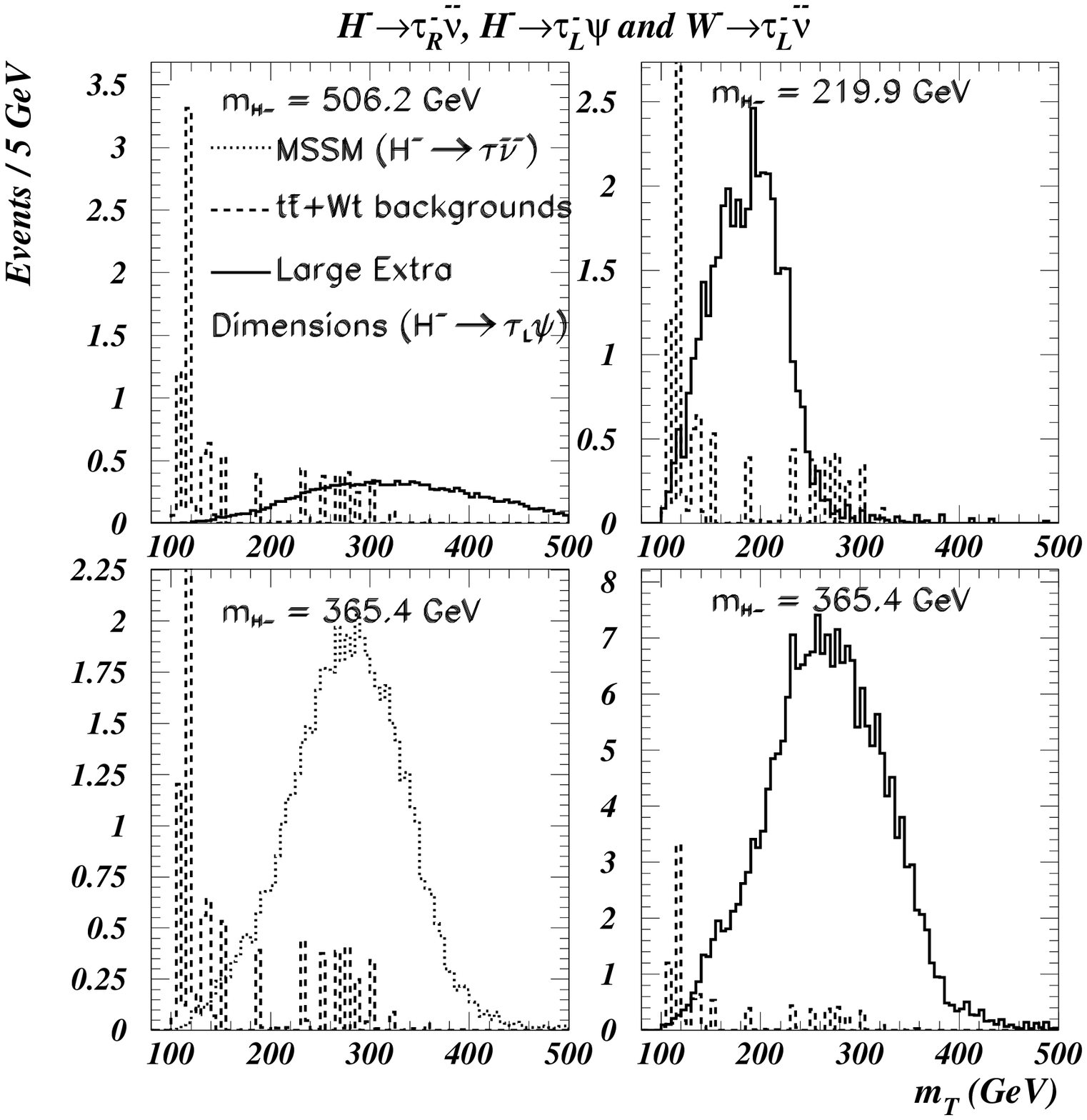} 
\caption{\small The reconstructions of the transverse mass of the signal in MSSM, the signal  
in models with a singlet neutrino in large extra dimensions and of the 
backgrounds, for an integrated luminosity of 100~fb$^{-1}$. The observation of the signal in the transverse mass 
distribution would not be sufficient to identify the model: the $\tau^\pm$ polarization effects 
must be explored further.}  
\label{fig:led_tnu_bgd} 
\end{center}  
\end{figure} 
Nevertheless, there are still some differences in the kinematics of the signal and of the backgrounds: the different transverse mass bounds and the increasingly harder $\tau$-jet and missing transverse momenta as the Higgs boson mass increases.
\begin{table}[!htbp] 
\begin{center}  
\caption{\small \label{tab:table3}The expected 
signal-to-background ratios and significances calculated for an integrated luminosity 
of 100~fb$^{-1}$ (one experiment). See Table~\ref{tab:table1} for the parameters used for 
Sig.-1, Sig.-2 and Sig.-3. In all the cases considered, the signal can be observed at 
the LHC with significances in excess of 5-$\sigma$ at high luminosity.} 
\begin{tabular}{cccc}  
\hline\hline
 & Sig.-1 & Sig.-2 & Sig.-3 \\ 
\hline
 Signal events  & 41 & 215 & 16 \\ 
 $t\bar{t}$ & 7 & 7 & 7 \\  
 $Wt$ & 3 & 3 & 3 \\   
 Total background & 10 & 10 & 10 \\ 
 $S/B$ & 4.1 & 21.5 & 1.6 \\  
 $S/\sqrt{B}$ & 13.0 & 68.0 & 5.1 \\    
\hline\hline
\end{tabular}
\end{center}  
\end{table}
\begin{figure}[!htb] 
\epsfysize=10truecm 
\begin{center} 
\epsffile{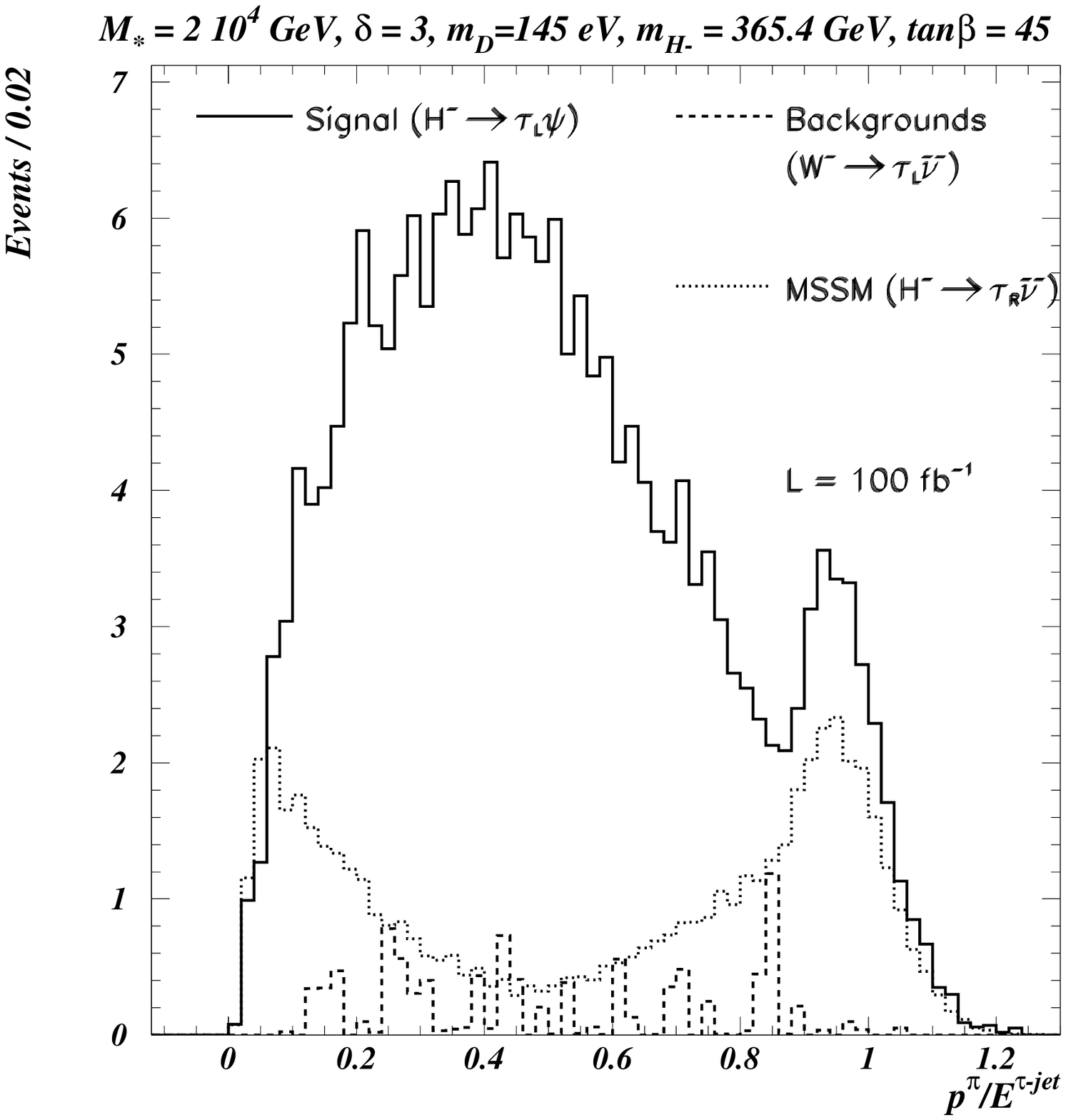} 
\caption{\small The distribution of the ratio of the charged pion track momentum in one  
prong $\tau$ decay to the $\tau$-jet energy for $m_A=350$~GeV, 
$\tan\beta=45$, $M_*= 20$~TeV, $\delta=3$ and $m_\nu=0.05$~eV, and for an integrated luminosity of 100~fb$^{-1}$. In the 2HDM-II, this ratio would peak near 0 and 1 as shown while in other models, the actual distribution of this ratio would depend on the polarization asymmetry since 
both left and right handed $\tau$'s would contribute. In the case shown, the 
asymmetry is $\sim 1$ and the ratio peaks near the center of the 
distribution.} \label{fig:350_45_145_1}  
\end{center}  
\end{figure} 
 The reconstructed transverse mass for the signal and the backgrounds are shown in Figure~\ref{fig:led_tnu_bgd} and the expected signal-to-background ratios and the signal significances in Table~\ref{tab:table3}.  The reconstruction of the transverse mass is not enough to distinguish between the MSSM and 
the singlet neutrinos in large extra dimensions. The differences in these two 
scenarios are best seen in the distribution of $p^\pi/E^{\tau-jet}$, the 
fraction of the energy carried by the charged track which is shown in 
Figure~\ref{fig:350_45_145_1}.  In the MSSM, this 
distribution peaks near 0 and 1 while in $H^-\rightarrow\tau_L^-\psi$ from 
large extra dimensions and in the backgrounds, this distribution peaks in the 
center. The backgrounds are relatively very small, and as concluded 
in~\cite{assa4,ritva}, the discovery reach is limited by the signal size 
itself. Therefore the observation of a signal in the transverse mass 
distribution and in the distribution of the fraction of the energy carried by 
the charged track should help determine whether the scenario is MSSM or not.  
 
\subsection{Conclusions} 

We studied the observability of the channel $H^-\rightarrow\tau_L^-\nu$ in models with a singlet neutrino in large extra dimensions at the LHC. Although the observation of a signal in the transverse mass distribution can be used to claim discovery of the charged Higgs boson, it is 
insufficient to pin down the scenario that is realized. Additionally, by 
reconstructing the fraction of the energy carried  by the charged track in the 
one-prong $\tau^\pm$ decay, it is possible to claim whether the scenario is the ordinary 2HDM or not. The further measurement of the polarization asymmetry might provide a distinctive evidence for models with singlet neutrinos in large extra dimensions.

\section{$\mathbf{H^\pm}$ Decays into SUSY Particles} 
\label{sec:Hp-to-susy}

Thus far, all analyzes have implicitly assumed that the SUSY counterparts of ordinary particles had 
masses much higher than $m_{H^\pm}$. However, lowering the typical SUSY mass scale may induce new 
interactions among $H^\pm$ bosons and several of the sparticles, so that the former may be abundantly 
produced in the decay of the latter (e.g., from gluinos and squarks, see \cite{mono}) or, 
alternatively, new Higgs boson decay channels into SUSY particles may well open at a profitable rate 
\cite{PAP1}. We will defer the study of the first scenario to section~\ref{sec:SUSYtoHpm}. Here, 
we will investigate in some depth the second possibility, focusing on the `intermediate' 
$\tan\beta$ region (say, between 3 and 10) left uncovered by the SM decay channels (see 
Figure~\ref{fig:figure2})\footnote{Depending on the actual rate of the new SUSY decay channels in 
the complementary $\tan\beta$ areas (less than three or larger than ten), some rescaling to the 
discovery reaches via the SM modes of Figure~\ref{fig:figure2} may be needed. This will be 
addressed in Ref.~\cite{preparation}.}.

\subsection{Motivation}

It was demonstrated in \cite{PAP1} that the decays of a charged Higgs boson into a chargino and a 
neutralino could probe regions of the MSSM
parameter space where decays to SM particles, such as
$b\bar{t}$ or $\tau^- \bar{\nu}_{\tau}$ yield no significant signal --- see Figure~\ref{fig:figure2}.  In particular,
intermediate values of $\tan\beta$ between $\sim 3$ and
$\sim10$ were in part accessible via
$H^{\pm} \rightarrow
\widetilde{\chi}_1^{\pm} \widetilde{\chi}_{\{2,3\}}^0$
modes resulting in three lepton final states (where leptons mean
electrons or muons), a hadronically reconstructed top quark
(from $gg \rightarrow \bar{b}tH^-$, $gb \rightarrow tH^-$
and their charge conjugate production processes) plus substantial
missing transverse momentum (from neutralino and chargino decays
to the stable lightest neutralino, i.e., the lightest
supersymmetric particle or LSP).
\par
We refer to the charginos and neutralinos collectively as ``inos'', 
which are the mass eigenstate mixtures of the electroweak (EW) gauginos and 
higgsinos. We expand on the parton level analysis of~\cite{PAP1}:  firstly, by studying the signal in a full event generator environment with an improved background analysis that includes potential
MSSM background processes (the previous study  considered
only SM backgrounds);  secondly, by further investigating
the possible role of on-shell or off-shell sleptons (the
supersymmetric partners of the leptons) in the signals.
As noted in the previous study, if there is a light slepton, then
the leptonic branching ratios (BRs) of the inos can be significantly
enhanced (especially those of $\widetilde{\chi}_2^0$,
$\widetilde{\chi}_3^0$).
Since both inos and sleptons play key roles in the signal process,
a considerable number of MSSM input parameters are relevant.
We seek to scan this expanse of parameter space,
at least at the parton level, identifying parameter points or
series of points that merit dedicated event generator analyzes.

To correctly delineate the portions of parameter space that can
potentially yield viable signals, it is necessary first to know
where experimental constraints cut off otherwise favorable zones.
LEP 2 experiments have yielded a number of bounds on MSSM
particles and parameters that impinge directly on our preferred
signal regions.  Among these, the most crucial are \cite{W1LEP2}:
the mass limit on the lighter chargino, taken as
$m_{\widetilde{\chi}_1^{\pm}} \, > \, 103\, \hbox{GeV}$
(which allows for the possibility of a light sneutrino resulting in
negative interference from
the then-significant $t$-channel in chargino pair production);
the mass limits on the sleptons,
taken as
$m_{\tilde{e}_1} \ge 99.0\, \hbox{GeV}$,
$m_{\tilde{\mu}_1} \ge 91.0\, \hbox{GeV}$,
$m_{\tilde{\tau}_1} \ge 85.0\, \hbox{GeV}$
and $m_{\tilde{\nu}} \ge 43.7\, \hbox{GeV}$
(the last being from studies at the $Z^0$ pole);
and the regions excluded by searches for signals of the type
$e^+e^- \rightarrow Z^{0*} h$ and $e^+e^- \rightarrow A h$.
For the numerical limits just given, it is
assumed that
$m_{\widetilde{\chi}_1^{\pm}} - m_{\widetilde{\chi}_1^0}$,
$m_{\tilde{\ell}_1} - m_{\widetilde{\chi}_1^0} \ge 5\, \hbox{GeV}$.
The constraint from Higgs boson production is somewhat vague owing to
an estimated uncertainty in the expected mass of the light Higgs boson
of $2$--$3\, \hbox{GeV}$ from un-calculated higher order
corrections \cite{highorder} and up to $5\, \hbox{GeV}$
from the error in the measurement of the top quark mass,
$m_t = 174.3 \pm 3.2 \pm 4.0\, \hbox{GeV}$ \cite{topmass}.
A small shift in the light Higgs boson mass translates into a substantial
shift in the location of the bound seen in the $\tan\beta$ versus
$m_A$ plane.  Other LEP 2 limits which generally should be less
restrictive than those just mentioned are also incorporated into our
analysis.

There are also other processes where charged Higgs bosons (or $A$, to
whose mass that of the $H^{\pm}$ is closely tied) enter as virtual
particles at the one-loop level.  These include neutral meson mixing
($K^0 \bar{K}^0$, $D^0 \bar{D}^0$, or $B^0\bar{B}^0$) \cite{FCNCpap,looppap},
$Z^0 \rightarrow b \bar{b}$ ($R_b$) \cite{looppap,EWprecis}, and
$b \rightarrow s \gamma$ decays \cite{FCNCpap,looppap,EWprecis,bsgam},
the last of which, where restrictions on $m_{H^{\pm}}$ are linked to a
number of MSSM variables, notably including the masses of the lighter
chargino and the stops, is generally thought to be the most constraining
\cite{EWprecis}.  This constraint may well disqualify regions of the
parameter space where our signal is strong and otherwise allowed; however,
the applicability of these low energy constraints is unclear
due to a variety of factors (see \cite{PAP1})
and thus they will not be included in this analysis (though we do
choose our stop parameters with an eye towards attempting to evade
potential $b \rightarrow s \gamma$ bounds).

\subsection{Parameter Space Exploration}

We begin in Figure~\ref{fig:HtoSUSY1} (top plot) with a look at the raw 
cross-section for $gg \rightarrow \bar{b}tH^-$, $gb \rightarrow tH^-$ and 
their charge conjugate production processes in the $\tan\beta$  vs $m_A$ plane. Other MSSM parameters are fixed at the values noted in the figure caption. Here we correctly take into account the subtraction needed to avoid
double counting between the $2\rightarrow 2$ and the $2\rightarrow 3$
production processes~\cite{subtraction,mor-roy}.
The charged Higgs boson mass is calculated including radiative corrections as
contained in ISAJET~\cite{isajet} and the
CTEQ4L~\cite{cteq} structure function set is employed.

In Figure~\ref{fig:HtoSUSY1} (bottom plot), we fold in the BRs of a charged Higgs
boson into an ino pair (restricted here to either
$\widetilde{\chi}_1^{\pm} \widetilde{\chi}_2^0$ or
$\widetilde{\chi}_1^{\pm} \widetilde{\chi}_3^0$)
multiplied by the BR of the ino pair into a trio
of charged leptons ${\ell}^+{\ell}^-{\ell}^{\prime\pm}$
(recall that ``$\ell$'' herein denotes either an electron or a muon), where
$\ell$ and ${\ell}^{\prime}$ may or may not be of the
same flavor.  
\begin{figure}[!htb]
\begin{center}
\epsfig{file=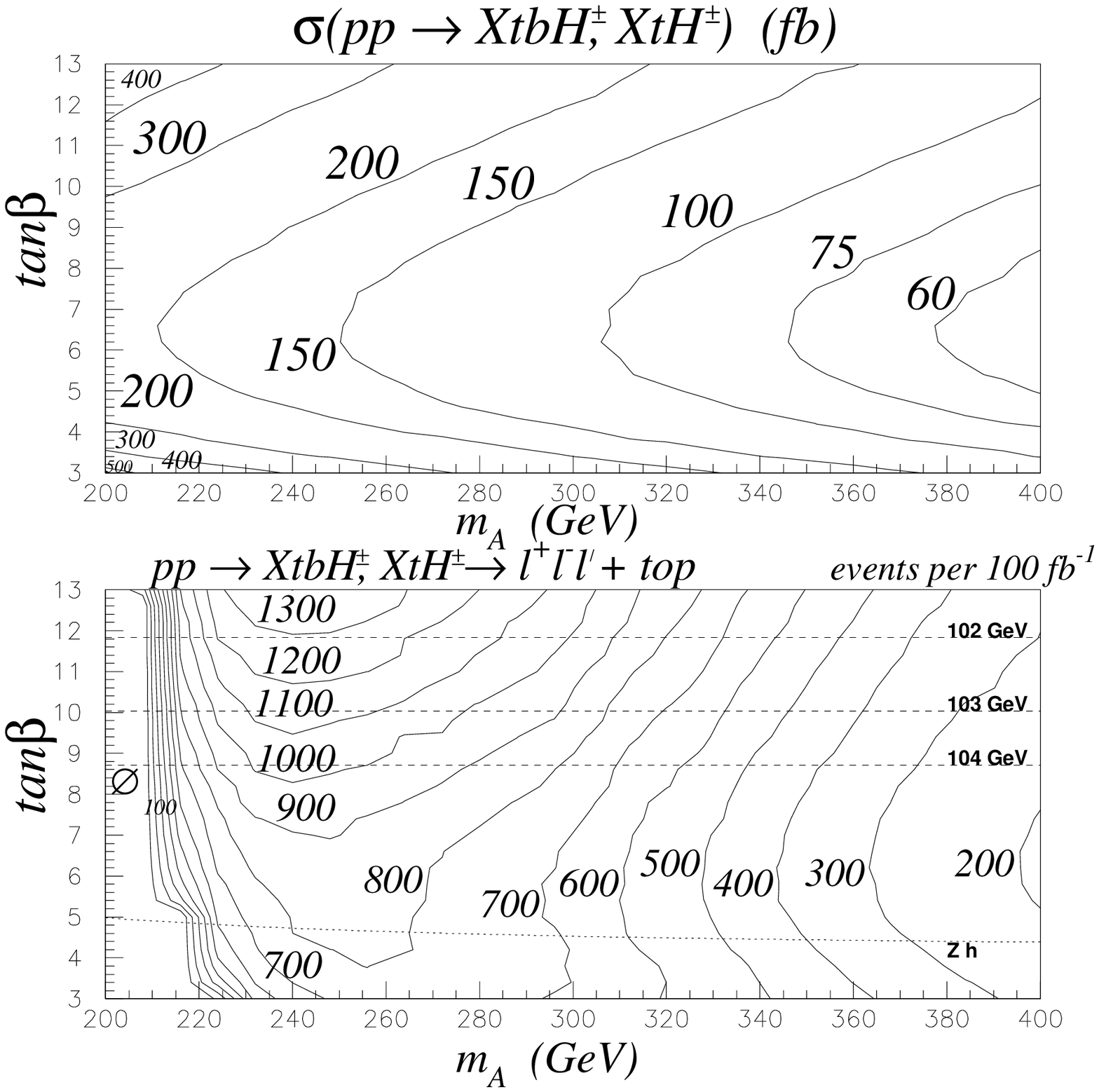,width=5in}
\caption{\small (top plot) Raw cross-section (in fb) for
$gg \rightarrow \bar{b}tH^-$, $gb \rightarrow tH^-$ and their charge
conjugate production processes in the $\tan\beta$  versus $m_A$ plane.
The proper subtraction factor to avoid double counting between the
$2\rightarrow 2$ and the $2\rightarrow 3$ production processes is
included. (bottom plot) Expected number of
$pp \rightarrow XtbH^{\pm}, XtH^{\pm} \rightarrow
{\ell}^+{\ell}^-{\ell}^{\prime\pm} \; + \; t$ events per
$100\, \hbox{fb}^{-1}$ at the LHC (with no cuts), assuming the charged
Higgs boson decays into $\widetilde{\chi}_1^{\pm}\widetilde{\chi}_2^0$
or $\widetilde{\chi}_1^{\pm}\widetilde{\chi}_3^0$. Here $M_{\scriptscriptstyle 2} = 200\, \hbox{GeV}$ and
$\mu = -115\, \hbox{GeV}$.  One-loop formul\ae\ as found in
\cite{isajet,thesis} are used to relate $m_{H^{\pm}}$ to $m_A$.
$m_t = 175\, \hbox{GeV}$ and $m_b = 4.25\, \hbox{GeV}$.
Other MSSM input parameters are:
$m_{\tilde{q}} = 1\, \hbox{TeV}$
for the first two generations,
$m_{\tilde{t}_{\scriptscriptstyle L}} = 600\, \hbox{GeV}$,
$m_{\tilde{t}_{\scriptscriptstyle R}} = 500\, \hbox{GeV}$,
$m_{\tilde{b}_{\scriptscriptstyle R}} = 800\, \hbox{GeV}$,
$A_t = 500\, \hbox{GeV}$, $A_b = 0$;
$m_{\tilde{\ell}_{\scriptscriptstyle R}} = 150\, \hbox{GeV}$,
$m_{\tilde{\ell}_{\scriptscriptstyle L}} = m_{\tilde{\nu}}^{soft} 
= 200\, \hbox{GeV}$ and $A_{\ell} = 0$ for all three generations.}
\label{fig:HtoSUSY1}
\end{center}
\end{figure}
In calculating the leptonic BRs of the inos
all conceivable decay chains are taken into
account\footnote{Including possible extra minor contributions arising
when a neutralino decays into a $\widetilde{\chi}_1^{\pm}$
(or a $W^{\pm}$) and one lepton and the $\widetilde{\chi}_1^{\pm}$
then decays into $\widetilde{\chi}_1^0$ along with the second lepton
(or $W^{\pm}$ decays leptonically).  These decay modes yield extra
neutrinos in addition to the final products of the main decay modes,
but should be experimentally indistinguishable.  Possible modes with
$\widetilde{\chi}_3^0 \rightarrow \widetilde{\chi}_2^0 X$
are also taken into account.  
}.
The plot shows that, assuming an integrated luminosity of
$100\, \hbox{fb}^{-1}$, hundreds to thousands of events are expected
for $m_A \le 400\, \hbox{GeV}$.  The possibility of extracting the SUSY signal
exists across the full range of allowed $\tan\beta$ values
and for $m_A$ (and $m_{H^{\pm}}$) $\lsim~400\, \hbox{GeV}$
provided the threshold for the
$H^{\pm} \rightarrow
\widetilde{\chi}_1^{\pm} \widetilde{\chi}_{\{2,3\}}^0$
decay is exceeded.
Higher values of $\tan\beta$ and $m_A
\simeq 240\,
\hbox{GeV}$ are optimal choices (quite different than when only the
raw production rate is considered).
The region from the top of the plot down to
the dashed curve marked as `$103\, \hbox{GeV}$' is excluded by the LEP 2
limit on the chargino mass, and the region below the dotted curve is
excluded by Higgs boson production.  As shown on the plot, the upper bound is
fairly sensitive to the $m_{\widetilde{\chi}_1^{\pm}}$ limit; and thus
in turn very dependent on the values chosen for other MSSM parameters,
in particular the higgsino mixing parameter $\mu$ and
$M_{\scriptscriptstyle 2}$\footnote{$M_{\scriptscriptstyle 1}$ and
$M_{\scriptscriptstyle 2}$ are the
$U(1)_{\hbox{\smash{\lower 0.25ex \hbox{${\scriptstyle Y}$}}}}$
and
$SU(2)_{\hbox{\smash{\lower 0.25ex \hbox{${\scriptstyle L}$}}}}$
gaugino masses, respectively; Grand Unified Theories (GUTs) predict
gaugino unification and
$M_{\scriptscriptstyle 1} =
\frac{5}{3}\tan\!^2{\theta}_{\scriptscriptstyle W}
M_{\scriptscriptstyle 2} \approx 0.5032~M_{\scriptscriptstyle 2}
\simeq \frac{1}{2}M_{\scriptscriptstyle 2}$,
as will be assumed in all numerical calculations.}, which
are chosen to be favorable to our signal in the plot.
Also, as mentioned in the previous section,
the exact location of the
latter bound is fairly loose due to uncertainty in the
mass of the light Higgs boson.

The expected number of events is shown again in Figure~\ref{fig:HtoSUSY2}, this time in the
$M_{\scriptscriptstyle 2}$ versus $\mu$ plane with $\tan\beta$ fixed
at $8$ and $m_A = 290\, \hbox{GeV} \;\,
(m_{H^{\pm}} \simeq 300\,\hbox{GeV})$.  The shaded region is excluded by
the LEP 2 bound on the chargino mass.   Again we see that hundreds to
thousands of events are possible in un-excluded regions of the parameter
space; but it is also apparent that small values of
$| \mu |$ are strongly preferred.  This is a serious restriction which
means that the preferred and un-excluded signal region is just beyond
that region probed by LEP 2 (and thus also the region expected to be
searched in the first phase of a future $e^+e^-$ linear collider).
\begin{figure}[!htb]
\begin{center}
\epsfig{file=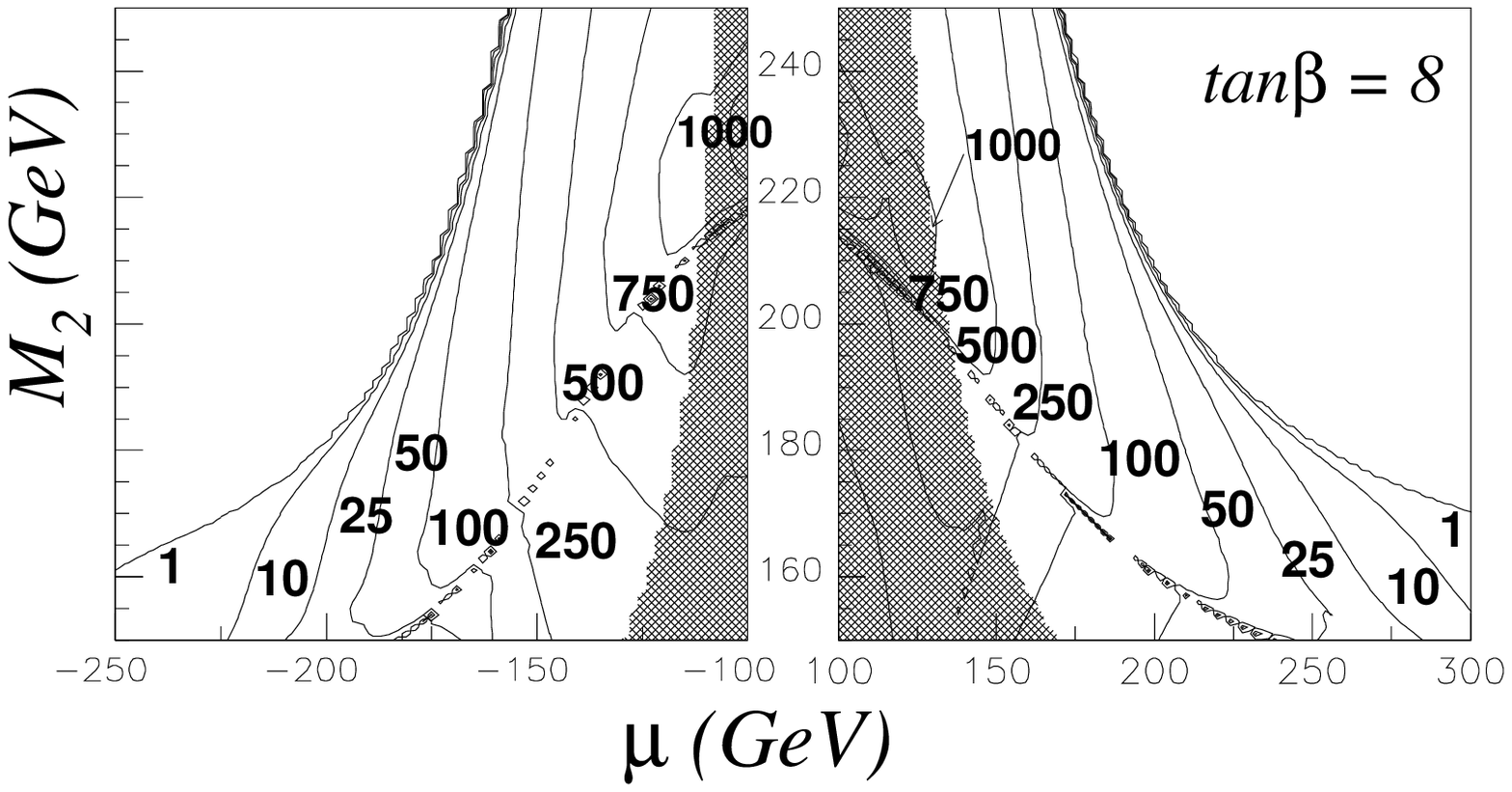,width=5in}
\caption{\small Expected number of
$pp \rightarrow XtbH^{\pm}, XtH^{\pm} \rightarrow
{\ell}^+{\ell}^-{\ell}^{\prime\pm} \; + \; t$ events per
$100\, \hbox{fb}^{-1}$ at the LHC (with no cuts), seen in the
$M_{\scriptscriptstyle 2}$ versus $\mu$ plane.
Here $\tan\beta = 8$ and $m_A = 290\, \hbox{GeV}~
(m_{H^{\pm}} \simeq 300\, \hbox{GeV})$.
Other inputs as in Figure~\ref{fig:HtoSUSY1}.}
\label{fig:HtoSUSY2}
\end{center}
\end{figure}

Figure~\ref{fig:HtoSUSY3} shows contours for the raw number of
three-lepton events
(${\ell}^{\pm}{\ell}^-{\ell}^+ \, + \, p_T^{miss} \, + \, t$)
expected at the LHC for $\sqrt{s} = 14\, \hbox{TeV}$ and
${\cal L} = 100\, \hbox{fb}^{-1}$
in the $M_2$ versus $m_{\tilde{\ell}_{\scriptscriptstyle R}}$ plane
before any selection cuts are applied.  The soft slepton mass spectrum
is fixed by 
$m_{\tilde{\ell}_{\scriptscriptstyle L}}
= m_{\tilde{\nu}}^{soft} = m_{\tilde{\ell}_{\scriptscriptstyle R}}
\, + \, 50\, \hbox{GeV}$
(soft slepton masses are assumed to be degenerate for all
three generations; the $A_{\ell}$'s are still kept equal to zero)
and $\mu$ is set to $-115\, \hbox{GeV}$.
All other inputs are as in the previous figures.
The shaded region is excluded by LEP 2 limits on
$m_{\widetilde{\chi}_1^{\pm}}$ (below) and $m_{\tilde{\ell}}$
(left)\footnote{There is a small ``sliver'' of allowed parameter space
that slices through the excluded region to the left.  This is where the
charged sleptons are only slightly more massive than the LSP
(taken as within $5\, \hbox{GeV}$) and thus can evade the LEP 2
searches.  Note though that then we expect to get soft leptons from
the slepton decay and so our signal probably disappears here.}.
We see that over a thousand events are possible.
The optimal spot (excluding the sliver region) is at
$(m_{\tilde{\ell}_{\scriptscriptstyle R}},M_2)
\approx (110.0~\hbox{GeV}, 195.5~\hbox{GeV})$
which boasts well over a thousand events.

\begin{figure}[!htbp]
\begin{center}
\epsfig{file=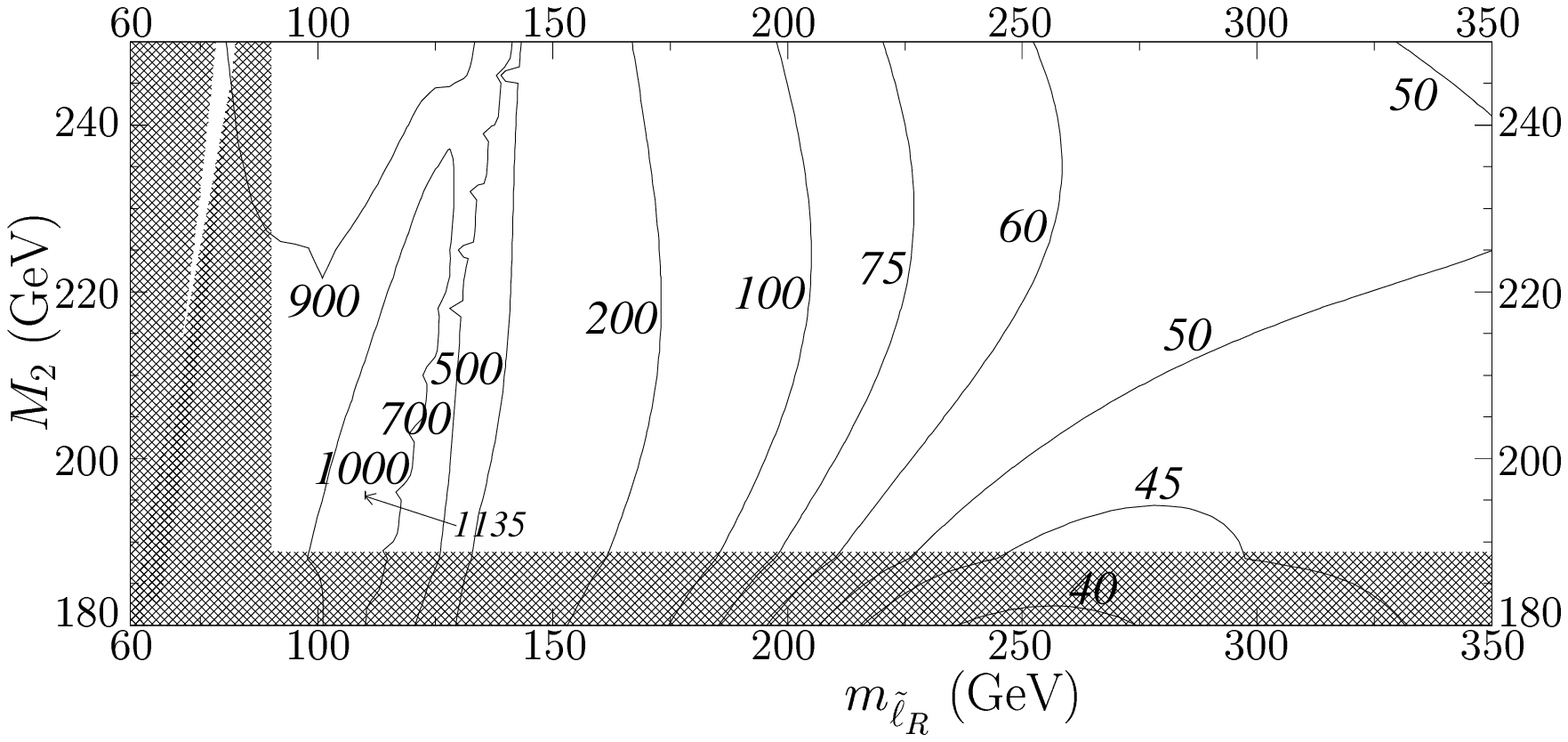,width=5in}
\caption{\small \small Expected number of
$pp \rightarrow XtbH^{\pm}, XtH^{\pm} \rightarrow
{\ell}^+{\ell}^-{\ell}^{\prime\pm} \; + \; t$ events per
$100\, \hbox{fb}^{-1}$ at the LHC (with no cuts), seen in the
$M_{\scriptscriptstyle 2}$ versus 
$m_{\tilde{\ell}_{\scriptscriptstyle R}}$ plane with $\mu$ fixed at
$-115\, \hbox{GeV}$. Other inputs as in Figure~\ref{fig:HtoSUSY2}.}
\label{fig:HtoSUSY3}
\end{center}
\end{figure}

\subsection{Analysis}

Here, we will show how all SM backgrounds can be completely
removed, leaving only MSSM processes as irreducible backgrounds in the
$3\ell+p^{{miss}}_T+t$ channel.
The relevant SM noise is constituted by $WZ^0$, $Z^0Z^0$, $b\bar{b}Z^0$,
$t\bar{t}$,
$t\bar{t}\ell^+\ell^-$ (consisting of both $t\bar{t}Z^0$ and
$t\bar{t}\gamma^*$, but not their interference), $t\bar{b}W$ and
$t\bar{b}W\ell^+\ell^-$ (again, via $Z^0$ and $\gamma^*$) production
and decay.  As for the MSSM backgrounds, one has to deal with
pair production of sleptons, gauginos or squarks/gluinos and
with neutral MSSM Higgs bosons produced in association with heavy quark
pairs (SM-like $t\bar{t}h$ production is found to merit attention while
associated production of the heavier neutral Higgs bosons is negligible).

We simulated the inclusive $H^\pm \to 3l$ signal and the MSSM backgrounds at the 5 points in the
intermediate $\tan\beta$ region of the MSSM parameter space listed in
Table~\ref{tab:table5}. These points were chosen in the favorable 
regions obtained in the previous section. In each case we have also chosen
$m_{\tilde{g}} = 700\, \hbox{GeV}$, $m_{\tilde{q}} = 1000\, \hbox{GeV}$,
$m_{\tilde{b}_{\scriptscriptstyle R}} = 800\, \hbox{GeV}$,
$m_{\tilde{t}_{\scriptscriptstyle L}} = 600\, \hbox{GeV}$,
$m_{\tilde{t}_{\scriptscriptstyle R}} = 500\, \hbox{GeV}$
and $A_t = 500\,  \hbox{GeV}$.
Rather large gluino and squark masses are chosen to preclude large
charged Higgs boson production rate from MSSM cascade decays
\cite{mono}, thus leaving the `direct' production modes of the
previous section as the only numerical relevant contributors at the LHC
\cite{hierarchy}.  Sleptons are chosen to be light in accordance with
the discussion in the last section.
\begin{table}[!htbp]
    \begin{center}
    \caption{\small Simulated MSSM parameter points.  All masses in GeV.
The event number is the parton-level result $H^\pm \to \chi_1^\pm\chi^0_{\{2,3\}} \to 3\ell X$ per 
$100\, \hbox{fb}^{-1}$.
$M_{\scriptscriptstyle 1} = \frac{1}{2}M_{\scriptscriptstyle 2}$ is
assumed.}
     {\begin{tabular}{lccccccc} \hline\hline
        Point  & $\tan\beta$ & $m_{H^{\pm}}$ & $\mu$ 
               & $M_{\scriptscriptstyle 2}$ 
               & $m_{\tilde{\ell}_{\scriptscriptstyle R}}$
               & $m_{\tilde{\ell}_{\scriptscriptstyle L}}$
               & events\\ \hline 
A &  $8$ & $250$ & $-115$ & $200$ & $120$ & $170$ & 1243\\ 
B & $10$ & $250$ & $-115$ & $200$ & $120$ & $170$ & 1521\\
C & $10$ & $300$ & $-115$ & $200$ & $120$ & $170$ & 1245\\
D & $10$ & $250$ & $+130$ & $210$ & $125$ & $175$ & 1288\\
E & $10$ & $300$ & $+130$ & $210$ & $125$ & $175$ & 1183\\
\hline\hline
      \end{tabular}}
\label{tab:table5}
    \end{center}
\end{table} 

HERWIG 6.3 \cite{herwig} is used to generate all hard
processes\footnote{An exception is the $t\bar t\gamma^*$ process,
which uses a set of separate subroutines \cite{me}.},
fragmentation and showering, with adoption of the default settings
\cite{herwig}.  The SUSY spectrum was obtained from ISASUSY 7.58~\cite{isajet} 
through the ISAWIG interface \cite{ISAWIG}. The detector
aspects were simulated using CMSJET 4.801 \cite{cmsjet}, which contains
a parameterized description of the CMS detector response; however,
since none of our selection cuts depend strongly on unique performance
factors of the CMS detector, we expect the analysis here to
roughly coincide with what  the ATLAS detector environment
would yield. The effects of pile-up at high luminosity running of the LHC have
not been included.

In order to distinguish between signal and background events, the
following selection criteria are applied:
\begin{description}
\item[\textbf{1.}] We require exactly three isolated leptons ($\ell=e$, $\mu$),
with $p_T \, > \, 20$, $7$ and $7\, \hbox{GeV}$, respectively,
all within $|\eta| \, < \, 2.4$.
The isolation criterion demands that there are no
charged particles with $p_T \, > \, 1.5\, \hbox{GeV}$ in a cone of
radius $R = 0.3$ radians centered around each
lepton track and that the sum of the transverse energy in the
ECAL (Electromagnetic Calorimeter of CMS) crystal towers between
$R = 0.05$ and $R = 0.3$ radians is smaller than $3\, \hbox{GeV}$.
\item[\textbf{2.}] We impose a $Z^0$-veto, i.e., rejection of all di-lepton pairs
with opposite charge and same flavor that have an invariant mass in the
range $m_Z \, \pm \, 10\, \hbox{GeV}$.
\item[\textbf{3.}] We select only events with three or four jets, all with
$p_T \, > \, 20\, \hbox{GeV}$
and $|\eta| \, < \, 4.5$.
\item[\textbf{4.}] Among these jets, we look for the combination of three that
are most probably coming from a top decay, by minimizing the
difference $m_{jjj}-m_{t}$, $m_{jjj}$ being the invariant mass of the
three-jet system.  These three jets must have $m_{jjj}$ in the range
$m_{t} \, \pm \, 35\, \hbox{GeV}$.
\item[\textbf{5.}] One of the jets must be $b$-tagged (the signed
significance of the transverse impact parameter must be
$\sigma(ip) = \frac{ip_{xy}}{\Delta ip_{xy}} \, > \, 2.5$).
The two other jets must have their invariant mass, $M_{jj}$,
in the range $m_W \, \pm \, 20\, \hbox{GeV}$.
\item[\textbf{6.}] The $p_T$ of the hardest lepton must be less than
$80\, \hbox{GeV}$\footnote{This reflects the relatively small mass
difference between the parent ino and the daughter ino (the latter
is typically the LSP).}.
\item[\textbf{7.}] The transverse missing momentum is in the following
interval: $50\, \hbox{GeV} \, < \, {p\!\!\!/}_T \, < \,
200\, \hbox{GeV}$.
\item[\textbf{8.}] The $p_T$ of the hardest jet is $< \, 180\, \hbox{GeV}$.
\item[\textbf{9.}] We exploit the `effective' transverse mass of
Ref.~\cite{PAP1}, by imposing
\newline
\centerline{
$M_{\rm{eff}} = \sqrt{2 p_T^{3\ell} {p\!\!\!/}_T (1-\cos{\Delta\phi})}
\, < \, 140\, \hbox{GeV}$
}
\newline
(where $p_T^{3\ell}$ is the three lepton transverse momentum and $\Delta\phi$ the azimuthal 
angle between $p_T^{3\ell}$ and ${p\!\!\!/}_T$).
\item[\textbf{10.}] In addition, the three-lepton invariant mass, $m_{3\ell}$,
 must be $< \, 120\, \hbox{GeV}$.
\end{description}
In Table~\ref{tab:table6} we summarize the signal and background events that 
\begin{table}[!htbp]
    \begin{center}
    \caption{\small Number of events after the cuts mentioned in the text at a luminosity of $100\, \hbox{fb}^{-1}$.}
     \begin{tabular}{lcccccc} \hline\hline
        Process          & 3$\ell$ events & $Z^0$-veto & 3,4 jets & $m_{jjj}\sim m_t$ & $M_{jj}\sim m_W^\dagger$  & others\\ \hline

    $t\bar{t}$             &    847      &  622  & 90 &   30      & 0    & 0                  \\
    $t\bar{t}Z^0$          &    244      &   34  & 13 &   5       & 0    & 0                  \\
    $t\bar{t}\gamma^*$     &    18       &   18  & 10 &   3       & 1    & 0                  \\
    $t\bar{t}h$            &    66       &   52  & 33 &   9       & 3    & 1                  \\
    $\tilde{\ell}\tilde{\ell}$ &    5007 &  4430 & 475& 112       & 2    & 0                    \\
    $\widetilde{\chi} \widetilde{\chi}$       &    8674      &  7047 & 1203 & 365 &19 &  3       \\
    $\tilde{q}$, $\tilde{g}$  & 37955    & 29484 &3507& 487  &100  & 0              \\    
    \hline
    $t H^+$ (point A)      &     251     &   241 & 80&   23     & 6 &   5            \\
    $t H^+$ (point B)      &     321     &   298 &118&   42     &13 &   9            \\
    $t H^+$ (point C)      &     279     &   258 &100&   36     &11 &   7            \\
    $t H^+$ (point D)      &     339     &   323 &121&   48     &13 &   9            \\
    $t H^+$ (point E)      &     291     &   278 &114&   40     &10 &   5            \\ \hline\hline
\end{tabular}
\label{tab:table6}
\newline
{\small
$^\dagger$Includes $b$-tagging efficiency for the third jet
 \phantom{aaaaaaaaaaaaaaaaaaaaaaaaaaaaaaaaaaaaaaaaaaaaaaaaa} }
\end{center}
\end{table}
remain after applying these criteria, assuming a luminosity of 
$100\, \hbox{fb}^{-1}$ and optimizing the selection using the 
$m_A = 250\, \hbox{GeV}$ signal.
Due to the aforementioned small mass differences in the ino sector, 
leptons coming from the signal events will often be soft, 
which explains much of the difference between the numbers in the 
``3$\ell$ events'' column of Table~\ref{tab:table6} and those from the parton-level
analysis in Table~\ref{tab:table5}.  Clearly, it is vital to have a low $p_T$ threshold
for accepting leptons.  Conservatively, here we have set this to 
$7\, \hbox{GeV}$ for both electrons and muons. 

Requiring three tightly isolated leptons in addition to a  
hadronically reconstructed top quark allows us to eliminate  
most of the SM backgrounds: $W^\pm Z^0$, $Z^0Z^0$, $b\bar{b}Z^0$ as well as
the initially large $t\bar{t}$ channel.
The $t\bar{t}\ell^+\ell^-$ background, resulting from $t\bar{t}Z^0$,
$t\bar{t}\gamma^*$ and $t\bar{t}h^0$ production can easily
mimic the signature of the signal. However, after applying the 
$Z^0$-veto, the previously dominant $t\bar{t}Z^0$ component of  this background 
becomes negligible compared to the intrinsic
MSSM backgrounds\footnote{The single-top
counterparts, $t\bar{b}W^-\ell^+\ell^-$, are sub-leading, as they
amount to a $\sim25\%$ correction to the double-top rates.}.
Among the latter, the channels that survive the selection are
squark/gluino and gaugino pair-production (in which we include
$\widetilde{\chi}\tilde{g}$ too) as well as the associated production of  
a light Higgs boson with a $t\bar{t}$ pair.
In gaugino-gaugino production, a gluon jet can fake one of the three
jets from the top decay, but in the next stage of the top reconstruction  
these events will always be rejected (see Table~\ref{tab:table6}).
On the other hand, squark/gluino and squark-gaugino background events
that pass the top reconstruction cut are likely to have a real top in 
the final state, or else a $bW^+$ pair.

The events that pass the top reconstruction cut can be distinguished from
the signal by noticing that MSSM cascade decays will typically produce
harder leptons, more and harder jets, as well as a larger amount of
missing $p_T$, with respect to the signal.
Also, one can further suppress these backgrounds by putting an upper
limit on the three lepton invariant mass and on the mass made up from
the $p_T$ of the three leptons and the ${p\!\!\!/}_T$.   In particular,
these last two variables are very effective if the mass of the Higgs boson
is just above the chargino+neutralino threshold, i.e., for
$m_A  \simeq 250\, \hbox{GeV}$.  For larger charged Higgs boson
masses, the lepton $p_T$ cut-off increases and the discriminative power
of this variable is reduced.


\subsection{Conclusions}

The inability to cover the intermediate $\tan\beta$ region by
exploiting charged Higgs boson decays into SM particles prompted us
to carefully investigate the potential offered by other channels.
In SUSY models such as the MSSM, it is natural to explore
to this end the interactions between charged Higgs bosons and the
inos. Assuming that squarks and gluinos are heavy enough so that the
production rates of $H^\pm$ states via cascade decays of sparticles are
negligible, the obvious place to look is the decay of a charged Higgs
boson itself into a chargino-neutralino pair.

The limited $X t H^-$ (and the charge conjugate) production rate
precludes exploration for large values of $m_{H^\pm}$; however,
for $m_{H^\pm}~\lsim~300\, \hbox{GeV}$, a signal could well be observed
above the background, provided that:
(1) $\mu$ and $M_{\scriptscriptstyle 2}$ are not much above the current
LEP restrictions from ino searches;
(2) sleptons are sufficiently light.
We are thus guided to a region of parameter space where both the
ino and the Higgs boson spectra will be accessible at the LHC,
making it a quite reasonable choice for initial phenomenological studies.
(A more refined analysis encompassing a wider span of the MSSM
parameter space will be ready in due course \cite{preparation}.)

Before closing, we would like to remark on possible improvements in
our analysis.   One such improvement would be the inclusion of the
leptonic decays of the top, in which case one can look for a signature
with four leptons and a $b$-jet in the final state \cite{thesis}\footnote{In
fact, it should be noted that both leptonic and hadronic decays of top
quarks (and of inos) have been allowed in the generated events of the
present analysis, although the former have negligible impact because of
the enforced top mass reconstruction procedure. }.
After applying a reasonable set of cuts on the lepton momenta, we find the
SM noise can be completely eliminated.
Furthermore, rejecting events with more than two jets and limiting
the $E_T^{miss}$ eliminate the background from squark/gluino pair production.
Ino (again, including $\widetilde{\chi}\tilde{g}$) and slepton pair
production as well as associated production of a light neutral Higgs boson
with a heavy quark pair then constitute the principal backgrounds,
though the former are largely reduced by requiring one $b$-tagged
jet.

Another extension under development is the study of charged Higgs boson
decays including the higher ino mass eigenstates
($\widetilde{\chi}_2^{\pm}$ and
$\widetilde{\chi}_4^0$) for $m_{H^{\pm}}~\gsim~300\, \hbox{GeV}$.
For the highest mass inos, the number of possible decay chains ending in
the LSP can quickly multiply.  Such decay modes may have preferred
regions of parameter space for $\mu$ and $M_{\scriptscriptstyle 2}$
quite different from the cases presented here.

Although limited to some restricted region of parameter space,
we have managed with our present analysis to cover some portion of the
elusive intermediate $\tan\beta$ region in the MSSM; and we regard our
results thus far as encouraging enough to look further into
these more exotic charged Higgs boson decay modes \cite{preparation}.

\section{$\mathbf{H^\pm}$ from SUSY Cascade Decays at the LHC}
\label{sec:SUSYtoHpm}

In this section, we analyze the cascade decays of the scalar quarks and gluinos of the
MSSM, which are abundantly produced at the LHC, into heavier charginos and 
neutralinos which then decay into the lighter ones and charged Higgs 
particles. We show that these decays can have substantial branching fractions. 
The production rates of these Higgs bosons can be much larger than those from 
the direct production mechanisms, in particular for intermediate values of 
$\tan \beta$. An event generator analysis shows that the detection of 
$H^\pm$ bosons produced through this mechanism is possible. 

\subsection{Motivation} 

As previously recalled, for masses $m_{H^\pm} >m_t$, the two production mechanisms with potentially
sizeable cross sections at the LHC are the $2\to 3$ and $2\to 2$ processes --- shown in 
Figure~\ref{fig:figure1} --- which have to be properly combined to avoid 
double counting \cite{subtraction,mor-roy,reviewLHC,Jaume}.  
However, the cross sections are rather small: even for the extreme values, $\tb =2$ and
$40$, they hardly reach the level of a picobarn for a charged Higgs boson mass
$m_{H^\pm}=200$ GeV. For intermediate values of $\tb$ and/or larger $H^\pm$
masses, the cross sections are too small for these processes to be useful. For
instance, for $\tb=10$, the cross section is below the level of a few femtobarn
for $m_{H^\pm} \gsim 250$ GeV. The other mechanisms for $H^\pm$ production at
hadron colliders give even smaller cross sections \cite{reviewLHC,hierarchy}. 

In a recent paper, Ref.~\cite{mono}, it has been shown that there is a
potentially large source of the $H^\pm$ bosons of the MSSM at the LHC: the
cascade decays of squarks and gluinos, which are abundantly produced in $pp$
collisions, thanks to their strong interactions. These squarks and gluinos can
decay into the heavy charginos and neutralinos and if enough phase space is
available, the latter particles could then decay into the lighter
charginos/neutralinos, and $H^\pm$ bosons, with substantial branching ratios.

In this section, we summarize the production of $H^\pm$ particles through
these cascade decays at the LHC and describe a Monte Carlo simulation
which shows that these final states can be possibly detected in some regions 
of the MSSM parameter space.  

\subsection{$\mathbf{H^\pm}$ Bosons from Cascade Decays in the MSSM} 

At the LHC the total squark and gluino production cross section is $\sigma
(\tilde{q}+\tilde{g}) \sim 110$ (3) pb for $m_{\tilde{g}} \sim m_{\tilde{q}}
\sim 0.5$ (1) TeV leading  to a large, $ \sim 3 \cdot 10^{7}$ to $10^{6}$,
number of events with an accumulated luminosity of ${\cal L} \sim 300$
fb$^{-1}$.  These squarks and gluinos can decay into the heavy charginos and
neutralinos, $\chi_2^\pm, \chi_3^0$ and $\chi_4^0$ with significant branching
fractions, a few ten percent. If enough phase space is available, the latter
particles could then decay into the lighter charginos/neutralinos, $\chi_1^\pm,
\chi_1^0$ and $\chi_2^0$, and $H^\pm$ bosons, with branching ratios of the
order of a few ten percent, again. A key point here, is that the coupling of
the Higgs bosons to chargino and neutralino states is maximal for
higgsino--gaugino mixed states \cite{Haber}, while the gauge boson couplings to
neutralinos are maximal for higgsino--like states. In the gaugino--like or
higgsino--like regions, this results into the dominance of the decays of the
heavier charginos and neutralinos into the lighter ones and Higgs bosons
compared to the same decays with gauge boson final states.  

The total number of charged Higgs particles produced at the end of the chain 
\begin{eqnarray} 
pp \to \tilde{g} \tilde{g}, \tilde{q} \tilde{q}, \tilde{q} \tilde{g}  \to 
\chi_2^\pm, \chi_3^0, \chi_4^0 + X \to   \chi_1^\pm, \chi_2^0, \chi_1^0 + 
H^\pm +X 
\end{eqnarray} 
could be rather large (of the order of a few 10.000 to a few 100.000 events for
the high--luminosity option) in favorable regions of the parameter space. The
interesting and important point to note is that the rate of $H^\pm$
production does not depend very crucially on $\tan\beta$ unlike in the other
mechanisms mentioned above --- Note also that $H^\pm$ bosons could be
searched for, if kinematically possible, in the direct decays of heavy third
generation squarks into their lighter partners or in direct gluino three--body
decays. This is illustrated below, in two scenarii with the intermediate value
$\tb=10$ and where the universality of the gaugino masses has been assumed at
the GUT scale, leading to the relation $M_2 \simeq 2M_1 \simeq M_3/3 \simeq 
m_{\tilde{g}}/3$ at the weak scale.

\begin{figure}[!htbp]
\begin{center}
\vspace*{-1.2in}
\epsfig{file=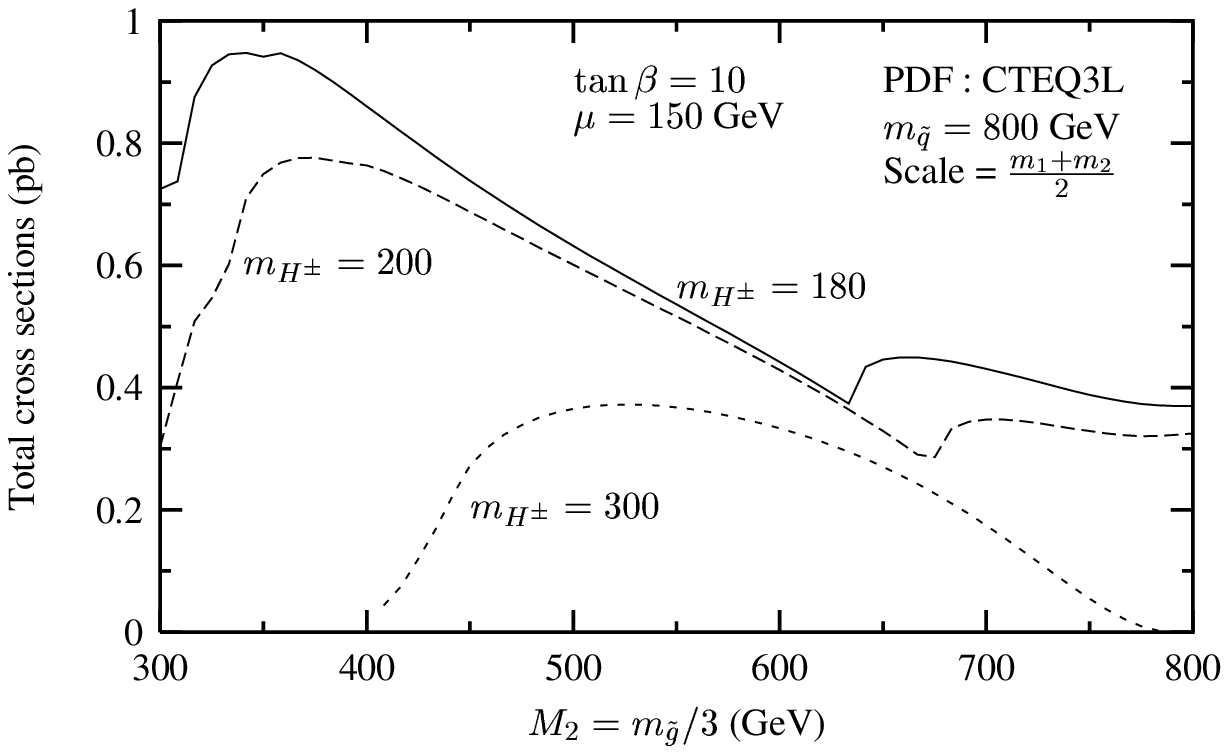,width=5in}
\vspace*{-4.2in}
\caption{Cross sections times branching ratios for gluinos decaying into 
squarks and squarks decaying through cascades into $H^\pm$ bosons as
functions of $M_2$ with $\mu=150$ GeV and $\tb=10$. $m_{\tilde{q}}=900$ GeV 
and $m_{\tilde{g}} =3M_2$.}
\label{fig:scn2_uni}
\end{center}
\end{figure}

Scenario 1 (Figure~\ref{fig:scn2_uni}): Here gluinos (with $m_{\tilde{g}}=3M_2$) are heavier than
squarks ($m_{\tilde{q}}=900$ GeV) and therefore $\tilde{g} \to \tilde{q} q$
occurs 100\% of the time. The higgsino mass parameter has been chosen to be
small, $\mu=150$ GeV, so that all squarks --- in particular those of the first two
generations --- will mainly decay into the heavier charginos and neutralinos which
are gaugino like with masses $m_{\chi_2^+} \sim m_{\chi_4^0} \sim 2
m_{\chi_3^0} \sim M_2$. For large enough $M_2$, there is then enough phase
space for the decay of the heavier gauginos into the lighter higgsino states,
with masses $m_{\chi_1^+} \sim m_{\chi_1^0} \sim m_{\chi_2^0} \sim |\mu|$, and
$H^\pm$ bosons to occur. For small $M_2$ values, the states $\chi_{3,4}^0$ and
$\chi_{2}^+$ are not heavy enough for the decays into $H^\pm$ bosons to occur,
in particular for large $m_{H^\pm}$. When these decays are allowed, $\sigma
\times {\rm BR}(\to H^\pm)$ values of the order of 1 pb for $m_{H^\pm} \sim
180$ GeV and 0.3 pb for $m_{H^\pm} \sim 300$ GeV can be reached.  For
increasing values of $M_2$, the gluino mass increases and the cross sections for
associated squark and gluino and gluino pair production drop and $\sigma \times
{\rm BR}(\to H^\pm)$ decreases accordingly; at some stage, only the cross
section for squark production survives $m_{\tilde{q}}$ being fixed.  The
decrease of $\sigma \times {\rm BR}(\to H^\pm)$ with increasing $M_2$ is also
due to the more suppressed phase space for $\tilde{q} \to q' \chi_2^\pm,
q\chi_4^0$ since for large $M_2$, $m_{\chi_4^0} , m_{\chi_2^\pm} \sim M_2$. For
even larger $M_2$ values, $M_2 \gsim 650$ GeV, the channel $\chi_3^0 \to H^\pm
\chi_1^\mp$ opens up, and since the phase space is more favorable, because
$m_{\chi_3^0} \sim M_2/2$, $\sigma \times {\rm BR}(\to H^\pm)$ increases again.

\begin{figure}[!htbp]
\begin{center}
\vspace*{-1.2in}
\epsfig{file=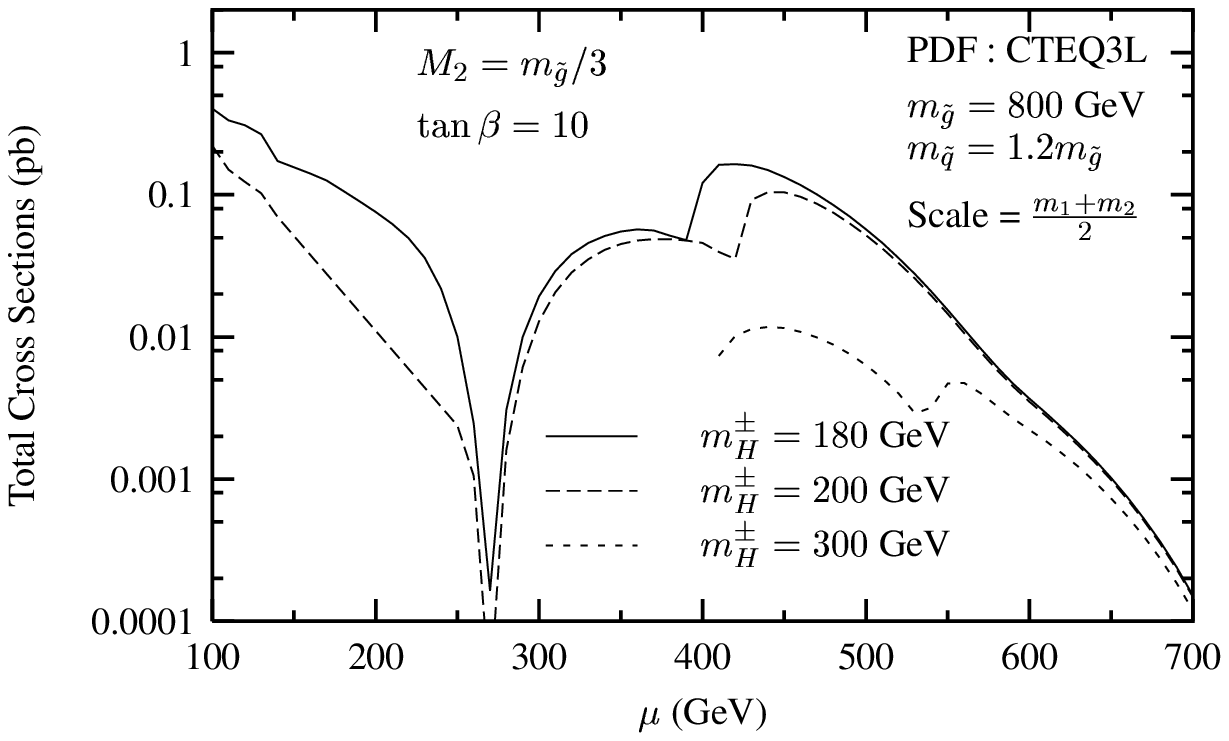,width=5in}
\vspace*{-4.2in}
\caption{Cross sections times branching ratios for squarks decaying into 
gluinos with the gluinos decaying through cascades into $H^\pm$ bosons, as
as functions of $\mu$ for $\tb=10$. $m_{\tilde{g}}=800$ GeV and 
$m_{\tilde{q}}=1.2m_{\tilde{g}}$.}
\label{fig:final_uni}
\end{center}
\end{figure}

Scenario 2 (Figure~\ref{fig:final_uni}): Here the scalar partners of light quarks are heavier than
stops which are heavier than gluinos, $m_{\tilde{q}} =1.2 m_{\tilde{g}}$ with
$m_{\tilde{g}}\sim 3M_2 =800$ GeV. Gluinos will decay mainly into three--body
final states mediated by the exchange of top squarks which have a smaller
virtuality. Note that the cross sections for squark and gluino production are
constant and the variation of $\sigma \times {\rm BR}(H^\pm)$ is only due to
the variation of the branching ratios BR$(\tilde{g} \to \chi_{3,4}^0 qq,
\chi_2^\pm qq')$ and   BR$(\chi_{3,4}^0, \chi_2^\pm \to \chi_{1}^\pm H^\mp,
\chi_{1,2}^0 H^\pm)$. One sees that $\sigma \times {\rm BR}(H^\pm)$ is
relatively large for small values of $\mu$ and $m_{H^\pm}$, when the
gaugino--like heavy $\chi$ states are light enough for the decays $\chi_{4}^0
\to \chi_1^\pm H^\mp$ and $\chi_{2}^\pm \to \chi_{1,2}^0 H^\pm$ to occur.  In
the mixed region, $\mu \sim M_2$, the mass difference between the heavy and
light $\chi$ states are too small to allow for decays in $H^\pm$ bosons. For
large values of $\mu$, $\sigma \times {\rm BR}(H^\pm)$ increases to reach
values of the order of $\sim 0.1$ pb for $m_{H^\pm} \sim 200$ GeV (in
particular when the additional channels $\chi_3^0 \to \chi_1^\pm H^\mp$ open
up) before it drops out because of the gradually closing phase space for the
decays $\tilde{g} \to q\bar{q} \chi^0_{3,4}, qq' \chi_2^\pm$. 

Thus, large samples of $H^\pm$ bosons can be produced in these SUSY cascade
decays.  In the regions of $m_{H^{\pm}}$ and $\tan\beta$ values that we are
interested in, the dominant decay modes of the charged Higgs bosons are
\cite{hdecay,hdecay1} 
$H^+ \ra t \bar{b}$ ($\sim 90\%$ sufficiently above the $tb$
threshold)  followed by $H^+ \ra \tau^+ \nu_\tau$ ($\sim 10\%$). In the
simulation which we present below, we will focus on the latter decays which are
easier to detect in the jet environment of the LHC.  

\subsection{An Event Generator Analysis}

Selecting the one-prong hadronic decays of these tau leptons will allow us to exploit
the tau polarization effects in our analysis.  Therefore, the signature of
$H^\pm$ bosons produced in SUSY cascade decays consists of one hard $\tau$--jet
plus additional hard jets (often $b$-jets) accompanied by a large amount of
missing energy due to the presence of the lightest neutralinos and the
neutrinos. 

The main SM processes leading to the same signature are top pair
production, $pp \ra t \bar{t}$, with one top decaying hadronically and the
other one decaying leptonically, and QCD $W^\pm\,+\,jets$ production with $W^\pm \ra
\tau^\pm \nu_\tau $.  As we will argue below, these SM backgrounds can be efficiently
suppressed.  However, a more difficult task will be to distinguish the $H^\pm$
signal from other SUSY cascade decay processes. As was discussed above, all
squark and gluino production in colliders will end up in lightest neutralinos
and fermions through cascade decays via gauginos and sleptons.  Events in which
these fermions are taus will mimic the signal.  Hence, a good understanding of
the nature of these SUSY backgrounds will be needed prior to a search for
charged Higgs bosons in cascade decays.

We performed a Monte Carlo simulation of the production signal and the main
backgrounds for the following scenario: $m_{H^{\pm}}=200$ GeV, $\tan\beta =
10$, $\mu = 450$ GeV, $M_{2} = 2M_1= 200$ GeV, $m_{\tilde{g}} = 800$ GeV,
$m_{\tilde{q}} = 1.2 m_{\tilde{g}}$ and $m_{\tilde{l}} = 300$ GeV.  The signal
and background events were generated with PYTHIA 6.152 \cite{pythia}.  To
account for the tau polarization, PYTHIA was interfaced with the TAUOLA
\cite{tauola} package.  The detector aspects were simulated using CMSJET 4.801
\cite{cmsjet}, which contains a parameterized description of the CMS detector
response. The effects of pile--up at high luminosity running of the LHC have
not been included. The features that will allow  to distinguish the 
$H^{\pm}$ signal from the SM and SUSY backgrounds are summarized below.

In the left--hand panel of Figure~\ref{fig:norma}, the $E_T^{\rm miss}$ distribution,
normalized to the number of events, is shown for the $t\bar{t}$ background, the
SUSY cascade background and the charged Higgs boson signal. Demanding a very
large $E_T^{\rm miss}$ in the events allows us to eliminate the $t\bar{t}$
background.  Also part of the SUSY cascade background can be suppressed
relative to the signal due to a slight excess of missing energy in cascade
decays including charged Higgs bosons. Similarly, making a hard requirement on
the $E_T$ of the hardest jet in the event will help in the background
rejection.  In the right--hand panel of Fig.~\ref{fig:norma}, this $E_T$ distribution is
shown and it is clear that the SM processes can be efficiently suppressed. 

\begin{figure}
  \begin{center}
  \vspace{-5.0mm}
  \epsfig{file=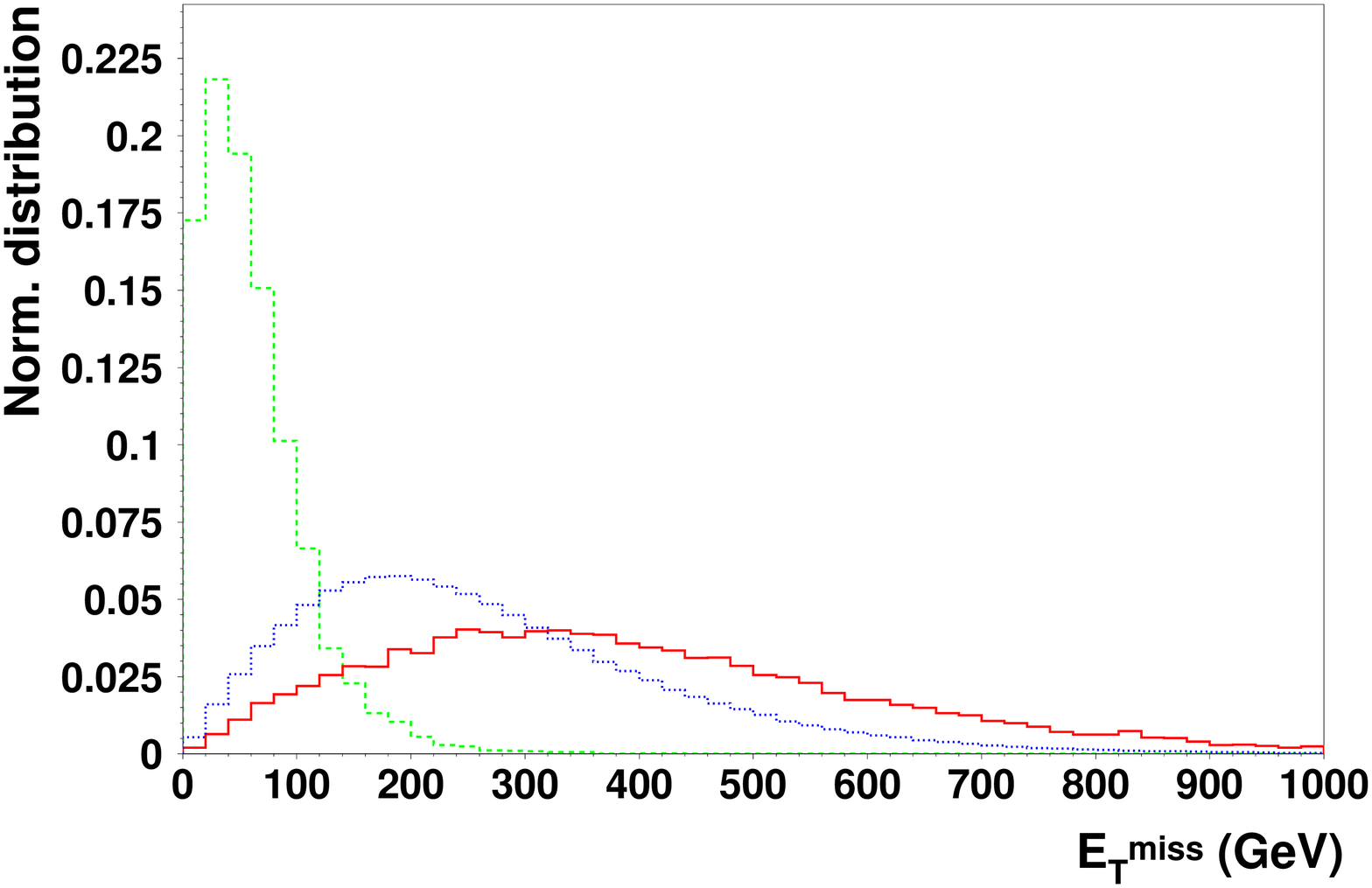, width=70mm, height=75mm}
  \epsfig{file=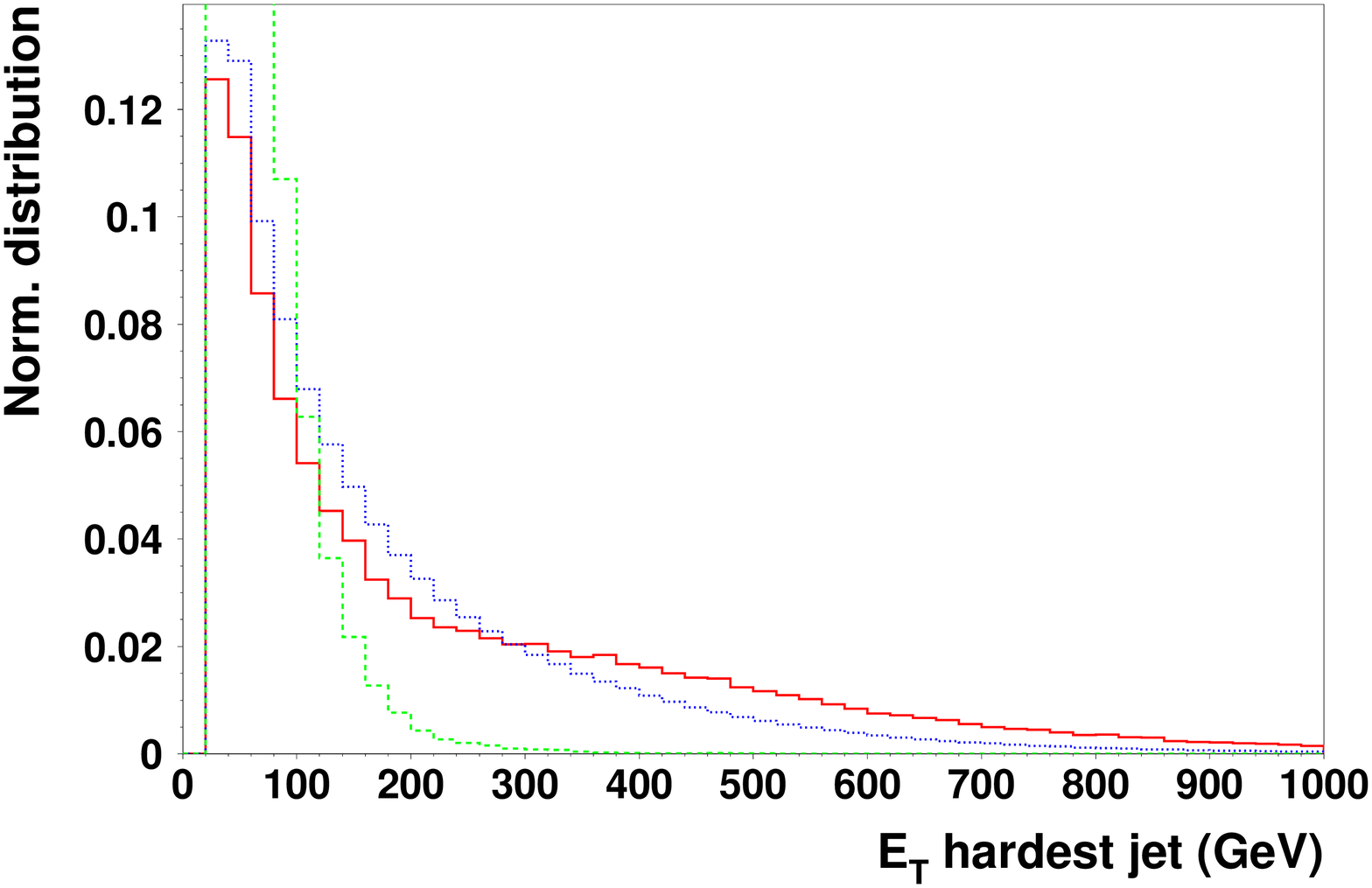, width=70mm, height=75mm}
  \vspace{-6.0mm}
\caption{{\small Normalized distribution of $E_T^{miss}$ (left) and 
normalized $E_T$ distribution of the hardest jet in the event (right) 
for the $t\bar{t}$ background (green--light dashed line), the SUSY cascade 
background (blue--medium line) and the charged Higgs signal (red--dark 
line).}}  
  \label{fig:norma} 
  \end{center}
  \vspace{-0.5cm}
\end{figure}

After eliminating the SM backgrounds, the more difficult task remains to
discriminate the $H^{\pm}$ bosons from the other particles in the SUSY cascade
decays. $\tau$--leptons in the cascade backgrounds originate from charginos and
neutralinos in the intermediate state. In our scenario, charginos decay
predominantly into $W^\pm$ bosons and neutralinos into $Z^0$ bosons. Taus coming
from these particles will typically have a softer spectrum than the ones coming
from a 200 GeV charged Higgs boson. Therefore we impose a lower limit on the
transverse energy of the $\tau$-jet, $E_T^{\rm \tau-jet}$ $>$ 120 GeV, which is
well above the mass of the $W^\pm$ and $Z^0$ bosons. As an upper limit on $E_T^{\rm
\tau-jet}$, we chose the mass of the charged Higgs boson.

The significant presence of $W^\pm$'s in the SUSY cascade background also needs to
exploit the tau polarization effects \cite{roy} explained in section~\ref{sec:mass}: 
by selecting events in which the fraction of the $\tau$-jet transverse momentum 
carried by the charged pion is large, the SUSY background involving $W^\pm$ bosons can 
be suppressed relative to the signal.  The tau leptons in the background coming from 
either $\tilde{\tau}$ and $Z^0$ decays or  MSSM neutral Higgs boson (in particular
$H$ and $A$) decays cannot be suppressed this way. 

These considerations lead us to the following selection criteria in order to
distinguish between signal and background events: 

$i)$ The transverse missing energy $E_T^{miss}$ in the event should be larger 
than 300 GeV.  

$ii)$ The hardest jet in the event should have $E_T$ $>$ 400 GeV.  

$iii)$ Events with more than five jets are rejected.  

$iv)$ We require exactly one hadronically decaying tau (1-prong) i.e. we
demand a narrow jet within $|\eta| \, < \, 2.5$ which should contain a hard
charged track with $p_T \, > \, 5$ GeV in a cone of $\Delta R$ = 0.15 radians
around the calorimeter jet axis, and it should be isolated i.e. no
charged tracks with $p_T \, > \, 2$ GeV are allowed in a cone of $\Delta R$ =
0.4 radians around the axis.  

$v)$ The $E_T$ of the $\tau$-jet, defined as the $E_T$ reconstructed in a cone
of $\Delta R$ = 0.4 radians around the jet axis, should be in the interval:
$120\, \hbox{GeV} \, < \, E_T^{\tau-jet} \, < \, 200\, \hbox{GeV}$.  

$vi)$  More than 80\% of the $\tau$-jet transverse energy should be carried by
the charged track: see the requirement~(\ref{eq:pfrac}) in section~\ref{sec:mass}. 

Events that satisfy conditions ($i$) and ($ii$) can be efficiently triggered on
using the jet and missing energy triggers.  If the above selection criteria are
applied, we obtain the following numbers for an integrated luminosity of 100
fb$^{-1}$: 17 events for the $H^{\pm}$ boson signal and 21 events for the SUSY
cascade background, while the $t\bar{t}$ background is completely suppressed. 
This results in a significance $S/\sqrt{B}$ of 3.5, meaning that a
5$\sigma$-observation of $H^{\pm}$ bosons produced in SUSY cascade decays could
be made with about 2 years of high luminosity data of the LHC, assuming the
above physics scenario and provided the SUSY background processes are well
understood.  

\subsection{Conclusions} 

We have shown that charged Higgs boson production from cascade decays of
strongly interacting SUSY particles can occur with large rates, in favorable
domains of the MSSM parameter space. This is true, in particular for
intermediate values of $\tb$ where the standard production processes are
disfavored because of the smallness of the $H^-tb$ Yukawa coupling.  We have
shown that the SM background to the cascade production can be
efficiently suppressed. By exploiting the characteristics of the $\tau$-jet in
the final state, the $H^\pm$ signal can be made visible above the other SUSY
cascade decays. 

Our strategy to distinguish the $H^{\pm}$ bosons from other cascade processes
depends on prior knowledge of the properties of the SUSY background. Therefore,
before any reasonable search for charged Higgs bosons in cascade decays can
take place, the general nature of SUSY cascade decays should be sufficiently
understood in order to select the kinematical region in which the signal is
enhanced with respect to the SUSY background.  Due to the complexity of the
signature and the dependence on calorimetric energy resolutions, these results
would need to be confirmed by a more detailed simulation. However, our current
findings point towards the conclusion that there exists a potential for
observing charged Higgs bosons produced in SUSY cascade decays at the LHC,
provided the nature of the SUSY background is well understood. 

A similar analysis, dealing with the production of the neutral Higgs particles
of the MSSM through the cascade decays of squarks and gluinos is under way
\cite{future}. 

\section{Summary}
\label{sec:sumup}

We have investigated the feasibility of detecting various signatures of the charged Higgs boson in 
order to provide a complete coverage for $H^\pm$ searches in upcoming experiments at the Tevatron 
and at the LHC. 

In the threshold region, i.e., for $m_{H^\pm}\sim m_t$, the $gg\rightarrow tbH^\pm$ process has been 
used to correctly account for the $H^\pm$ production and decay phenomenology in this region of 
parameter space instead of the usual narrow width approximation. It is found that a significant 
charged Higgs boson signal can be extracted in the channel $H^\pm \to \tau^\pm\nu_\tau$ after an 
integrated luminosity of 15~fb$^{-1}$ at the Tevatron Run~2 while the factorization approach used 
in the narrow with approximation would not predict such a favorable signal. The case of the LHC, 
which is further complicated by the potential problem of double counting when calculating the 
inclusive cross section, is currently being studied.

In models with singlet neutrinos in large extra dimensions, the process $H^-\rightarrow\tau^-_L
\psi+c.c.$ --- which is completely suppressed in the 2HDM --- can be enhanced thanks to the large 
number of Kaluza-Klein states of the right handed bulk neutrino. Such a signal can be observed at 
the LHC with significances exceeding 5-$\sigma$. However, in order to identify the scenario that is 
realized (2HDM-II or large extra dimensions) both the transverse charged Higgs mass and the fraction 
of the energy carried away by charged tracks in the one-prong $\tau$ decays must be reconstructed. 
Further evidence for large extra dimensions would come the measurement of the polarization asymmetry.

In the intermediate $\tan\beta$ region, the charged Higgs boson decays to SM particles do not yield 
any significant discovery potential. To cover this region of the parameter space, charged Higgs 
decays into chargino-neutralino pairs have been studied. It is demonstrated that, by searching for a 
three-lepton final state with a top quark and a large missing energy resulting from the 
$gg \to tbH^\pm$ where the charged Higgs decays to $\tilde{\chi}^\pm_1\tilde{\chi}^0_2$ or 
$\tilde{\chi}^\pm_1\tilde{\chi}^0_3$, a significant charged Higgs signal could be extracted for 
intermediate $\tan\beta$ values and $m_{H^\pm} \lsim 300$~GeV. Further analysis is in progress 
aiming to extend the coverage to higher charged Higgs boson masses.

A significant source of charged Higgs boson production is the cascade decays of SUSY particles 
(squarks and gluinos) also sensitive to the intermediate $\tan\beta$ region where, as previously 
mentioned, the SM charged Higgs boson productions and decays yield no discovery potential. The 
subsequent decay of the charged Higgs boson into the $\tau$-lepton has been studied taking advantage 
of the polarization effects in the $\tau$-jet final state. It is determined that a significant 
charged Higgs boson signal can be observed through these cascade decays of SUSY particles provided 
the SUSY background is well understood.

The charged Higgs boson mass can be determined in $H^\pm\rightarrow tb$ and 
$H^\pm\rightarrow\tau^\pm\nu_\tau$ where the precisions range from 0.5\% at $\sim$ 200~GeV to 1.8\% 
at $\sim$ 500~GeV for an integrated luminosity of 300~fb$^{-1}$. By measuring the rate of 
$H^\pm\rightarrow\tau^\pm\nu_\tau$, $\tan\beta$ can be determined with precisions ranging from 
7.4\% at $\tan\beta=20$ to 5.4\% at $\tan\beta=50$ for an integrated luminosity of 300~fb$^{-1}$ and 
assuming a 10\% uncertainty on the luminosity. 

The studies discussed here are the necessary continuation of previous work and in the process, 
reveal the enormous potential of the charged Higgs boson:
\begin{itemize}
 \item To understand the structure of the Higgs sector through the determination of $m_{H^\pm}$ and 
$\tan\beta$ and to probe the decoupling limit of the MSSM, thus distinguishing between the SM and 
the MSSM, particularly in the $H^\pm \to \tau^\pm\nu_\tau$ channel. Indeed, the scope of the 
parameter space covered by this channel is comparable to the reach of the $A/H \to \tau\tau$ channel 
in the neutral Higgs sector. 
\item To provide evidence --- or not --- of large extra dimensions, the existence 
of which, especially at the electroweak scale, would constitute the solution to the outstanding 
hierarchy problem.
\item To explore the SUSY particle arena and thus providing various signatures whose detection 
will constitute evidence of the existence of these exotic particles.
\end{itemize}

\section*{Acknowledgments}

The authors express gratitude to E.~Richter-W\c{a}s, D.~Cavalli, K.~Jakobs and D.~Denegri for 
helpful comments and discussions. K.A.~Assamagan thanks K.~Agashe for helpful 
correspondence. M.~Bisset is grateful to the  U.S. National Science Foundation for support under 
grant INT-9804704. F.~Moortgat is supported by the Fund for Scientific Research, Flanders (Belgium), A.K.~Datta 
is a MNERT fellow and M.~Guchait was supported by CNRS. This work was partially performed at Les 
Houches Workshop: ``Physics at TeV Colliders'', 21 May -- 1 June 2001. We thank the organizers for 
the invitation and for their effort.
}

\end{document}